%% file: draft.tex
\documentclass[a4paper,11pt]{article}
\usepackage{jheppub} 
\usepackage[figuresright]{rotating}
\usepackage{lineno}
\usepackage[utf8]{inputenc}
\usepackage{newunicodechar}
\newunicodechar{−}{-}

\arxivnumber{2601.01503} 
\usepackage{multirow}

\newcommand{\ee}{e^+e^-}

\newcommand{\tagmbc}{M_{\rm BC}^{\rm tag}}
\newcommand{\sigmbc}{M_{\rm BC}^{\rm sig}}
\newcommand{\mev}{\,\mathrm{MeV}}

\newcommand{\mevcc}{\,\mathrm{MeV}/c^2}
\newcommand{\gev}{\,\mathrm{GeV}}
\newcommand{\gevc}{\,\mathrm{GeV}/c}

\newcommand{\lcp}{\Lambda_{c}^{+}}
\newcommand{\lcm}{\bar{\Lambda}_{c}^{-}}

\newcommand{\modea}{pK_{S}^{0}}
\newcommand{\modeb}{pK^{-}\pi^+}
\newcommand{\modec}{pK_{S}^{0}\pi^0}
\newcommand{\moded}{pK_{S}^{0}\pi^+\pi^-}
\newcommand{\modee}{pK^{-}\pi^+\pi^0}
\newcommand{\modef}{p\pi^+\pi^-}
\newcommand{\modeaa}{\Lambda\pi^+}
\newcommand{\modebb}{\Lambda\pi^+\pi^0}
\newcommand{\modedd}{\Lambda\pi^+\pi^-\pi^+}
\newcommand{\modeaaa}{\Sigma^{0}\pi^+}
\newcommand{\modeccc}{\Sigma^{+}\pi^0}
\newcommand{\modeddd}{\Sigma^{+}\pi^+\pi^-}

\newcommand{\bmodea}{\bar{p}K_{S}^{0}}
\newcommand{\bmodeb}{\bar{p}K^{+}\pi^-}
\newcommand{\bmodec}{\bar{p}K_{S}^{0}\pi^0}
\newcommand{\bmoded}{\bar{p}K_{S}^{0}\pi^-\pi^+}
\newcommand{\bmodee}{\bar{p}K^{+}\pi^-\pi^0}
\newcommand{\bmodef}{\bar{p}\pi^-\pi^+}
\newcommand{\bmodeaa}{\bar{\Lambda}\pi^-}
\newcommand{\bmodebb}{\bar{\Lambda}\pi^-\pi^0}
\newcommand{\bmodedd}{\bar{\Lambda}\pi^-\pi^+\pi^-}
\newcommand{\bmodeaaa}{\bar{\Sigma}^{0}\pi^-}
\newcommand{\bmodeccc}{\bar{\Sigma}^{-}\pi^0}
\newcommand{\bmodeddd}{\bar{\Sigma}^{-}\pi^-\pi^+}
 \newcommand{\BESIIIorcid}[1]{\href{https://orcid.org/#1}{\hspace*{0.1em}\raisebox{-0.45ex}{\includegraphics[width=1em]{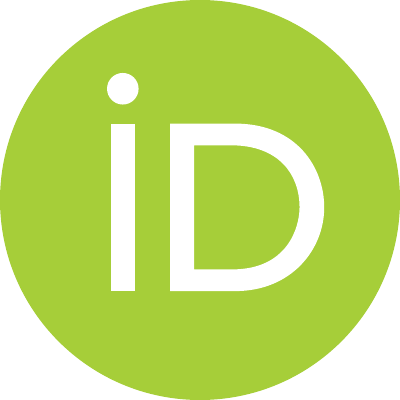}}}}

\title{\bf\boldmath Measurements of the absolute branching fractions of the $\lcp$ hadronic decays}
\author{The BESIII collaboration}
\collaboration{\includegraphics[height=17mm,angle=90]{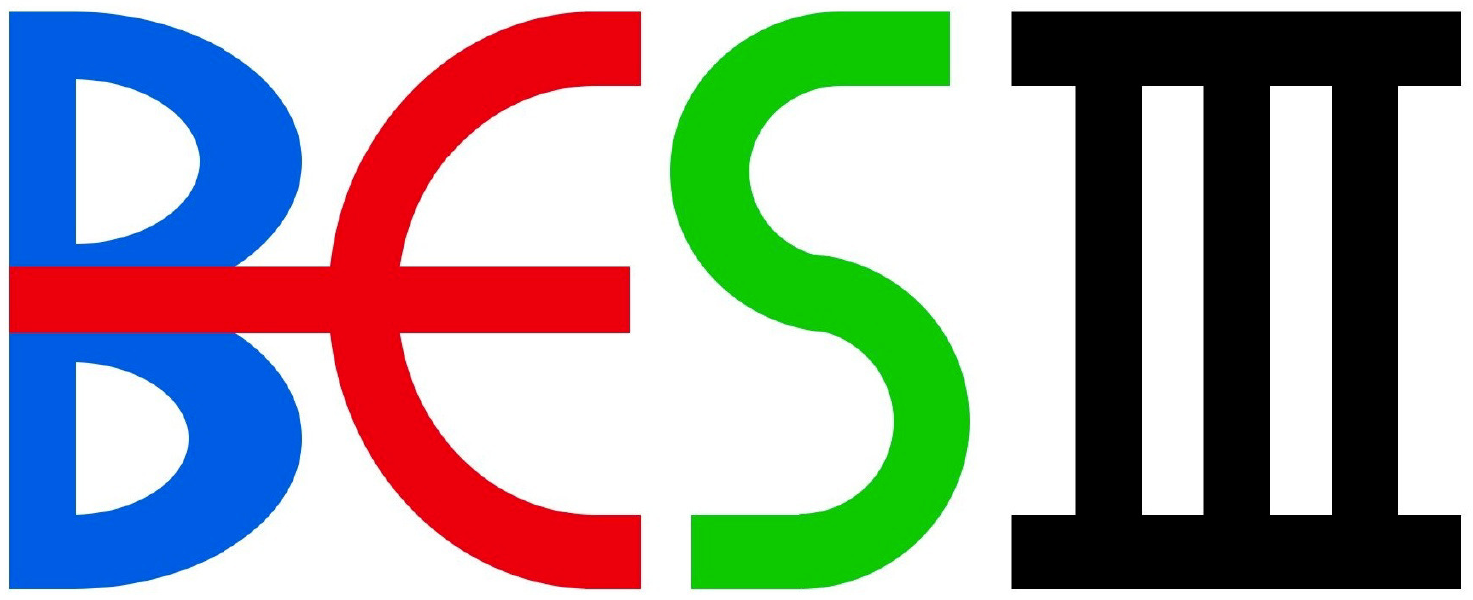}}
\emailAdd{besiii-publications@ihep.ac.cn}
\abstract{Based on 4.5 fb$^{-1}$ of $e^+e^-$ collision data collected at center-of-mass energies between 4599.53 MeV and 4698.82 MeV with the BESIII detector, 
the absolute branching fractions of twelve $\lcp$ hadronic decay modes are measured with a double-tag technique. 
A global least-square fit is implemented simultaneously among different decay modes at different energy points. 
This paper gives the most precise results on the branching fractions of different decay modes to date, with precision improved by a factor of 2 to 3. Among them, the branching fraction of $\lcp\to\modeb$ is determined to be $(6.61\pm0.11\pm0.12)\%$, where the first uncertainty is statistical and the second is systematic.
In addition, the $e^+e^-\to\Lambda_c^+\bar{\Lambda}_c^-$ Born cross sections and the effective form factors ($|G_{\rm eff}|$) at different energy points have been determined with the highest precision to date.}

\begin{document}
\maketitle
\flushbottom
\section{Introduction}

The lightest charmed baryon, $\lcp$, is the cornerstone of charmed baryon spectra.
Improved knowledge of $\lcp$ decays will aid the understanding of strong and weak interactions in the charm sector~\cite{BESIII:2020nme}.
The nonfactorizable contributions play an essential role in charmed baryon decays. 
There are various approaches to describe the nonfactorizable effects, including the covariant confined quark model~\cite{Korner:1992wi,Ivanov:1997ra}, 
the pole model~\cite{Cheng:1991sn,Xu:1992vc,Cheng:1993gf,Xu:1992sw,Zenczykowski:1993jm,Sharma:1998rd}, the current algebra~\cite{Cheng:1993gf,Sharma:1998rd,Korner:1978ec,Uppal:1994pt}
and the SU(3) flavor symmetry~\cite{Geng:2019xbo,Sharma:1996sc,Geng:2017esc,Geng:2017mxn,Geng:2018plk,Lu:2016ogy,Hsiao:2019yur,He:2018joe}.
Using $e^+e^-$ collision data corresponding to an integrated luminosity of
586.9~pb$^{-1}$ collected at the center-of-mass~(c.m.) energy of \mbox{$\sqrt{s}= 4599.53$} MeV in 2014, BESIII has measured the absolute branching fractions~(BFs) for a series of decay modes~\cite{Li:2021iwf,Li:2025nzx,BESIII:2015bjk,BESIII:2015ysy,BESIII:2016ozn}. These results have drawn a lot of attention, and stimulated the development of theoretical studies in the charmed baryon decays~\cite{Cheng:2021qpd,Li:2021iwf}.

Examples where the precision of known BFs have an impact include the determination of Cabibbo–Kobayashi–Maskawa matrix elements $|V_{c(u)b}|$. The LHCb experiment uses exclusive charmed baryon decays to measure the ratio of $|V_{cb}/V_{ub}|$~\cite{LHCb:2015eia}, where the BF of the $\lcp\to\modeb$ decay contributes to the dominant systematic uncertainty.
Similarly, the uncertainty due to the branching fraction of $\lcp$ also plays a dominant role in studies of excited-charm and bottom baryons that decay into final states involving $\lcp$, as observed by the LHCb, Belle, and BaBar experiments~\cite{LHCb:2022vns,Belle:2017jrt,LHCb:2020iby,BaBar:2009eml,Belle:2008xmh,Belle:2018yob}.

The BESIII experiment also reported the near-threshold Born cross section measurement of $e^+e^-\to \Lambda_c^{+}\bar{\Lambda}_c^{-}$ and the BF of $\lcp\to\modeb$ in Ref.~\cite{BESIII:2023rwv}. 
The results show no enhancement around 4630 MeV and the charmonium state Y(4630) is not found. 
The effective form factor ($|G_{\rm eff}|$) is also reported, however, unlike the nucleon case, no oscillatory behavior is observed in the effective form factor of $\Lambda_c^{+}$.
In addition, the data sample collected by the BESIII detector is now approximately eight times larger than that used in the previous measurement~\cite{BESIII:2015bjk}. 

In this paper, we report the updated measurement of the BFs of 11 Cabibbo-favored decay modes $\lcp\to \modea$, $\modeb$, $\modec$, $\moded$, $\modee$, $\modeaa$, $\modebb$, $\modedd$, $\modeaaa$, $\modeccc$, $\modeddd$ and one Cabibbo-suppressed decay mode $\lcp\to\modef$ using 4.5~fb$^{-1}$ of $e^+e^-$ collision data collected at c.m.~energies between 4599.53 MeV and 4698.82 MeV with the BESIII detector~\cite{BESIII:2022ulv}. Both single tags (ST), where one $\Lambda_c$ decay is reconstructed in one of the 12 signal channels with no requirement on the rest of the event, and double tags (DT), where two $\Lambda_c$ decays are reconstructed to any combination of signal channels, are used in the analysis.
 The BFs are then determined by 
 \begin{equation}
    \mathcal{B}_j= \frac{N^{\rm
DT}_{-,j}}{\Sigma_{i}N_{i}^{\rm ST}\cdot(\epsilon_{ij}^{\rm
DT}/\epsilon_{i}^{\rm ST})\cdot \mathcal{B}_{\rm int}},
 \end{equation}
where the indices $i$ and $j$ denote the ST and DT decay modes,
respectively; $\mathcal{B}_{\rm int}$ denotes the BF of
intermediate state decays; $N^{DT}_{-,j}$ represents the total
DT signal yields summing over 12 ST modes; $\epsilon_{ij}^{\rm DT}$,
$N_{i}^{\rm ST}$ and $\epsilon_{i}^{\rm ST}$ represent DT
efficiencies, ST yields, and ST efficiencies, respectively.
The $e^+e^-\to\Lambda_c^+\bar{\Lambda}_c^-$ Born cross
sections and $|G_{\rm eff}|$ are also updated in this analysis, based
on the determined yields of the produced $\lcp\lcm$ pairs.
A global fit, similar to that in Ref.~\cite{BESIII:2022xne} is
performed simultaneously at all the energy points, which
constrains the total number of $\Lambda_c^+\bar{\Lambda}_c^-
$ pairs and BFs among the DT~\cite{MARK-III:1985hbd} and ST
yields of the $\lcp$ at different energy points. Throughout this
paper, charge-conjugate modes are implicitly included.
\section{BESIII Detector and Monte Carlo Simulation}
\label{mc}

The BESIII detector~\cite{BESIII:2009fln} records symmetric $e^+e^-$ collisions provided by the BEPCII storage ring~\cite{Yu:2016cof}, which operates in the c.m. energy range from 1.84~$\gev$ to 4.946~$\gev$~\cite{BESIII:2020nme}. In this analysis data were collected at seven energy points between 4600 and 4700$~\mev$. Details of the integrated luminosity at each energy point are found in Ref.~\cite{BESIII:2022ulv}.

The cylindrical core of the BESIII detector covers 93\% of the full solid angle and consists of a helium-based multilayer drift chamber~(MDC), a time-of-flight system~(TOF) 
and a CsI(Tl) electromagnetic calorimeter~(EMC), which are all enclosed in a superconducting solenoidal magnet providing a 1.0~T magnetic field.
The solenoid is supported by an octagonal flux-return yoke instrumented with resistive plate counter muon identification modules interleaved with steel.
The charged-particle momentum resolution at $1\gevc$ is $0.5\%$, and the specific ionization energy loss (d$E/$d$x$) resolution is $6\%$ for electrons from Bhabha scattering.
The EMC measures photon energies with a resolution of $2.5\%$ ($5\%$) at $1\gev$ in the barrel (end cap) region.
The time resolution in the TOF barrel region is 68~ps, while that in the end cap region is 60~ps~\cite{Guo:2017sjt,Li:2017jpg,Cao:2020ibk}. This end cap resolution was achieved in 2015 after an upgrade. The data at 4600~$\mev$ was taken prior to the upgrade and have an end cap time resolution of 110 ps. All other data used here benefit from the improved resolution of the upgrade.

Simulated data samples are produced with a {\sc geant-4}~\cite{GEANT4:2002zbu} based Monte Carlo (MC) package,
which includes the geometric description~\cite{Huang:2022wuo} of the BESIII detector and the detector response, and are used to determine detection efficiencies and estimate backgrounds (BKGs). 
The simulation includes the beam energy spread and initial state radiation in the $e^+e^-$ annihilation modeled with the event generator {\sc kkmc} ~\cite{Jadach:2000ir,Jadach:1999vf}. 
The inclusive MC samples include the production of $\lcp\lcm$ pairs, open charm processes, the Initial State Radiation~(ISR) production of vector charmonium(-like) states, and the continuum Quantum Chromodynamics  processes $e^+e^-\to q \bar{q}$($q = u,d,s$). 
The known decay modes are modeled with {\sc evtgen}~\cite{Lange:2001uf,Ping:2008zz} using BFs taken from the Particle Data Group~\cite{ParticleDataGroup:2024cfk}, 
and the remaining unknown charmonium decays are modeled with \mbox{\sc lundcharm}~\cite{Chen:2000tv,Yang:2014vra}.
Final state radiation from charged final state particles is incorporated using {\sc photos}~\cite{Richter-Was:1992hxq}.
For the MC production of $e^+e^-\to\lcp\lcm$, the c.m. energy-dependent Born cross section measured by BESIII~\cite{BESIII:2023rwv} is fed into the generator and the ISR effect is taken into account.
The angular distribution of $e^+e^-\to\lcp\lcm$ is generated as $1+\alpha\cos^2\theta_{\lcp}$, where $\theta_{\lcp}$ is the polar angle between the $\lcp$ and the positron beam in the c.m. frame, and $\alpha$ is the angular parameter of $\lcp$ production, which are different at the seven c.m. energy points~\cite{BESIII:2023rwv}.
The ST MC sample, where one $\lcp$ decays to a signal decay mode and the decay of the $\lcm$ is not reconstructed, is used to extract the signal shape.
Both the $\lcp$ and $\lcm$ decay to signal modes in the signal MC sample which is used to determine the DT efficiency and model the signal shape. For the multi-body decays, the signal mode is modeled by the amplitude analysis results~\cite{BESIII:2022udq}.
For the two-body decays, the angular parameters are considered in MC generation~\cite{LHCb:2022sck,BESIII:2019odb,BESIII:2025zbz}.

\section{Event Selection}

The $\lcm$ candidates with the charged track selection follow the same criteria as in Ref.~\cite{BESIII:2022xne}. 
Charged tracks detected in the MDC are required to be within a polar angle ($\theta$) range of $|\rm{cos\theta}|<0.93$, where $\theta$ is defined with respect to the $z$-axis,
which is the symmetry axis of the MDC.
For charged tracks not originating from $K_S^0$ or $\Lambda$ decays, the distance of closest approach to the interaction point (IP) 
must be less than 10\,cm
along the $z$-axis, $|V_{z}|$,  
and less than 1\,cm
in the transverse plane, $|V_{xy}|$.

Particle identification (PID) for charged tracks combines measurements of ionization energy loss in the MDC ($dE/dx$) and the flight time in the TOF to form likelihoods $\mathcal{L}(h)~(h=p,K,\pi)$ for each hadron $h$ hypothesis.
Charged tracks are identified as protons, kaons and pions when $\mathcal{L}(p)>\mathcal{L}(\pi)$ and $\mathcal{L}(p)>\mathcal{L}(K)$, $\mathcal{L}(K)>\mathcal{L}(\pi)$, $\mathcal{L}(\pi)>\mathcal{L}(K)$, respectively.

Photon candidates are identified using showers in the EMC. The energy deposited by each shower is required to be larger than 25\,MeV in the barrel region ($|\!\cos\theta|<0.80$), or 50\,MeV in the end cap region ($0.86<|\!\cos\theta|<0.92$).
To reject photons arising from beam background, electronic noise, and showers unrelated to the event, the difference between the EMC time and the event start time \cite{Guan:2013jua} is required to be within [0, 700] ns.
The $\pi^0$ candidates are reconstructed from photon pairs with an invariant mass $M(\gamma\gamma)$ within [0.115,0.150]~\,GeV/$c^2$.

Candidates for $K_S^0$ and $\bar{\Lambda}$ hadrons are selected by combining two oppositely charged tracks.
These tracks are required to have distances to the IP along the beam direction within 20\,cm.
The two daughter tracks are constrained to originate from a common decay vertex by requiring the $\chi^2$ of the vertex fit to be less than 100.
The decay vertex is required to be separated from the IP by a distance of at least twice the fitted vertex resolution.
For protons, the previously mentioned PID requirement is applied, while there is no PID requirement for charged pions.
The invariant masses of $\pi^+\pi^-$ and $\bar{p}\pi^+$ pairs are required to be within [0.487,0.511] \,GeV/$c^2$ and [1.111,1.121] \,GeV/$c^2$, respectively.

For the decay modes $\bar\Lambda_c^-\to$ $\bmodec$, $\bmoded$, $\bmodeddd$ and $\bmodef$, events with the invariant mass $M(\bar{p}\pi^+)$ within the region [1.11, 1.12] GeV/$c^2$ are rejected to suppress the intermediate $\bar{\Lambda}$ baryon contributions.
For the decay modes $\bar\Lambda_c^-\to$ $\bmodedd$, $\bmodeccc$, $\bmodeddd$ and $\bmodef$, the invariant mass of $\pi^+\pi^-$ or $\pi^0\pi^0$ pairs within the region [0.48, 0.52] GeV/$c^2$ are rejected to suppress the intermediate $K_S^0$ meson contributions.   
In addition, to remove the $\bar{\Sigma}^-$ contribution in the decay mode $\bmodec$, the invariant mass of $\bar{p}\pi^0$ pairs must fall outside of the range [1.17,1.20] GeV/$c^2$. 


The energy difference $\Delta E$ and beam-constrained mass $M_{\rm BC}$ are used to distinguish the $\Lambda_c^+$ signal and combinational background. They are defined as
$M_{\rm BC}=\sqrt{E_{\rm beam}^{2}/c^{4}-{p}^{2}_{\Lambda_c}/{c^2}}$, 
and $\Delta E= E - E_{\rm beam}$, where $E$ is the total reconstructed energy of the $\lcm$ candidate, $E_{\rm beam}$ is the average value of the $e^+$ and $e^-$ beam energies, and ${p}_{\Lambda_c}$ is the momentum of the $\lcm$ measured in the c.m. system of the $\ee$ collision. 
For multiple candidates in one event, only the one with minimum $|\Delta E|$ is chosen.
In order to further suppress the combinational backgrounds, the selected candidates are imposed with the same $\Delta E$ requirements as used in Ref.~\cite{BESIII:2022xne}, corresponding to approximately three times the resolutions.

Although the mass and $\Delta E$ requirements are imposed, misidentified and partial reconstruction backgrounds remain, which generate a non-flat component in the background.
For the decay modes $\bar\Lambda_c^-\to$ $\bmodeaa$, $\bmodebb$ and $\bmodeaaa$, the $\lcm\to\bar{\Lambda} e^- \bar{\nu}_e$ and $\lcm\to\bar{\Lambda} \mu^- \bar{\nu}_{\mu}$ background decays have a very small peaking contribution due to muon, electron and pion misidentification. 
Residual backgrounds where final states similar to the signal modes with an extra or missing photon can contribute to the partially reconstructed backgrounds. For example, the $\lcm\to\bmodeaaa$ and $\lcm\to\bar{\Sigma}^0\pi^-\pi^0$ decays can contribute to the $\lcm\to\bmodebb$ channel;
the $\lcm\to\bmodebb$ decay can contaminate the $\lcm\to\bmodeaaa$ channel;
the $\lcm\to\bmodeaaa$ decay can contribute to the $\lcm\to\bmodeaa$ channel;
the $\lcm\to\bmodea$ ($K_S^{0} \to \pi^0\pi^0$) decay could contaminate the $\lcm\to\bmodeccc$ channel.
For the $\lcm\to\bmodef$ decay, there are residual peaking backgrounds from the $\lcm\to\bmodea$ and $\lcm\to\bmodeaa$ channels.
\section{Extractions of the ST and DT yields}

\begin{figure*}[!hbt]
  \centering
  \includegraphics[width=1.0\textwidth]{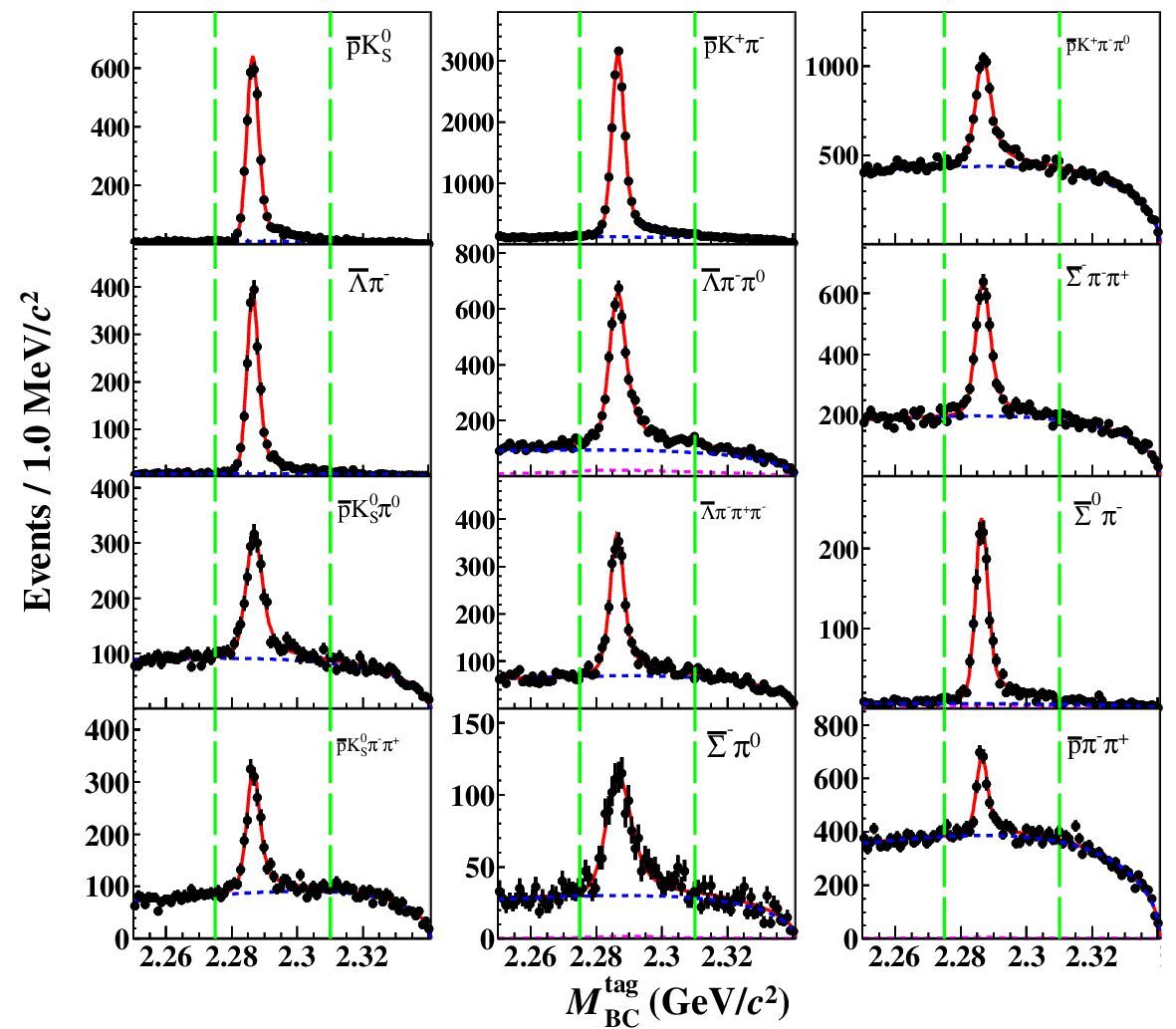}
 \caption{The fits to the $M^{\rm tag}_{\rm BC}$ distributions of the ST $\bar \Lambda^-_c$ candidates for various tag modes at $\sqrt{s}=4681.92~\mev$. 
     The points with error bars are data. The red line is the total fitted distribution, with the dashed-blue lines showing the ARGUS function and the dashed-magenta lines showing the residual non-flat background shape. The dashed-green lines denote the signal regions.}
 \label{fig:STfit}
\end{figure*}
 
  \begin{table}[htbp]
\centering
      \caption{The ST yields $N_{i}^{\rm ST}$ for different decay modes at different energy points, where the uncertainties are statistical only.}
  \begin{center}
  \scriptsize
  \begin{tabular}{l|c|ccccccc}
      \hline   \hline
       \multirow{2}{*}{Tag side}  & $\Delta E$ &\multicolumn{7}{c}{Energy point ($\mev$)} \\ \cline{3-9}
     &(MeV) & $4599.53$ & $4611.86$  & $4628.00$  & $4640.91$ & $4661.24$ & $4681.92$ & $4698.82$ \\ \hline
  $\textbf{$\bmodea$}$ & ($-21,18$) & $1273\pm37$ & $239\pm16$  & $1052\pm35$  & $1099\pm35$ &   $1115\pm35$  & $3370\pm61$   &  $958\pm33$   \\
  $\textbf{$\bmodeb$}$ &($-29,26$) & $6778\pm91$ & $1164\pm39$ & $5867\pm42$  & $6232\pm90$ &   $5920\pm86$  & $17480\pm148$ &  $5158\pm81$  \\
  $\textbf{$\bmodec$}$&($-49,34$)  & $603\pm35$  & $127\pm16$  & $596\pm39$   & $581\pm38$  &   $575\pm38$   & $1728\pm66$   &  $456\pm37$   \\
  $\textbf{$\bmoded$}$ & ($-34,31$) & $609\pm34$  & $106\pm16$  & $490\pm33$   & $516\pm33$  &   $532\pm32$   & $1504\pm59$   &  $458\pm32$   \\
  $\textbf{$\bmodee$}$&($-60,41$) & $2055\pm75$ & $356\pm33$  & $1574\pm69$  & $1610\pm78$ &   $1644\pm77$  & $4809\pm131$  &  $1319\pm63$  \\
  $\textbf{$\bmodeaa$}$&($-23,21$) & $746\pm28$  & $121\pm11$  & $664\pm28$   & $694\pm28$  &   $661\pm27$   & $2042\pm48$   &  $532\pm24$   \\
  $\textbf{$\bmodebb$}$&($-50,41$) & $1685\pm57$ & $294\pm22$  & $1404\pm52$  & $1561\pm52$ &   $1432\pm50$  & $4122\pm82$   &  $1260\pm48$  \\
  $\textbf{$\bmodedd$}$ & ($-40,36$) & $806\pm 42$ & $138\pm15$  & $584\pm33$   & $747\pm38$  &   $777\pm38$   & $2043\pm61$   &  $632\pm26$   \\
  $\textbf{$\bmodeaaa$}$&($-33,31$) & $502\pm24$  & $98\pm10$   & $401\pm23$   & $434\pm25$  &   $439\pm24$   & $1351\pm40$   &  $358\pm21$   \\
  $\textbf{$\bmodeccc$}$&($-67,32$) & $304\pm29$  & $71\pm10$   & $259\pm23$   & $289\pm24$  &   $295\pm26$   & $892\pm48$    &  $276\pm28$   \\
  $\textbf{$\bmodeddd$}$&($-40,32$) & $1182\pm51$ & $219\pm22$  & $989\pm46$   & $1049\pm52$ &   $1028\pm51$  & $3022\pm88$   &  $990\pm47$   \\
  $\textbf{$\bmodef$}$ &($-26,20$) & $528\pm44$  & $145\pm21$  & $475\pm50$   & $524\pm50$  &   $563\pm50$   & $1545\pm79$   &  $445\pm54$   \\ 
  \hline  \hline
   \end{tabular}
   \label{tab:deltaE_yields}

  \end{center}
  \end{table}

  \begin{table}[htbp]
       \caption{The ST efficiencies $\epsilon_{i}^{\rm ST}$(\%) for different decay modes at different energy points, where the uncertainties are statistical only.}
   \begin{center}
     \scriptsize
   \begin{tabular}{l|ccccccc}
       \hline \hline
       \multirow{2}{*}{Tag side}  &\multicolumn{7}{c}{Energy point ($\mev$)} \\ \cline{2-8}
      & $4599.53$ & $4611.86$  & $4628.00$  & $4640.91$ & $4661.24$ & $4681.92$ & $4698.82$ \\ \hline
       $\textbf{$\bmodea$}$    & $56.1\pm0.2$   & $53.9\pm0.5$  & $51.4\pm0.2$ & $50.6\pm0.2$ & $49.4\pm0.2$ & $48.4\pm0.1$ &$47.6\pm0.2$   \\
       $\textbf{$\bmodeb$}$    & $51.2\pm0.1$   & $50.7\pm0.2$  & $49.0\pm0.1$ & $48.3\pm0.1$ & $48.0\pm0.1$ & $47.0\pm0.1$ &$46.4\pm0.1$   \\
       $\textbf{$\bmodec$}$    & $22.8\pm0.2$   & $22.3\pm0.4$  & $20.6\pm0.2$ & $20.7\pm0.2$ & $19.8\pm0.2$ & $19.0\pm0.1$ &$18.6\pm0.2$   \\
       $\textbf{$\bmoded$}$    & $22.8\pm0.2$   & $21.5\pm0.4$  & $20.6\pm0.2$ & $20.9\pm0.2$ & $21.3\pm0.2$ & $20.4\pm0.1$ &$19.8\pm0.2$   \\
       $\textbf{$\bmodee$}$    & $19.8\pm0.1$   & $19.5\pm0.2$  & $18.7\pm0.1$ & $18.1\pm0.1$ & $18.0\pm0.1$ & $17.4\pm0.1$ &$17.2\pm0.1$   \\
       $\textbf{$\bmodeaa$}$   & $47.6\pm0.3$   & $45.7\pm0.6$  & $42.8\pm0.3$ & $42.4\pm0.3$ & $41.2\pm0.3$ & $40.6\pm0.2$ &$38.8\pm0.3$   \\
       $\textbf{$\bmodebb$}$   & $20.6\pm0.1$   & $19.3\pm0.2$  & $18.5\pm0.1$ & $18.4\pm0.1$ & $18.2\pm0.1$ & $17.6\pm0.1$ &$17.3\pm0.1$   \\
       $\textbf{$\bmodedd$}$   & $15.6\pm0.1$   & $13.5\pm0.3$  & $13.5\pm0.1$ & $14.2\pm0.1$ & $13.9\pm0.1$ & $14.0\pm0.1$ &$14.3\pm0.1$   \\
       $\textbf{$\bmodeaaa$}$  & $28.5\pm0.2$   & $26.4\pm0.6$  & $26.1\pm0.2$ & $25.3\pm0.2$ & $25.3\pm0.2$ & $24.9\pm0.1$ &$23.9\pm0.2$   \\
       $\textbf{$\bmodeccc$}$  & $22.8\pm0.3$   & $23.3\pm0.6$  & $22.9\pm0.3$ & $23.1\pm0.3$ & $22.7\pm0.3$ & $22.0\pm0.2$ &$21.6\pm0.3$   \\
       $\textbf{$\bmodeddd$}$  & $25.3\pm0.1$   & $24.9\pm0.3$  & $23.7\pm0.1$ & $23.5\pm0.1$ & $23.0\pm0.1$ & $22.2\pm0.1$ &$22.0\pm0.1$   \\
       $\textbf{$\bmodef$}$    & $60.5\pm0.3$   & $59.0\pm0.8$  & $55.4\pm0.4$ & $55.0\pm0.4$ & $56.4\pm0.4$ & $53.0\pm0.2$ &$50.9\pm0.4$   \\
       
   \hline\hline
    \end{tabular}
    \label{tab:STefficiency}
 
   \end{center}
     \end{table}

   \begin{table}[htbp]
   \begin{center}
      \caption{The DT yields $N_{i}^{\rm DT}$ for different decay modes at different energy points, where the uncertainties are statistical only.}
        \scriptsize
   \begin{tabular}{l|ccccccc}
       \hline \hline
        \multirow{2}{*}{Signal side}  &\multicolumn{7}{c}{Energy point ($\mev$)} \\ \cline{2-8}
      & $4599.53$ & $4611.86$  & $4628.00$  & $4640.91$ & $4661.24$ & $4681.92$ & $4698.82$ \\ \hline
       $\textbf{$\modea$}$   & $87\pm9$  & $20\pm4$ &$95\pm10$  & $93\pm10$  &$80\pm9$  & $260\pm15$   & $71\pm9$     \\
       $\textbf{$\modeb$}$   & $439\pm21$& $70\pm8$ &$407\pm21$  & $388\pm20$  &$396\pm21$  & $1110\pm35$   & $316\pm18$    \\
       $\textbf{$\modec$}$   & $47\pm7$  & $7\pm3$  &$37\pm6$  & $37\pm7$  &$47\pm7$  & $140\pm13$   & $35\pm5$     \\
       $\textbf{$\moded$}$   & $34\pm6$  & $4\pm1$  &$35\pm5$  & $35\pm6$  &$43\pm6$  & $116\pm10$   & $27\pm6$     \\
       $\textbf{$\modee$}$   & $125\pm9$  & $23\pm4$&$125\pm12$  & $114\pm11$  &$116\pm12$  & $337\pm20$   & $102\pm11$   \\
       $\textbf{$\modeaa$}$  & $60\pm7$  &  $10\pm2$&$45\pm7$  & $54\pm7$  &$46\pm6$  & $141\pm11$   & $40\pm5$    \\
       $\textbf{$\modebb$}$  & $105\pm10$ & $14\pm4$&$99\pm17$  & $104\pm10$  &$110\pm9$  & $283\pm17$   & $97\pm10$   \\
       $\textbf{$\modedd$}$  & $56\pm8$  & $14\pm4$ &$53\pm7$  & $47\pm7$  &$57\pm8$  & $145\pm12$   & $42\pm7$     \\
       $\textbf{$\modeaaa$}$ & $34\pm7$  & $10\pm2$ &$35\pm6$  & $46\pm7$  &$23\pm5$  & $114\pm10$   & $28\pm5$     \\
       $\textbf{$\modeccc$}$ & $21\pm5$  & $4\pm1$  &$27\pm5$  & $19\pm4$  &$19\pm5$  & $70\pm8$   & $19\pm4$     \\
       $\textbf{$\modeddd$}$ & $49\pm7$  & $9\pm3$  &$60\pm7$  & $49\pm7$  &$53\pm8$  & $154\pm11$   & $54\pm8$     \\
       $\textbf{$\modef$}$   & $42\pm7$  & $7\pm3$  &$39\pm6$  & $45\pm7$  &$28\pm6$  & $119\pm12$   & $41\pm6$      \\       
       
   \hline\hline
    \end{tabular}
    \label{tab:DTyields}
 
   \end{center}
   \end{table}

An extended unbinned maximum likelihood fit is performed to the $M_{\rm BC}$ distributions of the tag side ($M^{\rm tag}_{\rm BC}$) to obtain the ST yields.
In the fit, the signal shape is derived from the MC simulated shape convolved with a Gaussian function to model the resolution difference between MC and data. 
The parameters of the Gaussian function are allowed to vary in the fits. The non-flat background shape is extracted from the inclusive MC sample and its fraction relative to the signal is fixed from simulation.  
The ARGUS function is used to describe the remaining combinatorial background and the end-point is fixed to $E_{\rm beam}$. 
The ST yields are extracted within different $M^{\rm tag}_{\rm BC}$ signal regions depending on the c.m. energy.
The signal regions are defined as: (2275,~2300) $\mevcc$ at $\sqrt{s}=$ 4599.53 MeV; (2275,~2306) $\mevcc$ at $\sqrt{s}=$ 4611.86, 4628.00, 4640.91 MeV 
and (2275,~2310) $\mevcc$ at $\sqrt{s}=$ 4661.24, 4681.92, 4698.82 MeV.
The fit results are shown in Fig.~\ref{fig:STfit} for the energy point  $\sqrt{s}=4681.92$ MeV, with fit results for the other energy points found in Appendix A. 
The ST efficiency is estimated by the inclusive MC sample using the same procedure with data. The obtained ST yields and efficiencies for each energy point are listed in Table~\ref{tab:deltaE_yields} and Table~\ref{tab:STefficiency}, respectively.

The signal decays are reconstructed with the remaining tracks and showers against the ST $\lcm$ candidate.
If there are multiple candidates in one event, the one with minimum $|\Delta E|$ is retained using the same requirement as on the ST side.
Due to limited data, we combine the DT signal candidates over the twelve ST modes. 

\begin{figure}[htbp]
\centering
\includegraphics[width=0.6\textwidth]{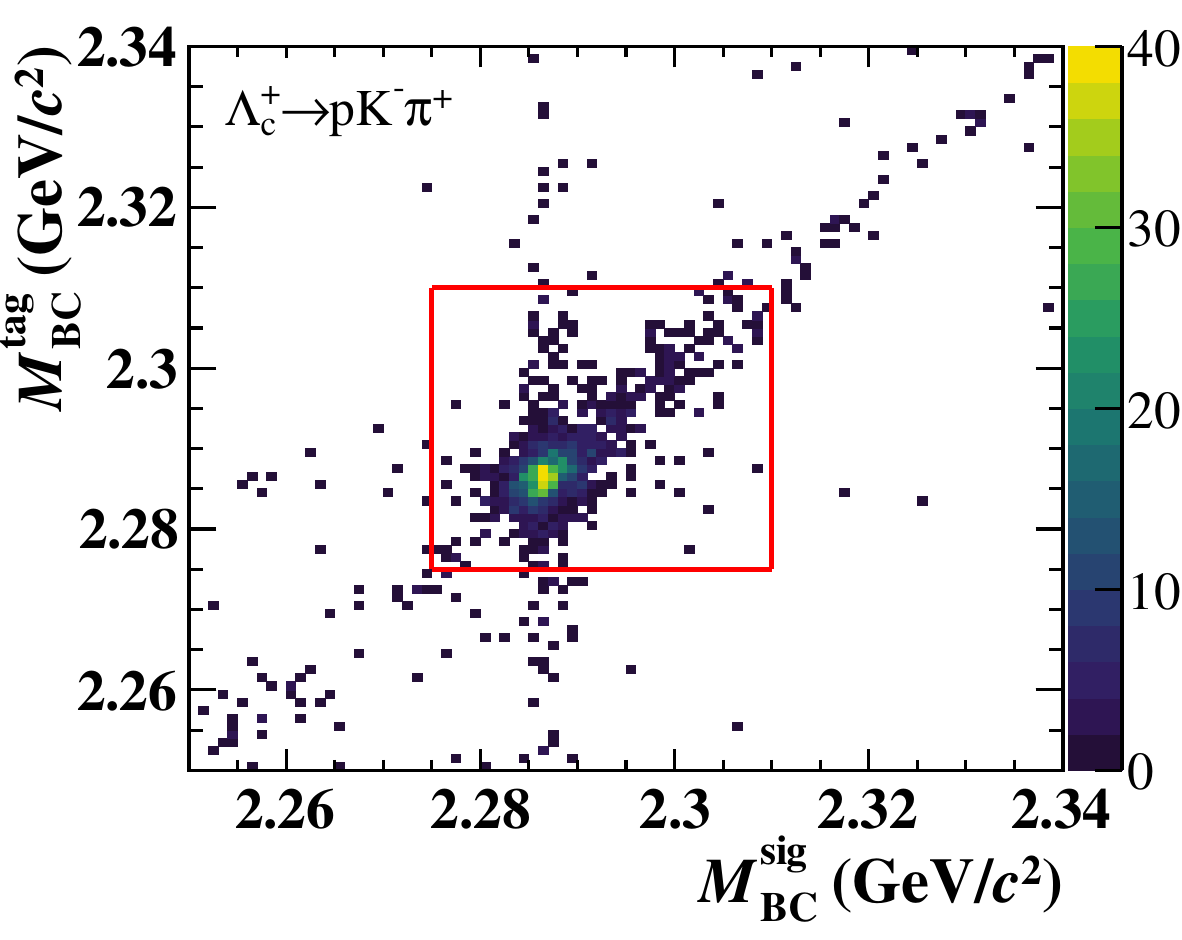}
\caption{The $M^{\rm tag}_{\rm BC}$ versus $M^{\rm sig}_{\rm BC}$ distribution of the signal decay mode $\Lambda_c^+\to\modeb$ in data at $\sqrt{s}=$ 4681.92~MeV.}
\label{2D_4680_pkpi}
\end{figure}

\begin{figure}[htbp]
\centering
\includegraphics[width=0.4\textwidth]{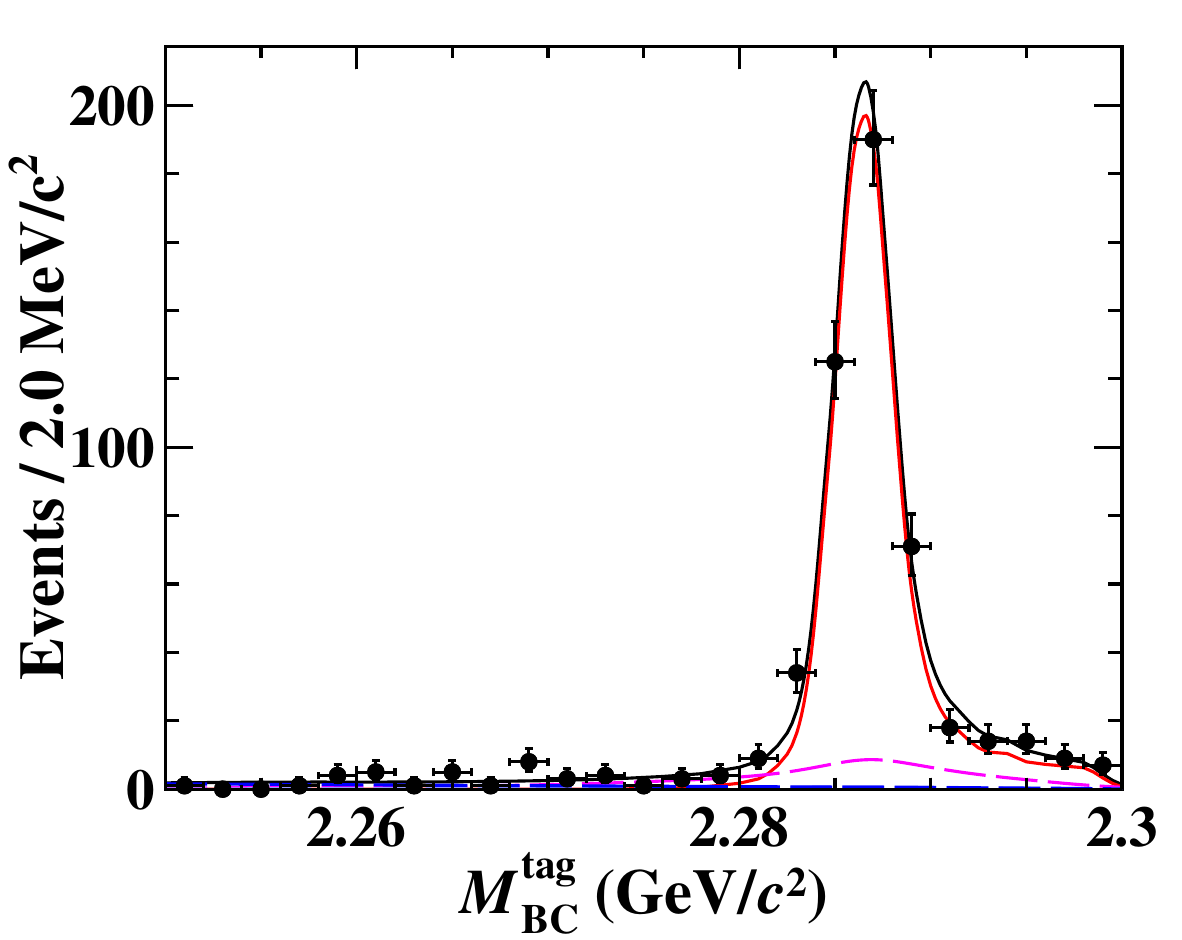}
\qquad
\includegraphics[width=.4\textwidth]{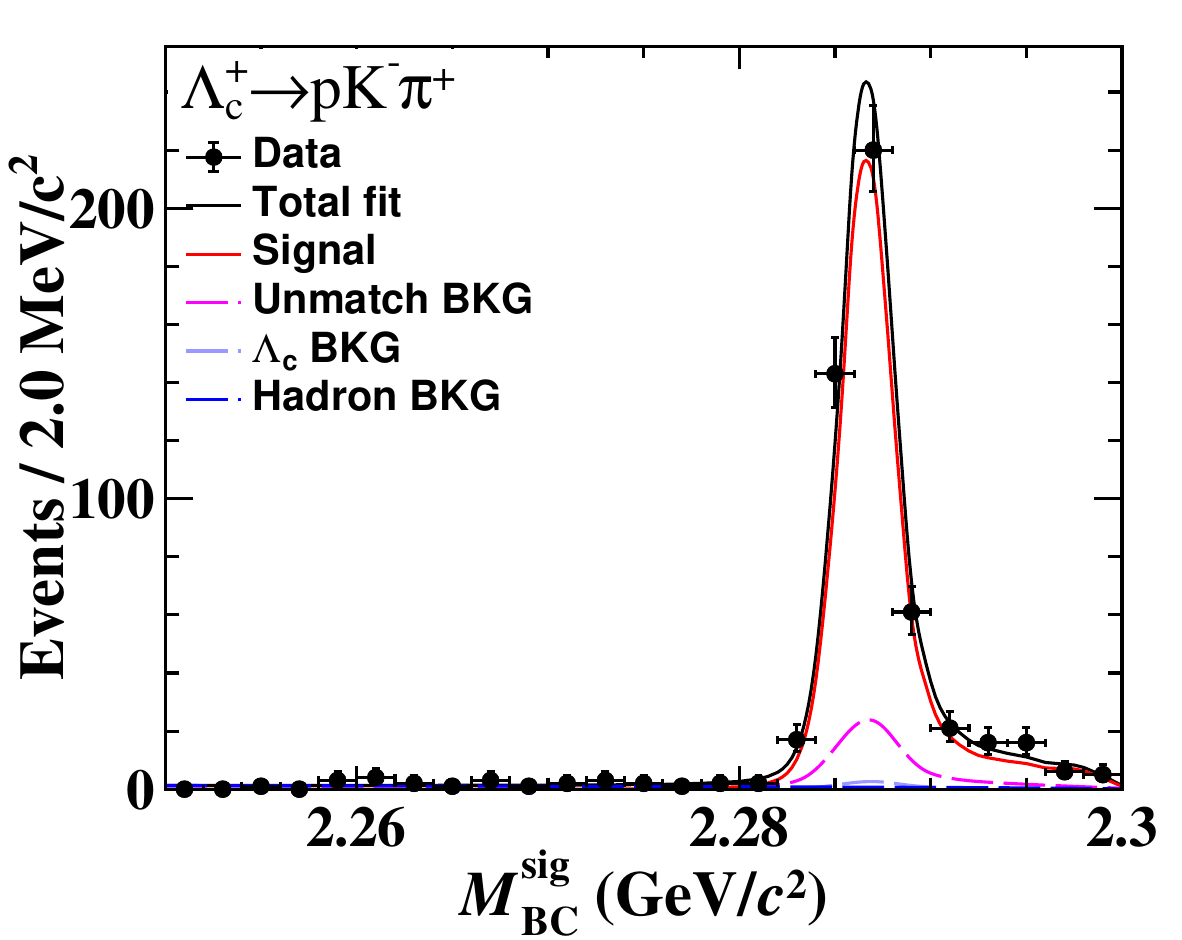}
\caption{The projections of the 2D fit of the accepted DT candidates for $\Lambda^+_c\to pK^-\pi^+$ at $\sqrt{s}= 4681.92~\mev$, on the $M_{\rm BC}^{\rm tag}$ and $M_{\rm BC}^{\rm sig}$ distributions.
The plots show the combined 12 tag modes.
The points with error bars are data. The total fit function is given in black, the red shows the matched signal shapes, and the dashed-pink line shows the unmatched signal shapes. The light blue shows the non-signal $\lcp\lcm$ shapes, and the dark blue shows hadronic background.}
\label{DT_4680_pkpi}
\end{figure}

A two-dimensional~(2D) unbinned likelihood fit is performed on the $M_{\rm BC}$ distribution of tag and signal sides ($\tagmbc$ versus $\sigmbc$) to extract the DT signal yield. 
As an example, the 2D distribution for the signal decay mode $\Lambda_c^+\to\modeb$ at $\sqrt s =$ 4681.92~MeV is shown in Fig.~\ref{2D_4680_pkpi}. 
For the signal MC sample, tracks are truth matched through comparison of the angle of the true and reconstructed tracks. Tracks with a truth matched angle less than $10^\circ$ are classified as the matched-signal component, while the remaining are treated as unmatched background.
In the fit to each $M_{\rm BC}$ distribution, the 2D signal shape is extracted from the matched-signal MC sample and convolved with a Gaussian resolution function.
The parameters of the Gaussian function are fixed based on the results of the fits to the $M_{\rm BC}^{\rm tag}$ distributions of the corresponding ST $\bar \Lambda^-_c$ candidates in data.
These values are taken from studies of different decay modes at various energy points. 
For the un-matched background, its shape is directly extracted from the signal MC sample. 
The fraction between the matched and un-matched components is fixed according to the signal MC simulation. 
As discussed, on the ST side, the residual partial reconstructed and misidentified backgrounds can also contribute to the signal side. These backgrounds can lead to a correctly reconstructed $\bar{\Lambda}_c^-$ on the tag side but a mis-reconstructed $\Lambda_c^+$ on the signal side, or vice versa. 
To account for this effect, the 2D background shape of non-signal $\lcp$ decays is extracted from the inclusive MC sample. The yield of this component is treated as a free parameter and determined by the fit. 
For hadronic backgrounds not associated with $\Lambda_c$ production, we employ an ARGUS function with its end-point fixed at the beam energy.
As an example, the projections of the 2D fit on the $M_{\rm BC}^{\rm tag}$ and $M_{\rm BC}^{\rm sig}$ distributions of the accepted DT candidates with $\lcp\to\modeb$ in data at $\sqrt{s}=$ 4681.92~$\mev$ are shown in Fig.~\ref{DT_4680_pkpi}.   
The results of the 2D fits for the remaining eleven channels and other energy points can be found in Appendix B.
The obtained DT yields are summarized in Table~\ref{tab:DTyields}.
The 12 $\times$ 12 efficiency matrix is evaluated based on the DT matched-signal MC sample and is presented in Table~\ref{tab:DTeff_4680}. 


\section{Systematic uncertainties}

Systematic uncertainties due to the selection of single tags primarily cancel due to the use of the DT technique. Sources of systematic uncertainties are mainly from the 
tracking and PID,  $\pi^0$, $K_S^{0}$ and $\Lambda$ reconstruction. Uncertainties also arise from the 2D $M_{\rm BC}$ fit, the models employed by the MC, the size of the MC samples and the background veto.

\begin{itemize}

\item 	\emph{Tracking, PID, $\Lambda$ and $\pi^0$ reconstruction}. 
The systematic uncertainties of tracking and PID for pion and proton are studied with the 
control sample of $J/\psi\rightarrow p\bar{p}\pi^+\pi^-$ decays.	
The tracking and PID efficiencies of charged kaons are studied with the control sample of $J/\psi\rightarrow K_S^0K^{\pm}\pi^{\mp}$ decays. 
The photon reconstruction efficiency is studied through the decay of $J/\psi \rightarrow \rho^0\pi^0$~\cite{BESIII:2010ank}.
The $K_S^0$ reconstruction efficiency is studied with the control samples of $J/\psi \rightarrow K^*(892)^{\pm}K^{\mp},K^*(892)^{\pm}\rightarrow K^0_S\pi^{\pm}$ and $J/\psi \rightarrow \phi K^0_SK^{\mp}\pi^{\pm}$ decays \cite{BESIII:2015jmz}. 
The $\pi^0$ reconstruction efficiency is studied with the process $\psi(3686)\to J/\psi\pi^0\pi^0$.							
For the $\Lambda$ reconstruction efficiency, the control samples of $J/\psi \rightarrow pK^- \bar{\Lambda}$ and $\Lambda\bar{\Lambda}$ are used.
The DT efficiency of signal mode is recalculated after reweighting the signal MC sample on an event-by-event basis according to the momentum and polar angle-dependent efficiency differences between data and MC simulation. 
The relative differences between the nominal and corrected BFs are taken as the systematic uncertainties.

\item 	\emph{2D fit model}.
Imperfect modeling of the shapes used in the 2D fit model leads to systematic uncertainties.
For the signal shape, the parameters of the smearing Gaussian function are varied within $1\sigma$. 
For the background shape, the end-point of ARGUS function is randomly varied within $\pm0.02$~MeV/$c^2$ around its nominal value. 
The kernel parameters of the MC background shape are changed from $1$ to $2$. 
We generate 1000 pseudo data sets and randomly vary the fit parameters for each of them.
The relative shift of the fitted yield distribution obtained from the pseudo data sets is assigned as the systematic uncertainty. 
\item 	\emph{MC model}.
The systematic uncertainty of MC model is estimated by varying the amplitude analysis model according to the error matrix and reweighting the detection efficiency. 
\item 	\emph{MC statistics}.
The uncertainty due to the finite sample size of the MC samples is propagated to the BFs of signal channels and considered as a systematic uncertainty.
\end{itemize}
The systematic uncertainties due to the application of mass window vetoes for intermediate-state suppression are found to be negligible.
The systematic uncertainties associated with the ST fits are evaluated by varying the fit conditions and are found to be negligible. 
For the unmatched component in extracting DT signal yields, modified matching criteria are applied and the resultant changes on the BF results are also negligible.
The summary of systematic uncertainties at $\sqrt{s}=$ 4681.92 MeV is listed in Table~\ref{tab:systematic_4680}. Those for the other energy points can be found in Appendix~\ref{sys}.  
The total uncertainties are then taken to be the quadratic sums of the individual items. 

  \begin{table*}[!htbp]
    \caption{The systematic uncertainties in BF measurements(\%) at $\sqrt s=$ 4681.92 MeV.}
            \scriptsize
    \begin{center}
    \begin{tabular}{l|ccccccccc}
        \hline \hline
  
        Source                &  Tracking  &  PID  & $K_S^0$  & $\Lambda$ & $\pi^0$ &2D fit & Signal model & MC statistics  & Intermediate BF \\  \hline
        $\textbf{$\modea$}$   & 0.4        & 0.1    & 0.8   &           &       & 0.6 & 0.5 & 0.5 &  0.1 \\
        $\textbf{$\modeb$}$   & 1.7        & 0.7    &       &           &       & 0.8 & 0.2 & 0.4 &      \\
        $\textbf{$\modec$}$   & 0.6        & 0.4    & 1.1   &           &  1.0  & 0.0 & 0.9 & 0.6 &  0.1 \\
        $\textbf{$\moded$}$   & 2.1        & 1.4    & 2.5   &           &       & 2.5 & 1.3 & 0.6 &  0.1 \\
        $\textbf{$\modee$}$   & 1.7        & 1.2    &       &           &  0.6  & 0.1 & 1.0 & 0.7 &      \\
        $\textbf{$\modeaa$}$  & 0.6        & 0.6    &       &   1.7     &       & 0.1 & 0.2 & 0.6 &  0.8 \\
        $\textbf{$\modebb$}$  & 0.6        & 0.4    &       &   0.4     &  1.1  & 1.0 & 1.4 & 0.7 &  0.8 \\
        $\textbf{$\modedd$}$  & 1.7        & 0.6    &       &   1.8     &       & 0.9 & 0.2 & 0.6 &  0.8 \\
        $\textbf{$\modeaaa$}$ & 0.6        & 0.6    &       &   1.3     &       & 3.5 & 1.3 & 0.7 &  0.8 \\
        $\textbf{$\modeccc$}$ & 0.4        & 0.1    &       &           &  2.5  & 2.4 & 0.1 & 0.5 &  0.1 \\
        $\textbf{$\modeddd$}$ & 1.9        & 1.2    &       &           &  0.4  & 1.9 & 1.1 & 0.5 &  0.1 \\
        $\textbf{$\modef$}$   & 2.0        & 0.9    &       &           &       & 1.8 & 0.5 & 0.5 &       \\
        \hline\hline
\end{tabular}
\label{tab:systematic_4680}
\end{center}
\end{table*}
\section{Global least-square fit}
A least-square fit is used to determine BFs of 12 $\lcp$ decays and the 7 yields of $\lcp\lcm$ pairs, $N_{\lcp\lcm}$, at each energy point. This fit accounts for statistical and systematic correlations among the different decay modes at different energy points. A global fitting strategy, similar to that used in
Refs.~\cite{Guan:2013hua,BESIII:2015bjk}, is adopted to extract
all parameters simultaneously.

The direct observables are the 12 ST yields, $N_i^{\rm ST}$, and
12 DT yields, $N_{-,j}^{\rm DT}$, collectively denoted by the
vector \textbf{n}. The corresponding efficiency-corrected yields
are represented by \textbf{c}, which are related to the observed
yields and efficiency through
\begin{equation}
\textbf{c} = \textbf{E}^{-1}\textbf{n},
\label{eq1}
\end{equation}
where \textbf{E} is the $N \times N$ signal efficiency matrix, with
$N$ being the total number of observables, the product of 24
signal yields (12 ST and 12 DT) and 7 energy points.
The expected values, $\tilde{\textbf{c}}$, are functions of the 12
BFs and 7 $N_{\Lambda_c^+\Lambda_c^-}$ parameters,
expressed as
$2N_{\Lambda_c^+\Lambda_c^-}\mathcal{B}_i$ and
$2N_{\Lambda_c^+\Lambda_c^-}\sum_i
\mathcal{B}_i\mathcal{B}_j$.

Based on the least square principle, the $\chi^{2}$ can be constructed as:
\begin{equation}
\chi^{2} \equiv (\textbf{c}-\tilde{\textbf{c}})^{T}\textbf{V}^{-1}_{\textbf{c}}(\textbf{c}- \tilde{\textbf{c}}), 
\label{eq2}
\end{equation}
where $\textbf{V}_{\textbf{c}}
=(\frac{\partial \textbf{c}}{\partial \textbf{n}})^{T}\textbf{V}_{\textbf{n}}\frac{\partial \textbf{c}}{\partial \textbf{n}} + (\frac{\partial \textbf{c}}{\partial \textbf{E}})^{T} \textbf{V}_{\textbf{E}} \frac{\partial \textbf{c}}{\partial \textbf{E}}$,
where $\textbf{V}_{\textbf{n}}$ and $\textbf{V}_{\textbf{E}}$ are the covariance matrix of $\textbf{n}$ and $\textbf{E}$, respectively. 
The variances of \textbf{E} arise from the systematic uncertainties.
For different energy points, the statistical uncertainty is uncorrelated. 
The tracking and PID related systematical uncertainties are correlated. The extracted BFs are listed in Table ~\ref{tab:globalfitresult}.
The correlations amongst the 19 fit parameters are listed in Table ~\ref{tab:correlation}.

The Born cross section of $e^+e^- \to \lcp\lcm$ at each energy point is determined as:
\begin{equation}
\sigma_{\rm Born}=\frac{N_{\lcp\lcm}}{\mathcal{L}_{\rm int}f_{\rm VP}f_{\rm ISR}} ,
\end{equation}
where $N_{\lcp\lcm}$ is listed in Table~\ref{tab:total numbers}, the factor $f_{\rm VP}$ is the vacuum polarization correction factor, and the ISR correction factor $f_{\rm ISR}$ is derived from the QED calculations~\cite{Jadach:2000ir} using {\sc kkmc} in an iterative procedure, in which the input Born cross-section line shape is updated in each iteration. The specific values of these correction factors are taken from Ref.~\cite{BESIII:2023rwv}. 
In the uncertainty estimation, the value of $N_{\lcp\lcm}$ has been determined in the global fit,
and the $\mathcal{L}_{\rm int}$ is taken from Ref.~\cite{BESIII:2022ulv}.
The uncertainties from  the factors $f_{\rm VP}$ and $f_{\rm ISR}$ are taken from Ref.~\cite{BESIII:2023rwv}.
The resulting Born cross sections are summarized in Table~\ref{tab:total numbers}.
\begin{table}[htbp]
  \caption{The globally determined BFs~(\%) obtained at the seven energy points, where the first uncertainties are statistical and the second systematic. Values from the PDG are provided for comparison~\cite{ParticleDataGroup:2024cfk}.}
  \label{tab:globalfitresult}
  \centering
  \begin{tabular}{l | r@{\,$\pm$\,}c@{\,$\pm$\,}l | r@{\,$\pm$\,}l}
    \hline \hline
    Signal mode & \multicolumn{3}{c|}{Global fit} & \multicolumn{2}{c}{PDG2025} \\ 
    \hline
    \textbf{$\modea$}   & 1.696 & 0.033 & 0.048 & 1.61 & 0.07 \\
    \textbf{$\modeb$}   & 6.61  & 0.11  & 0.12  & 6.35 & 0.25 \\
    \textbf{$\modec$}   & 2.186 & 0.057 & 0.048 & 1.99 & 0.12 \\
    \textbf{$\moded$}   & 1.883 & 0.040 & 0.070 & 1.62 & 0.11 \\
    \textbf{$\modee$}   & 4.89  & 0.10  & 0.11  & 4.52 & 0.28 \\
    \textbf{$\modeaa$}  & 1.322 & 0.027 & 0.026 & 1.31 & 0.05 \\
    \textbf{$\modebb$}  & 6.67  & 0.13  & 0.10  & 7.10 & 0.34 \\
    \textbf{$\modedd$}  & 4.092 & 0.091 & 0.096 & 3.67 & 0.26 \\
    \textbf{$\modeaaa$} & 1.449 & 0.034 & 0.025 & 1.29 & 0.05 \\
    \textbf{$\modeccc$} & 1.374 & 0.044 & 0.032 & 1.26 & 0.10 \\
    \textbf{$\modeddd$} & 4.58  & 0.10  & 0.10  & 4.54 & 0.20 \\
    \textbf{$\modef$}   & 0.499 & 0.016 & 0.010 & 0.47 & 0.02 \\
    \hline \hline
  \end{tabular}
\end{table}
The effective form factor is calculated with the Born cross section as 
\begin{equation}
|G_{\rm eff}|=\sqrt{\frac{\sigma}{\frac{\sigma_0}{3} (1+\frac{\kappa}{2}) }} ,
\end{equation}
where $\sigma_0 = 4\pi\alpha^2\beta C/s$, $C$ is the Coulomb factor~\cite{BESIII:2023rwv}, $\beta=\sqrt{1-\kappa},\kappa=4m^2c^4/s$, and $m$ is the known mass of the $\Lambda_c^+$ baryon. 
The obtained results are shown in Table~\ref{tab:total numbers}.
\begin{table}[!htbp]
  \caption{The integrated luminosities and vacuum polarization correction factors~$f_{\rm VP}$, and the ISR correction factors $f_{\rm ISR}$ at different energy points. The first uncertainties are statistical, and the second systematic.}
 \begin{center}
 \begin{tabular}{l|c|c|c}
     \hline \hline
$\sqrt s$($\mev$) & $\mathcal{L}_{\rm int}(\rm pb^{-1})$ & $f_{\rm ISR}$ & $f_{\rm VP}$ \\ \hline
4599.53  & $586.9\pm0.1\pm3.9$  & 0.7428 & 1.0547 \\
4611.86  & $103.7\pm0.1\pm0.6$  & 0.7657 & 1.0545 \\
4628.00  & $521.5\pm0.1\pm2.8$  & 0.7868 & 1.0544 \\
4640.91  & $551.6\pm 0.1\pm2.9$ & 0.8001 & 1.0544 \\
4661.24  & $529.6\pm0.1\pm2.8$  & 0.8193 & 1.0544 \\
4681.92  & $1667.4\pm0.2\pm8.8$ & 0.8422 & 1.0545 \\
4698.82  & $535.5\pm0.1\pm2.8$  & 0.8668 & 1.0545 \\
\hline\hline
    \end{tabular}
    \label{tab:total numbers_vp}
   \end{center}
   \end{table}
   \begin{table}[!htbp]
  \caption{The measured average Born cross sections and the effective form factors at each energy point, where the first and second uncertainties are statistical and systematic, respectively. The uncertainties of $N_{\lcp\lcm}$ are determined from a global fit performed simultaneously across all energy points; the values at different energies are correlated and not statistically independent.}
 \begin{center}
 \begin{tabular}{l|c|c|c}
     \hline \hline
$\sqrt s$($\mev$) &$N_{\lcp\lcm}$($\times 10^3$) &$\sigma$(pb)&$|G_{\rm eff}|(\times10^{-2})$\\ \hline
4599.53  & $98.0\pm1.8\pm0.5$  & $ 213.1 \pm 3.9  \pm 1.9$ & $53.8 \pm 0.5 \pm 0.2$ \\
4611.86  & $17.7\pm0.5\pm0.1$  & $ 211.5 \pm 6.0  \pm 2.0$ & $49.5 \pm 0.7 \pm 0.2$\\
4628.00  & $89.8\pm1.6\pm0.4$  & $ 207.6 \pm 3.7  \pm 1.8$ & $45.6 \pm 0.4 \pm 0.2$ \\
4640.91  & $96.1\pm1.9\pm0.5$  & $ 206.5 \pm 4.1  \pm 1.9$ & $43.5 \pm 0.4 \pm 0.2$\\
4661.24  & $94.3\pm1.8\pm0.5$  & $ 206.2 \pm 3.9  \pm 2.0$ & $41.2 \pm 0.4 \pm 0.2$\\
4681.92  & $284.3\pm4.9\pm1.3$ & $ 192.0 \pm 3.3  \pm 1.8$ & $38.1 \pm 0.4 \pm 0.2$ \\
4698.82  & $84.2\pm1.7\pm0.4$  & $ 172.0 \pm 3.5  \pm 2.2$ & $35.0 \pm 0.4 \pm 0.2$\\
\hline\hline
    \end{tabular}
    \label{tab:total numbers}
   \end{center}
   \end{table}
\section{Summary}

Based on $e^+e^-$ collision data with an integrated luminosity of 4.5 fb$^{-1}$ collected with the BESIII detector at 7 energy points between $\sqrt s=$ 4599.53 and 4698.82 MeV,
we measure the absolute BFs of twelve $\lcp$ hadronic decay modes.
The BFs measured in this work are improved by a factor of 2 to 3 compared to our previous measurements~\cite{BESIII:2015bjk}, and therefore supersede those reported in Ref.~\cite{BESIII:2015bjk}. 
The measured BF of $\lcp\to\modeb$ agrees well with the recent result reported in Ref.~\cite{BESIII:2023rwv}.
The obtained results help to improve and extend the current understanding of the dynamics of the charmed baryon hadronic decays and will benefit the Cabibbo–Kobayashi–Maskawa matrix element measurements. 
The cross sections and $\lcp$ effective form factors at different energy points are determined. The results are consistent with and more precise than our previous measurements~\cite{BESIII:2023rwv}, benefiting a comprehensive treatment of the correlations among systematic uncertainties.
However, the two measurements are not fully statistically independent, as they share data from the common decay mode $\lcp\to\modeb$.
Further studies of decay dynamics, including the amplitude analyses of three- and four-body decays and the decay asymmetry measurements will offer essential inputs for the bottom baryon decay asymmetry measurements to be performed in the future. 





\appendix
\section{ST fits}
\begin{figure*}[!hbt]
    \centering
    \includegraphics[width=1.0\textwidth]{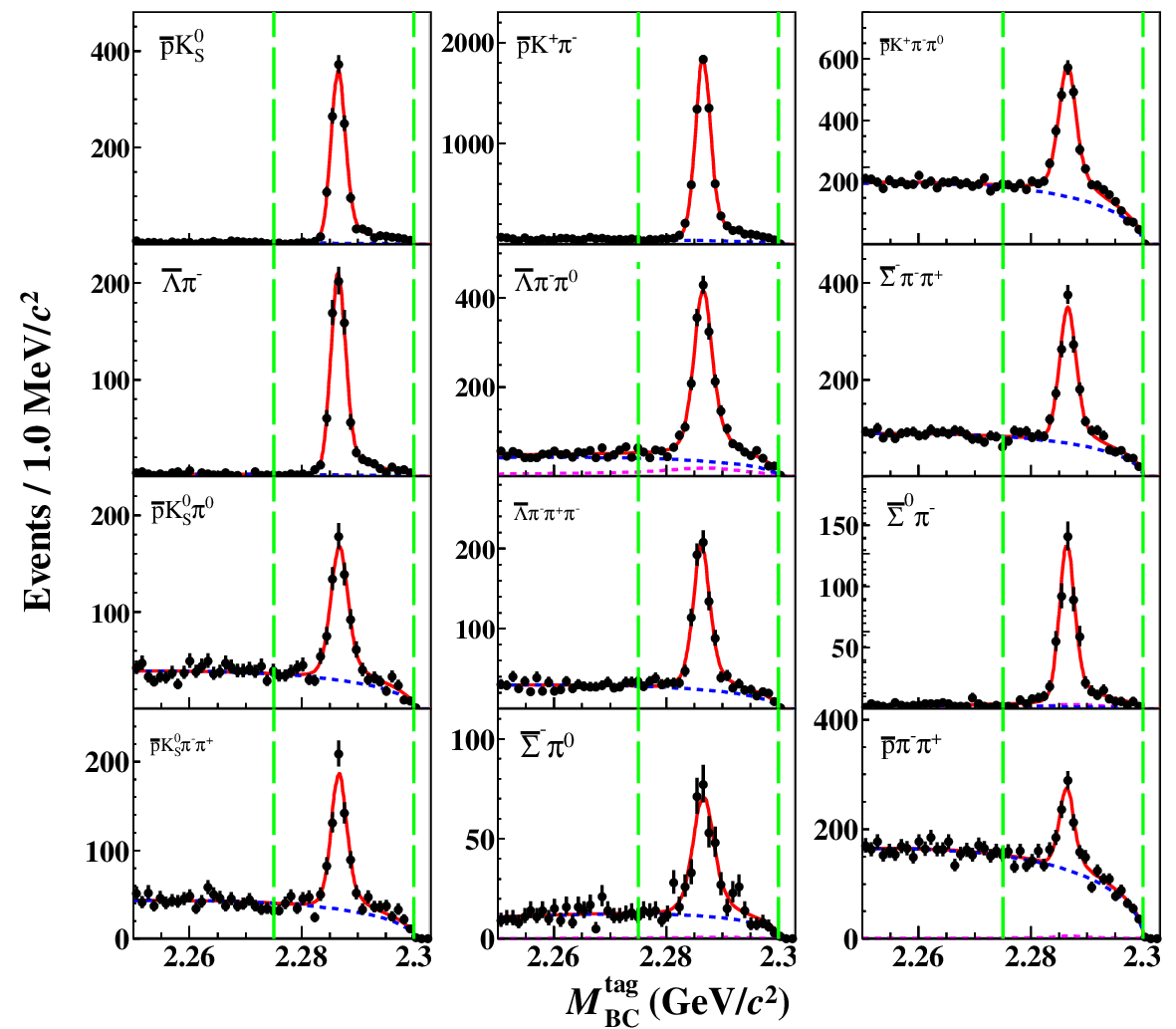}
   \caption{The fits to the $M^{\rm tag}_{\rm BC}$ distributions of the ST $\bar \Lambda^-_c$ candidates for different tag modes at $\sqrt{s}=4599.53~\mev$. 
     The points with error bars are data. The red line is the total fitted distribution, with the dashed-blue lines showing the ARGUS function and the dashed-magenta lines showing the residual non-flat background shape. The dashed-green lines denote the signal regions.
     }
   \label{fig:STfit_4600}
  \end{figure*}

\begin{figure*}[!hbt]
    \centering
    \includegraphics[width=1.0\textwidth]{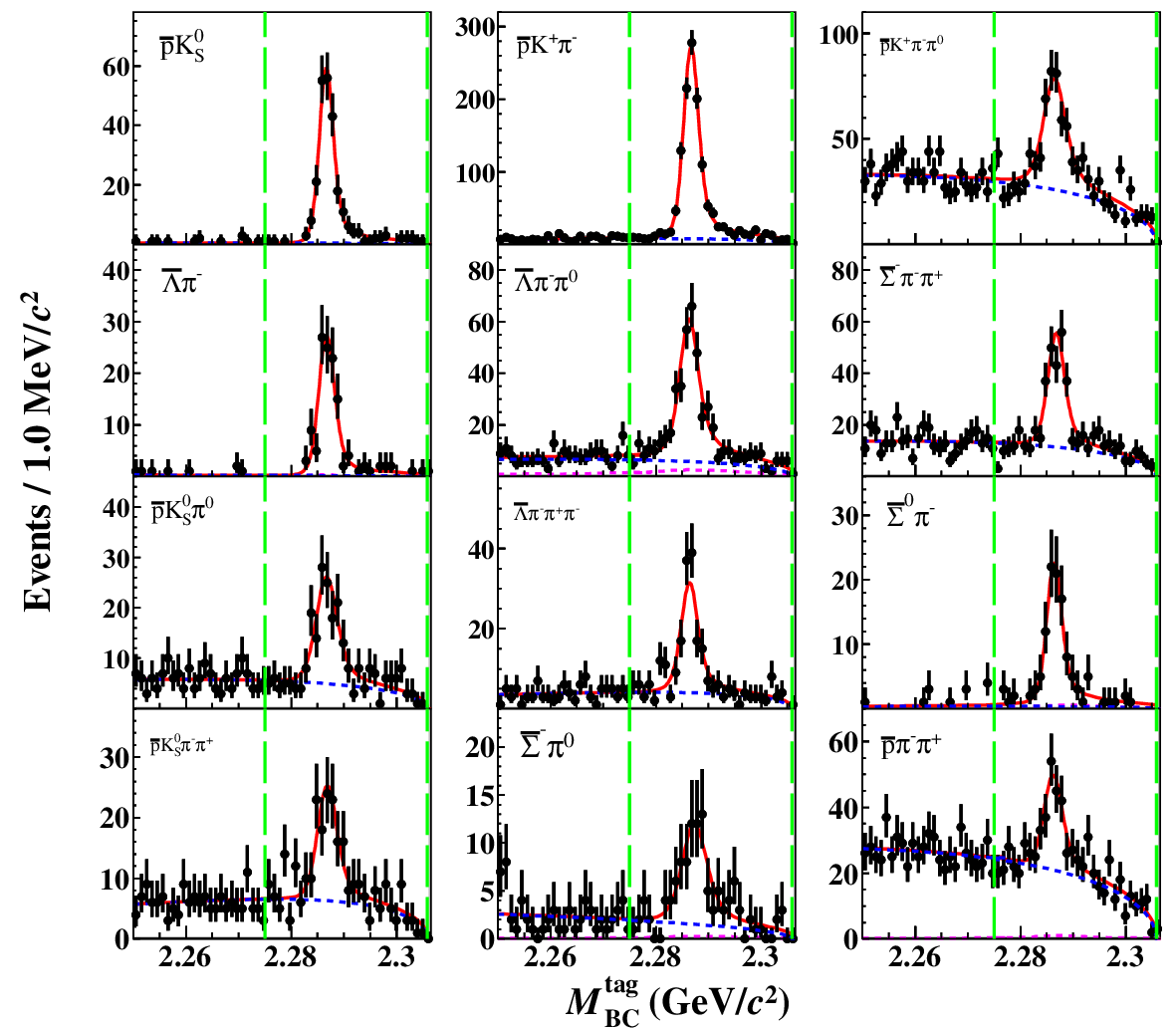}
   \caption{The fits to the $M^{\rm tag}_{\rm BC}$ distributions of the ST $\bar \Lambda^-_c$ candidates for different tag modes at $\sqrt{s}=4611.86~\mev$. 
     The points with error bars are data. The red line is the total fitted distribution, with the dashed-blue lines showing the ARGUS function and the dashed-magenta lines showing the residual non-flat background shape. The dashed-green lines denote the signal regions.
  }
   \label{fig:STfit_4612}
  \end{figure*}

\begin{figure*}[!hbt]
    \centering
    \includegraphics[width=1.0\textwidth]{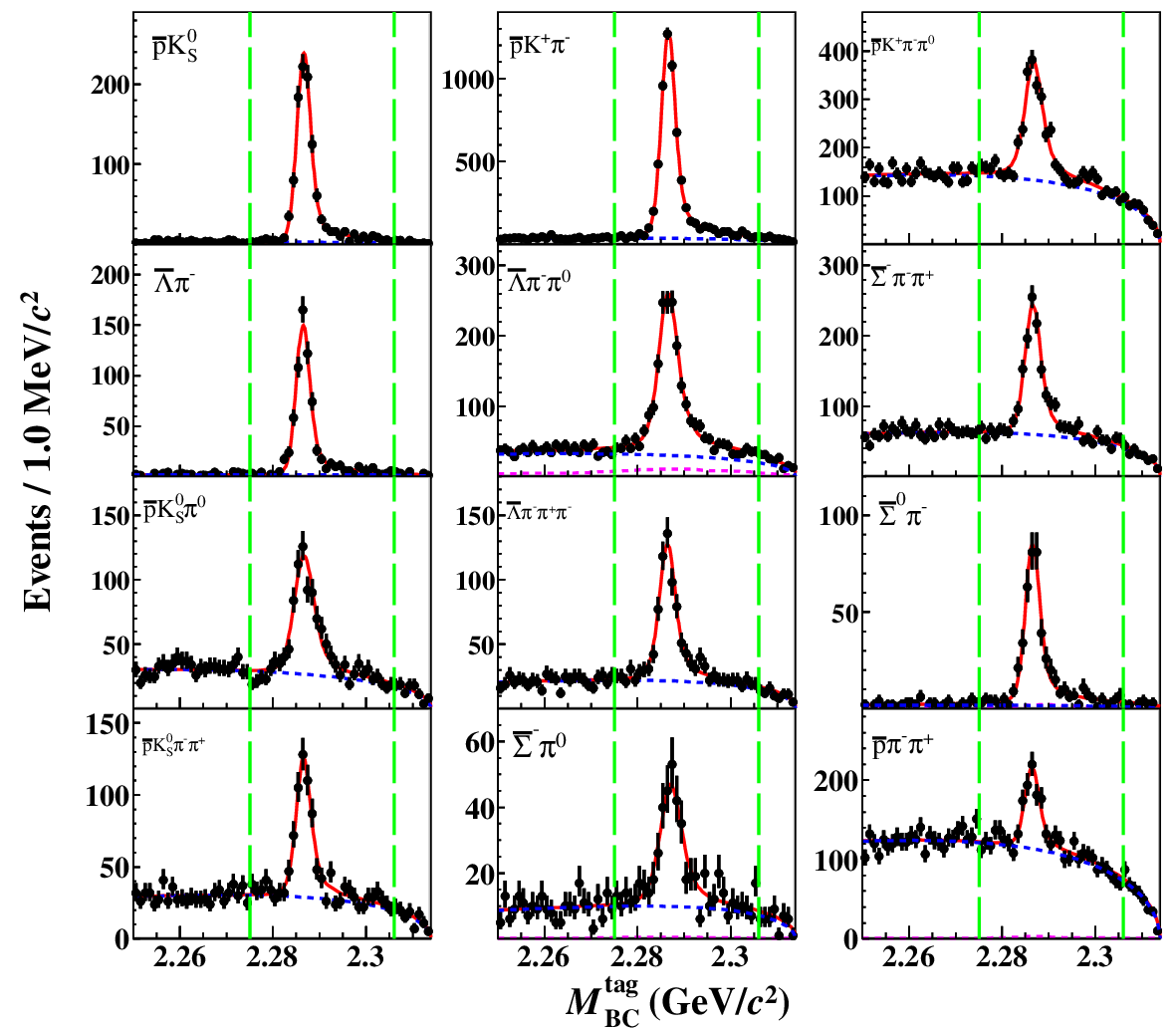}
   \caption{The fits to the $M^{\rm tag}_{\rm BC}$ distributions of the ST $\bar \Lambda^-_c$ candidates for different tag modes at $\sqrt{s}=4628.00~\mev$. 
     The points with error bars are data. The red line is the total fitted distribution, with the dashed-blue lines showing the ARGUS function and the dashed-magenta lines showing the residual non-flat background shape. The dashed-green lines denote the signal regions.}
   \label{fig:STfit_4628}
  \end{figure*}

\begin{figure*}[!hbt]
    \centering
    \includegraphics[width=1.0\textwidth]{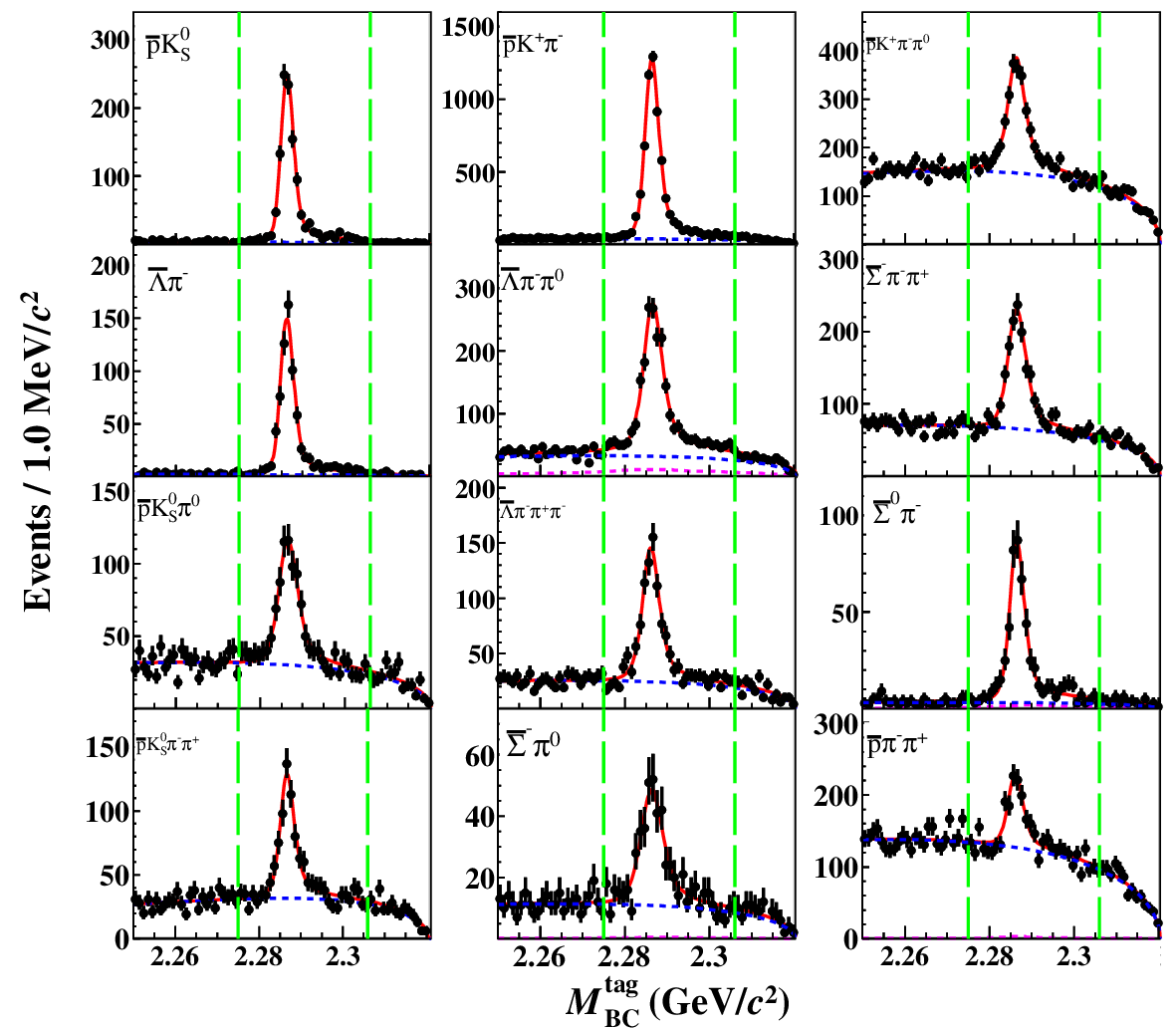}
   \caption{The fits to the $M^{\rm tag}_{\rm BC}$ distributions of the ST $\bar \Lambda^-_c$ candidates for different tag modes at $\sqrt{s}=4640.91~\mev$. 
     The points with error bars are data. The red line is the total fitted distribution, with the dashed-blue lines showing the ARGUS function and the dashed-magenta lines showing the residual non-flat background shape. The dashed-green lines denote the signal regions.
   }
   \label{fig:STfit_4640}
  \end{figure*}

\begin{figure*}[!hbt]
    \centering
    \includegraphics[width=1.0\textwidth]{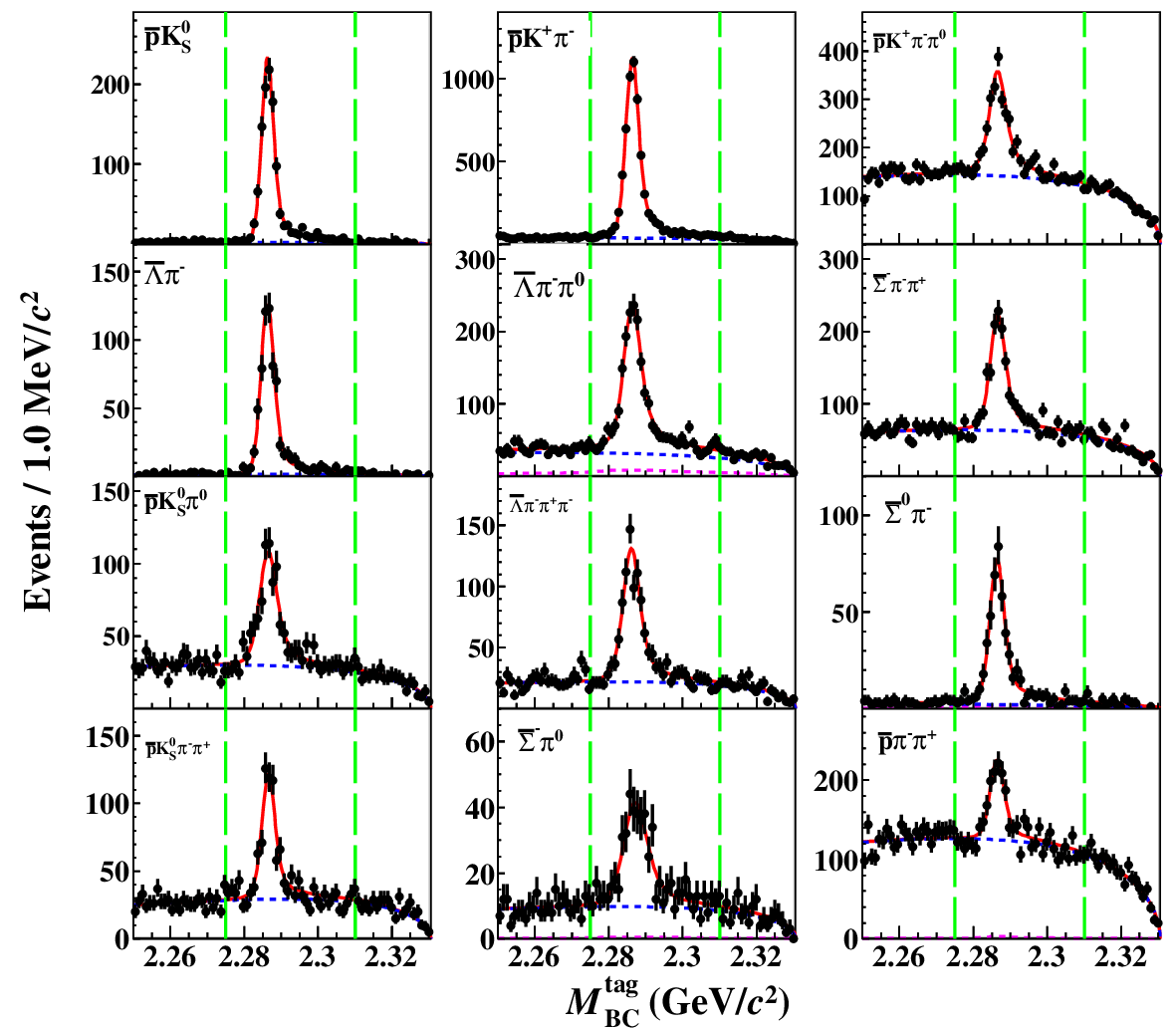}
   \caption{The fits to the $M^{\rm tag}_{\rm BC}$ distributions of the ST $\bar \Lambda^-_c$ candidates for different tag modes at $\sqrt{s}=4661.24~\mev$. 
     The points with error bars are data. The red line is the total fitted distribution, with the dashed-blue lines showing the ARGUS function and the dashed-magenta lines showing the residual non-flat background shape. The dashed-green lines denote the signal regions.
    }
   \label{fig:STfit_4660}
  \end{figure*}

\begin{figure*}[!hbt]
    \centering
    \includegraphics[width=1.0\textwidth]{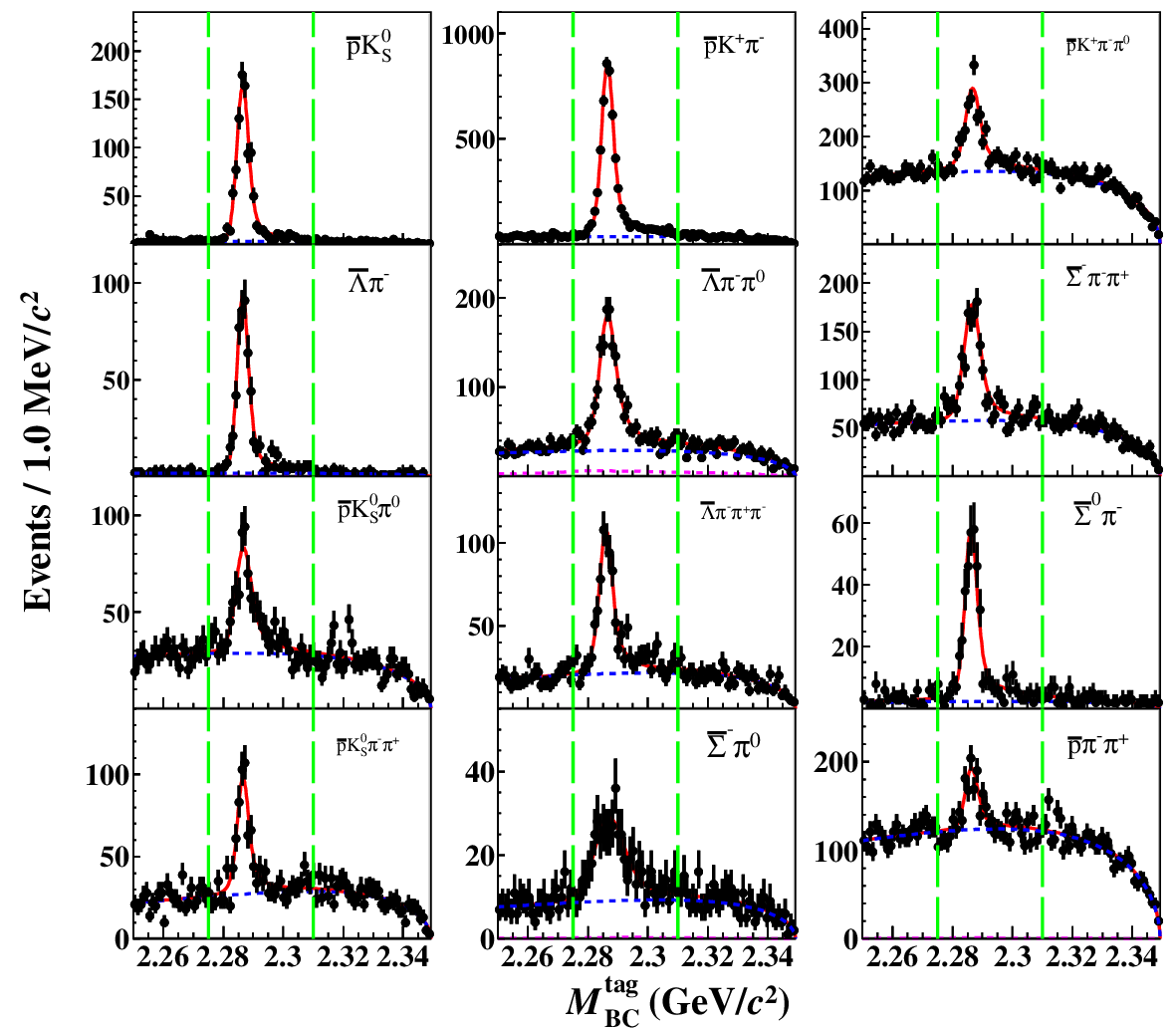}
   \caption{The fits to the $M^{\rm tag}_{\rm BC}$ distributions of the ST $\bar \Lambda^-_c$ candidates for different tag modes at $\sqrt{s}=4698.82~\mev$. 
     The points with error bars are data. The red line is the total fitted distribution, with the dashed-blue lines showing the ARGUS function and the dashed-magenta lines showing the residual non-flat background shape. The dashed-green lines denote the signal regions.
    }
   \label{fig:STfit_4700}
  \end{figure*}

Figures~\ref{fig:STfit_4600},~\ref{fig:STfit_4612},~\ref{fig:STfit_4628},~\ref{fig:STfit_4640},~\ref{fig:STfit_4660} and ~\ref{fig:STfit_4700} show the results of the fits to the $M^{\rm tag}_{\rm BC}$ distributions of the ST $\bar \Lambda^-_c$ candidates in data.

\clearpage
\section{DT fits}
\begin{figure}[!htbp]
  \begin{center}
  
  \includegraphics[width=0.24\textwidth]{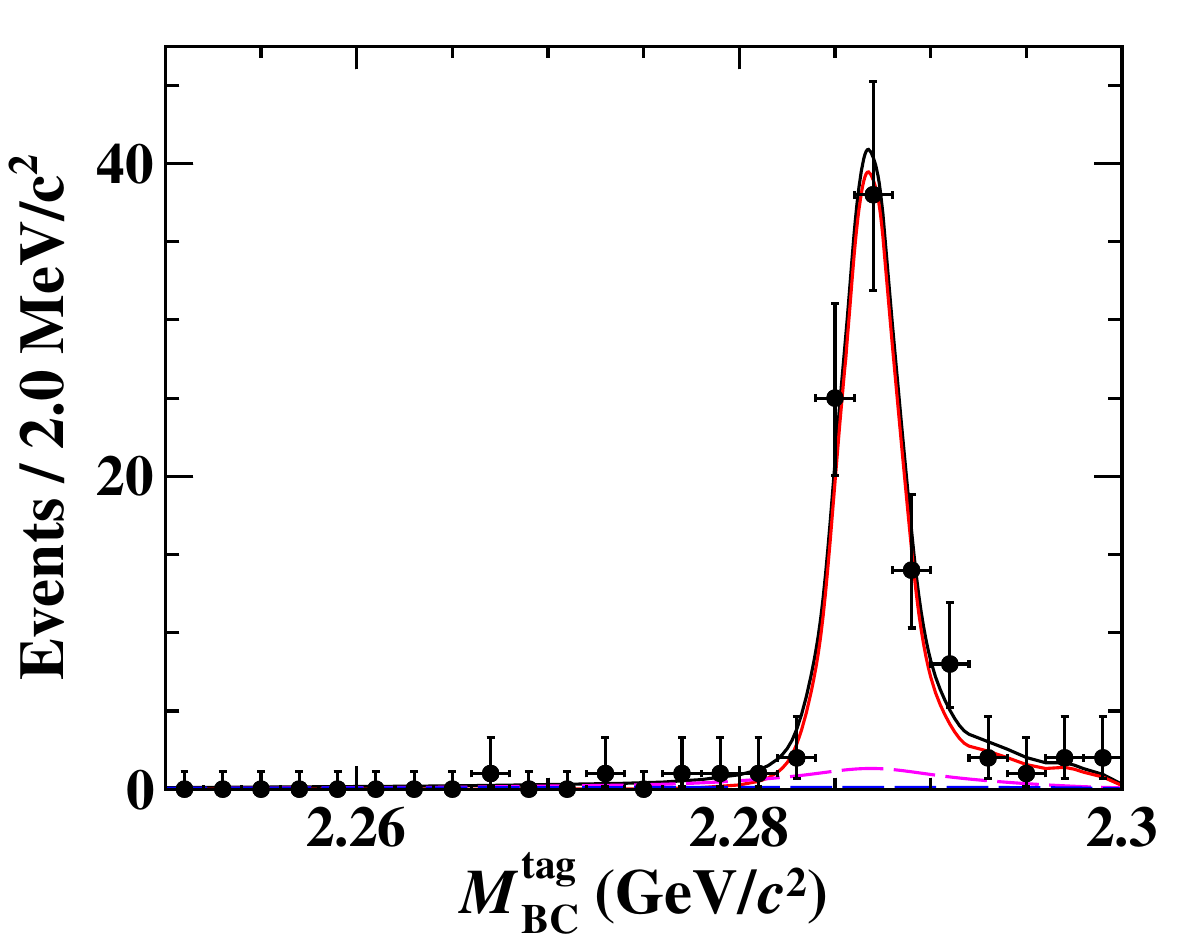}
  \includegraphics[width=0.24\textwidth]{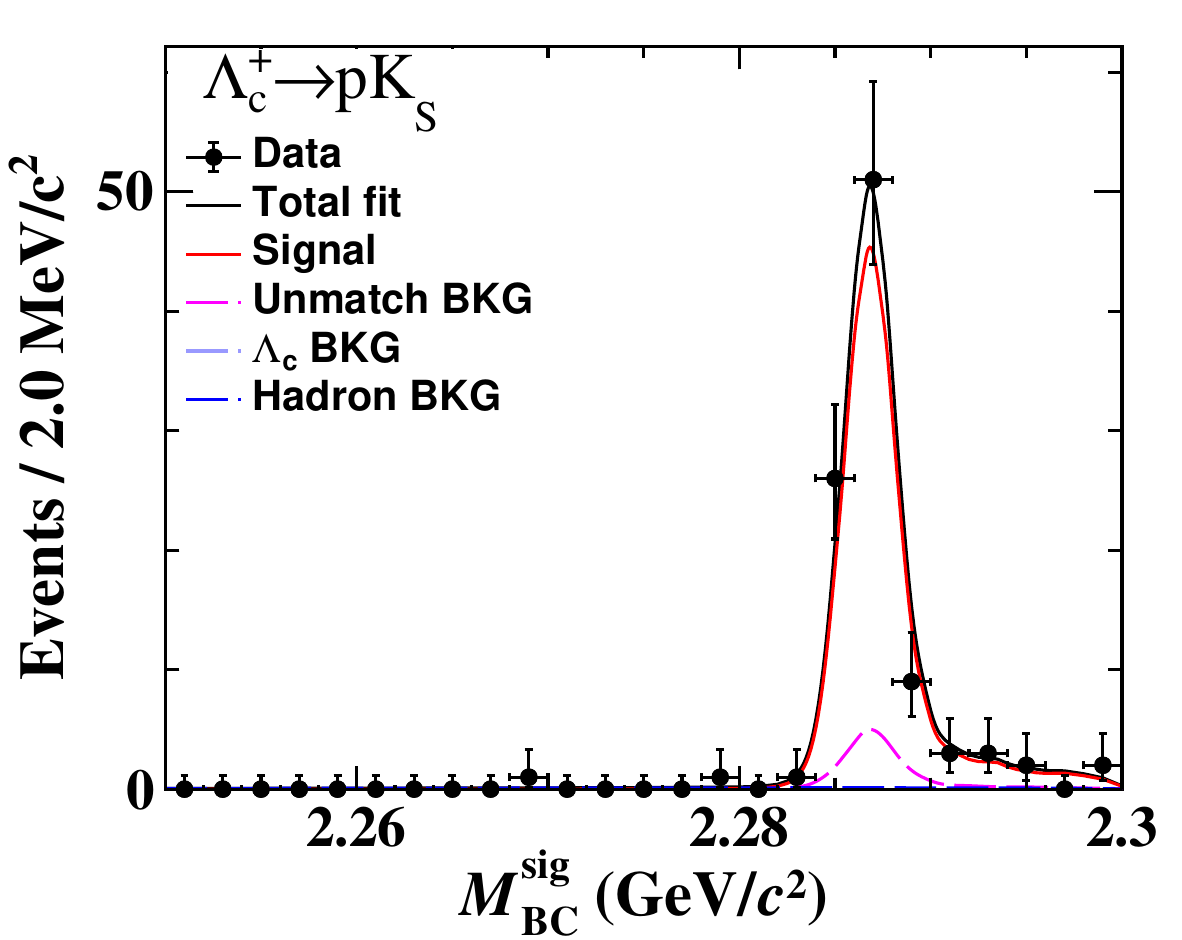}
  \includegraphics[width=0.24\textwidth]{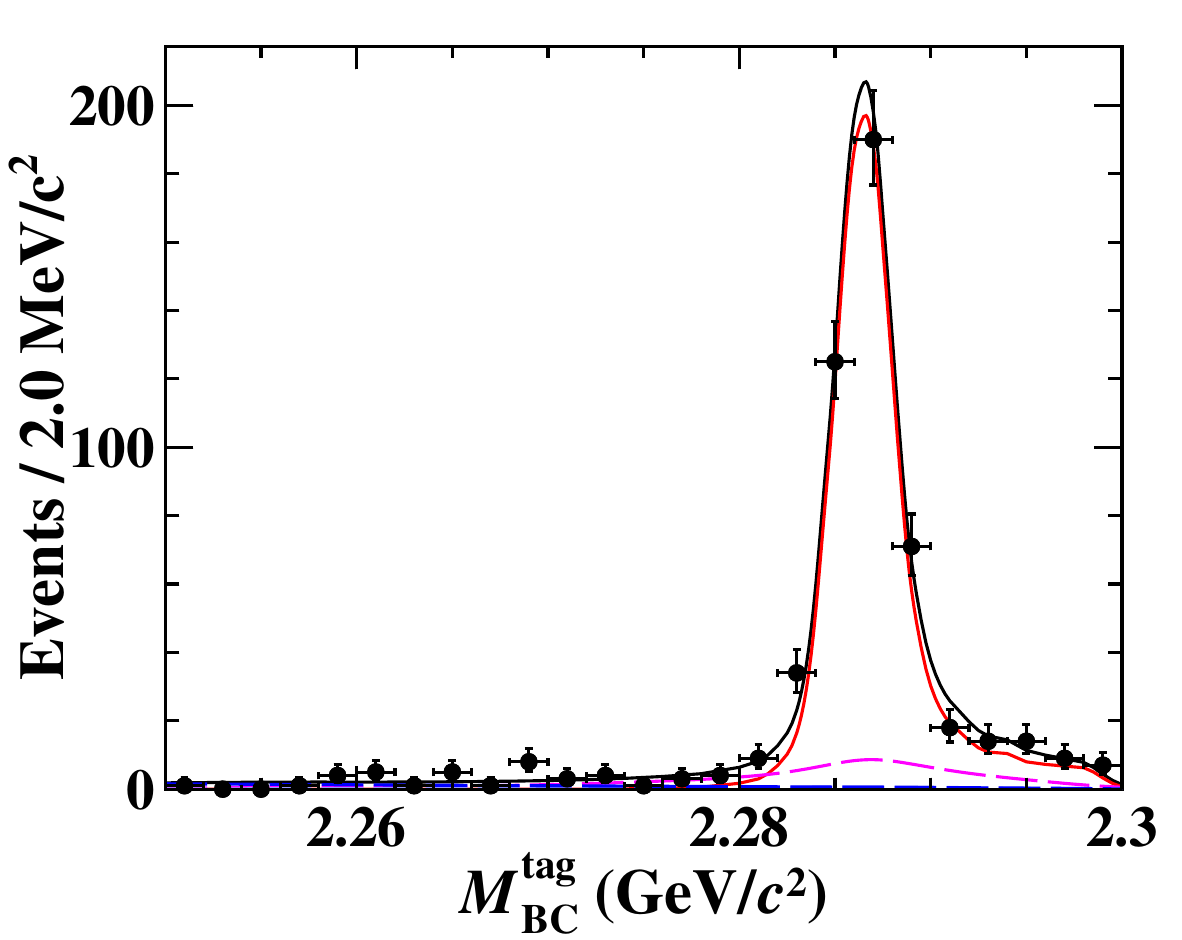}
  \includegraphics[width=0.24\textwidth]{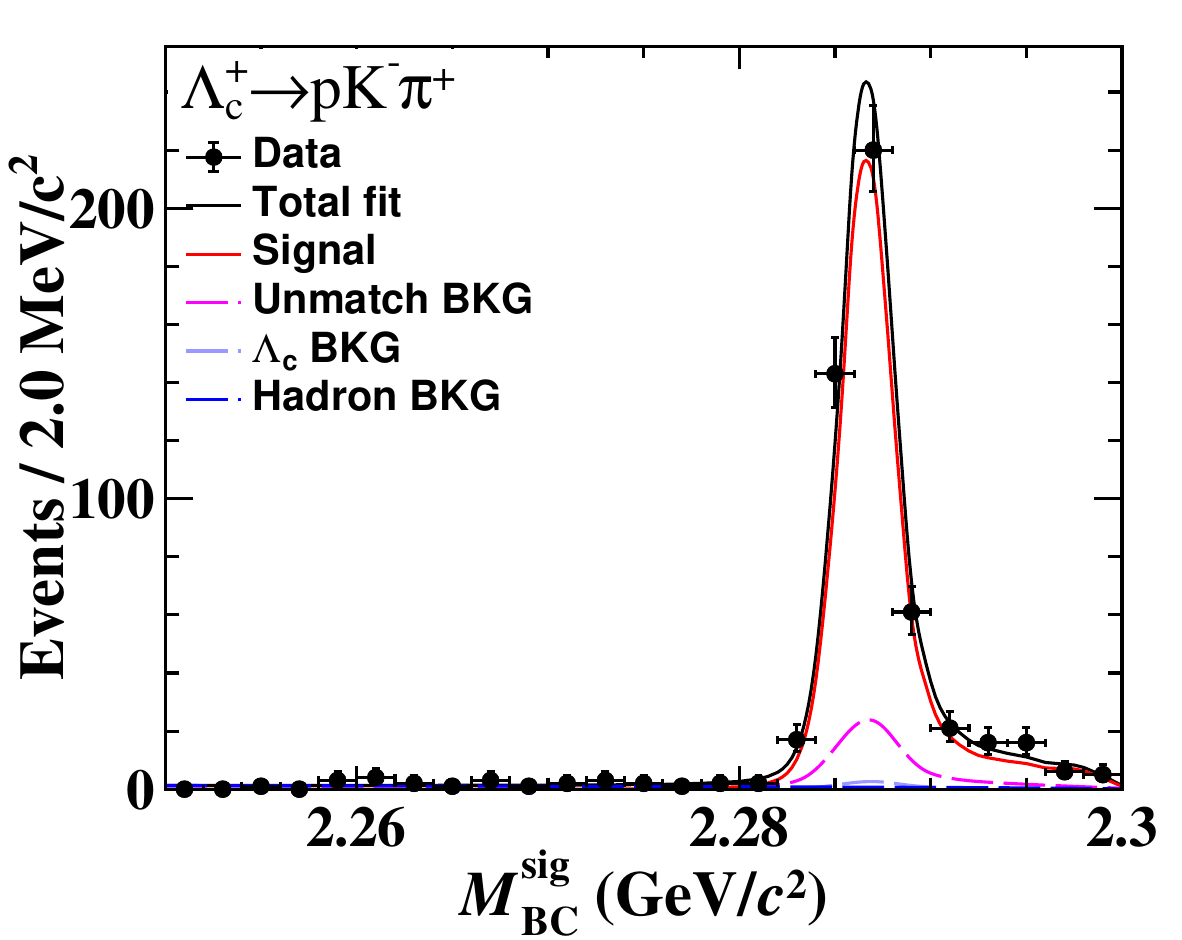}
  \includegraphics[width=0.24\textwidth]{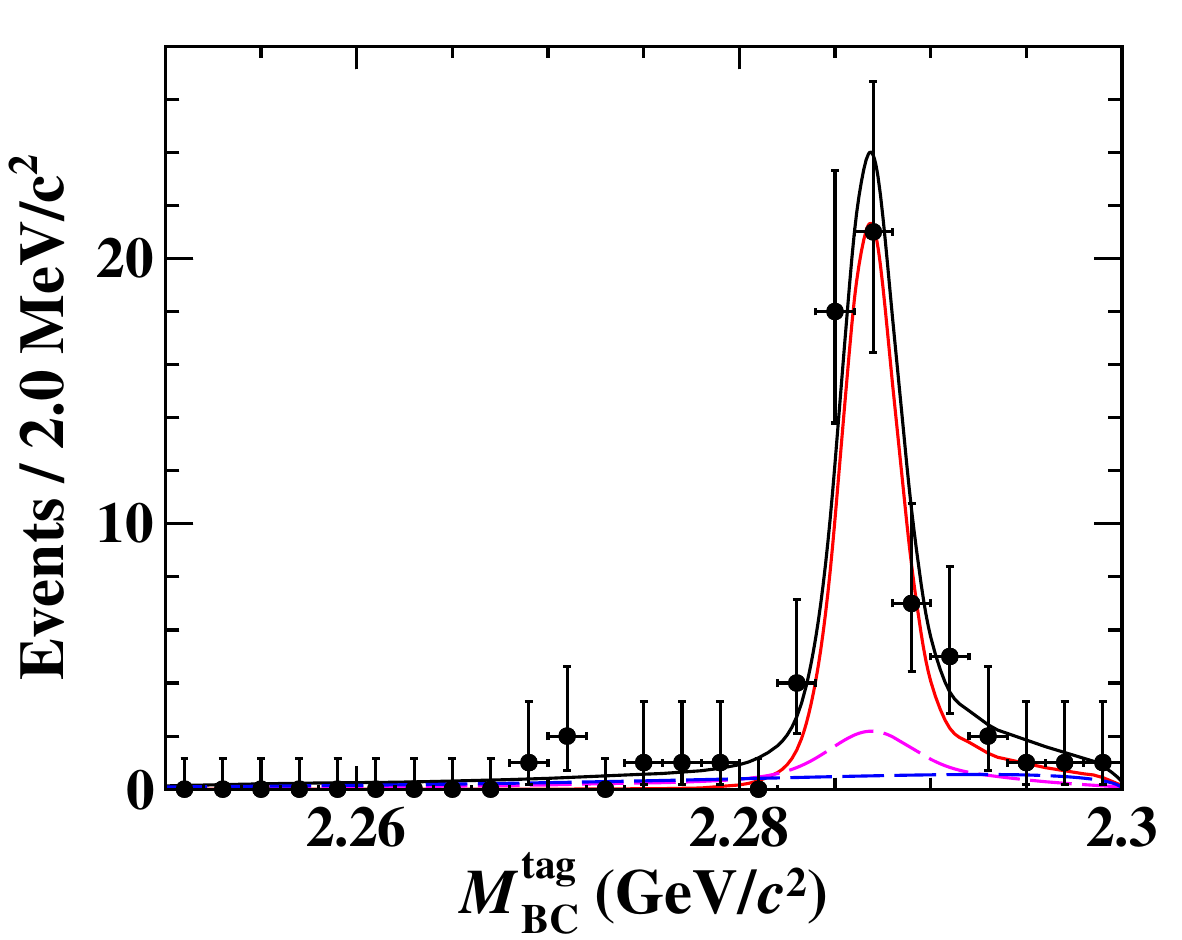}
  \includegraphics[width=0.24\textwidth]{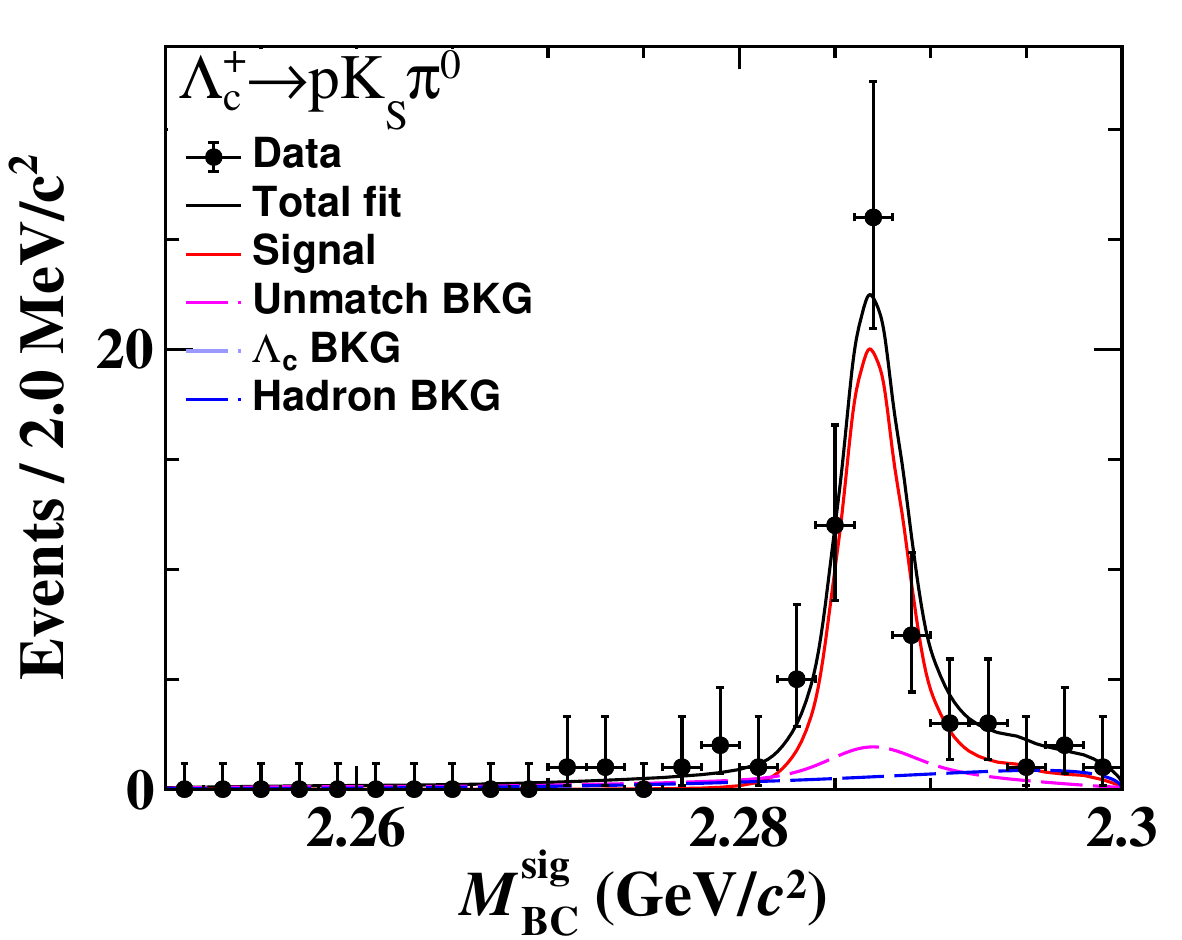}
  \includegraphics[width=0.24\textwidth]{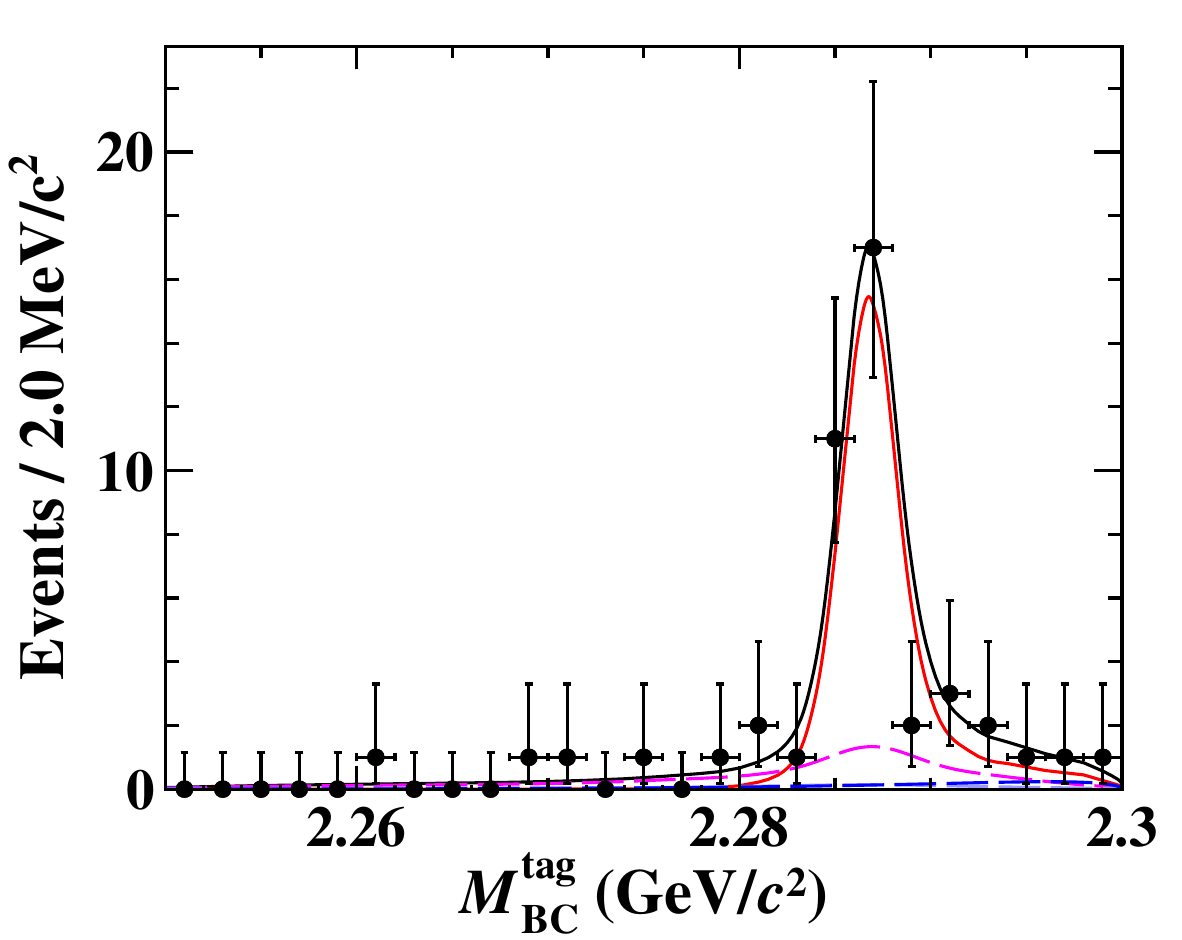}
  \includegraphics[width=0.24\textwidth]{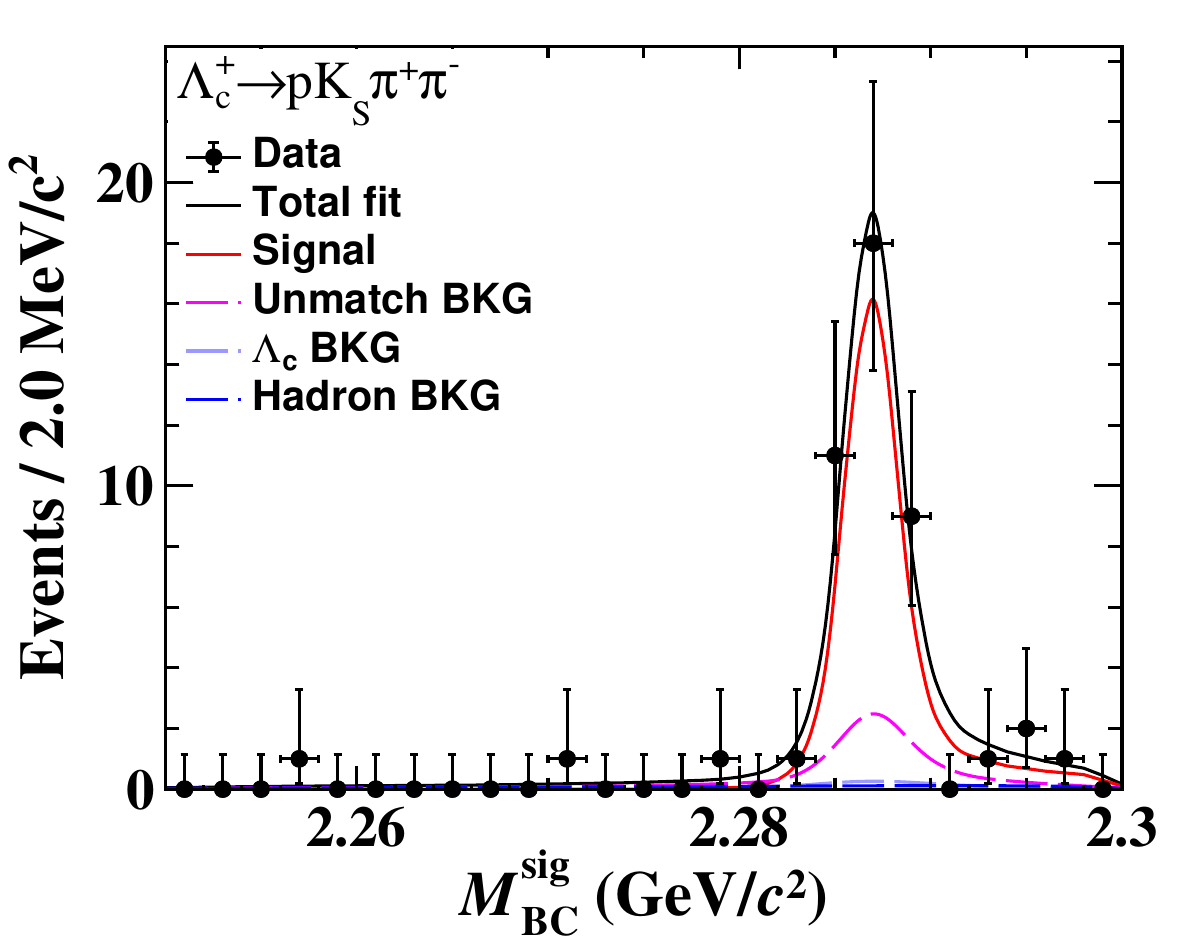}
  \includegraphics[width=0.24\textwidth]{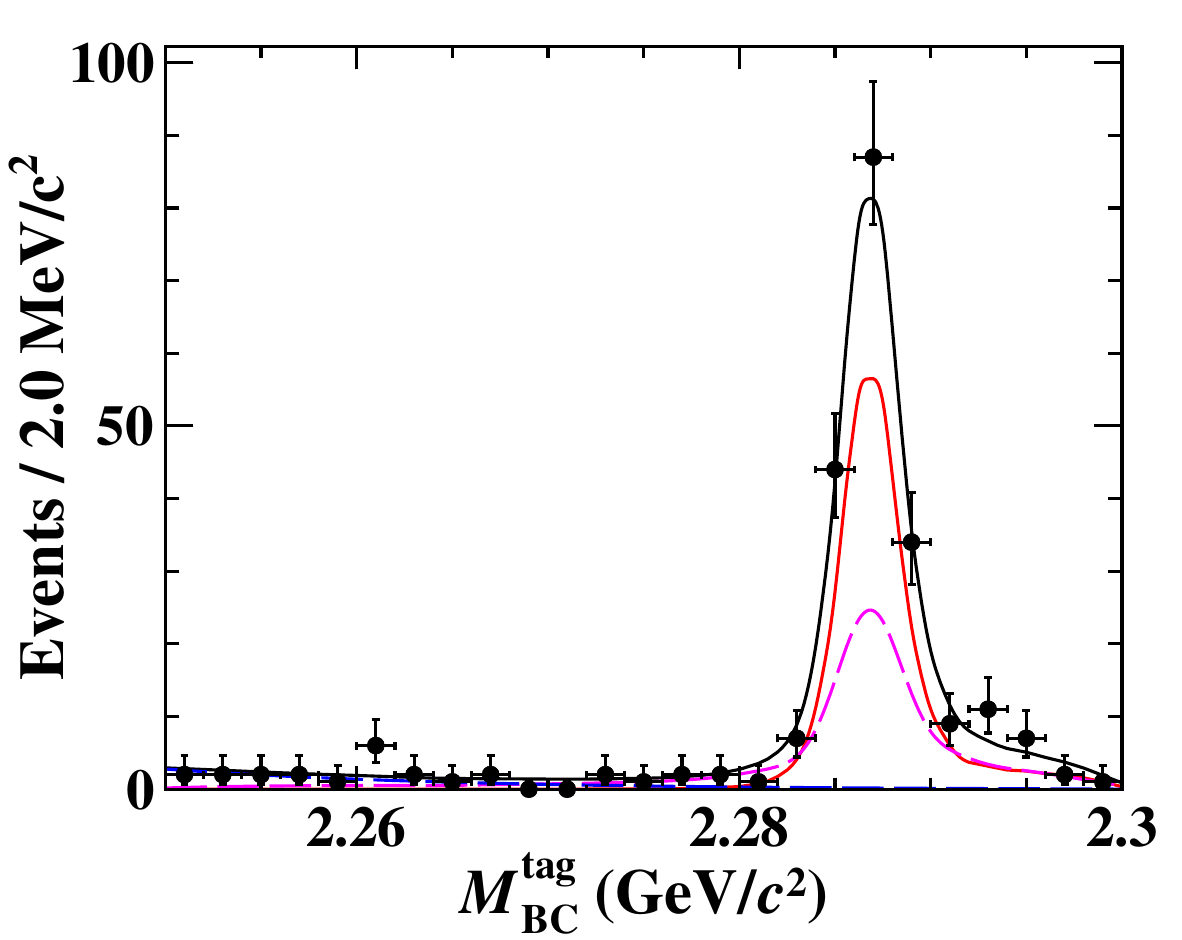}
  \includegraphics[width=0.24\textwidth]{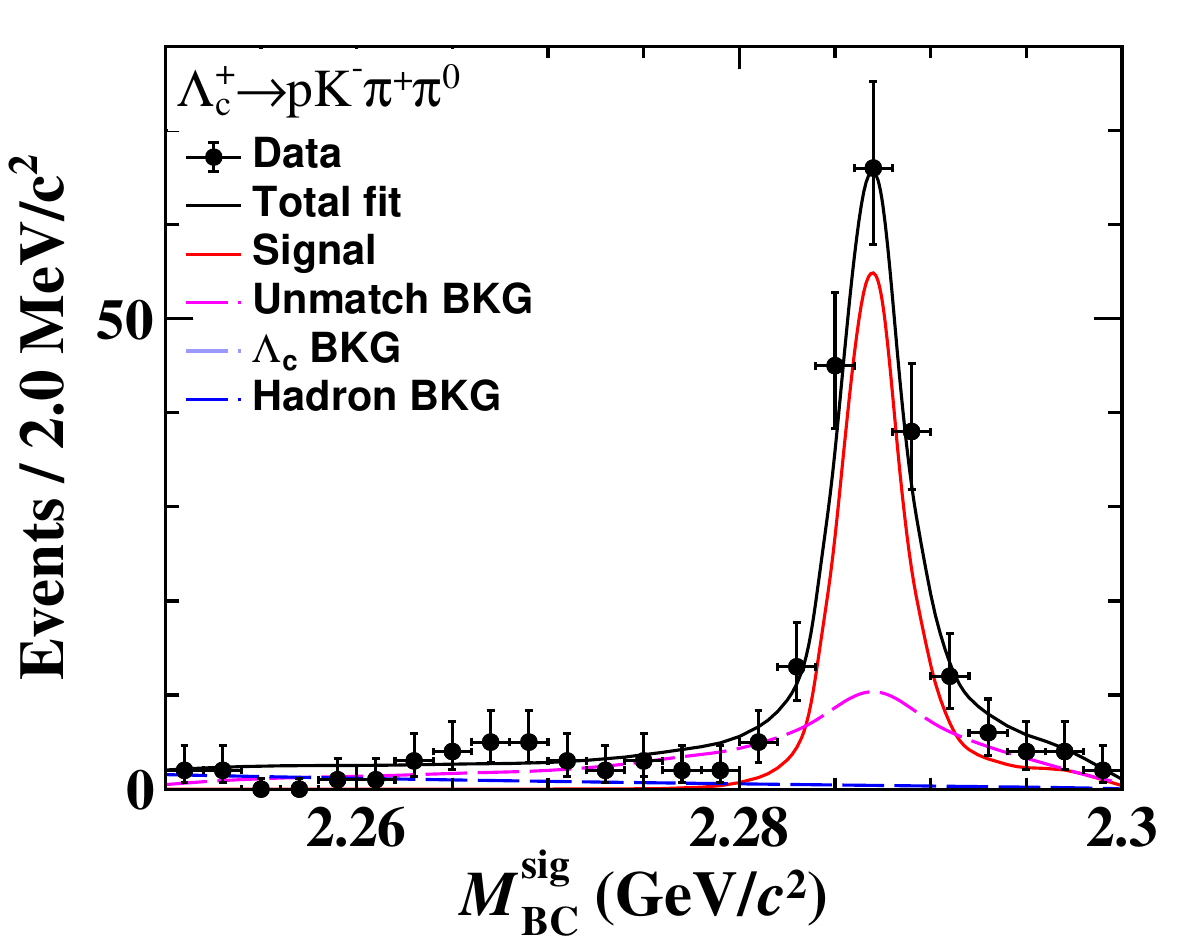}
  \includegraphics[width=0.24\textwidth]{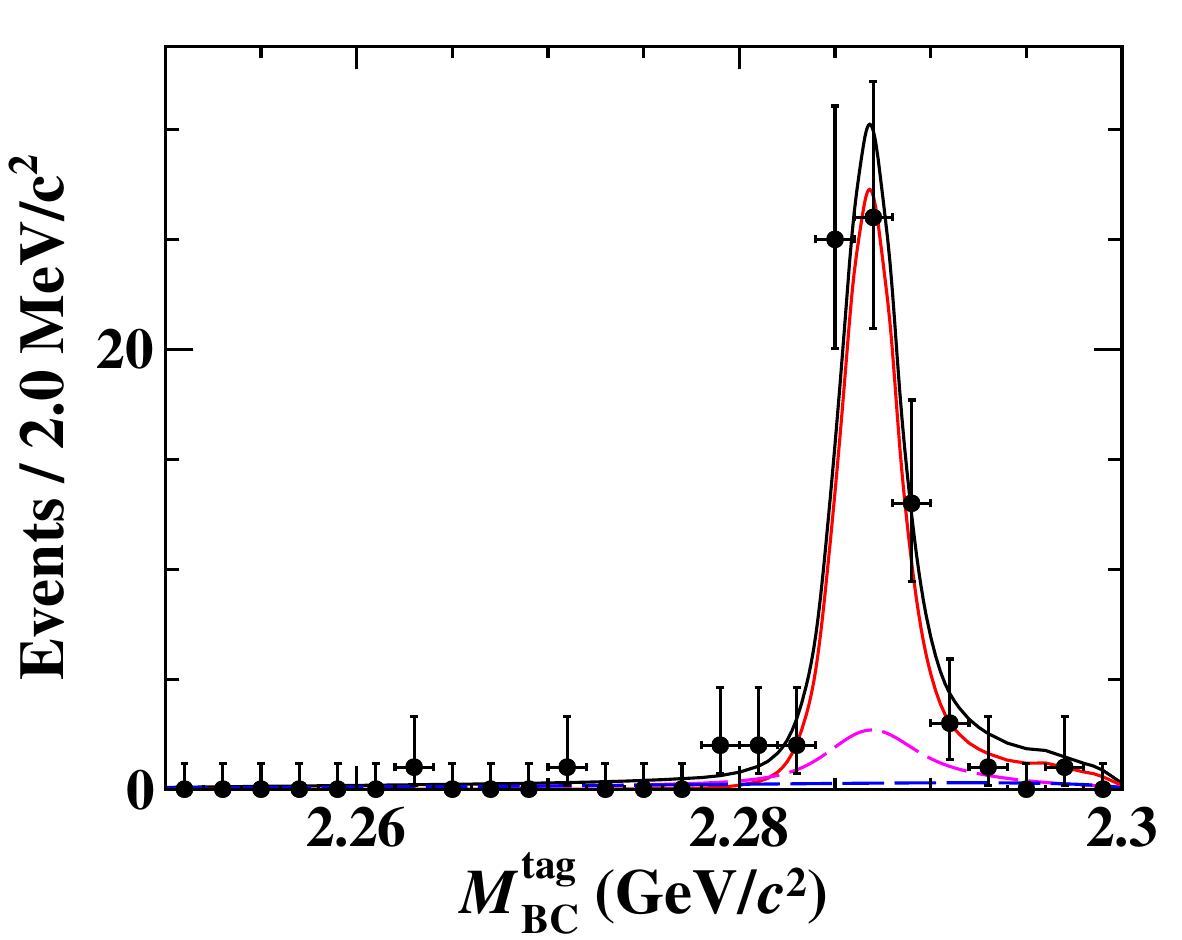}
  \includegraphics[width=0.24\textwidth]{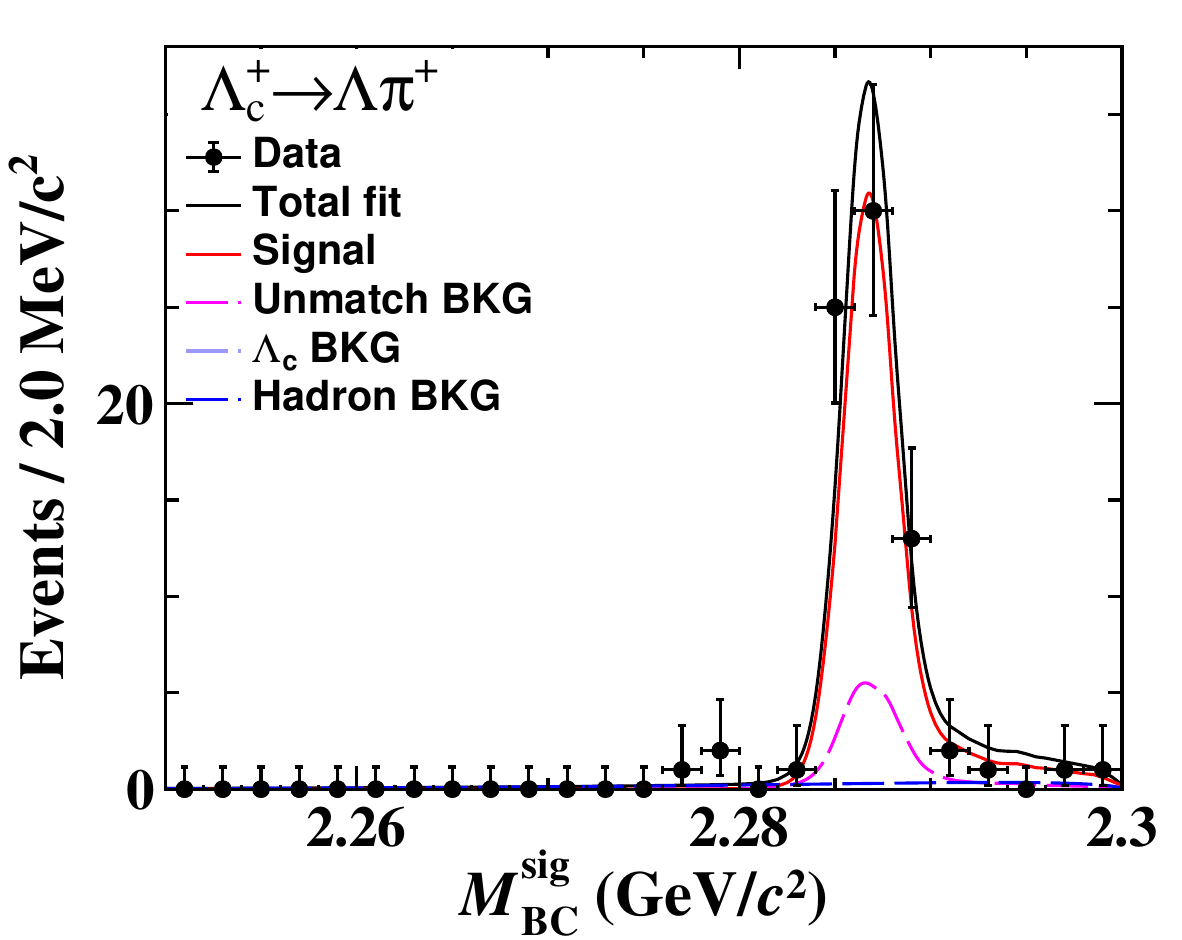}
  \includegraphics[width=0.24\textwidth]{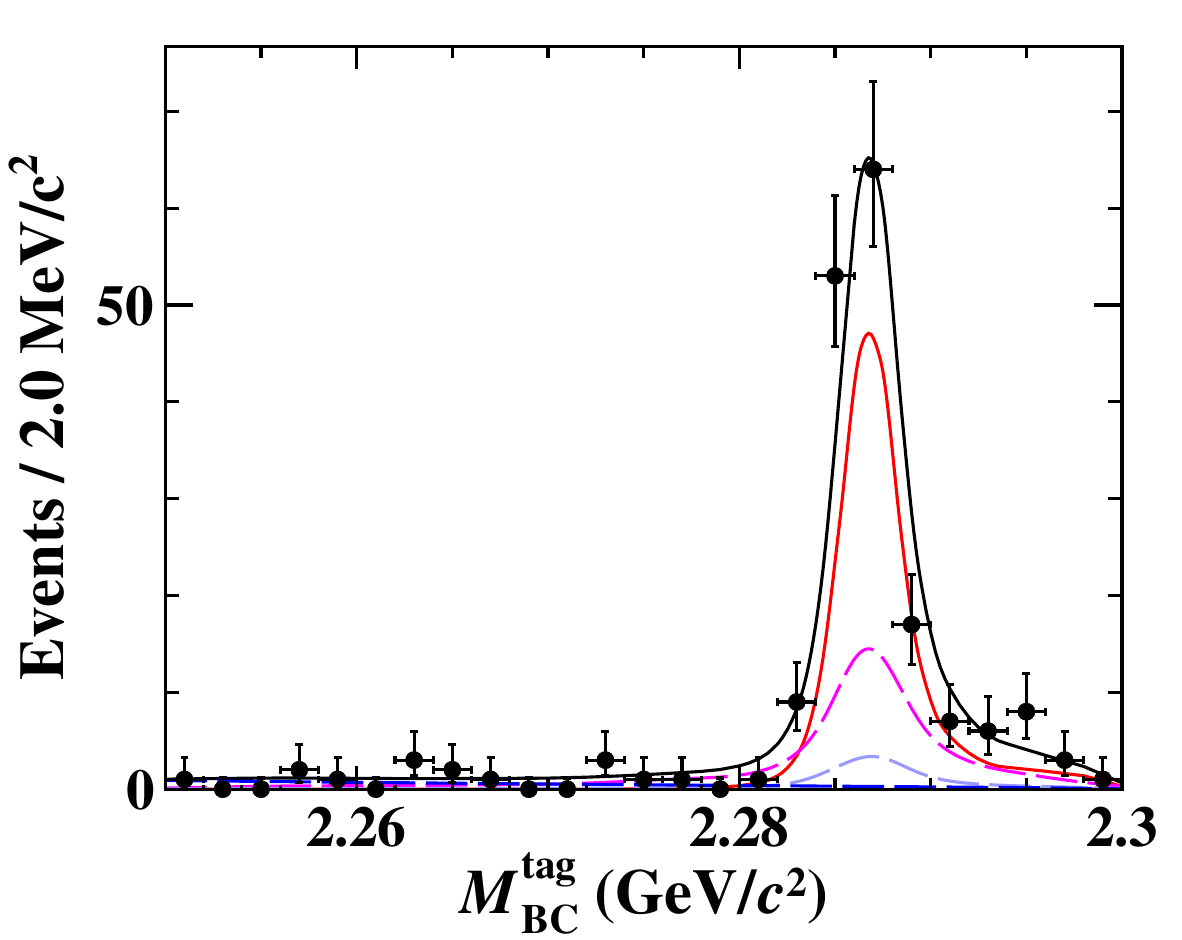}
  \includegraphics[width=0.24\textwidth]{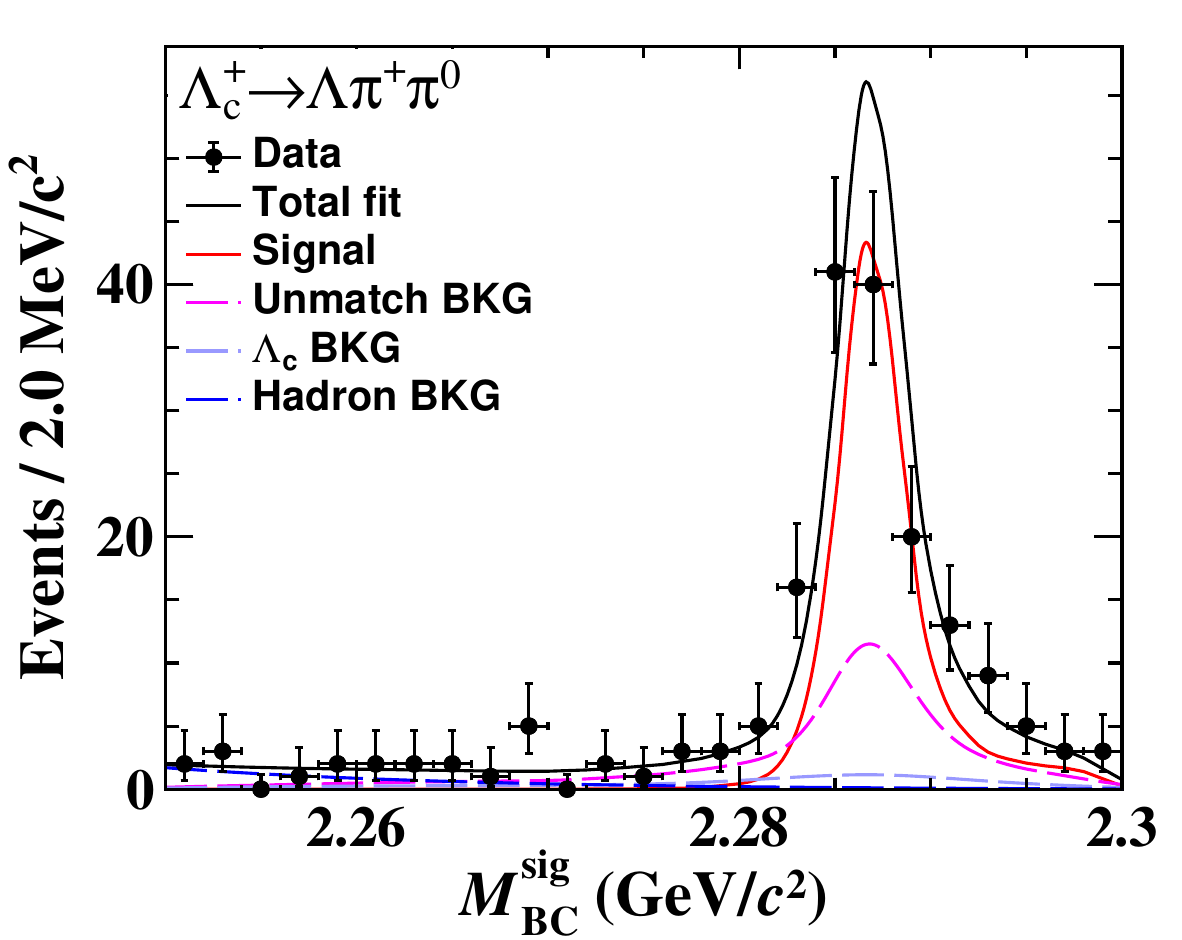}
  \includegraphics[width=0.24\textwidth]{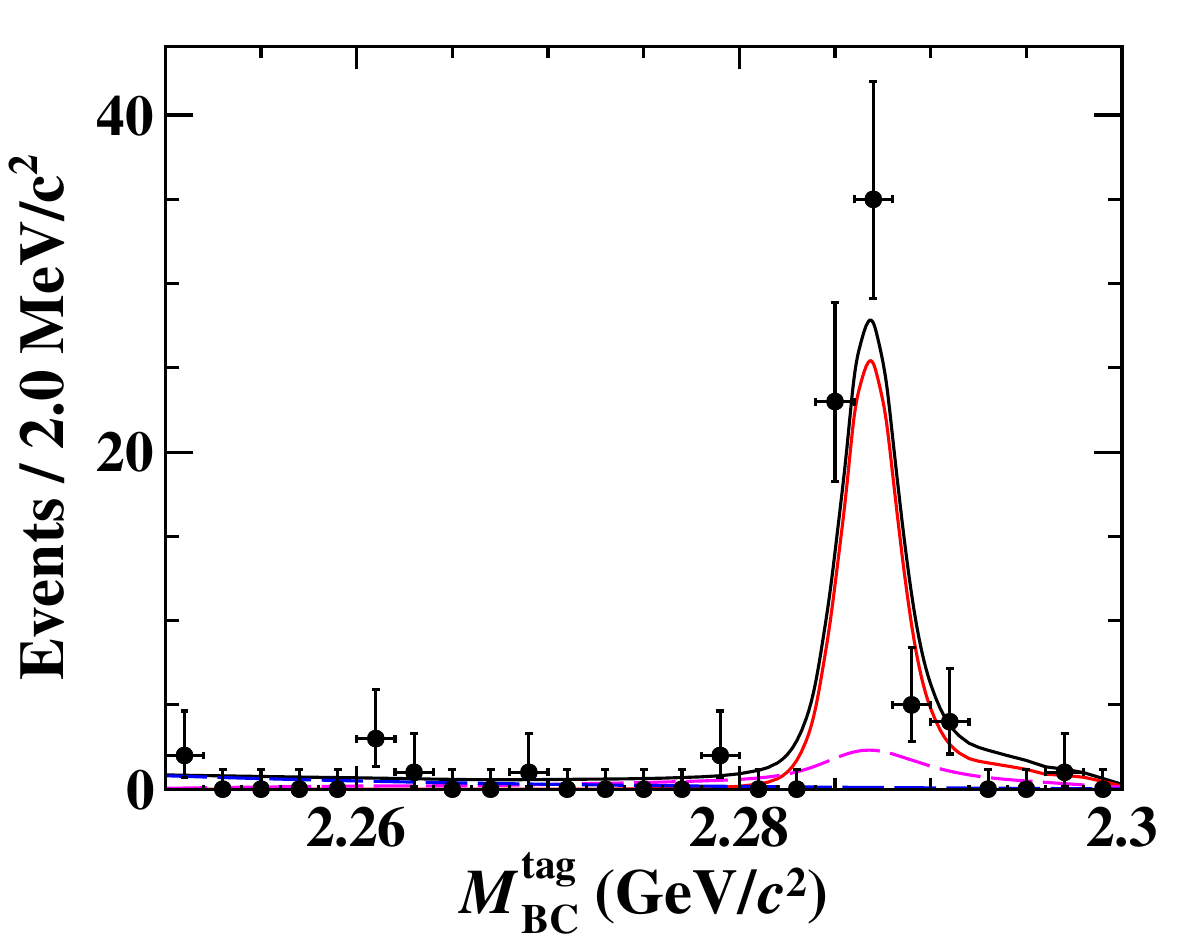}
  \includegraphics[width=0.24\textwidth]{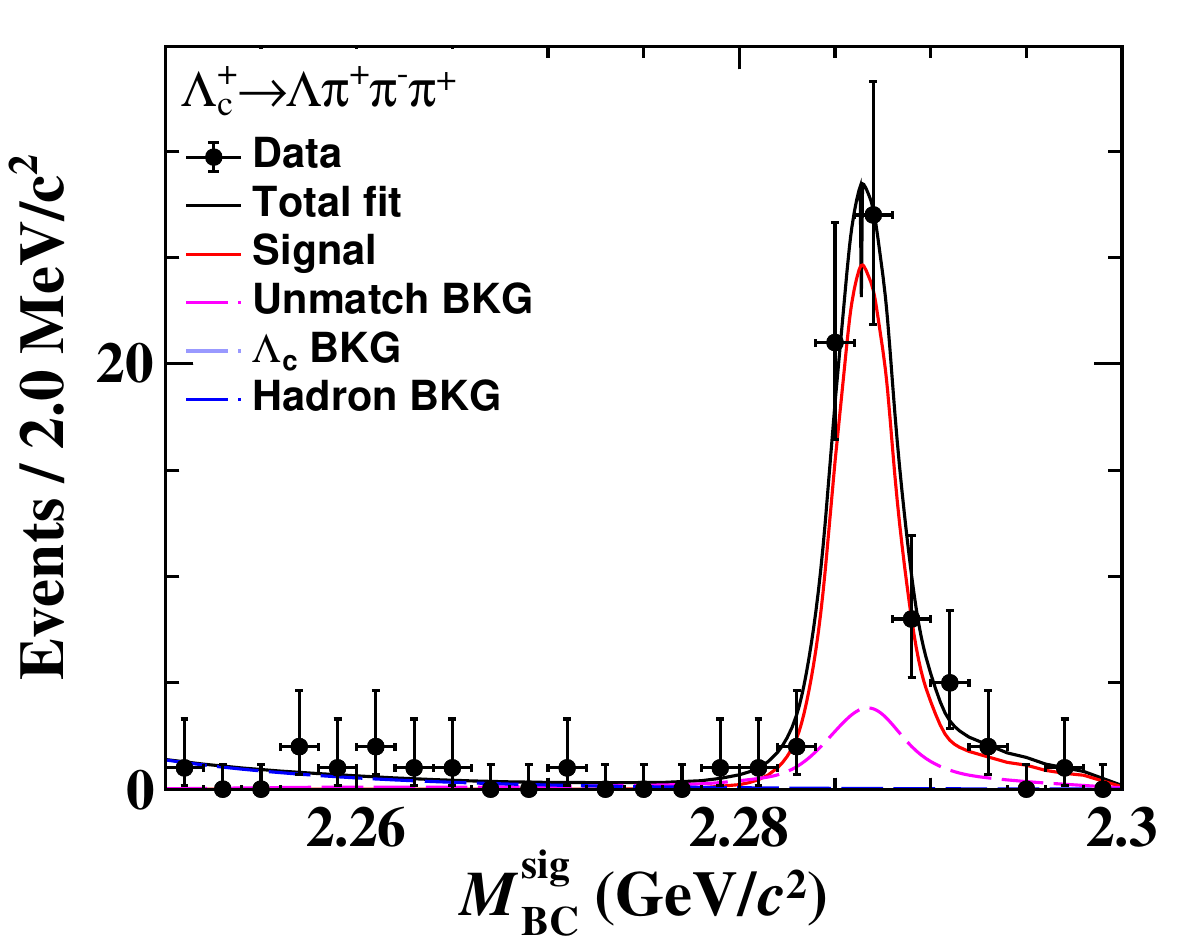}
  \includegraphics[width=0.24\textwidth]{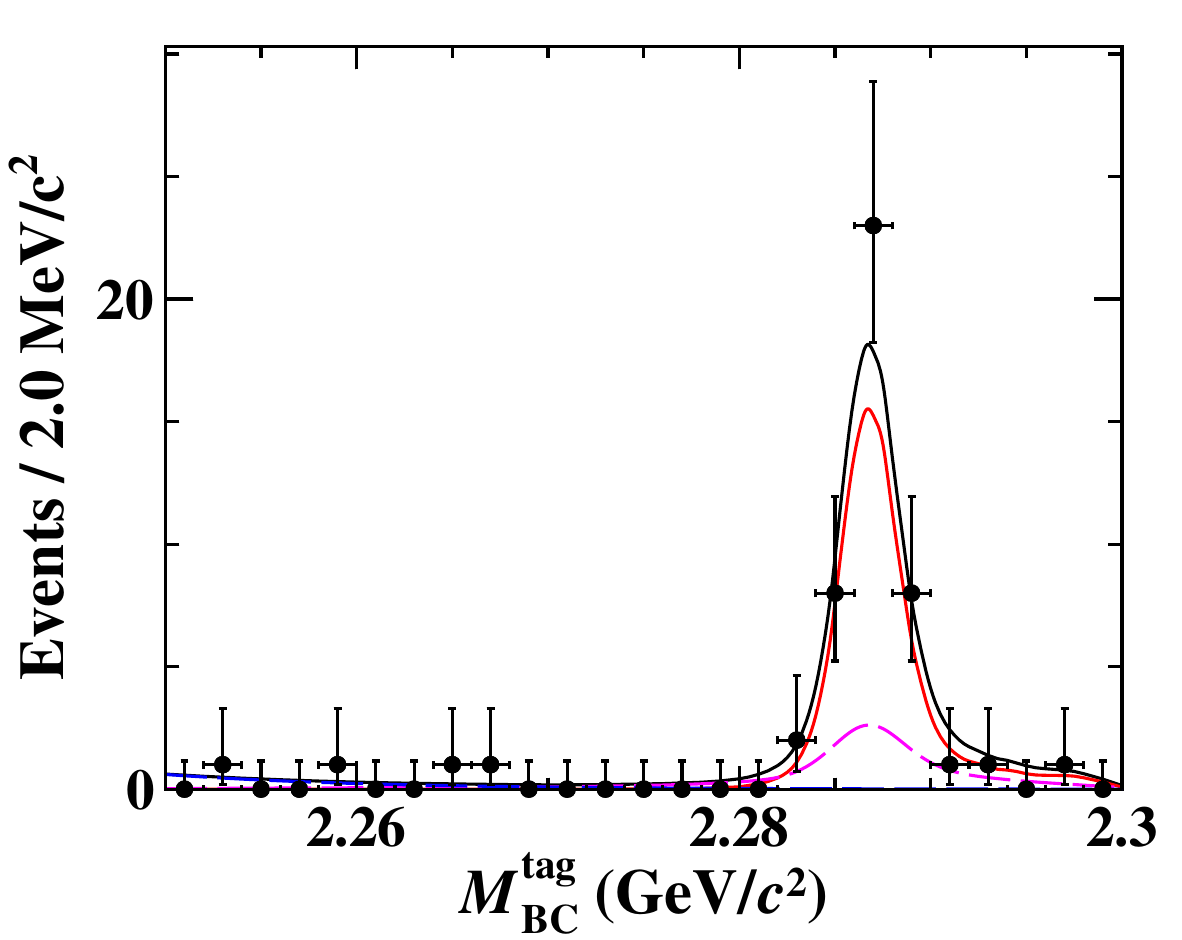}
  \includegraphics[width=0.24\textwidth]{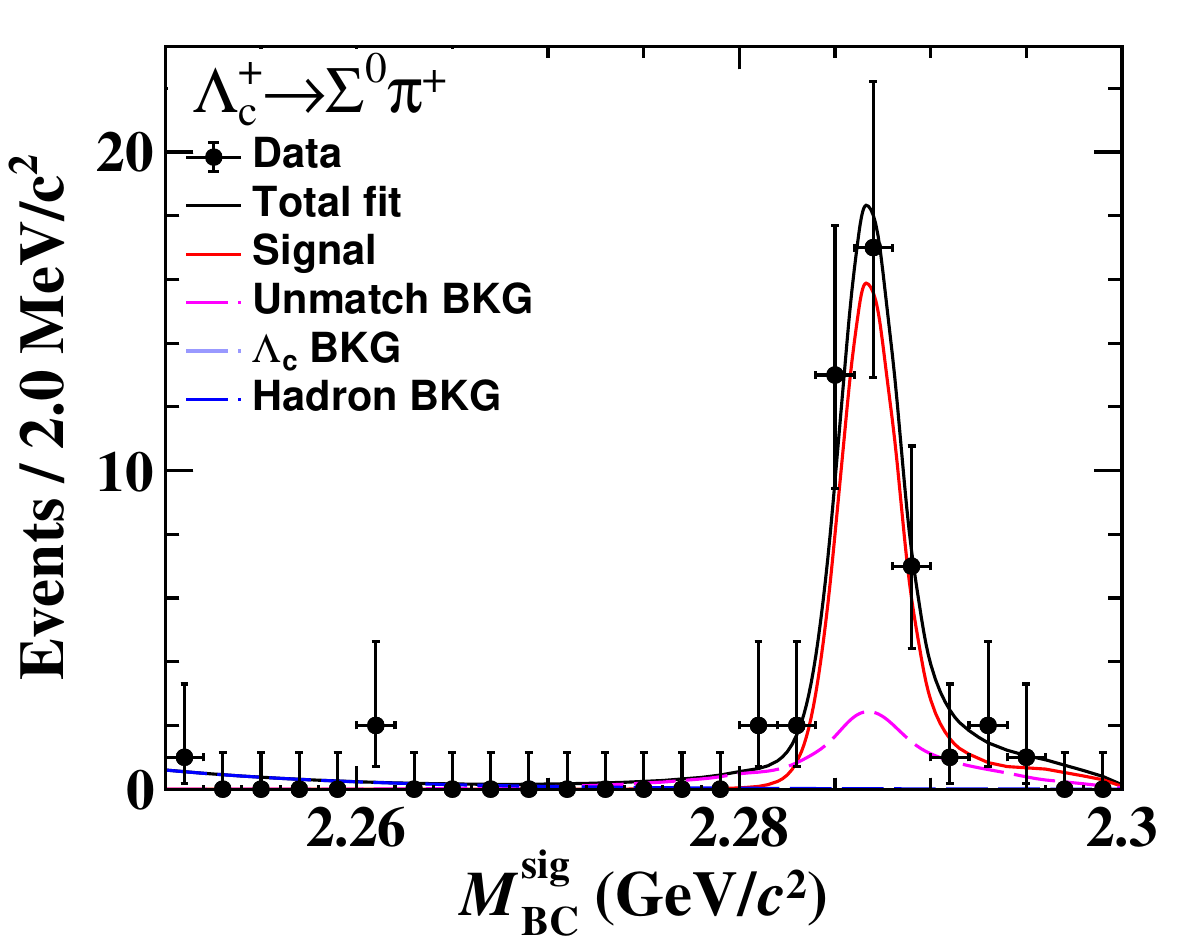}
  \includegraphics[width=0.24\textwidth]{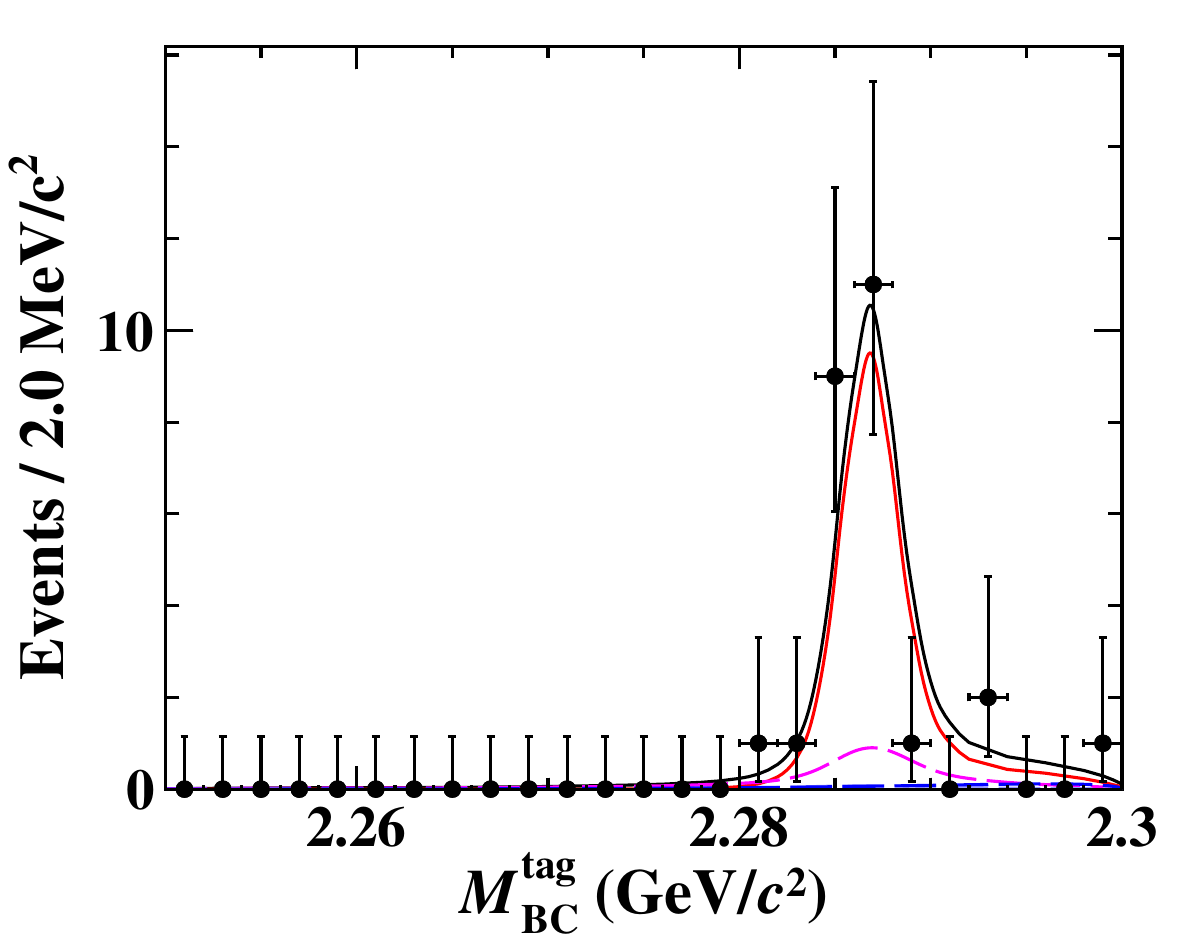}
  \includegraphics[width=0.24\textwidth]{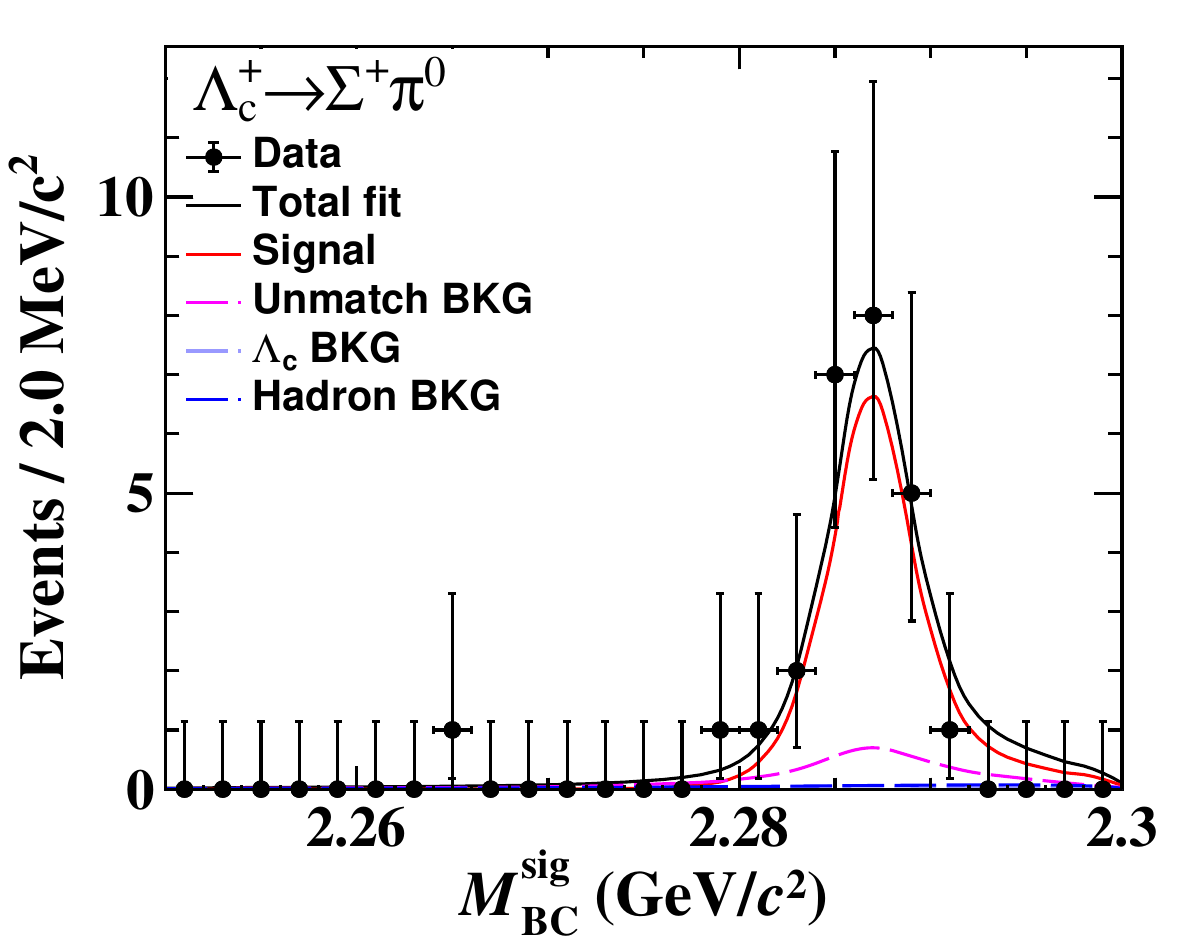}
  \includegraphics[width=0.24\textwidth]{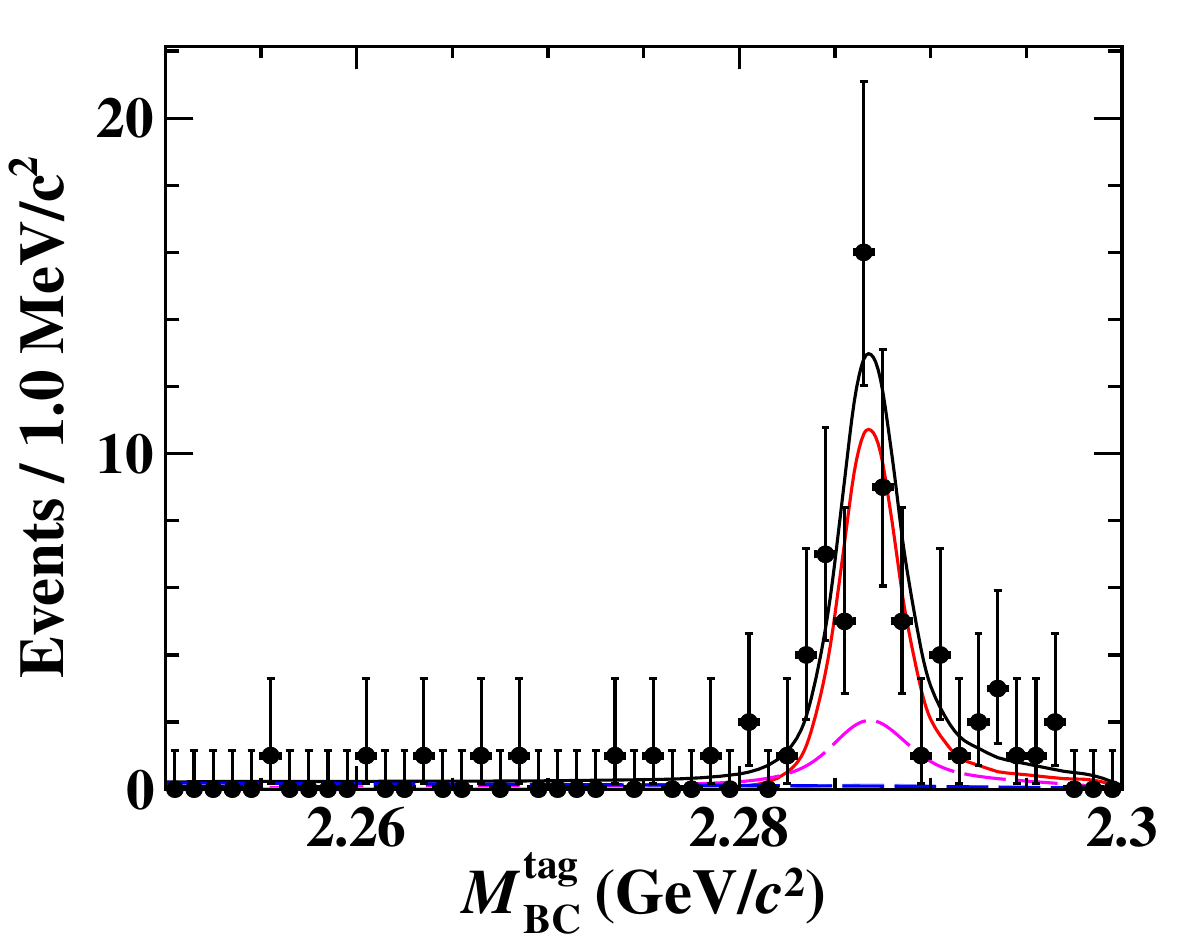}
  \includegraphics[width=0.24\textwidth]{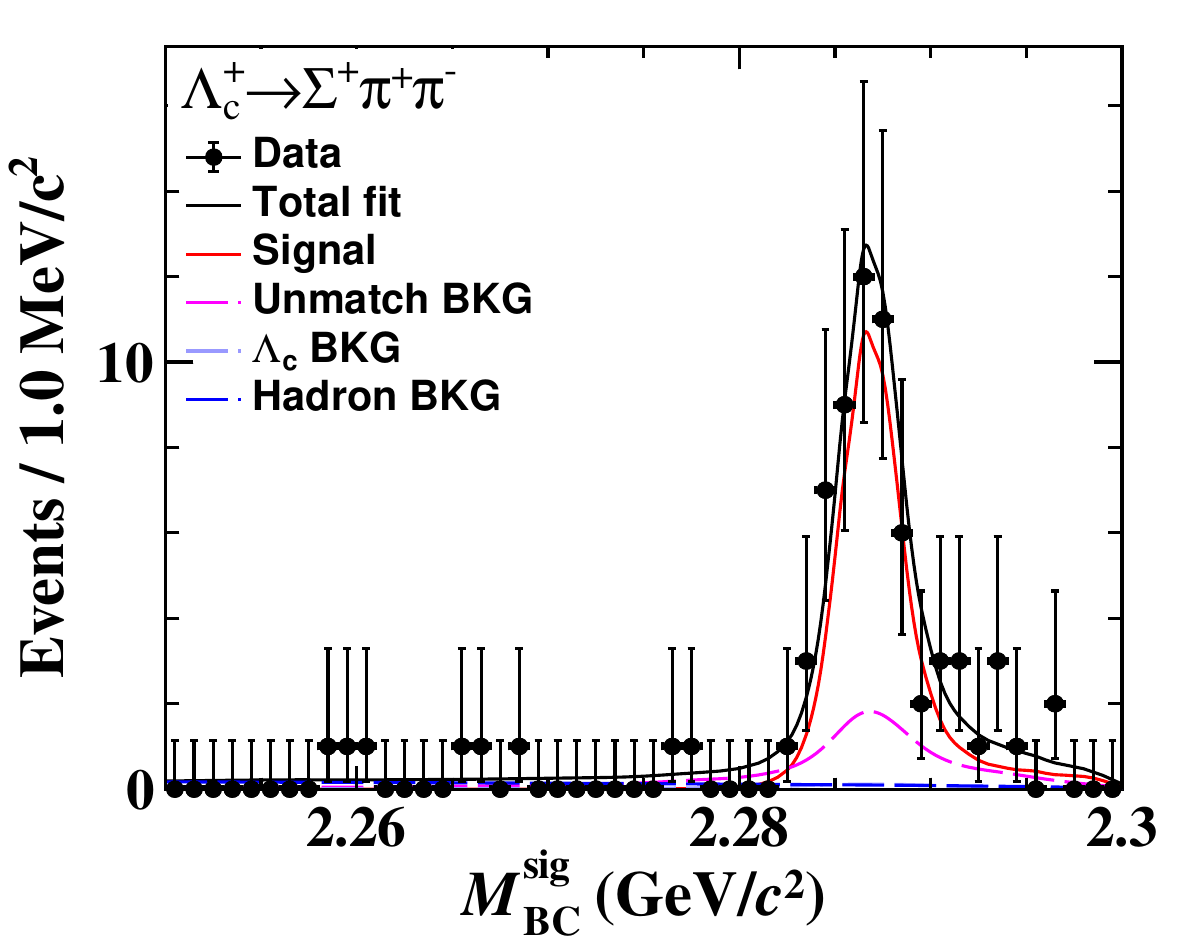}
  \includegraphics[width=0.24\textwidth]{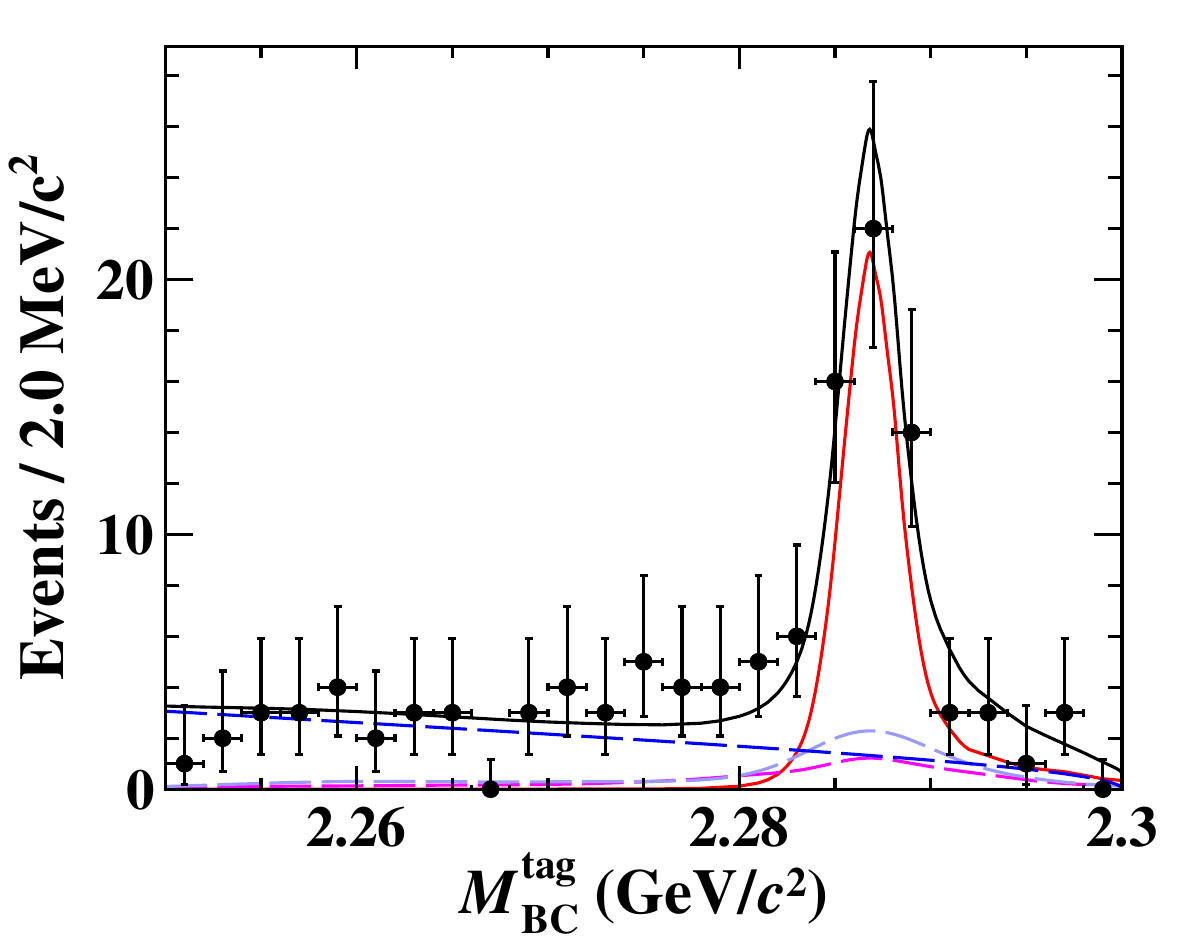}
  \includegraphics[width=0.24\textwidth]{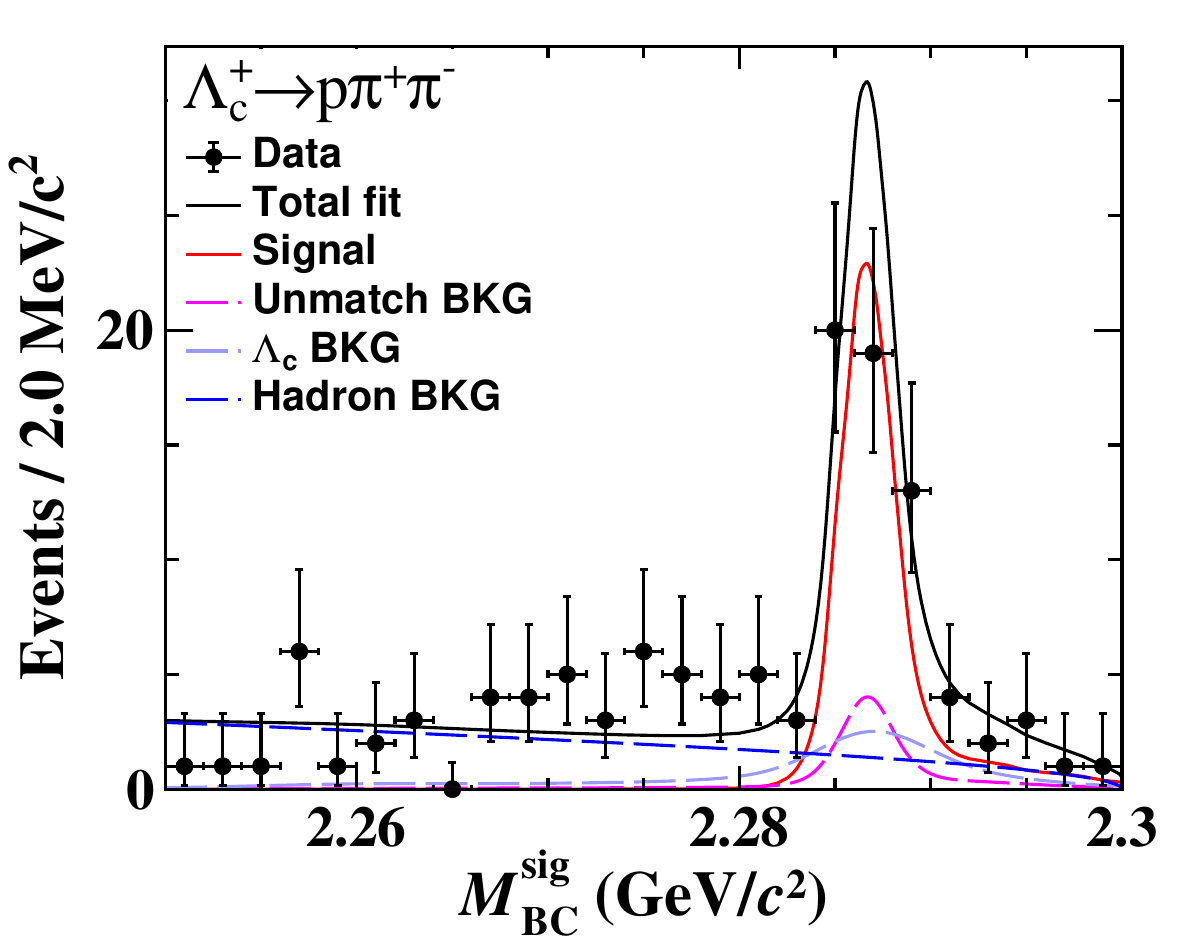}
     \vspace*{-0.5cm}
  \end{center}
\caption{The projections of the 2D fits on the $M_{\rm BC}^{\rm tag}$ and $M_{\rm BC}^{\rm sig}$ distributions of the accepted DT candidates at $\sqrt{s}=4599.53~\mev$. The plots in the first and third columns show the combined 12 tag modes for each signal mode.
The points with error bars are data, the black lines are the sum of fit functions, the red lines are the matched signal shapes, the pink dashed lines are the unmatched signal shapes, the lilac dashed lines are the non-signal $\lcp\lcm$ shapes, and the blue dashed lines are the ARGUS functions.}
\label{fig:DT_yield_4600}
\end{figure}

\begin{figure}[!htbp]
  \begin{center}
  
  \includegraphics[width=0.24\textwidth]{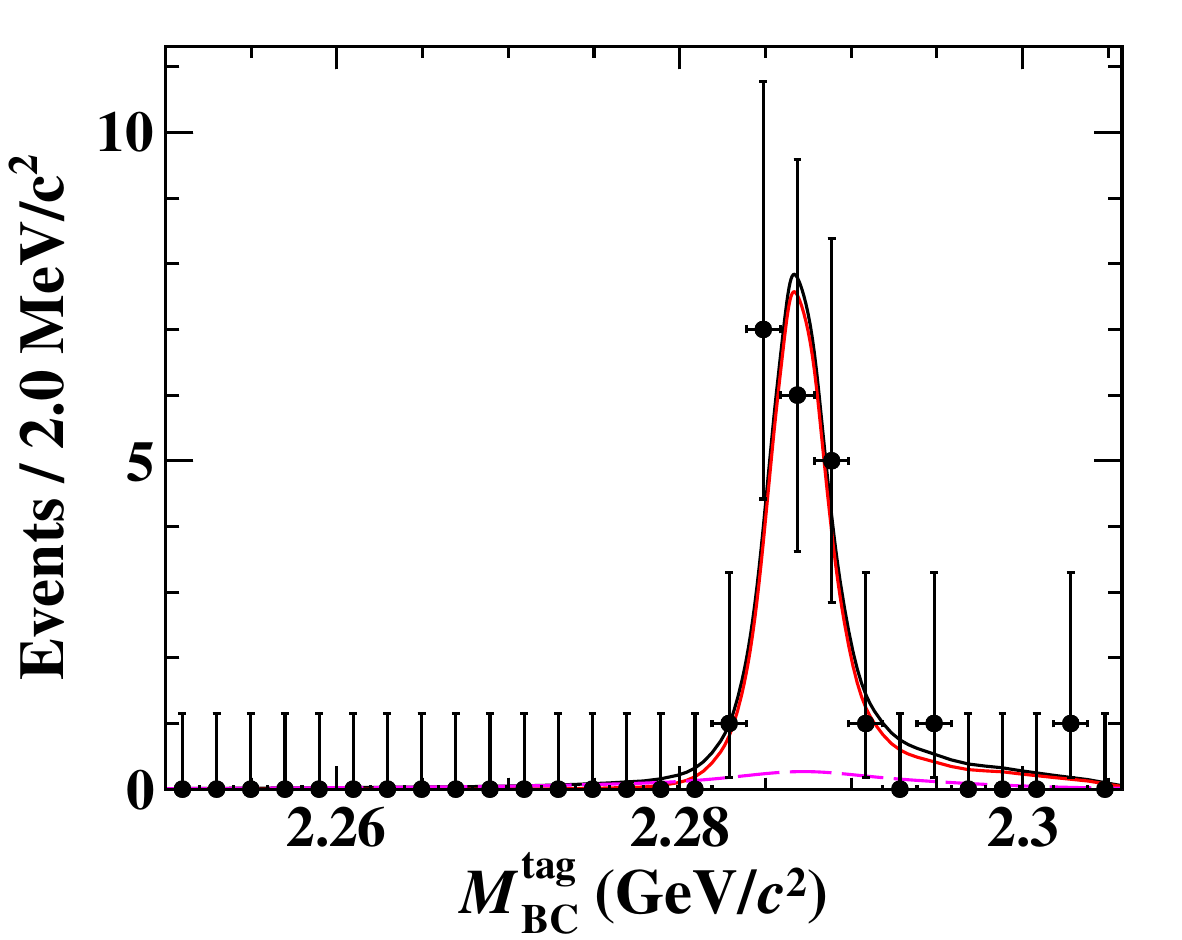}
  \includegraphics[width=0.24\textwidth]{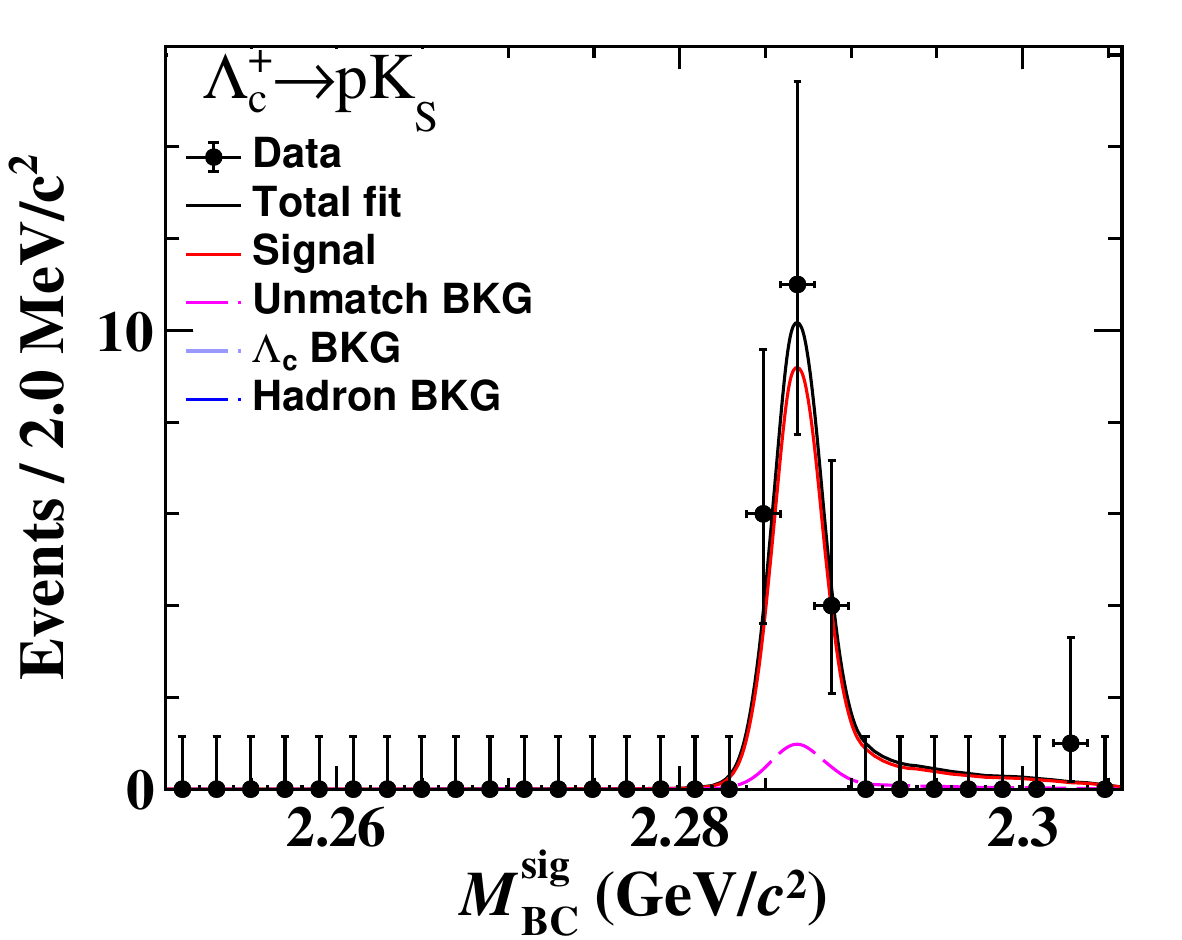}
  \includegraphics[width=0.24\textwidth]{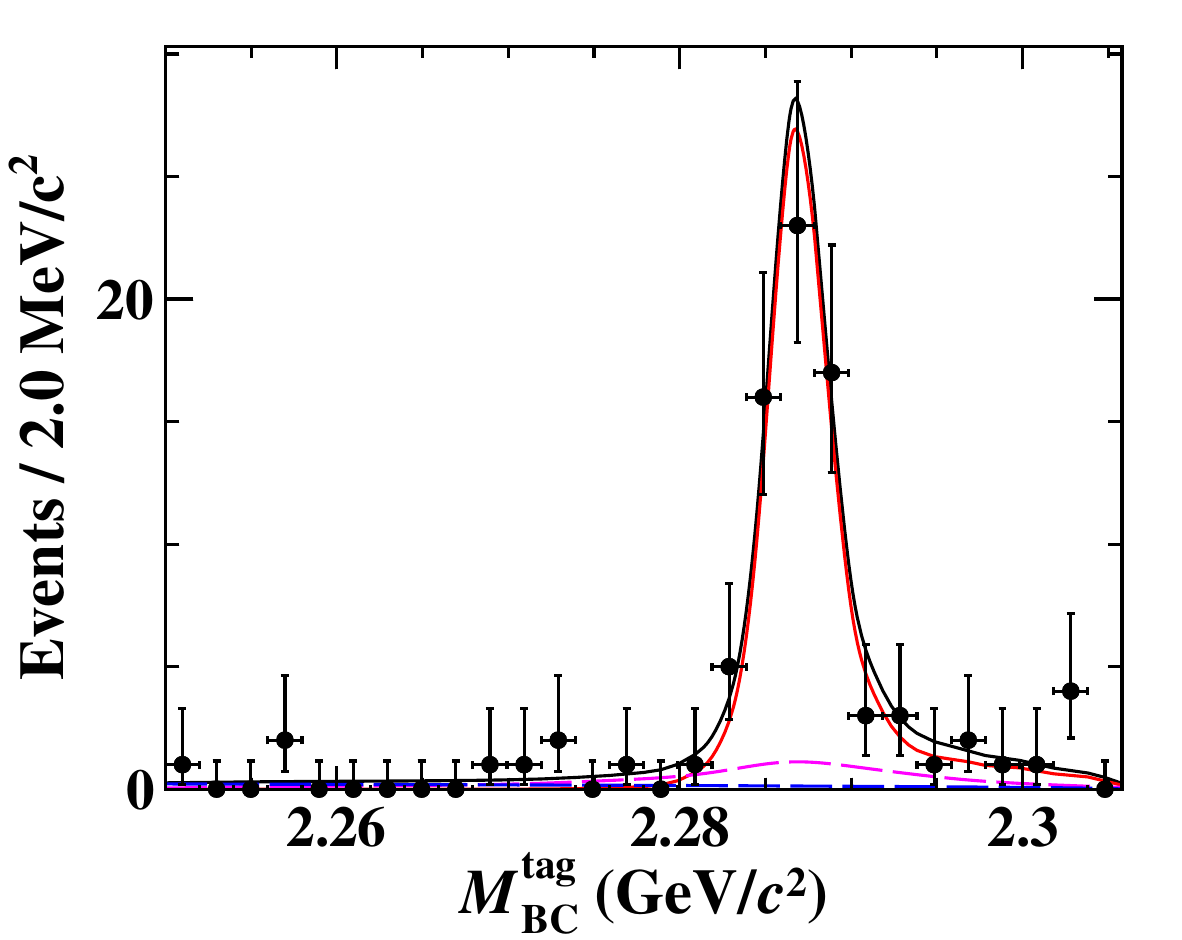}
  \includegraphics[width=0.24\textwidth]{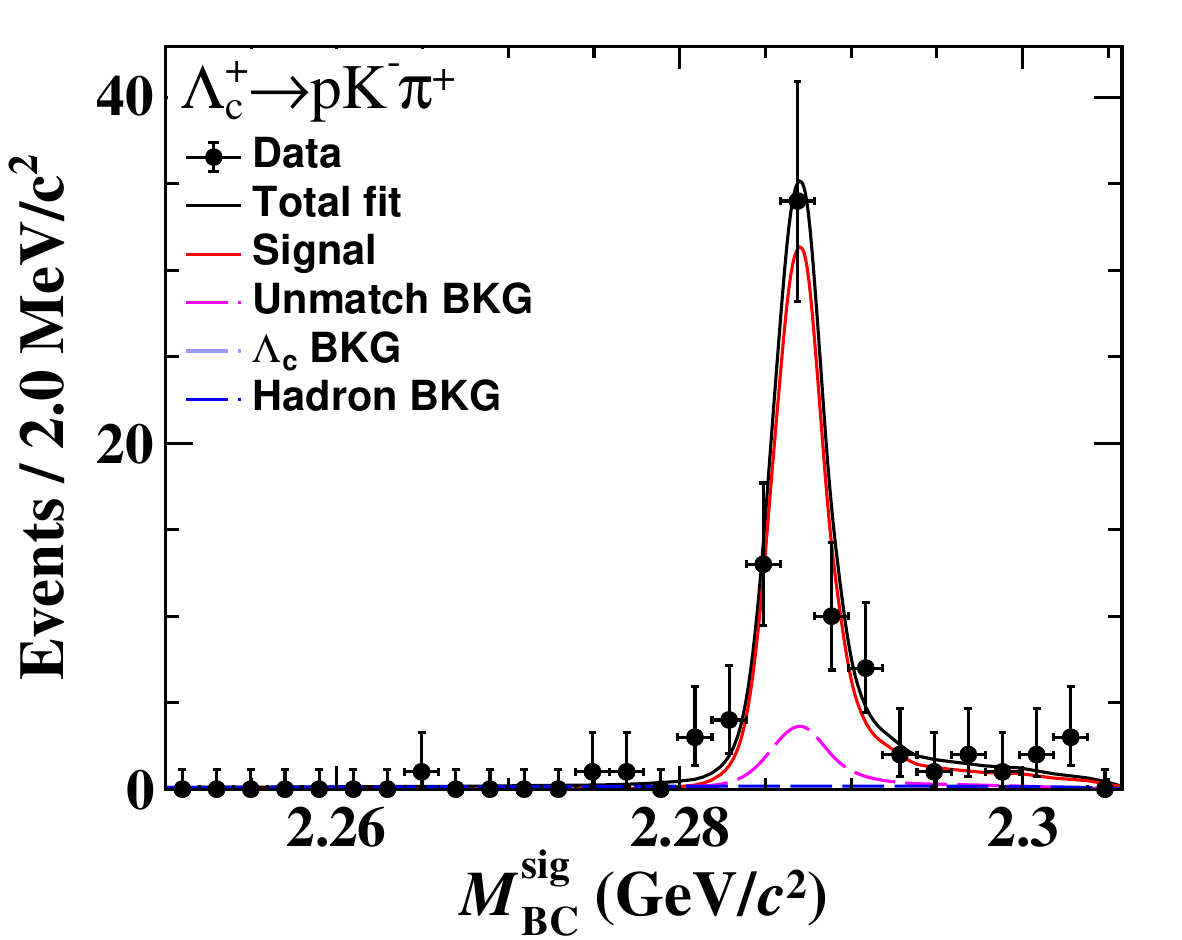}
  \includegraphics[width=0.24\textwidth]{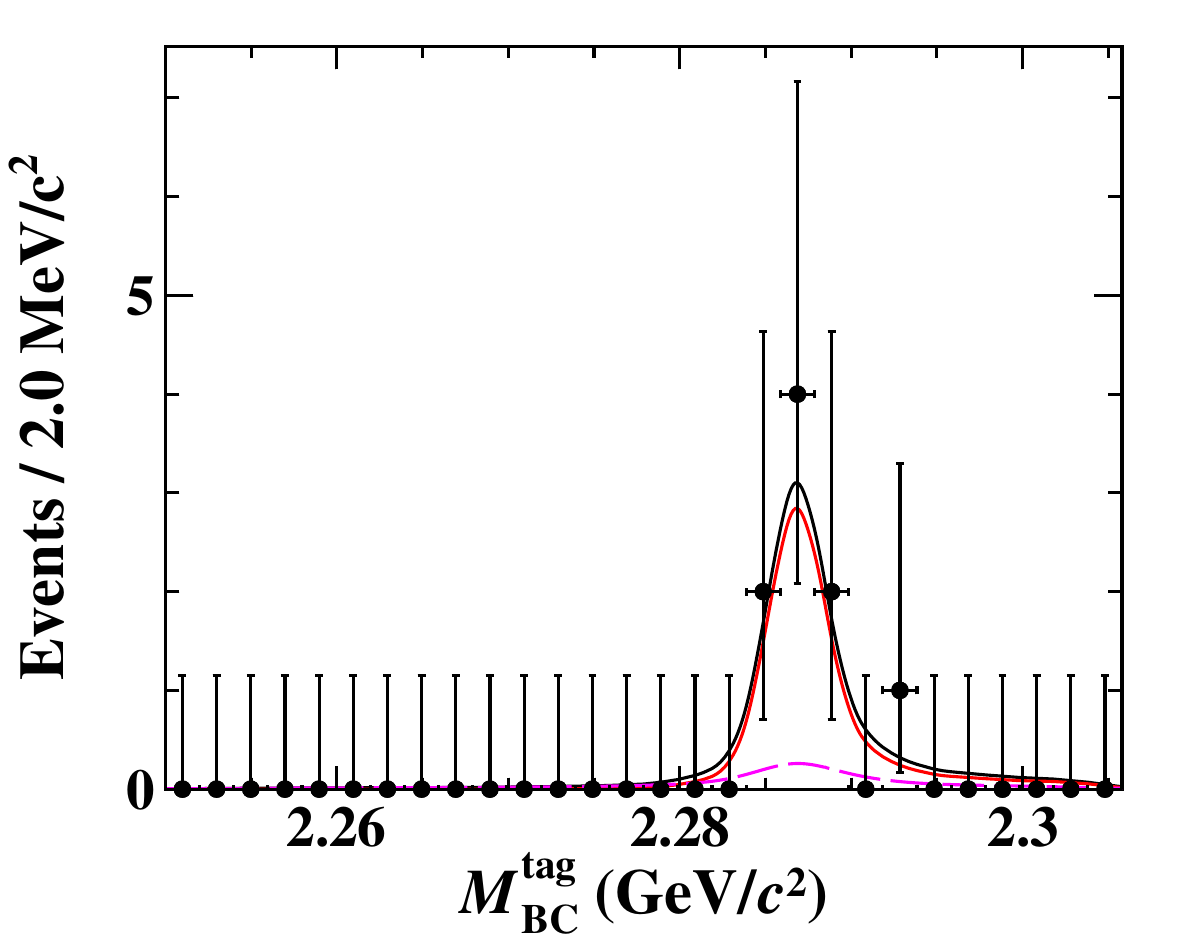}
  \includegraphics[width=0.24\textwidth]{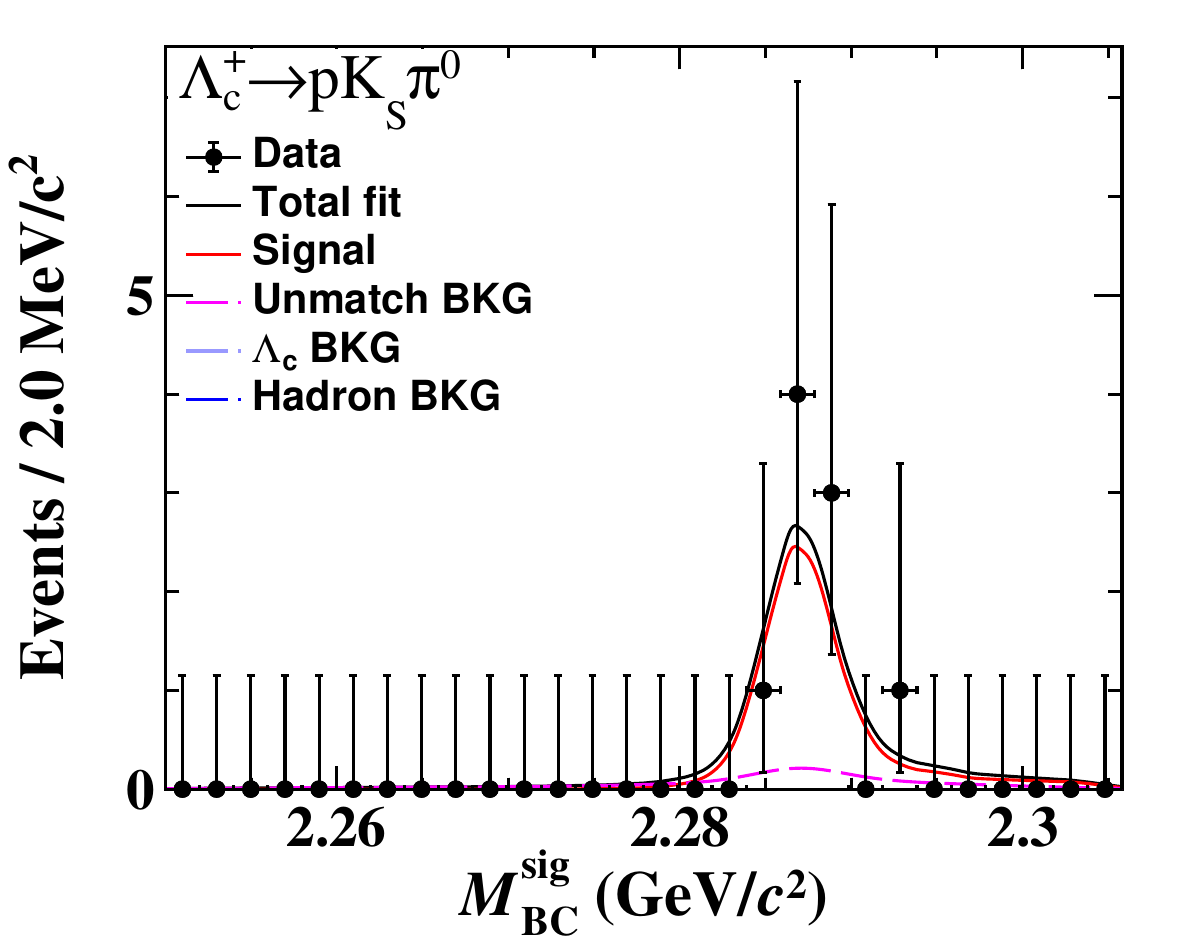}
  \includegraphics[width=0.24\textwidth]{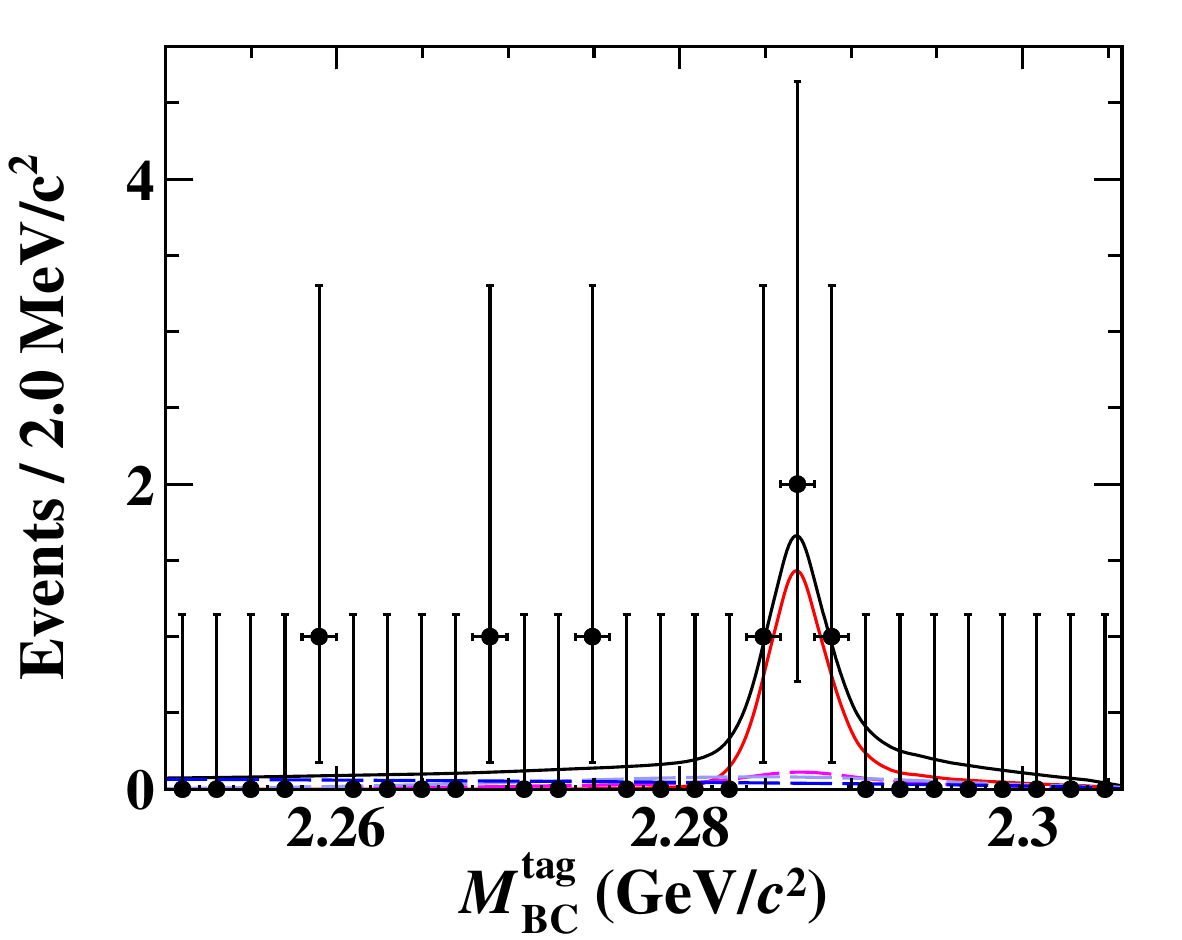}
  \includegraphics[width=0.24\textwidth]{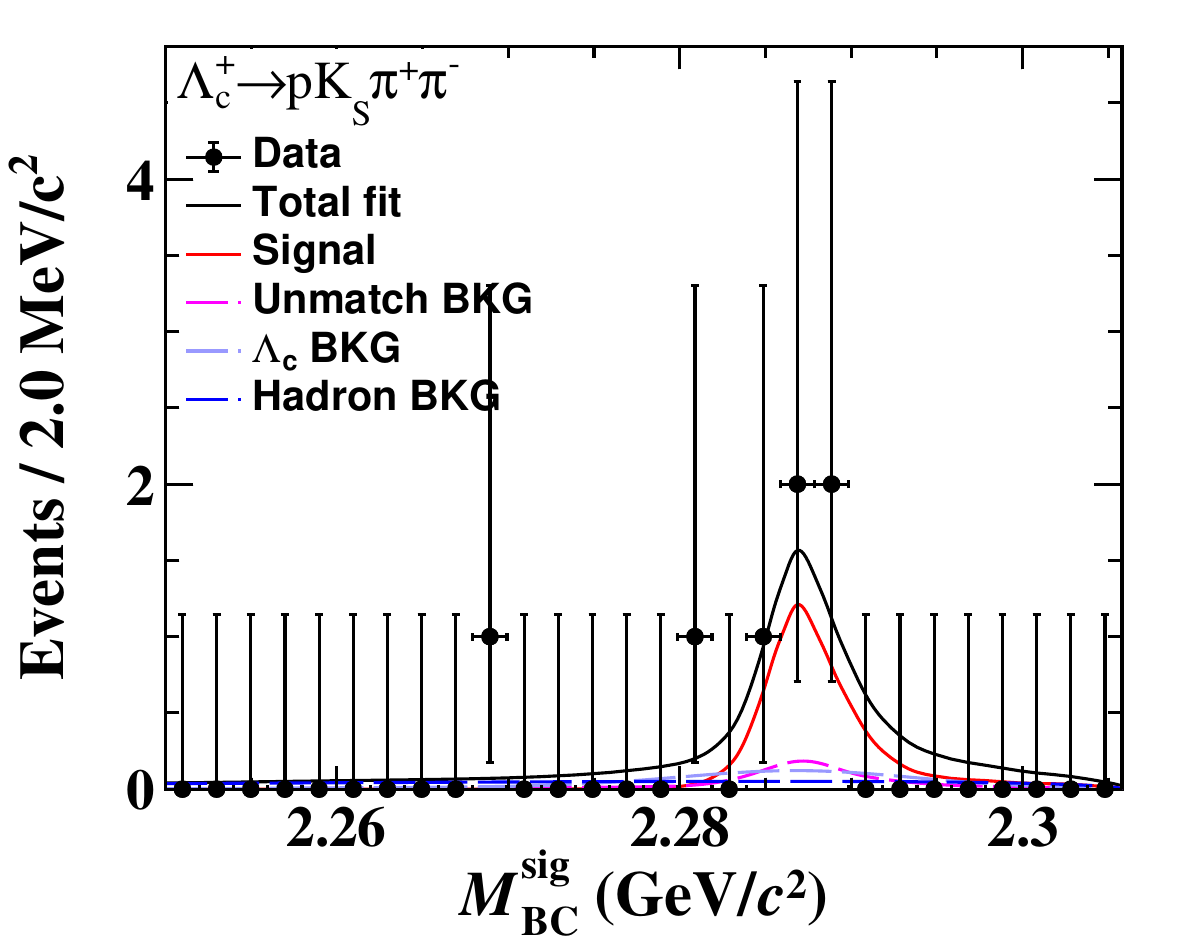}
  \includegraphics[width=0.24\textwidth]{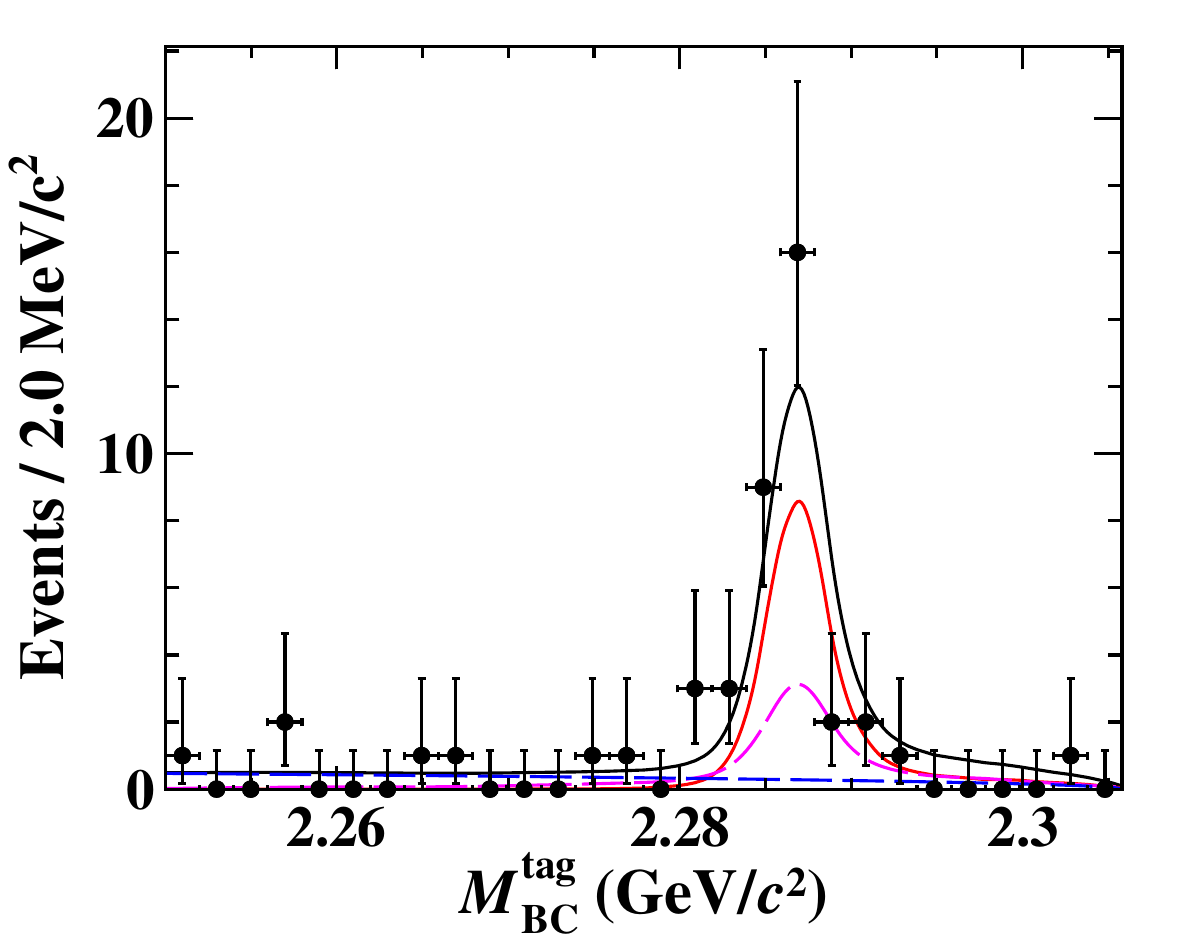}
  \includegraphics[width=0.24\textwidth]{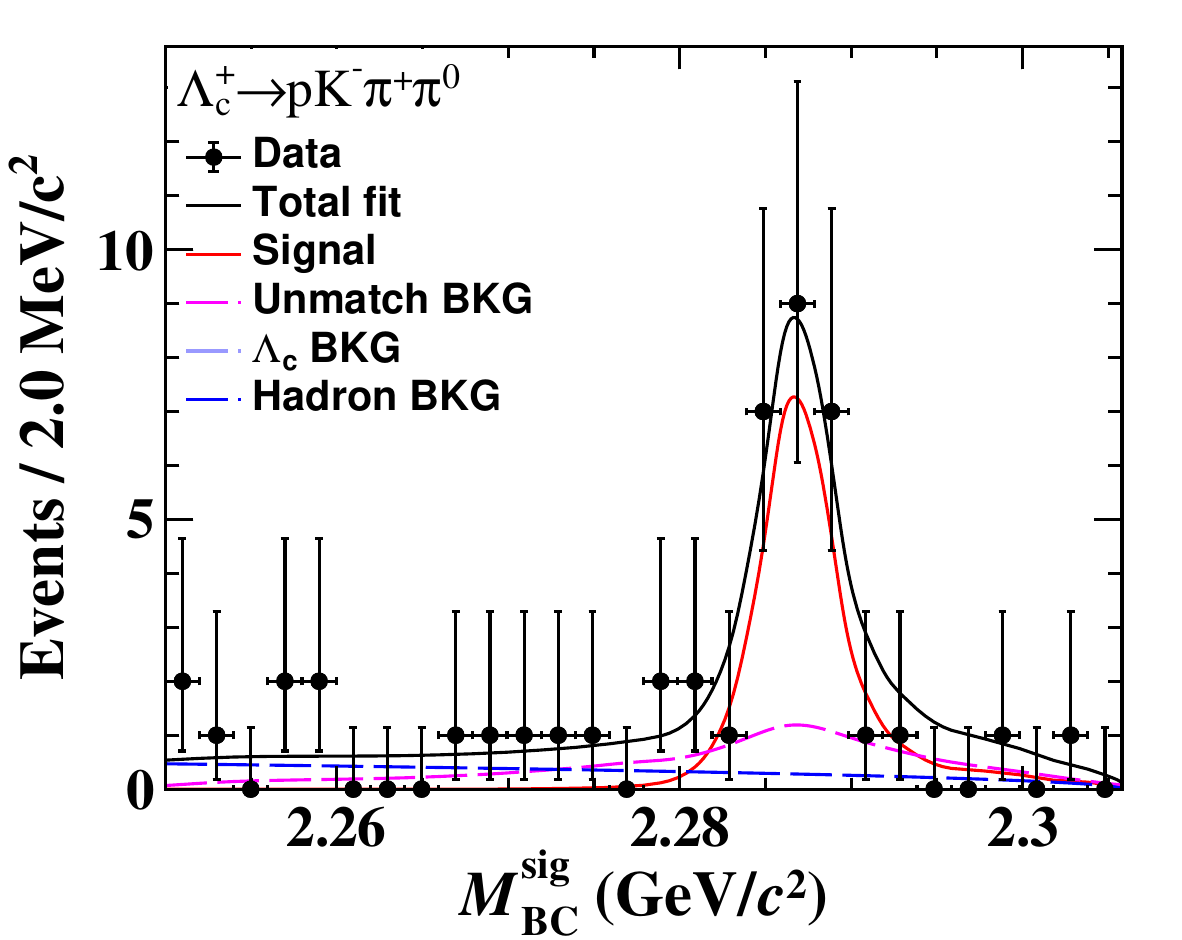}
  \includegraphics[width=0.24\textwidth]{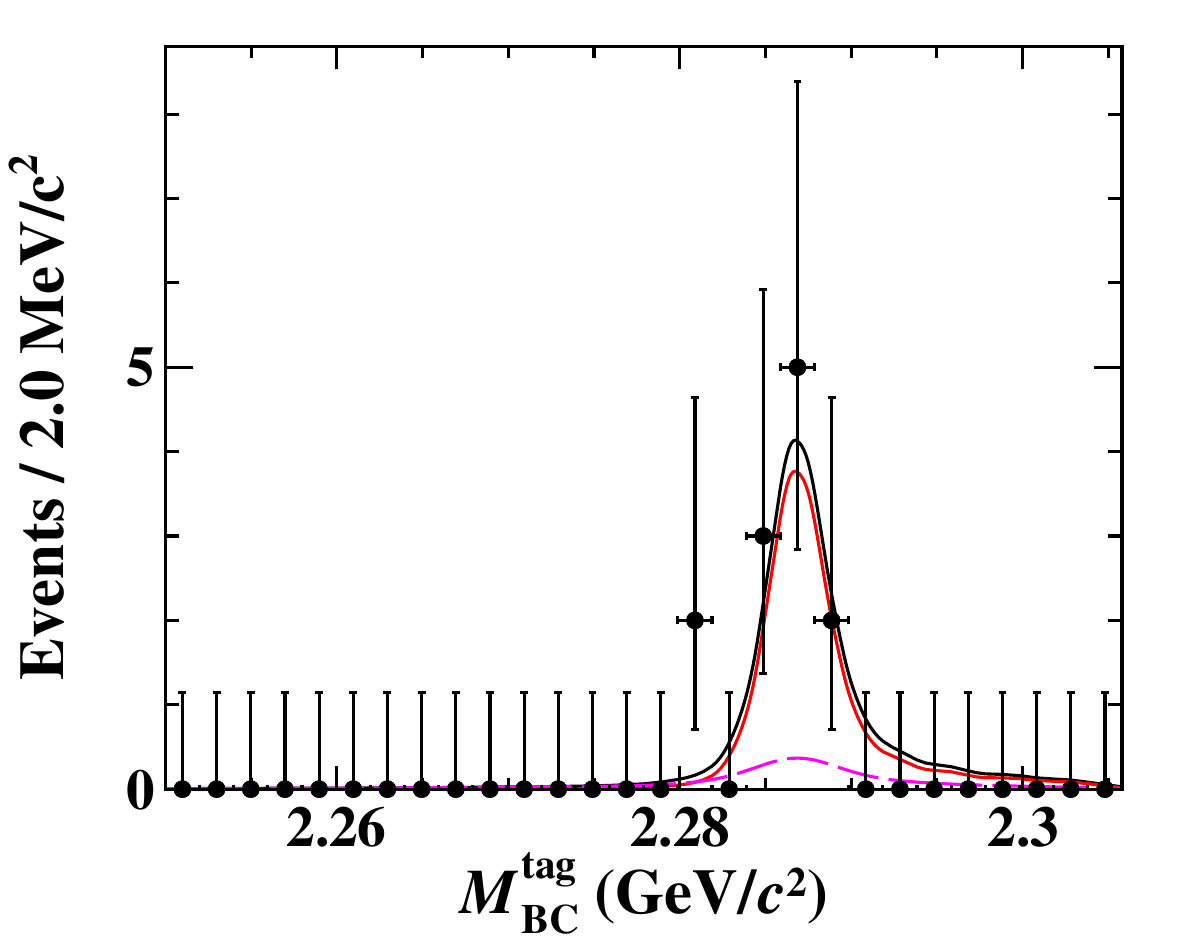}
  \includegraphics[width=0.24\textwidth]{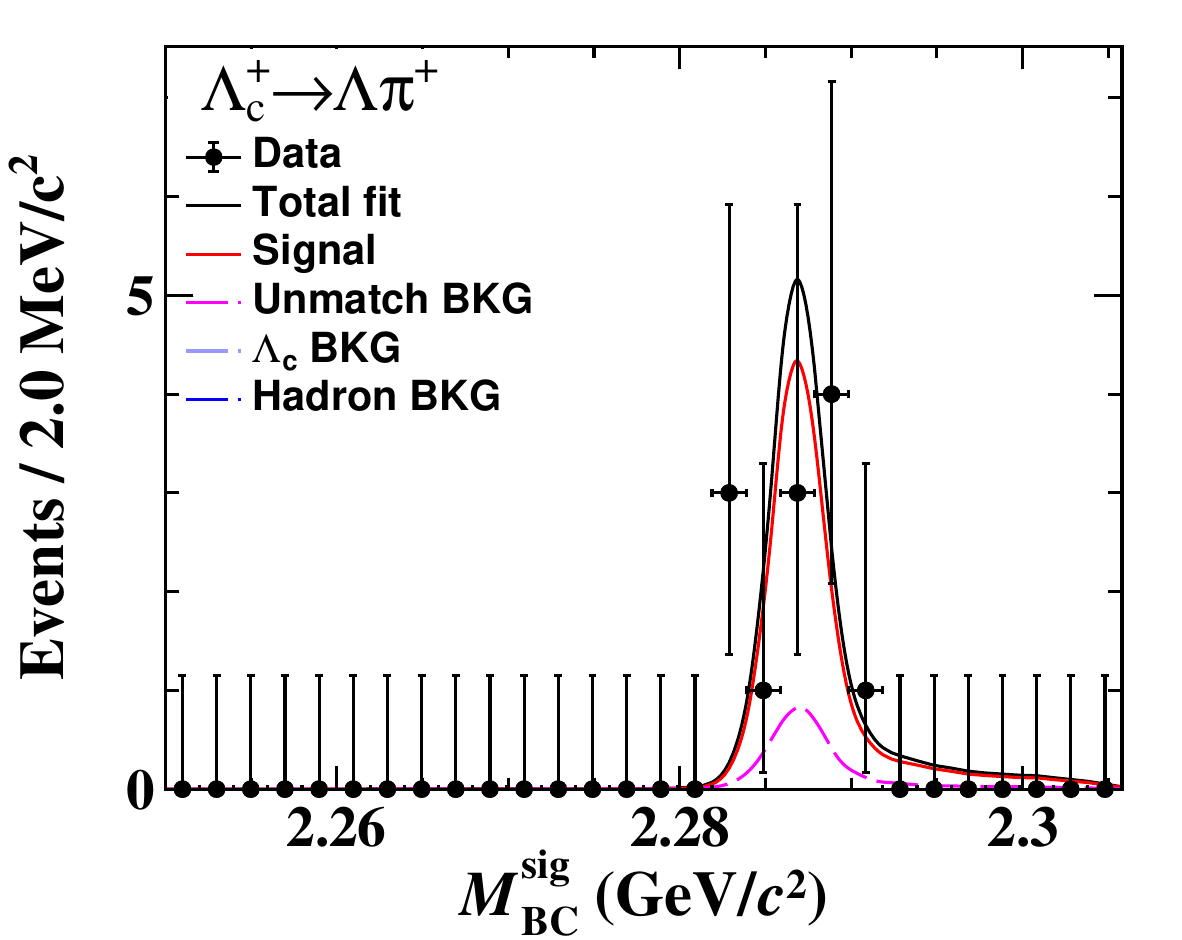}
  \includegraphics[width=0.24\textwidth]{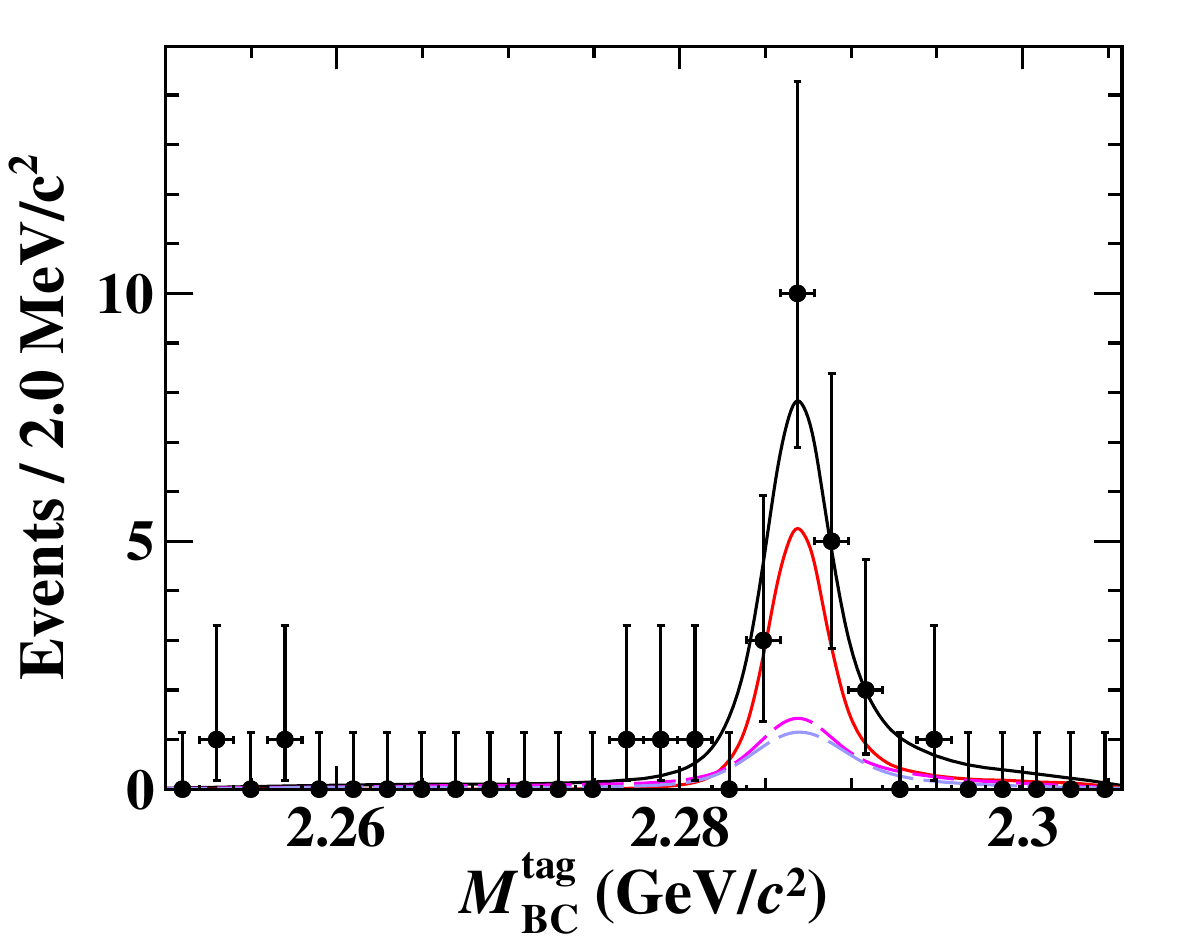}
  \includegraphics[width=0.24\textwidth]{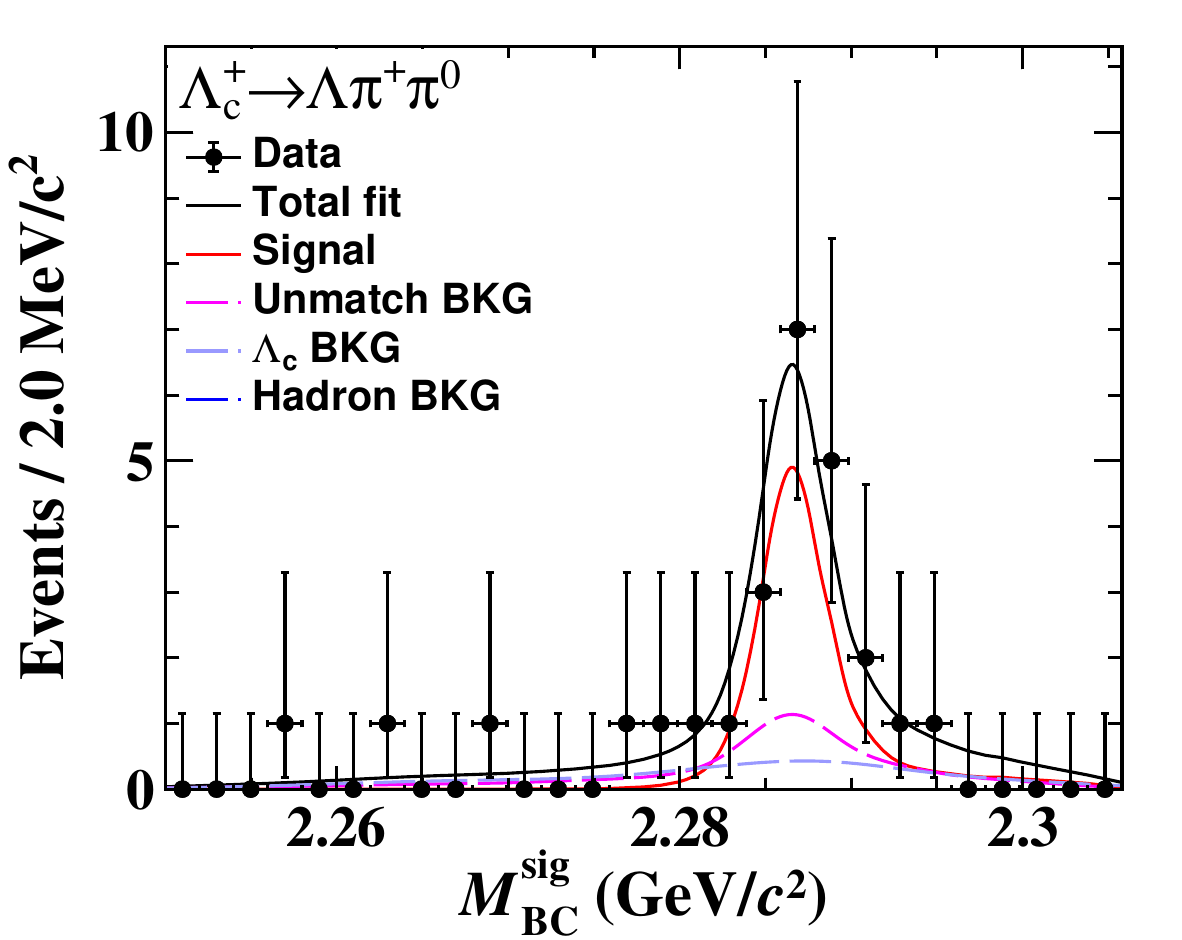}
  \includegraphics[width=0.24\textwidth]{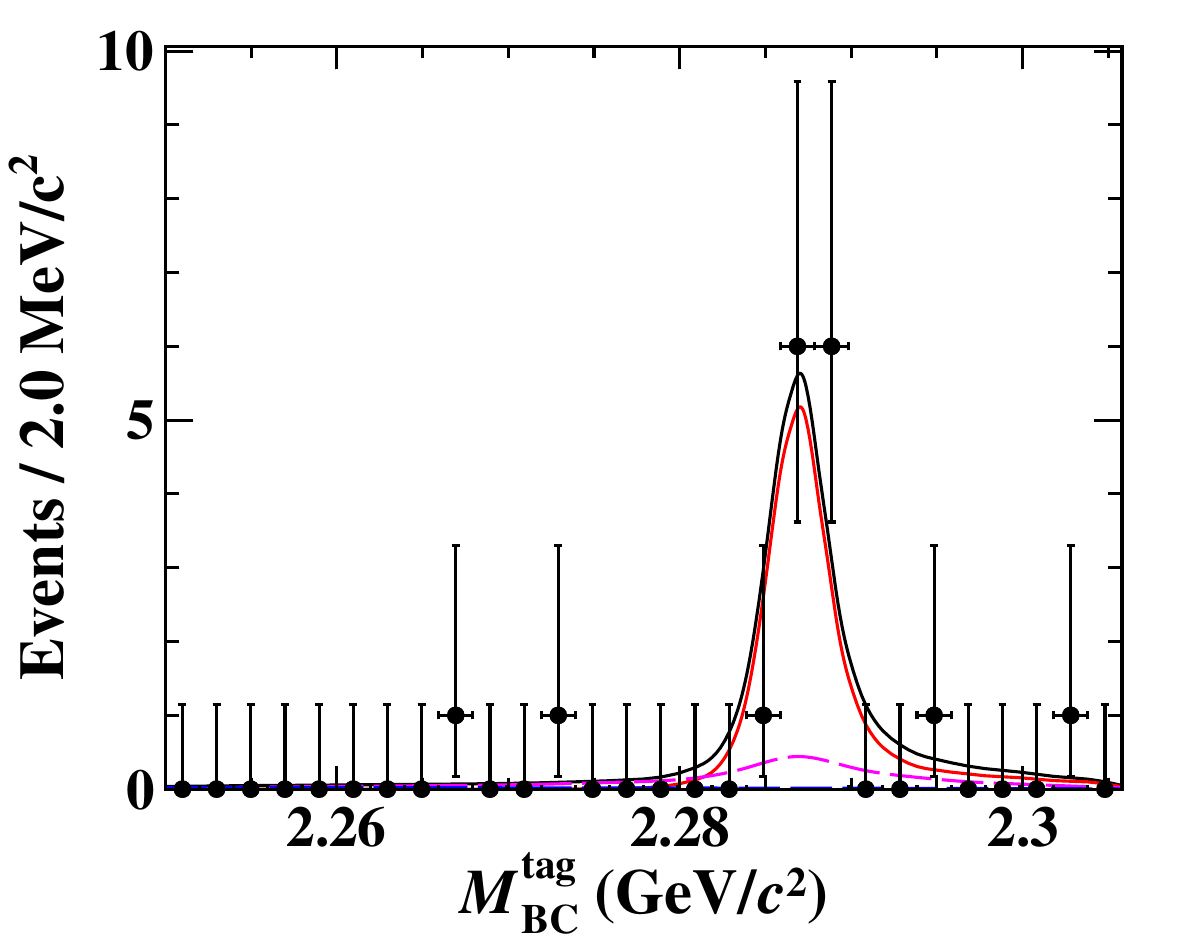}
  \includegraphics[width=0.24\textwidth]{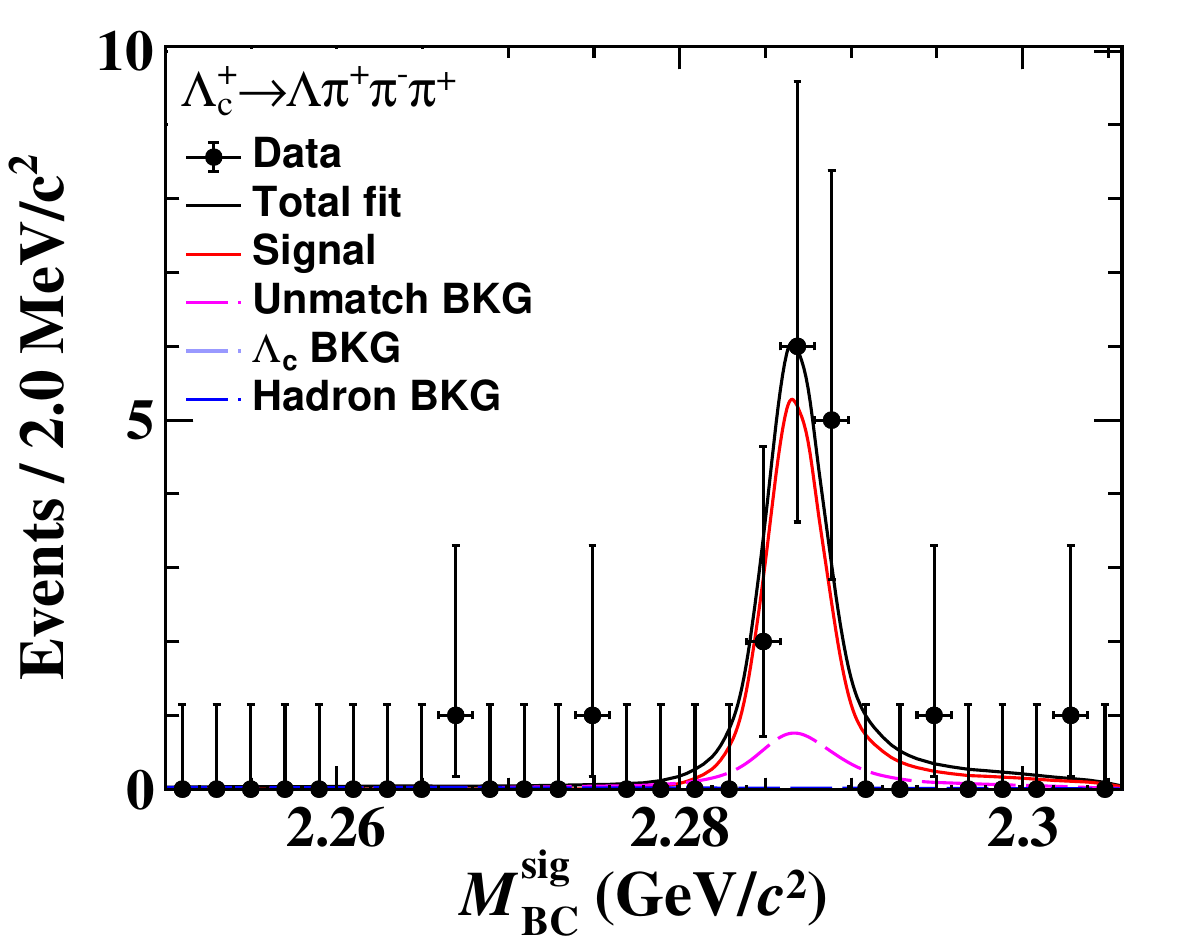}
  \includegraphics[width=0.24\textwidth]{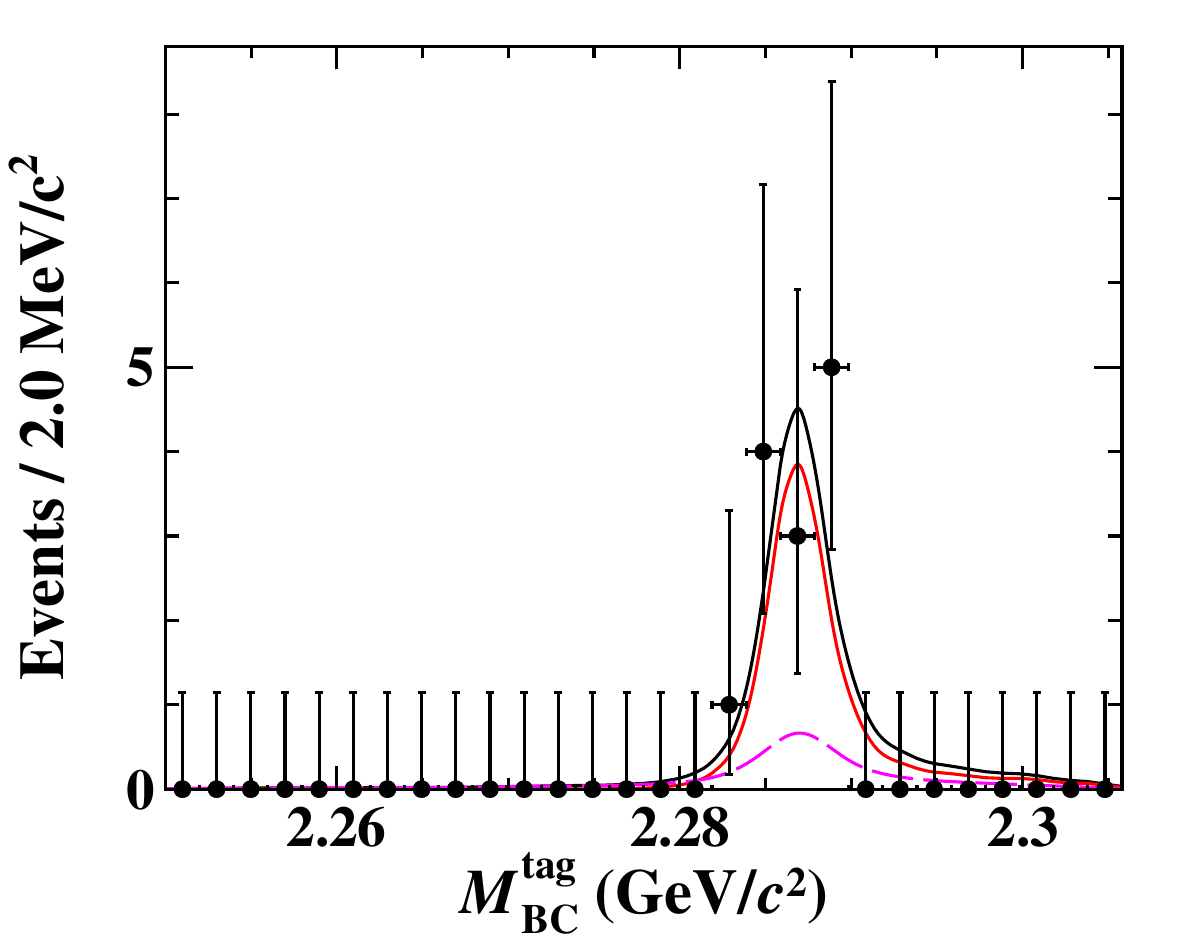}
  \includegraphics[width=0.24\textwidth]{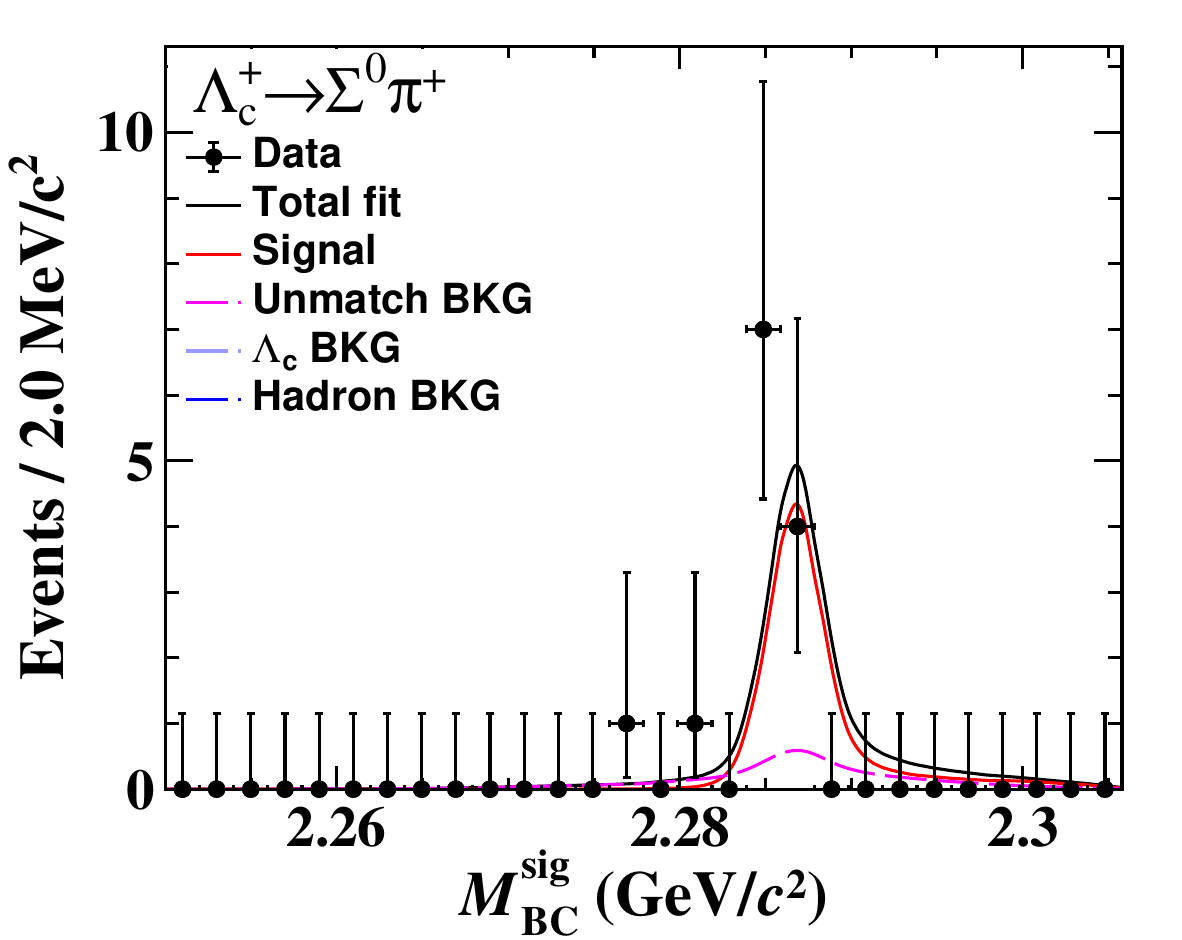}
  \includegraphics[width=0.24\textwidth]{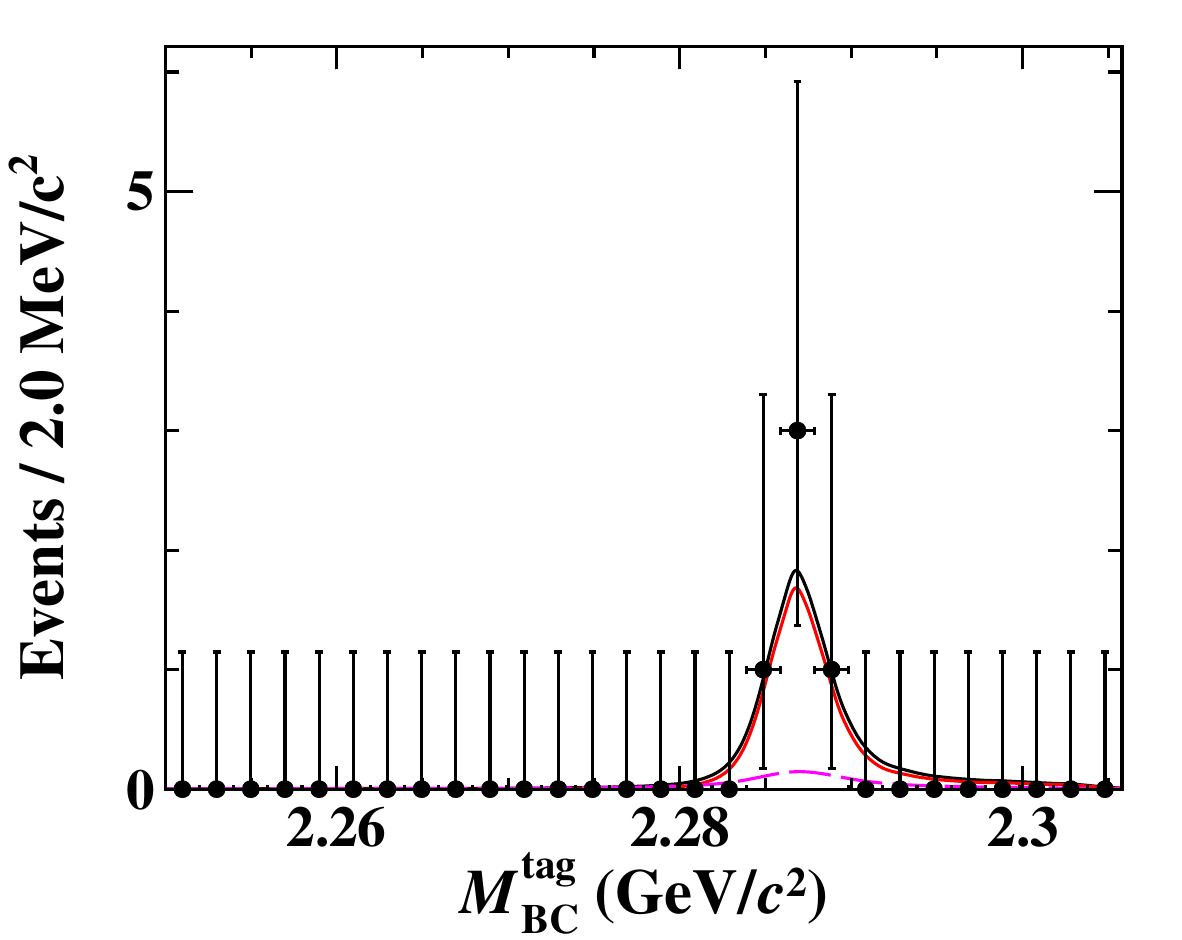}
  \includegraphics[width=0.24\textwidth]{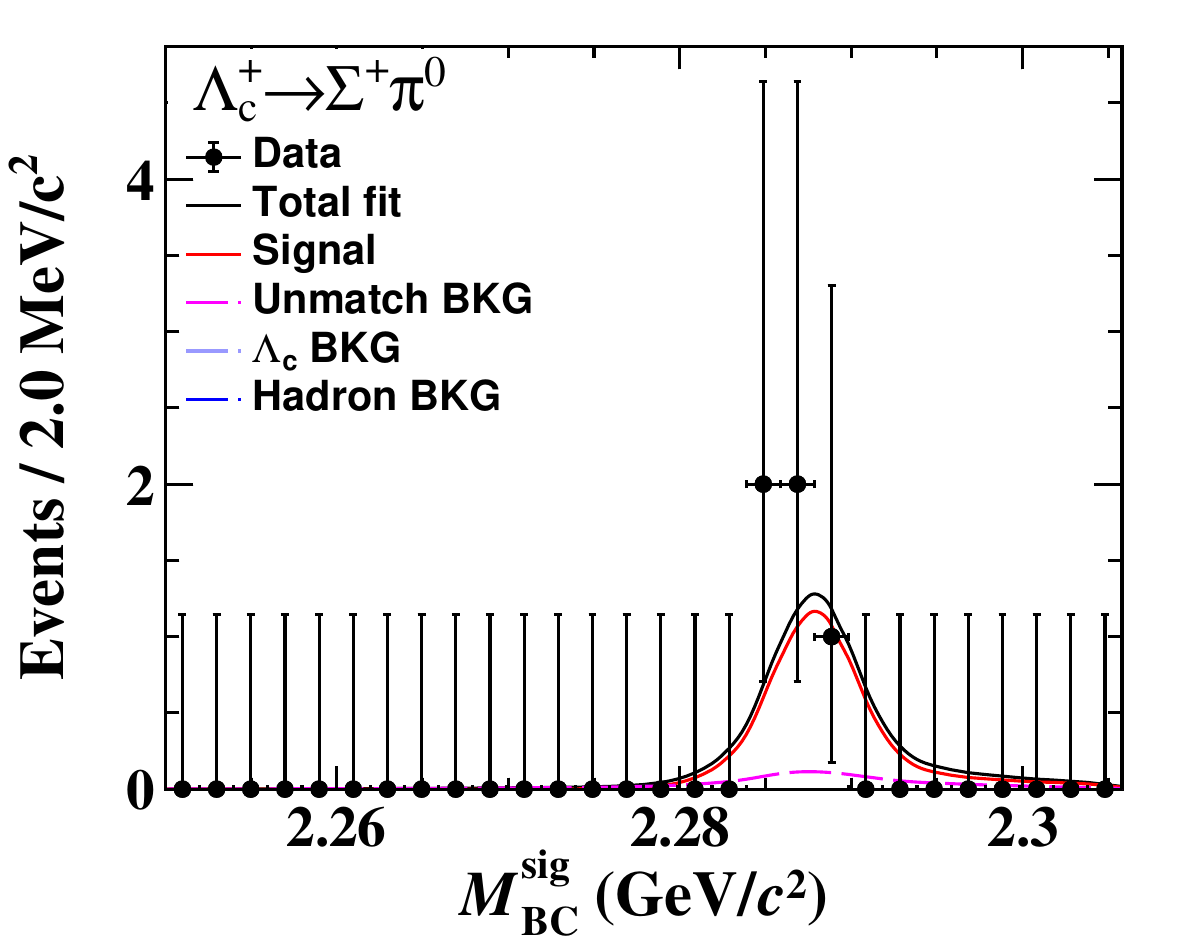}
  \includegraphics[width=0.24\textwidth]{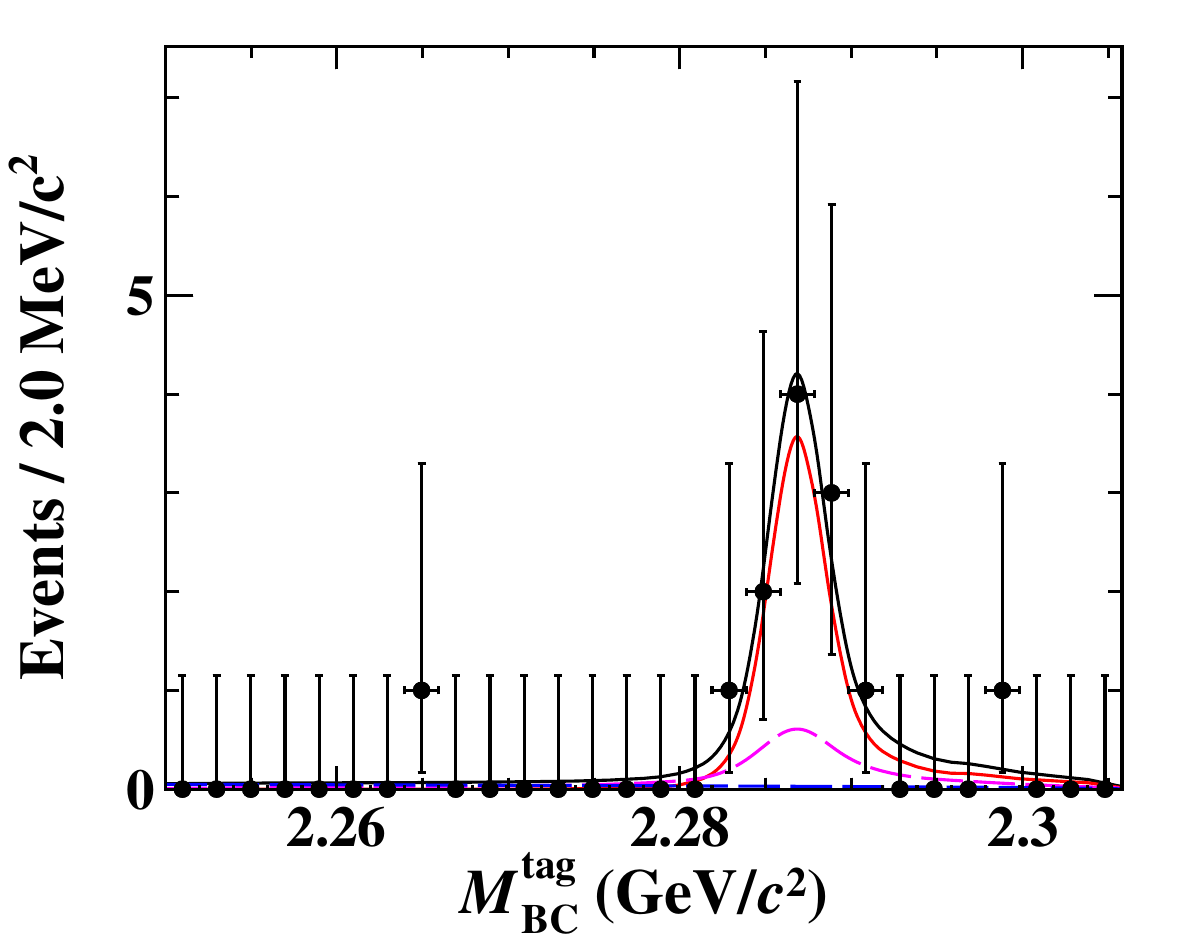}
  \includegraphics[width=0.24\textwidth]{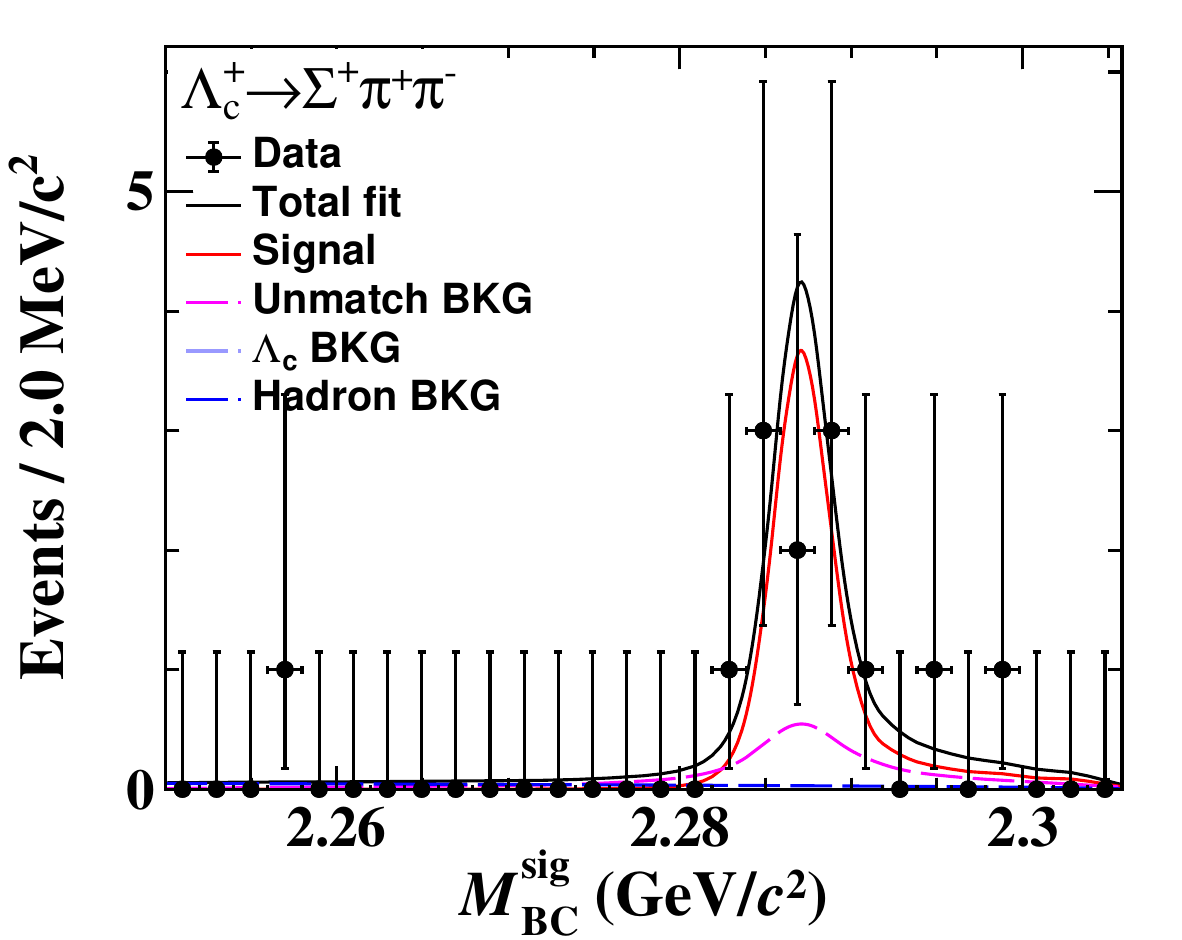}
  \includegraphics[width=0.24\textwidth]{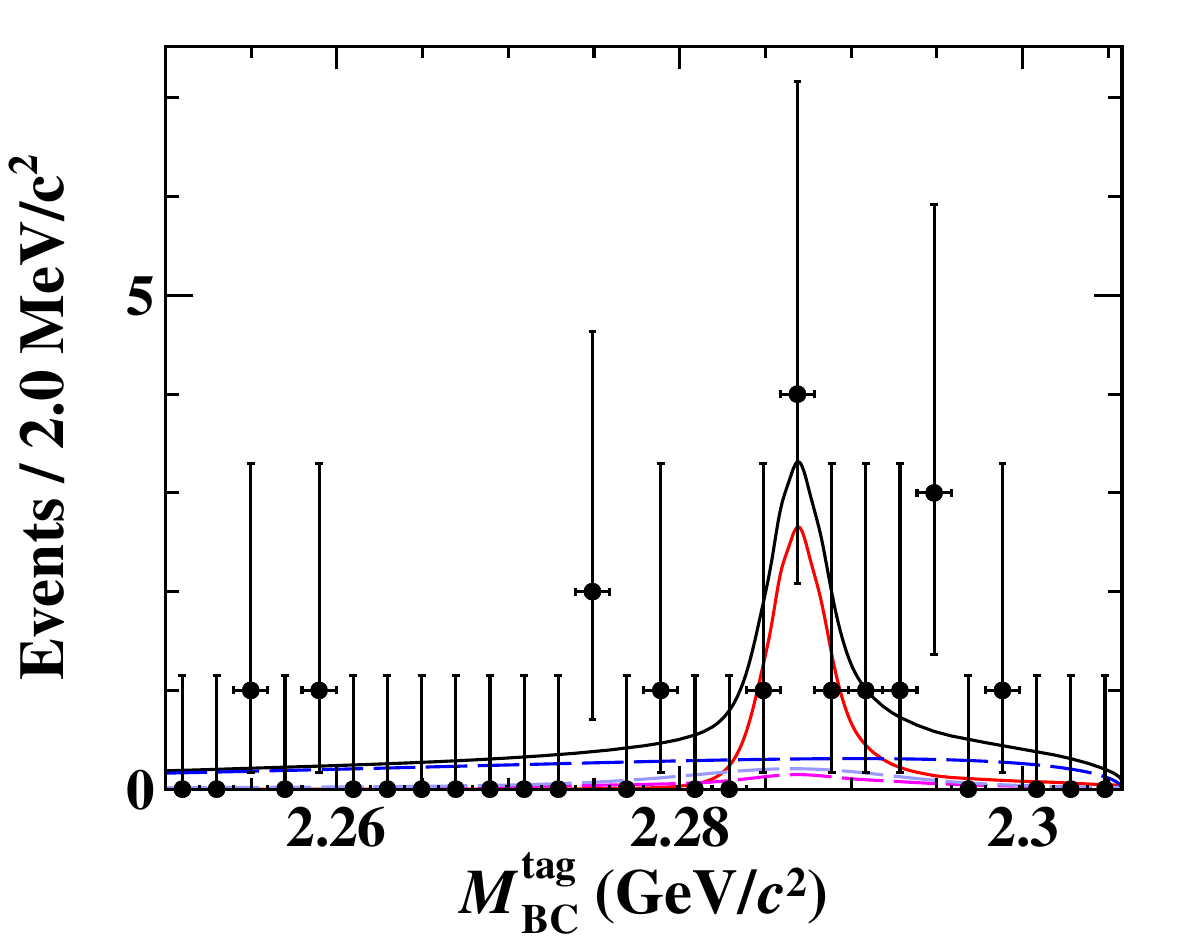}
  \includegraphics[width=0.24\textwidth]{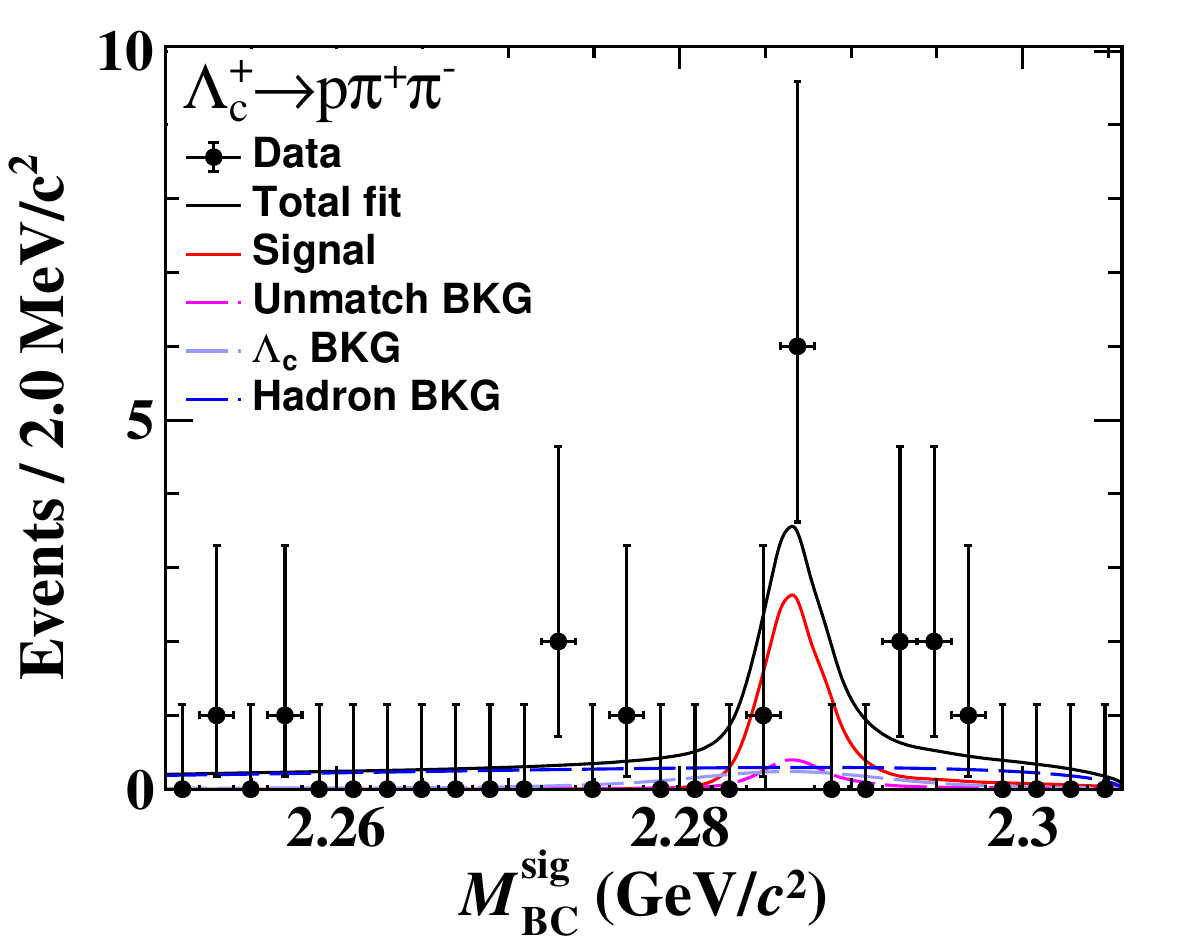}
     \vspace*{-0.5cm}
  \end{center}
\caption{The projections of the 2D fits on the $M_{\rm BC}^{\rm tag}$ and $M_{\rm BC}^{\rm sig}$ distributions of the accepted DT candidates at $\sqrt{s}=4611.86~\mev$. The plots in the first and third columns show the combined 12 tag modes for each signal mode. 
The points with error bars are data, the black lines are the sum of fit functions, the red lines are the matched signal shapes, the pink dashed lines are the unmatched signal shapes, the lilac dashed lines are the non-signal $\lcp\lcm$ shapes, and the blue dashed lines are the ARGUS functions.}
\label{fig:DT_yield_4612}
\end{figure}

\begin{figure}[!htbp]
  \begin{center}
  
  \includegraphics[width=0.24\textwidth]{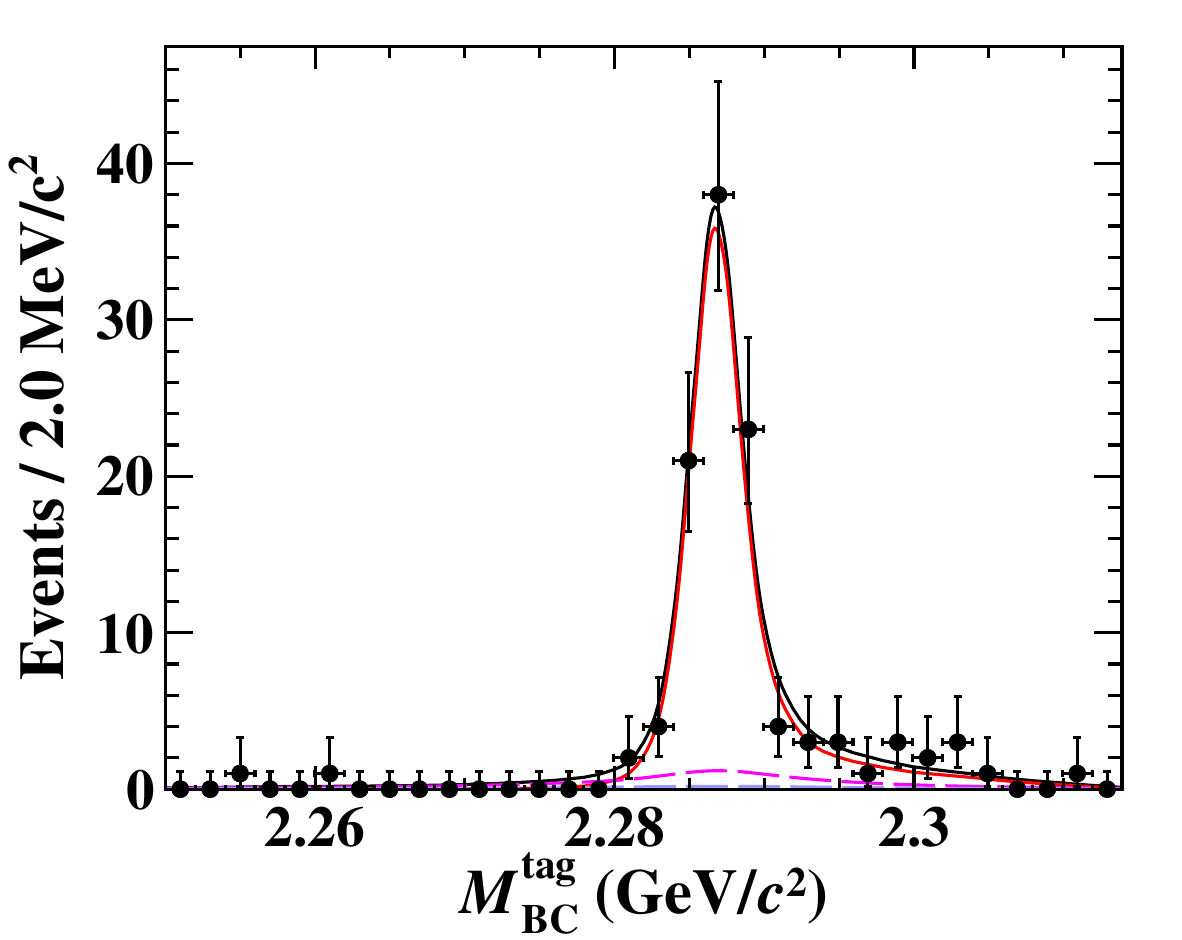}
  \includegraphics[width=0.24\textwidth]{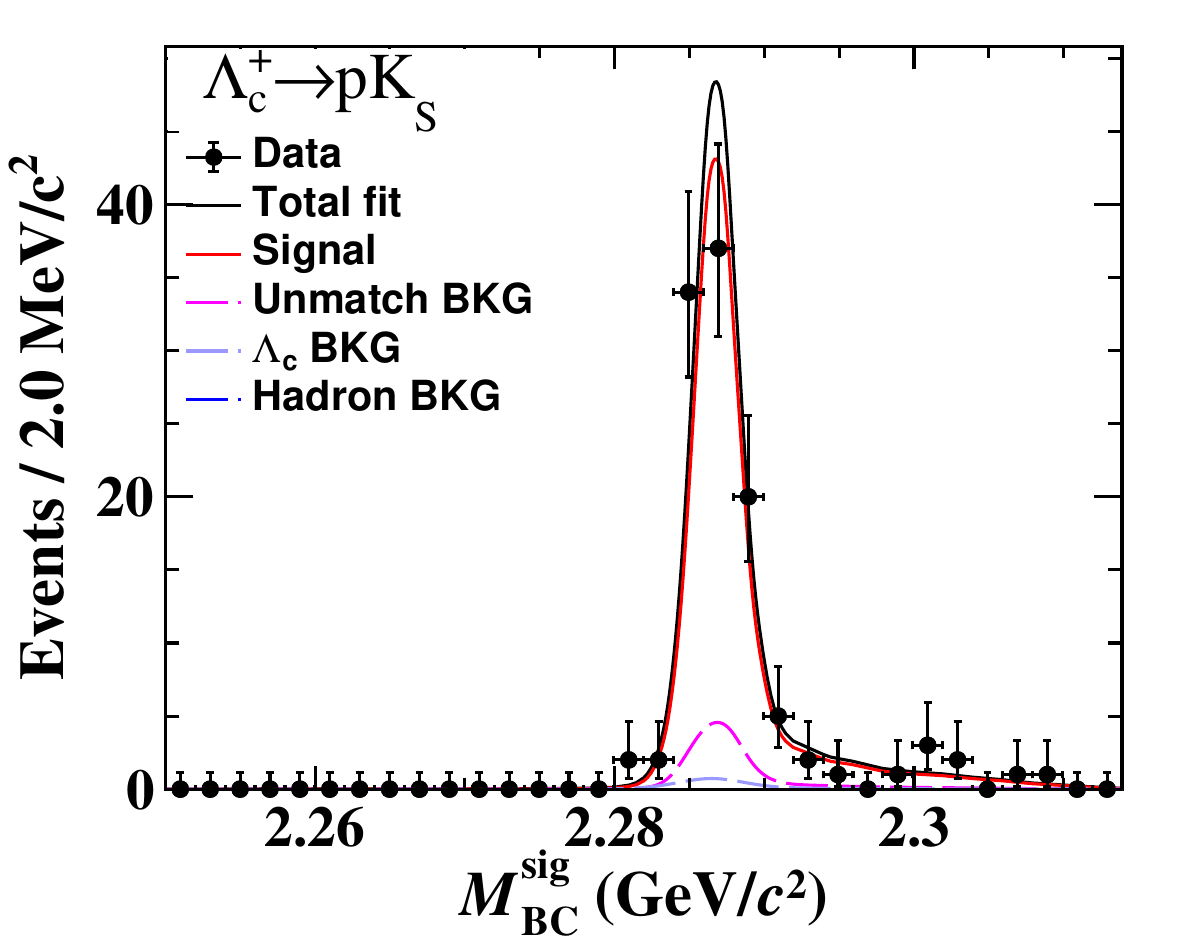}
  \includegraphics[width=0.24\textwidth]{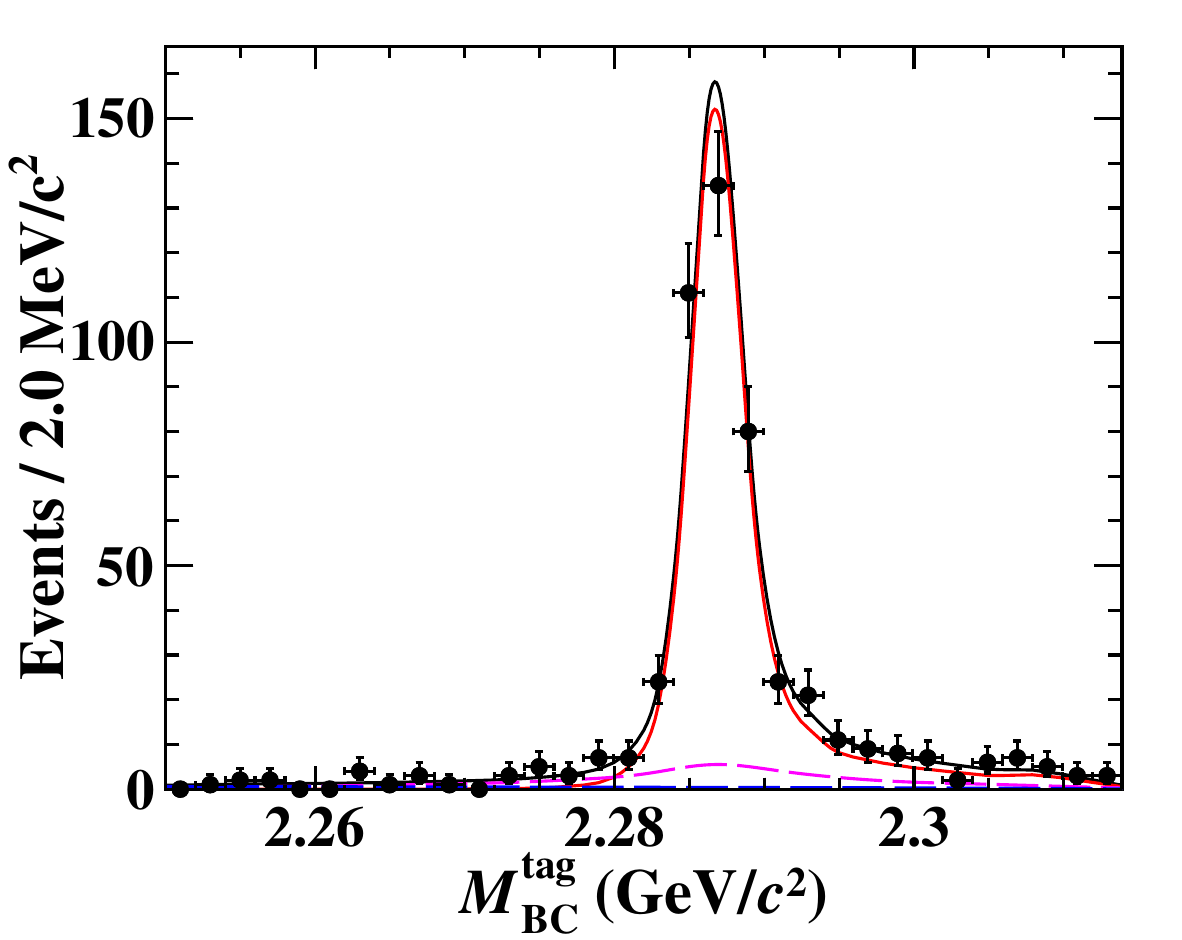}
  \includegraphics[width=0.24\textwidth]{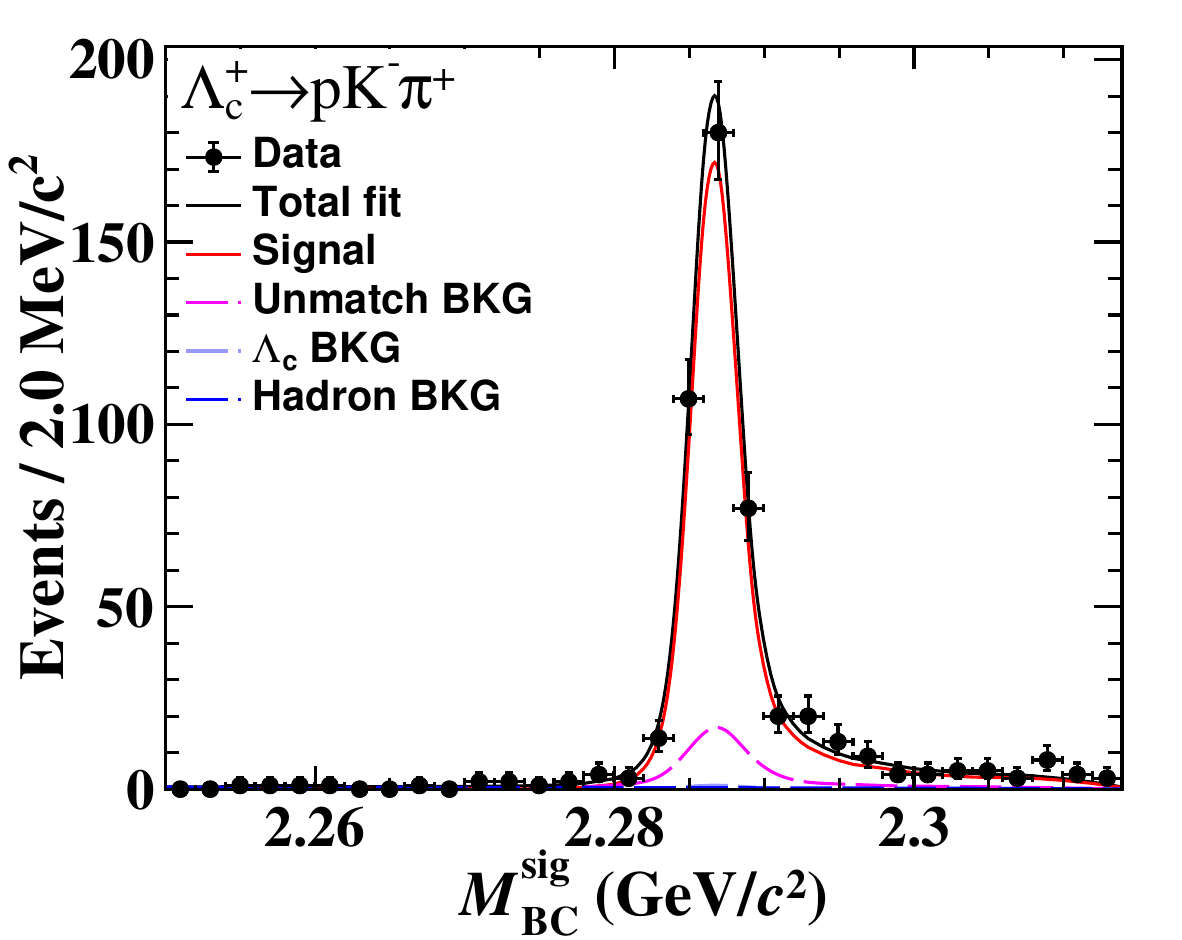}
  \includegraphics[width=0.24\textwidth]{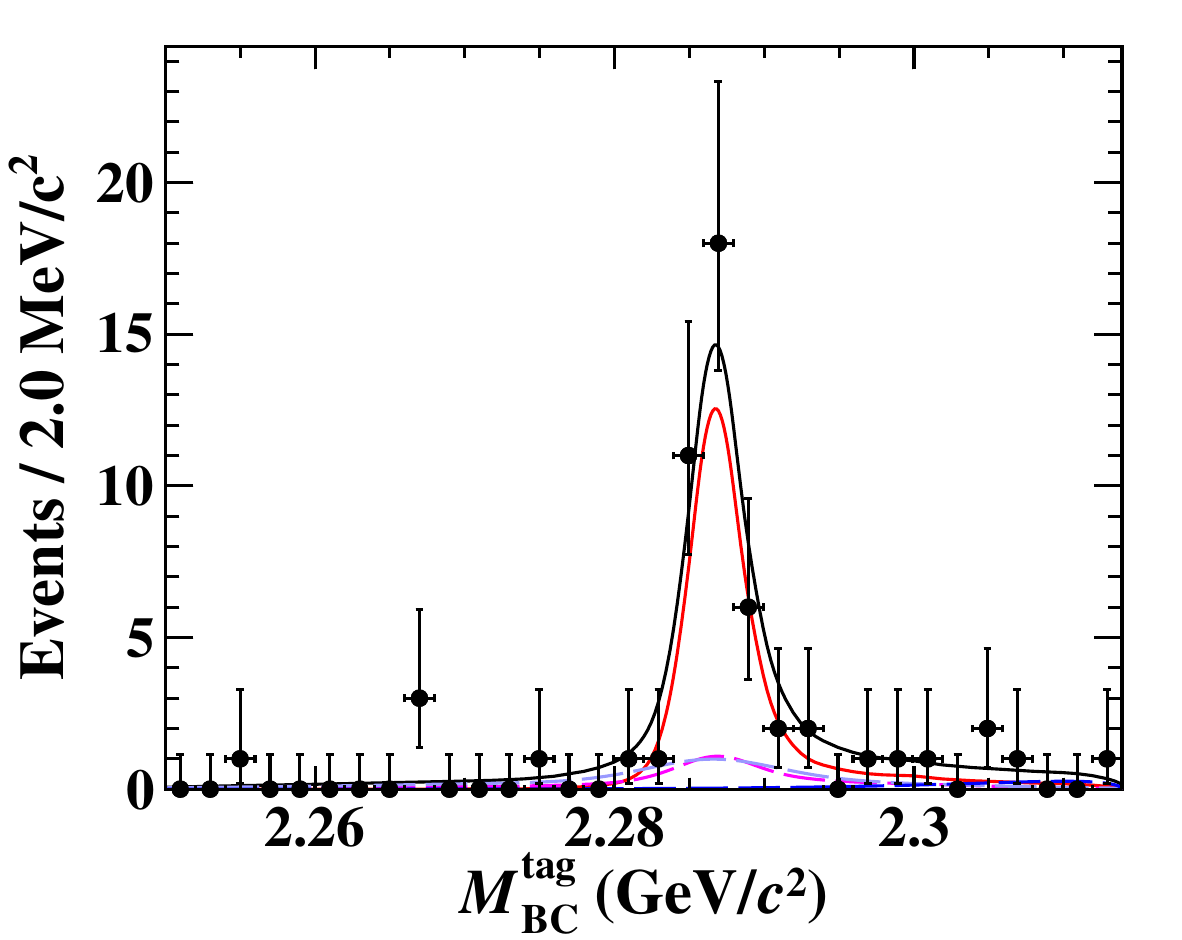}
  \includegraphics[width=0.24\textwidth]{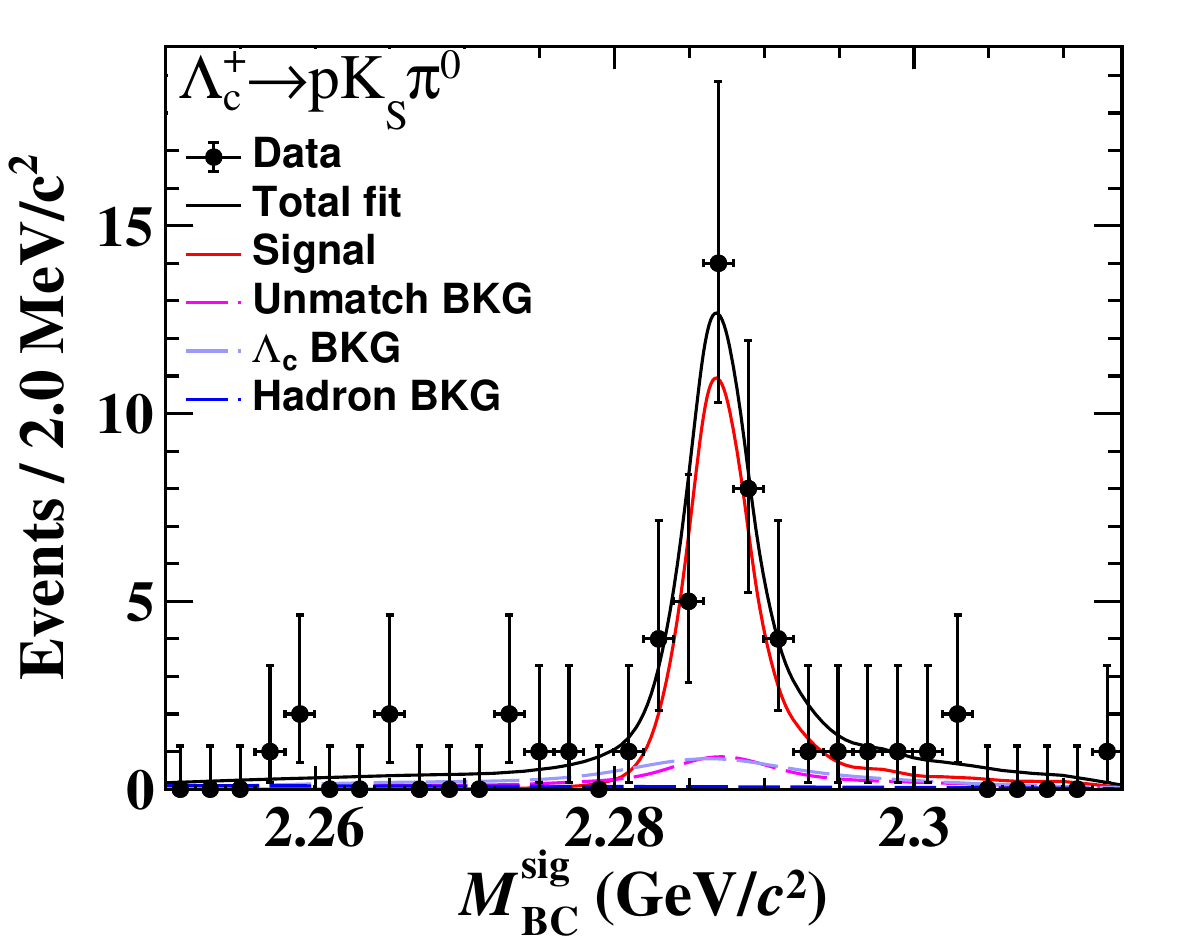}
  \includegraphics[width=0.24\textwidth]{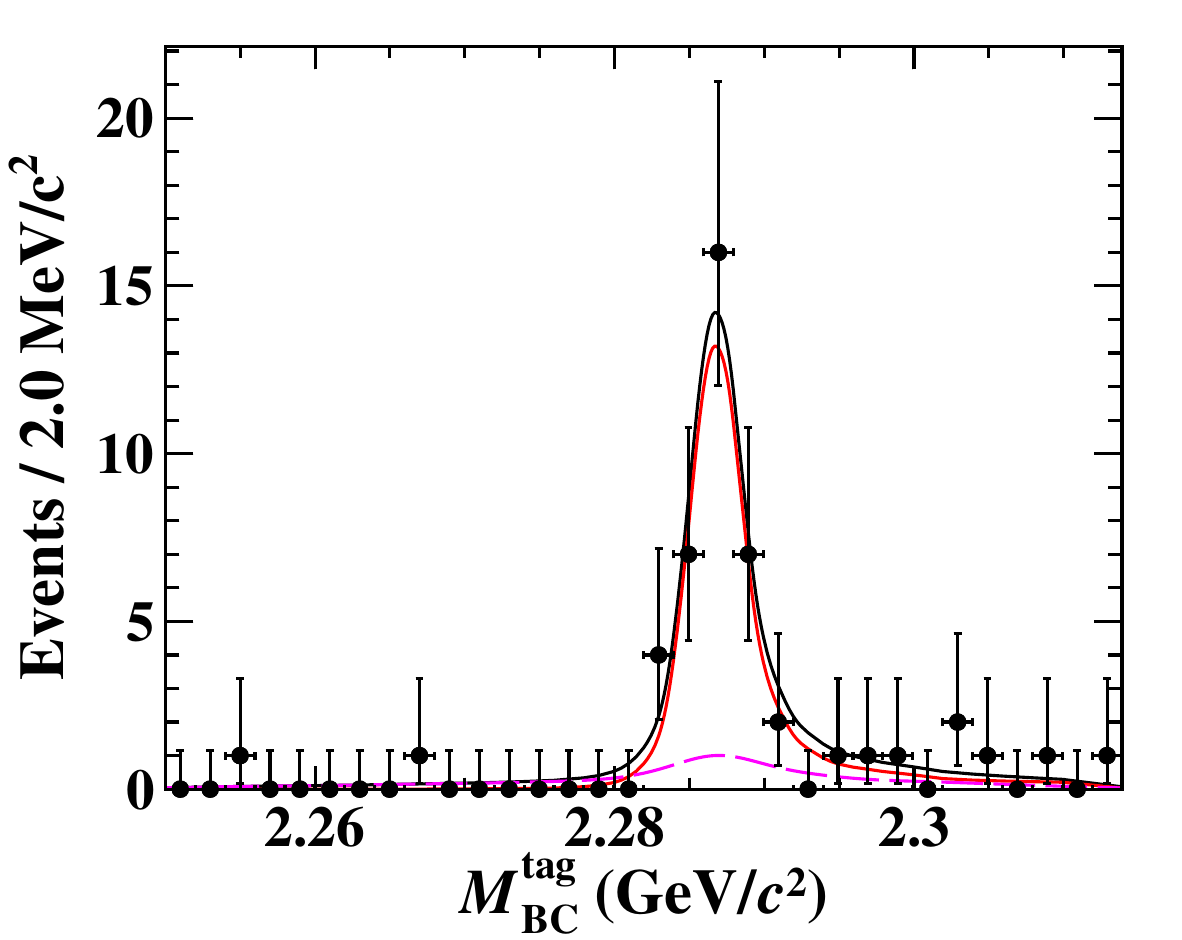}
  \includegraphics[width=0.24\textwidth]{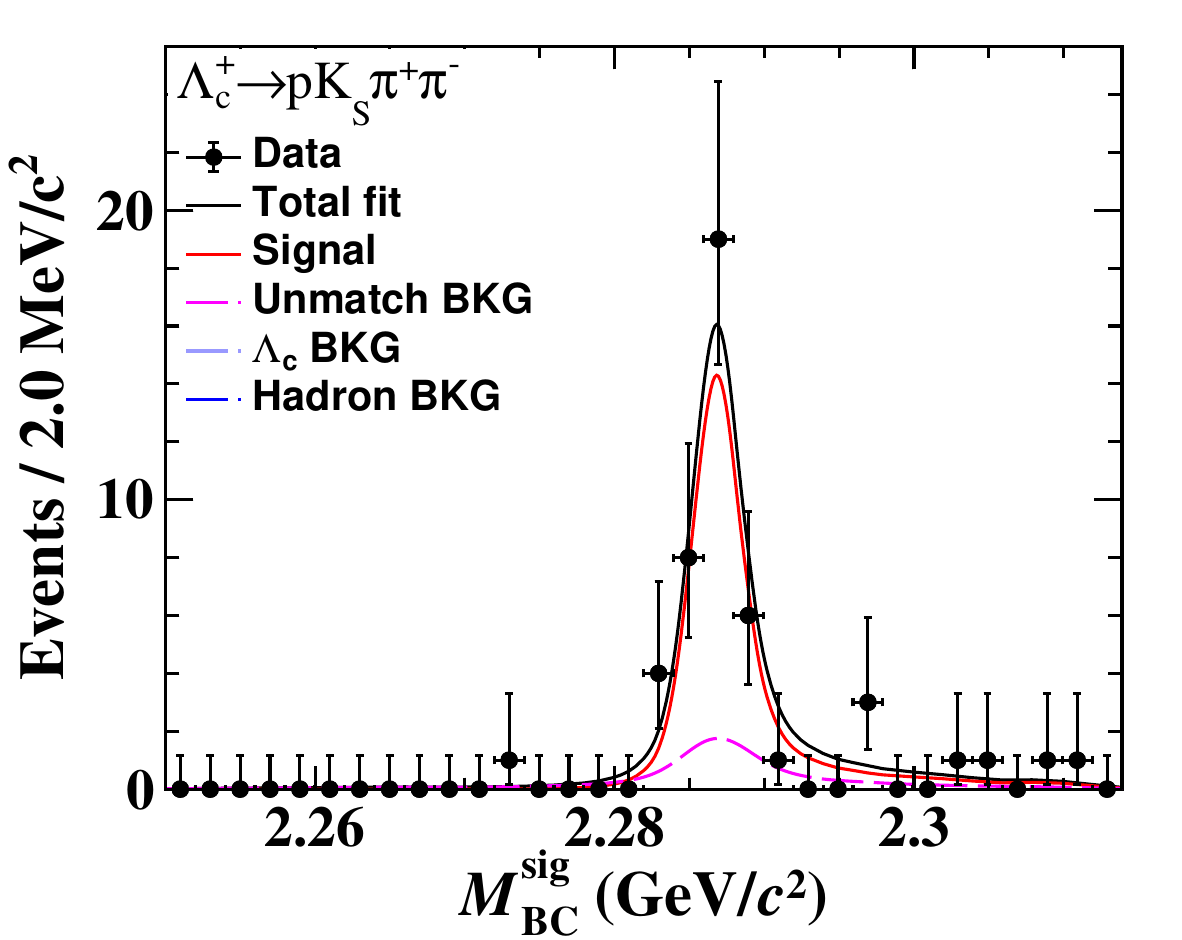}
  \includegraphics[width=0.24\textwidth]{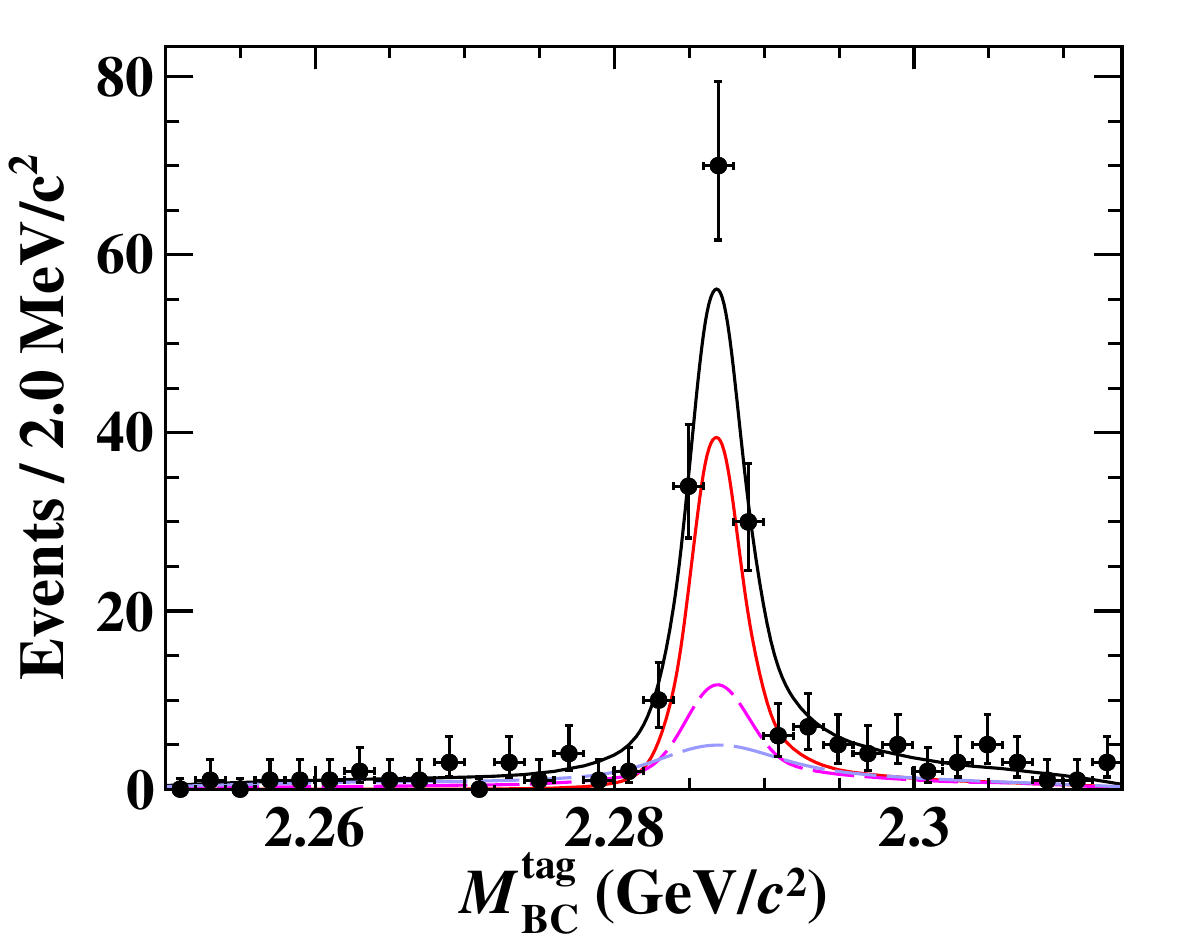}
  \includegraphics[width=0.24\textwidth]{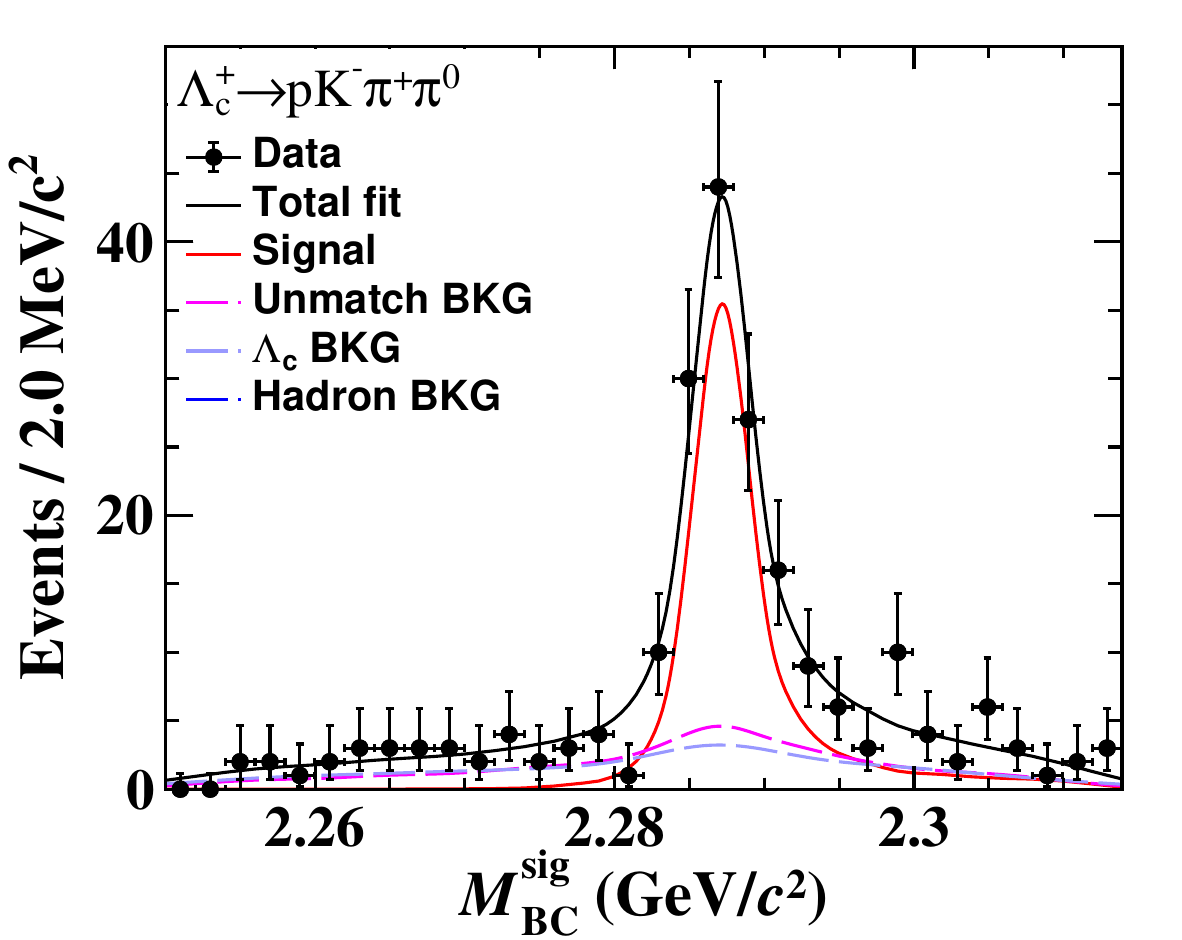}
  \includegraphics[width=0.24\textwidth]{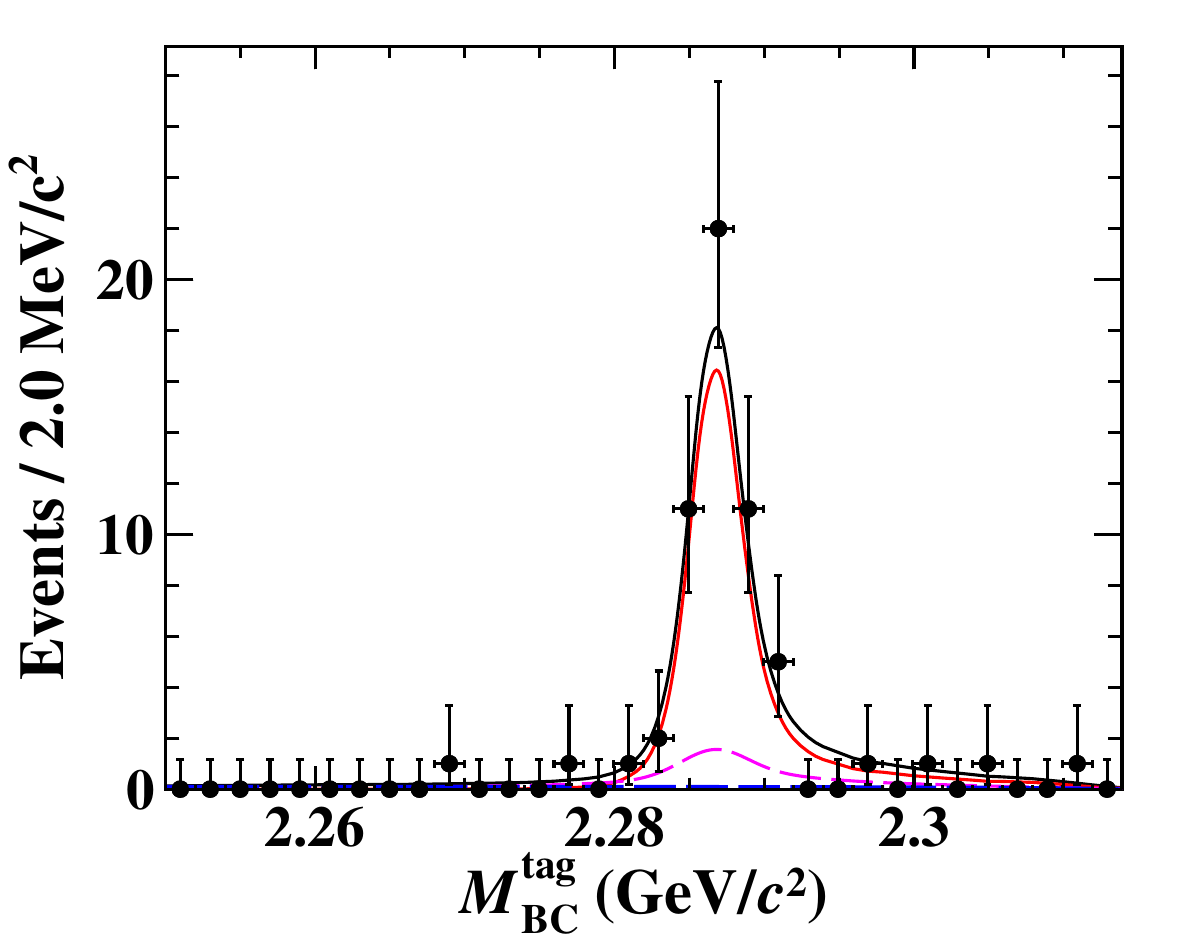}
  \includegraphics[width=0.24\textwidth]{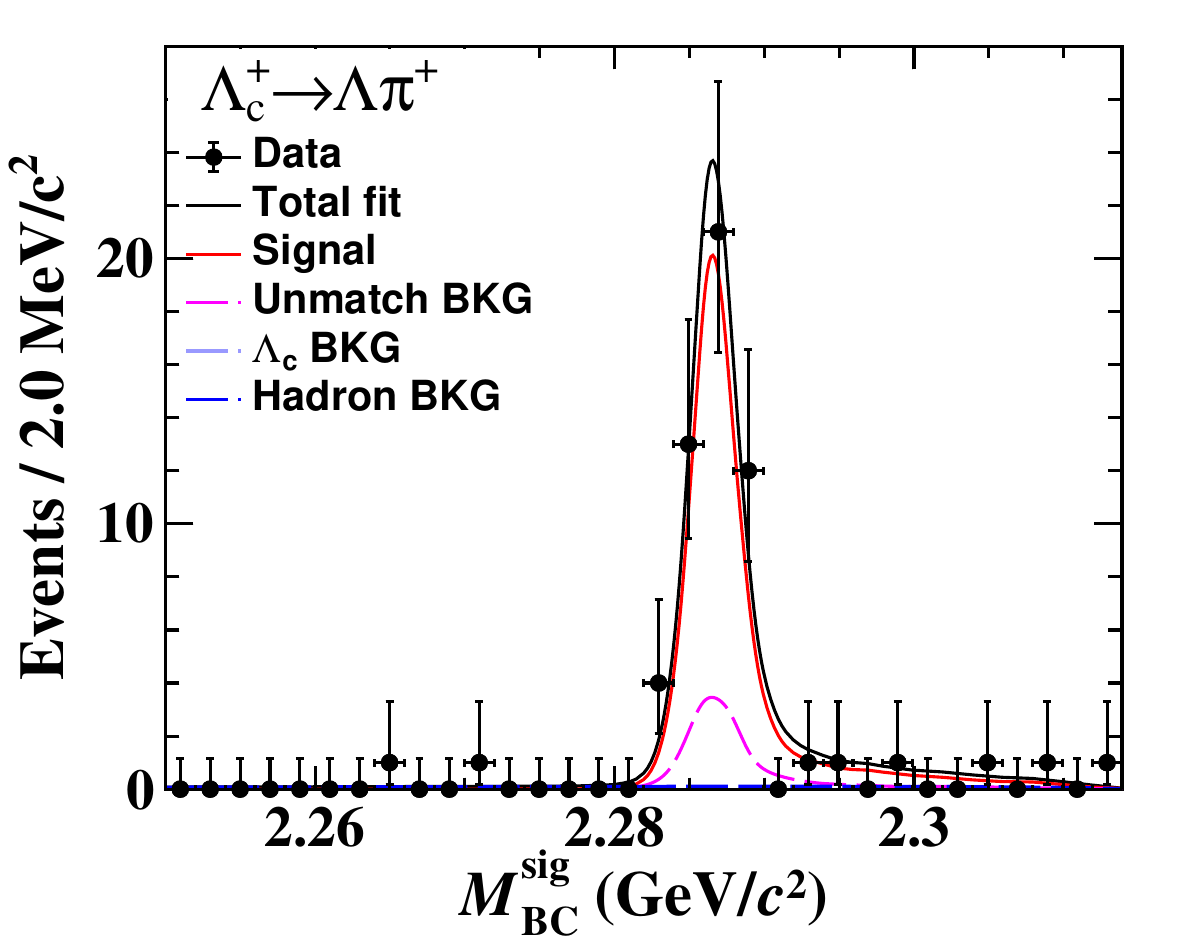}
  \includegraphics[width=0.24\textwidth]{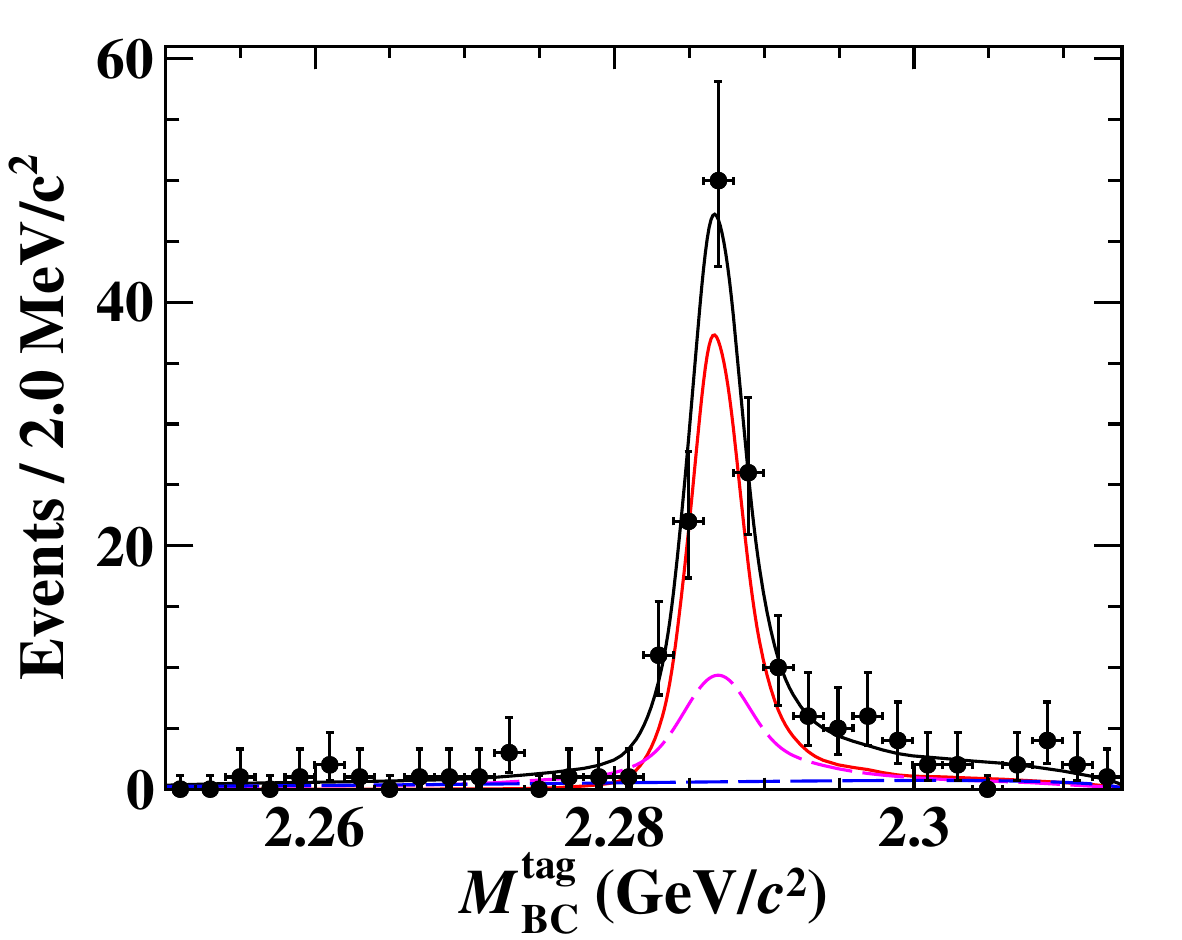}
  \includegraphics[width=0.24\textwidth]{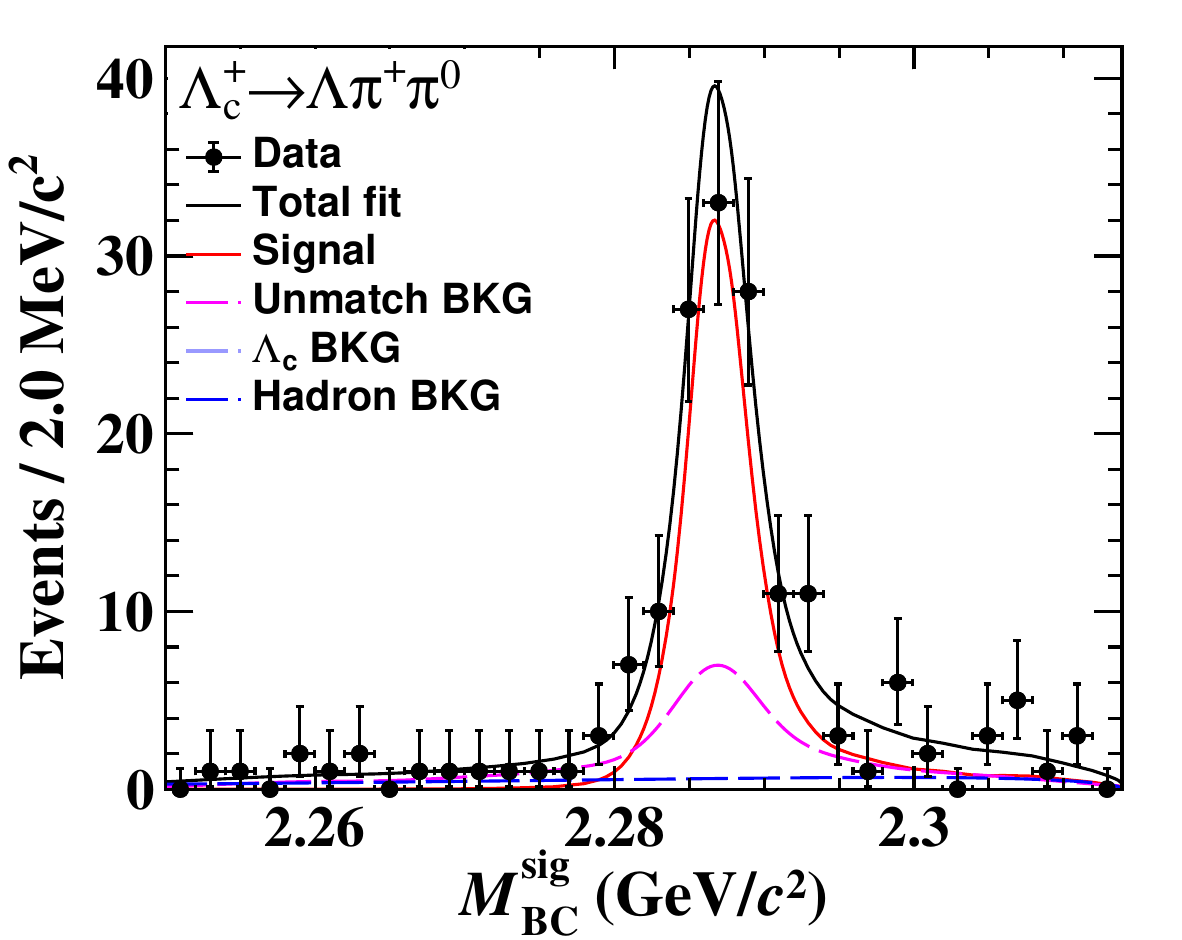}
  \includegraphics[width=0.24\textwidth]{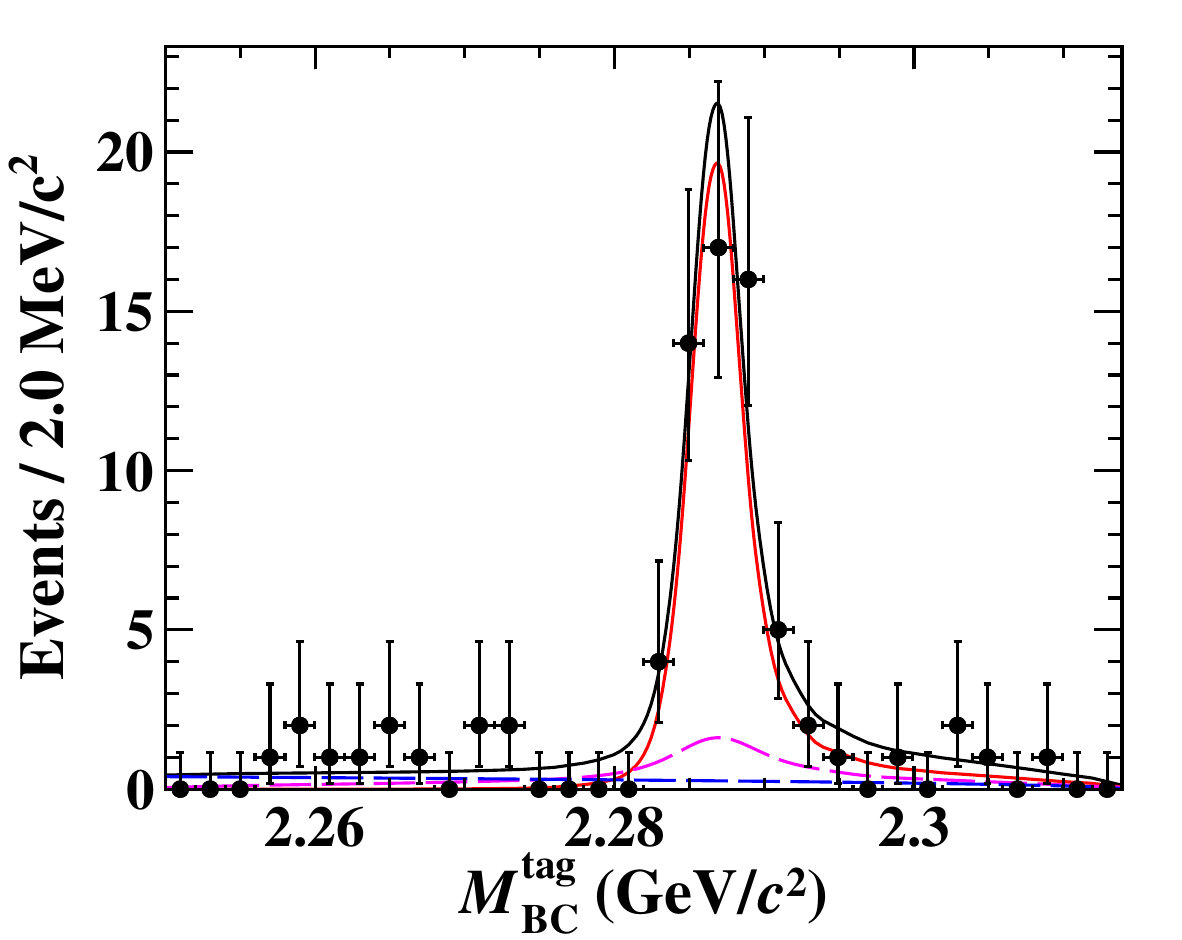}
  \includegraphics[width=0.24\textwidth]{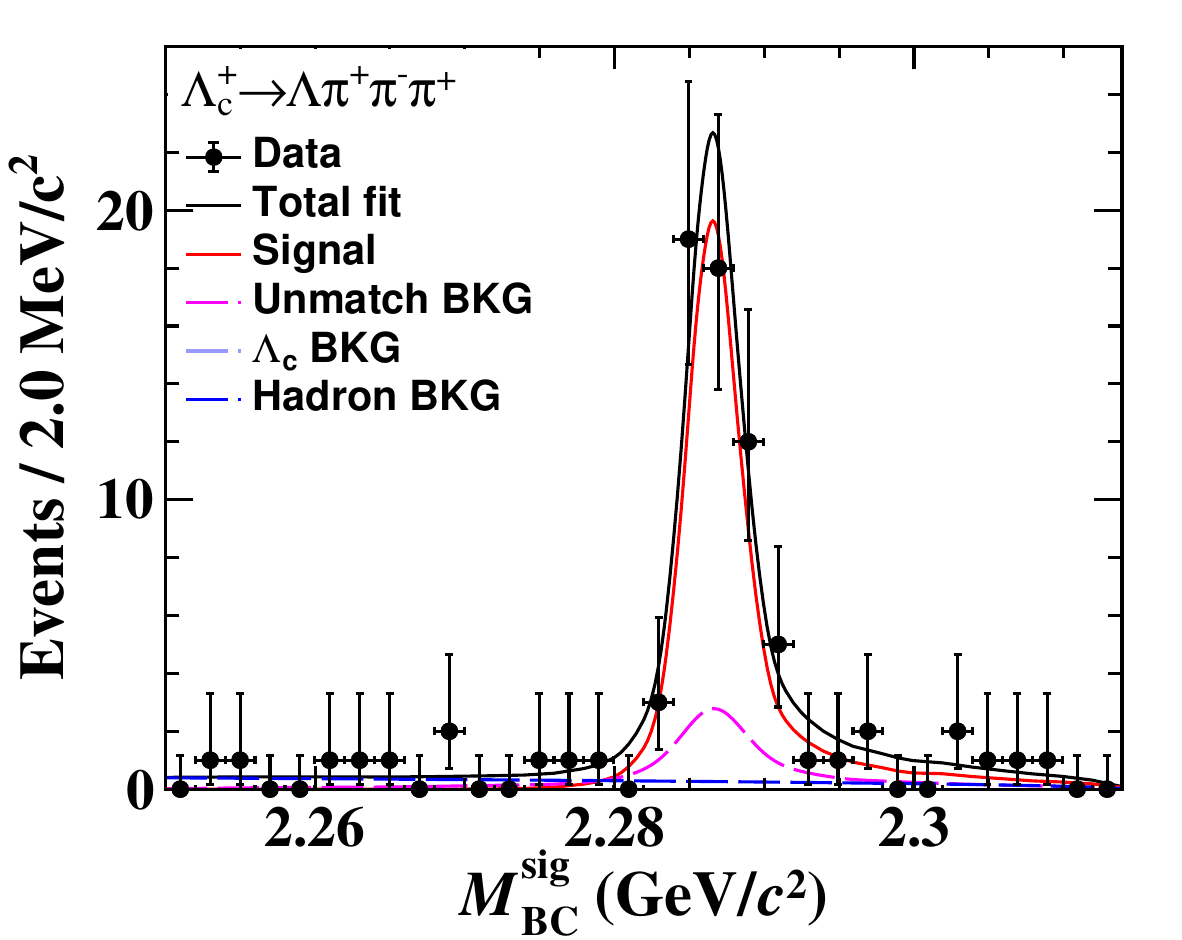}
  \includegraphics[width=0.24\textwidth]{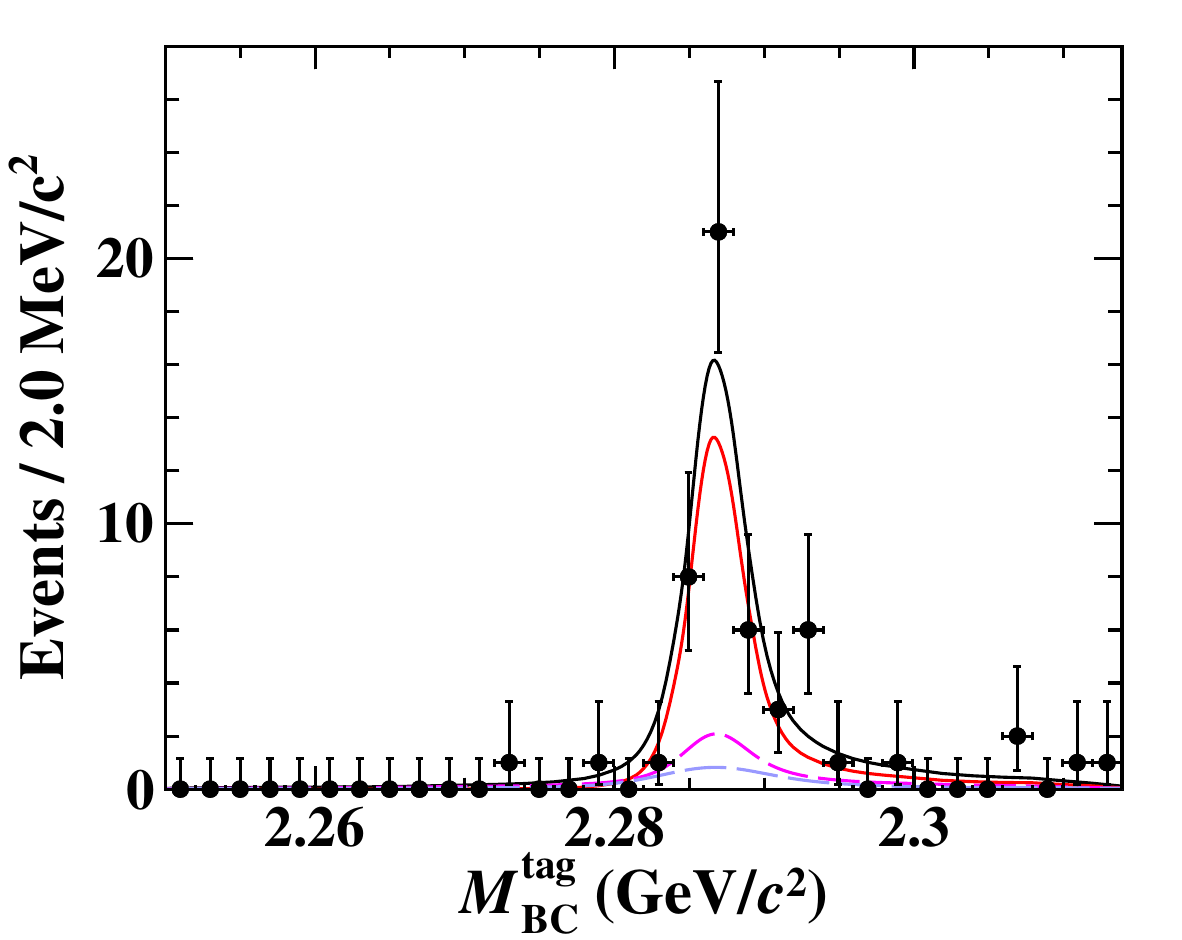}
  \includegraphics[width=0.24\textwidth]{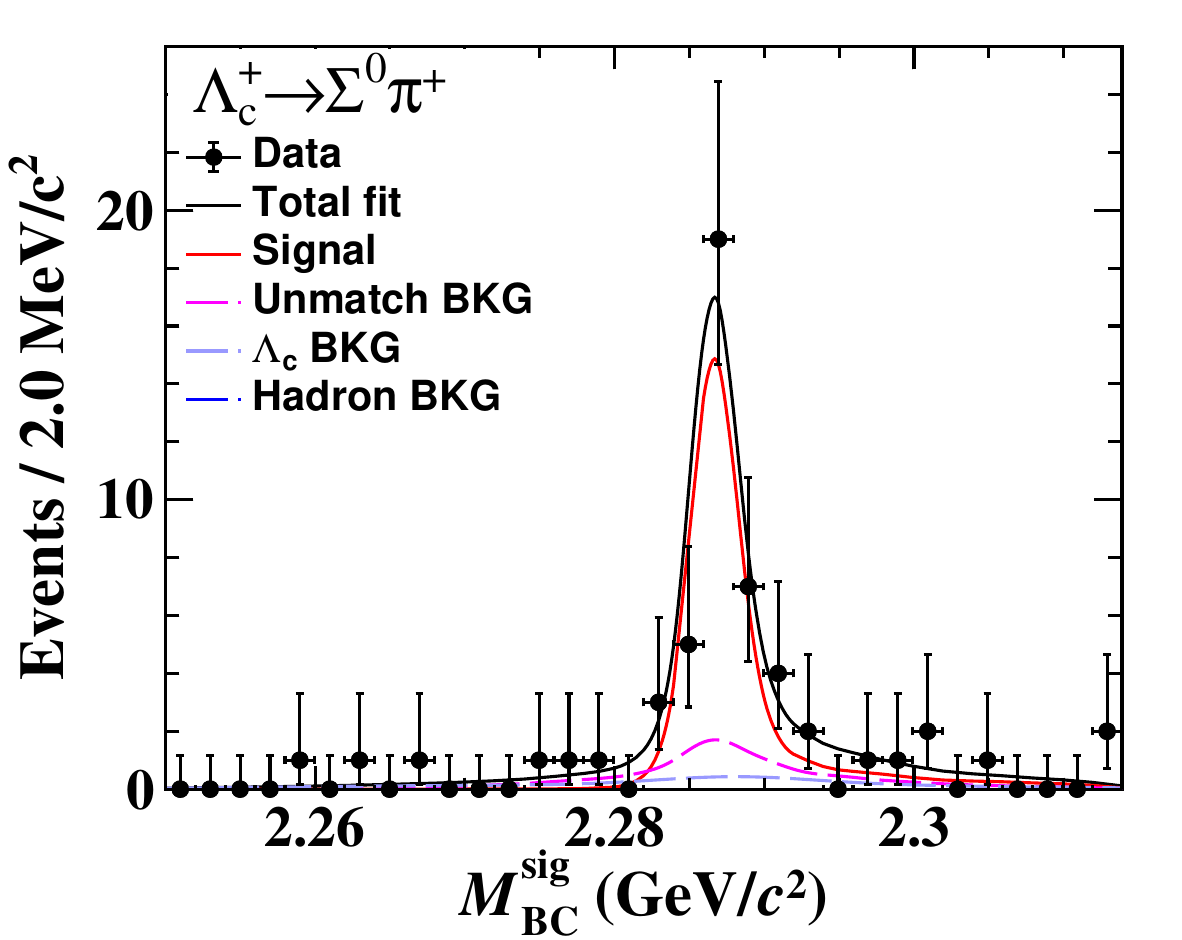}
  \includegraphics[width=0.24\textwidth]{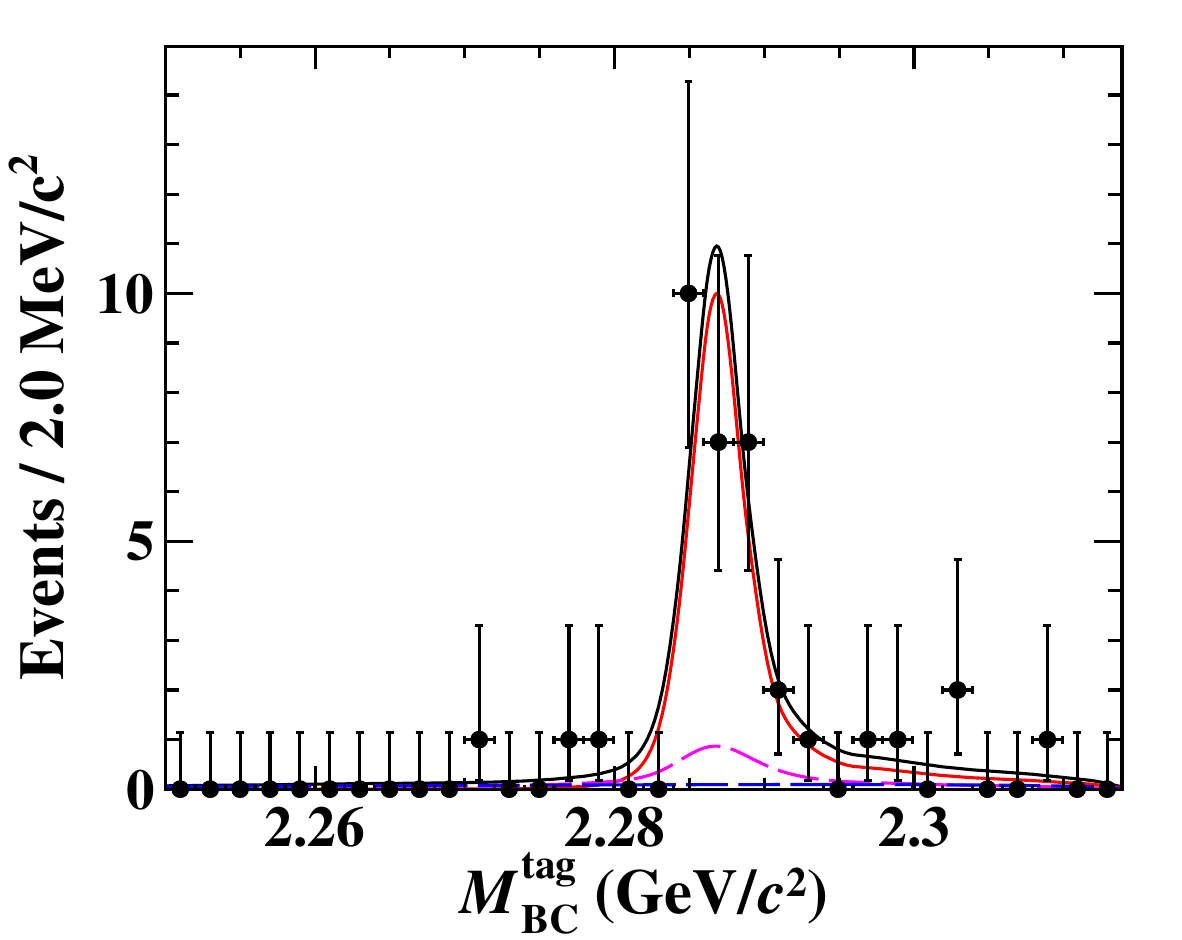}
  \includegraphics[width=0.24\textwidth]{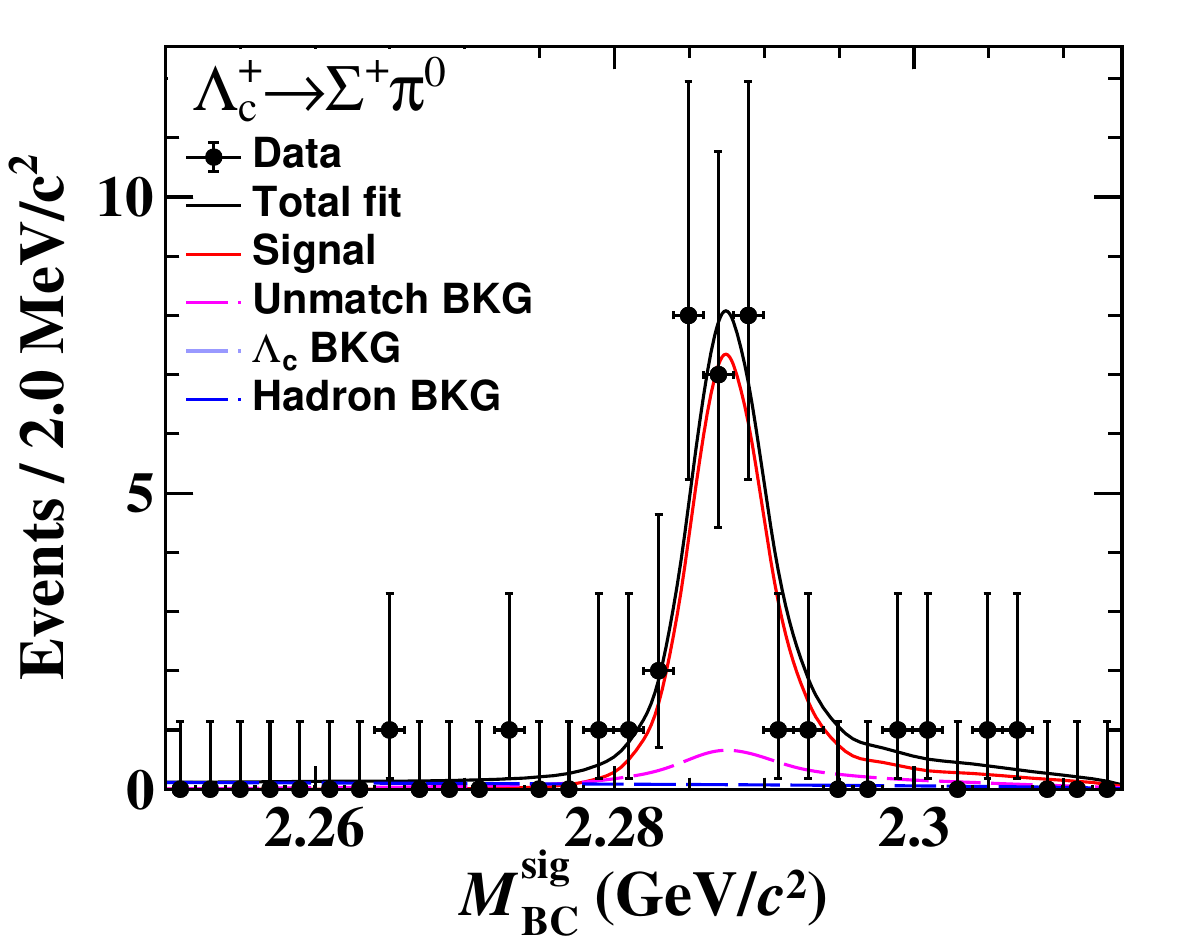}
  \includegraphics[width=0.24\textwidth]{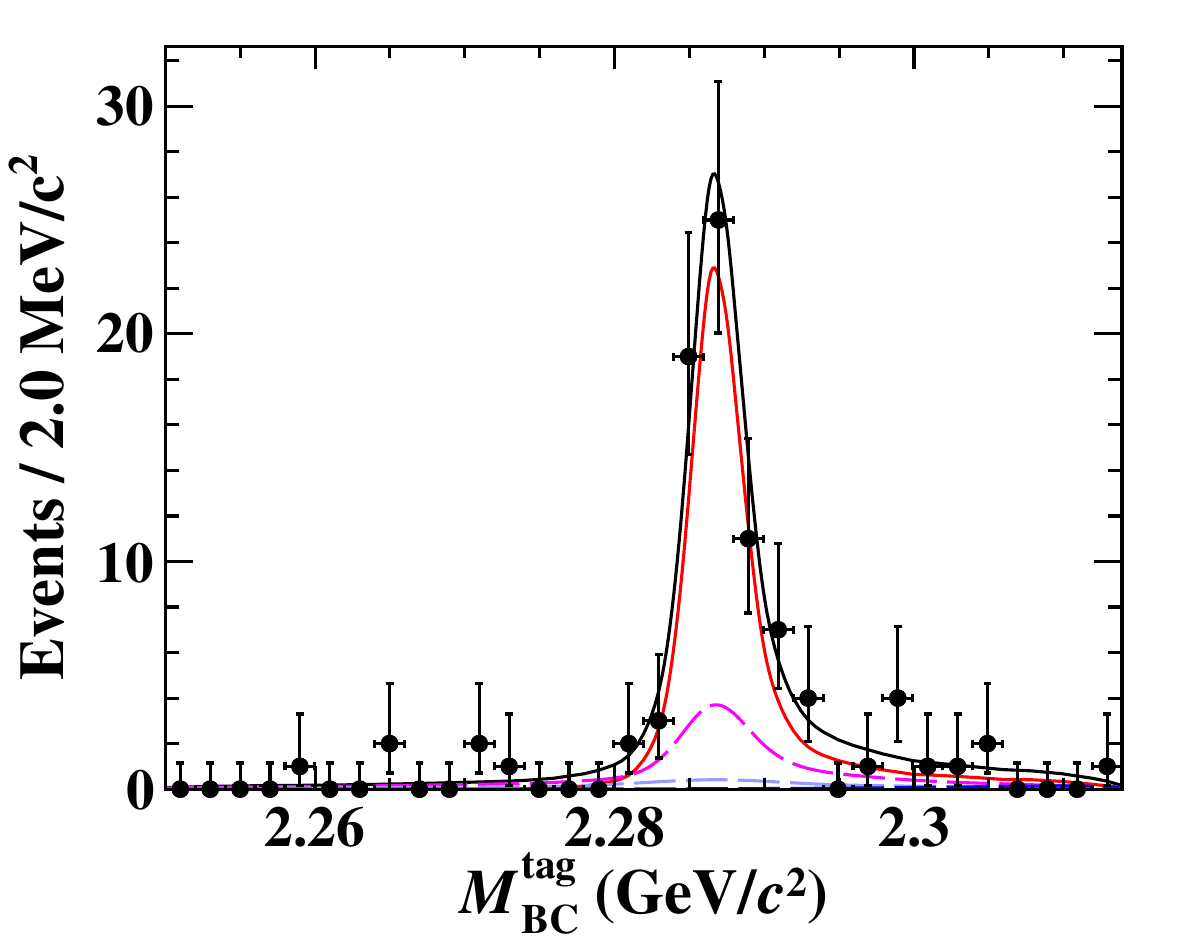}
  \includegraphics[width=0.24\textwidth]{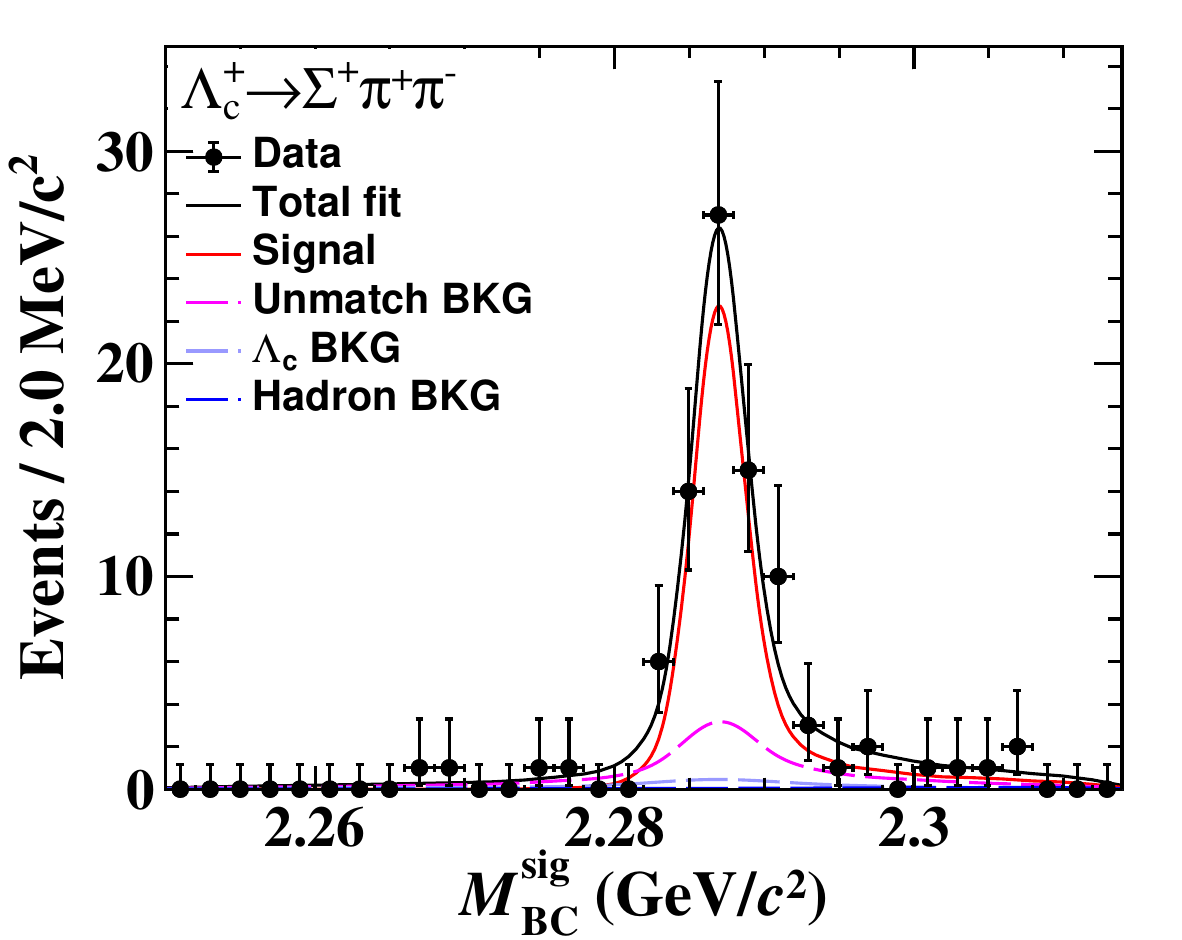}
  \includegraphics[width=0.24\textwidth]{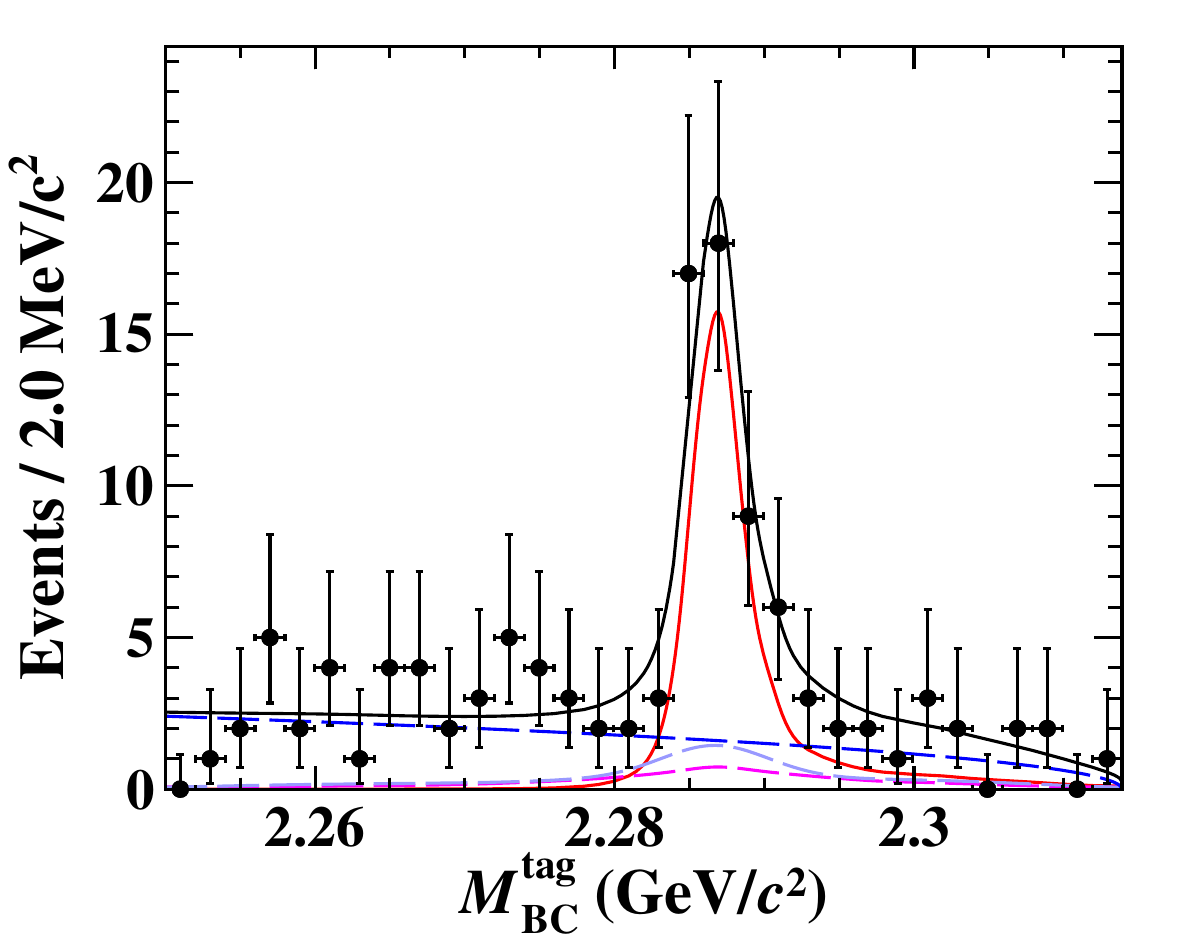}
  \includegraphics[width=0.24\textwidth]{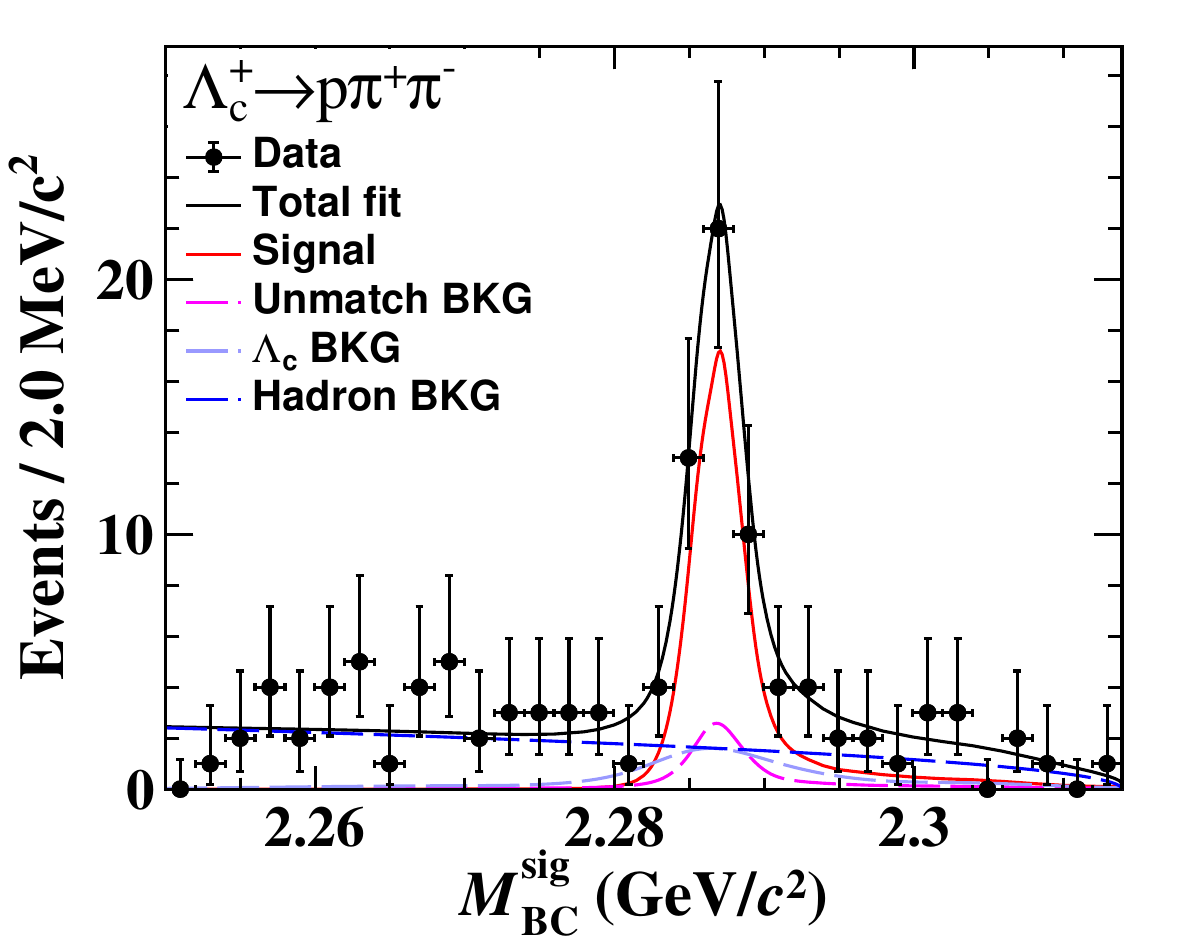}
     \vspace*{-0.5cm}
  \end{center}
\caption{The projections of the 2D fits on the $M_{\rm BC}^{\rm tag}$ and $M_{\rm BC}^{\rm sig}$ distributions of the accepted DT candidates at $\sqrt{s}=4628.00~\mev$. The plots in the first and third columns show the combined 12 tag modes for each signal mode. 
The points with error bars are data, the black lines are the sum of fit functions, the red lines are the matched signal shapes, the pink dashed lines are the unmatched signal shapes, the lilac dashed lines are the non-signal $\lcp\lcm$ shapes, and the blue dashed lines are the ARGUS functions.}
\label{fig:DT_yield_4626}
\end{figure}

\begin{figure}[!htbp]
  \begin{center}
  
  \includegraphics[width=0.24\textwidth]{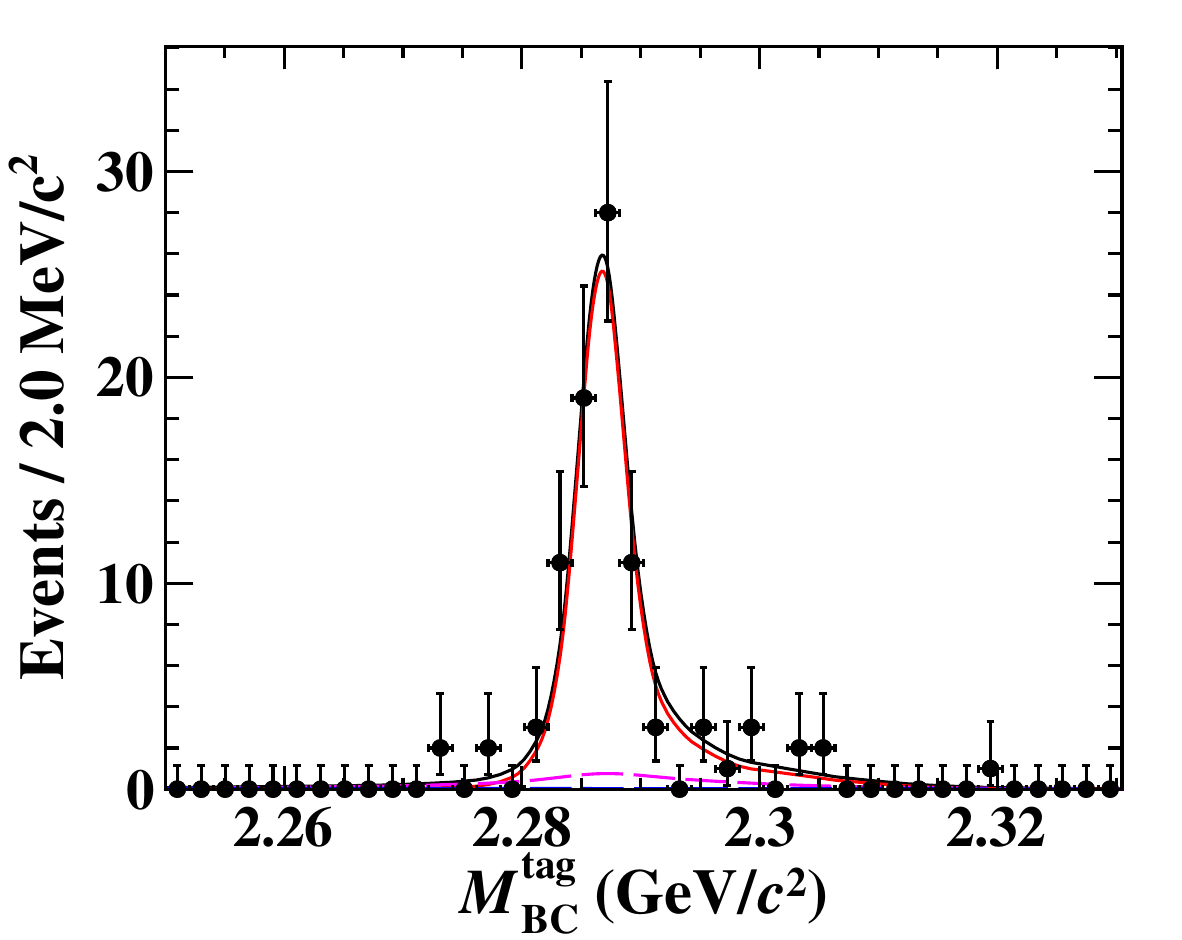}
  \includegraphics[width=0.24\textwidth]{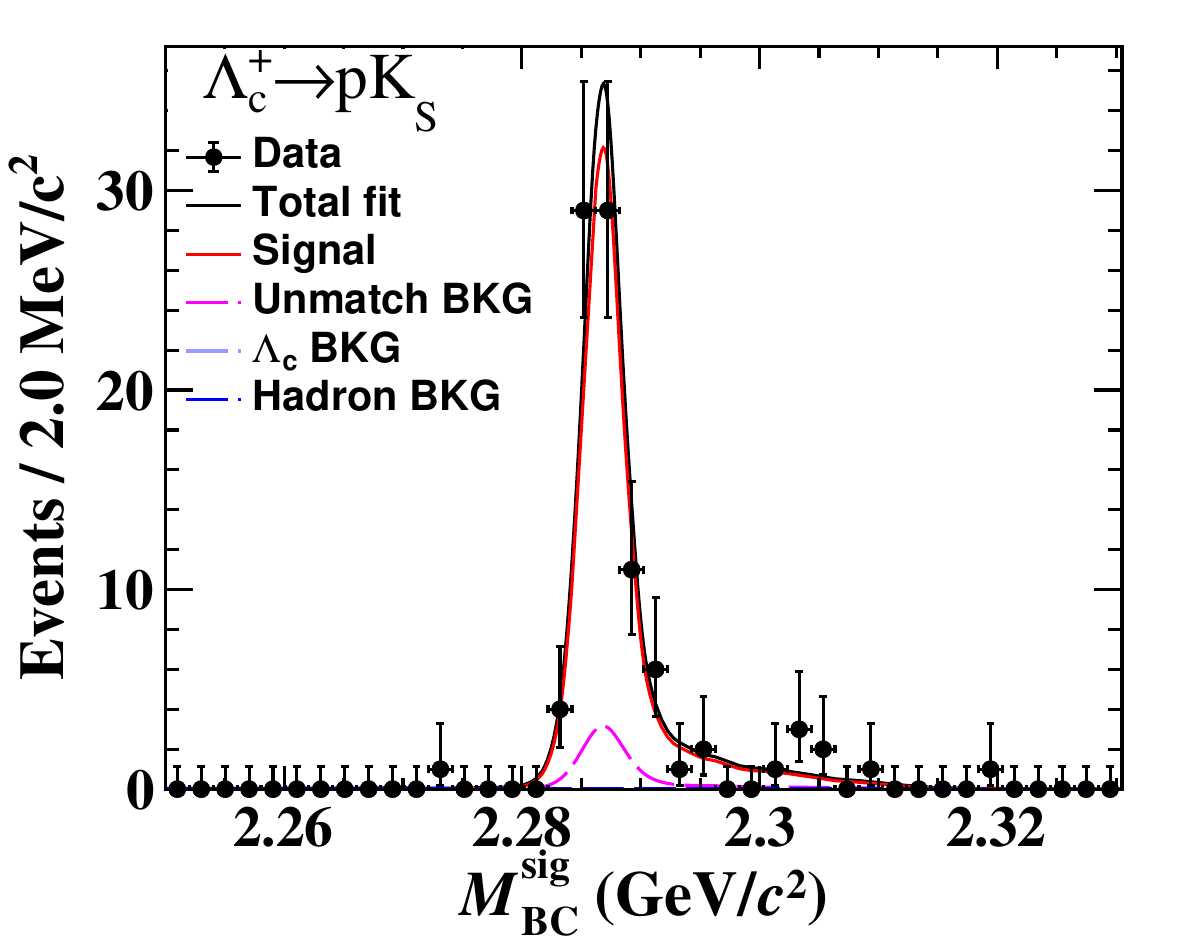}
  \includegraphics[width=0.24\textwidth]{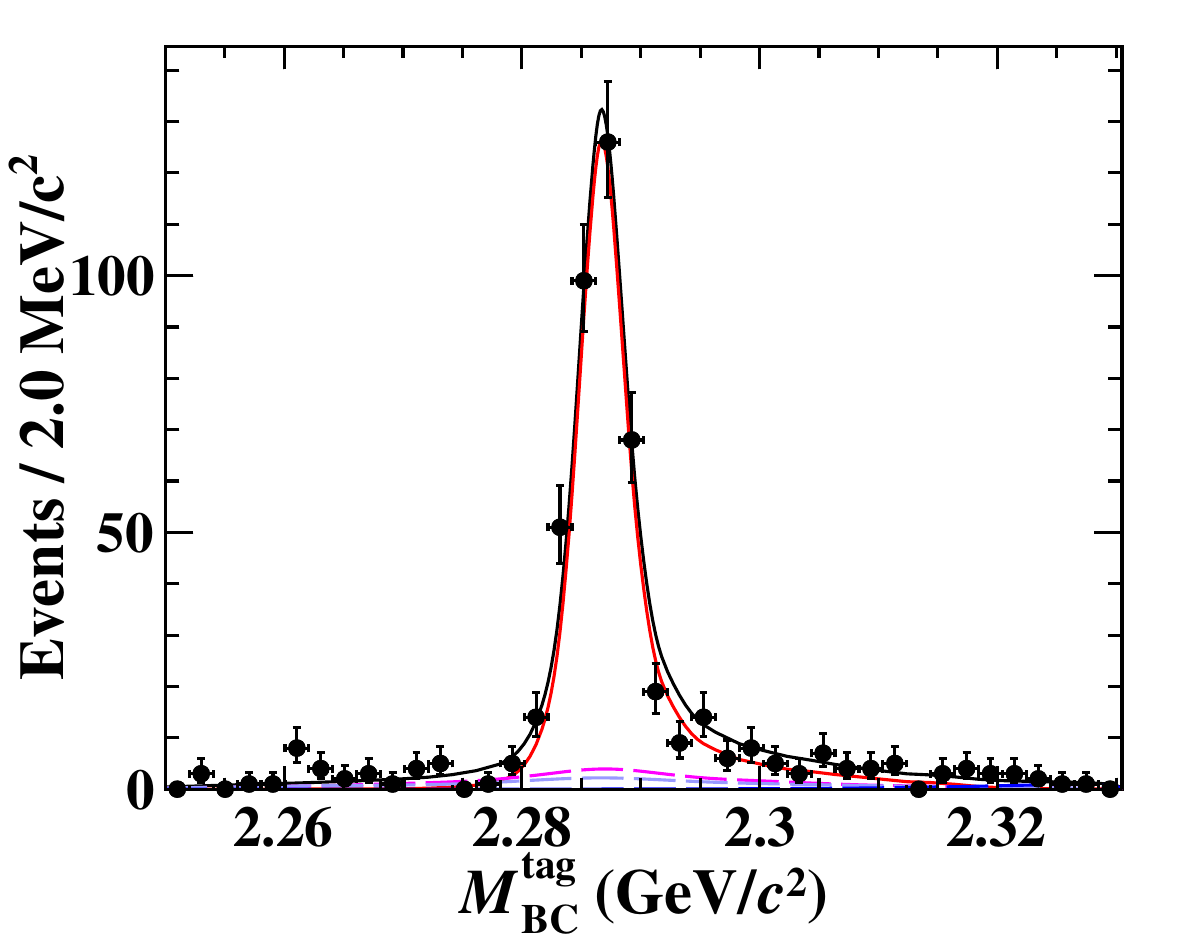}
  \includegraphics[width=0.24\textwidth]{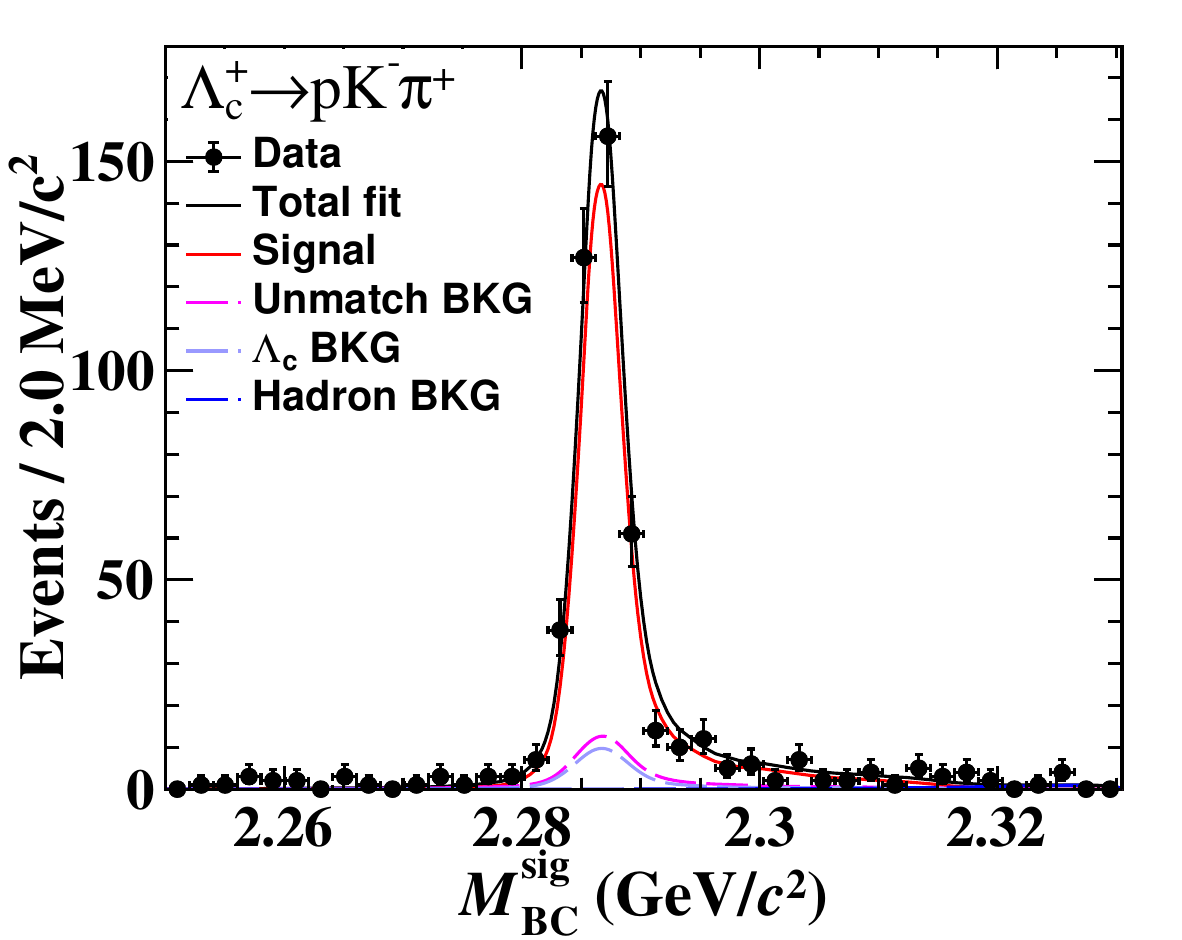}
  \includegraphics[width=0.24\textwidth]{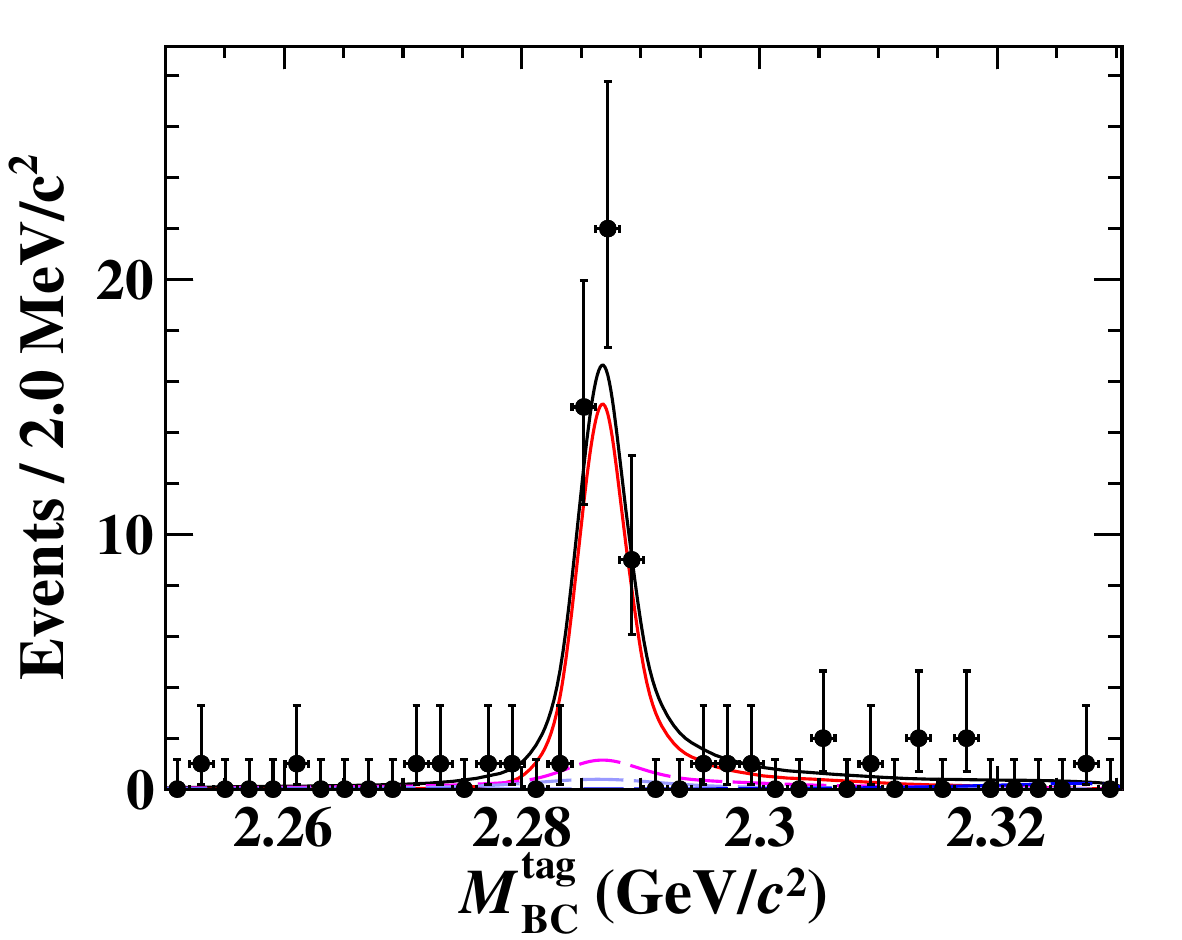}
  \includegraphics[width=0.24\textwidth]{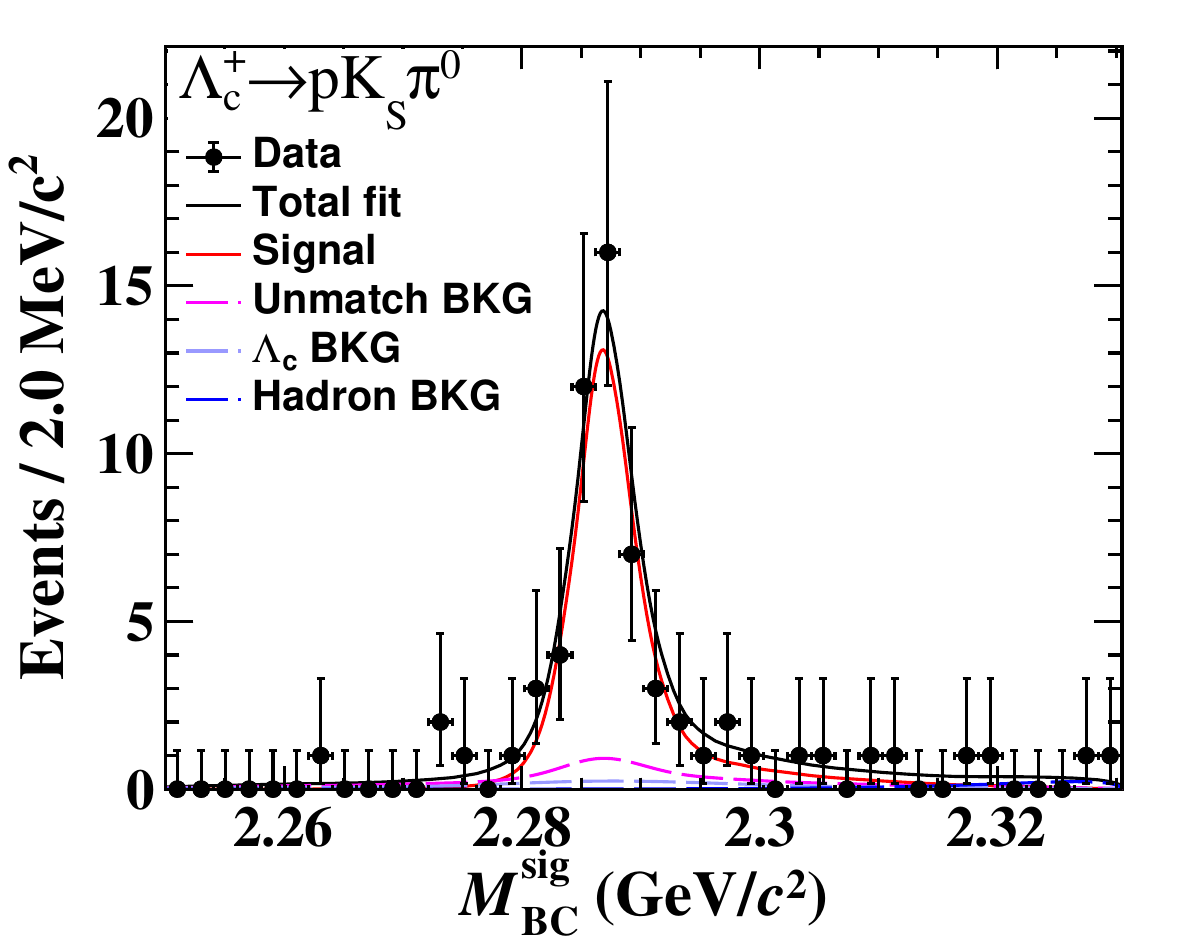}
  \includegraphics[width=0.24\textwidth]{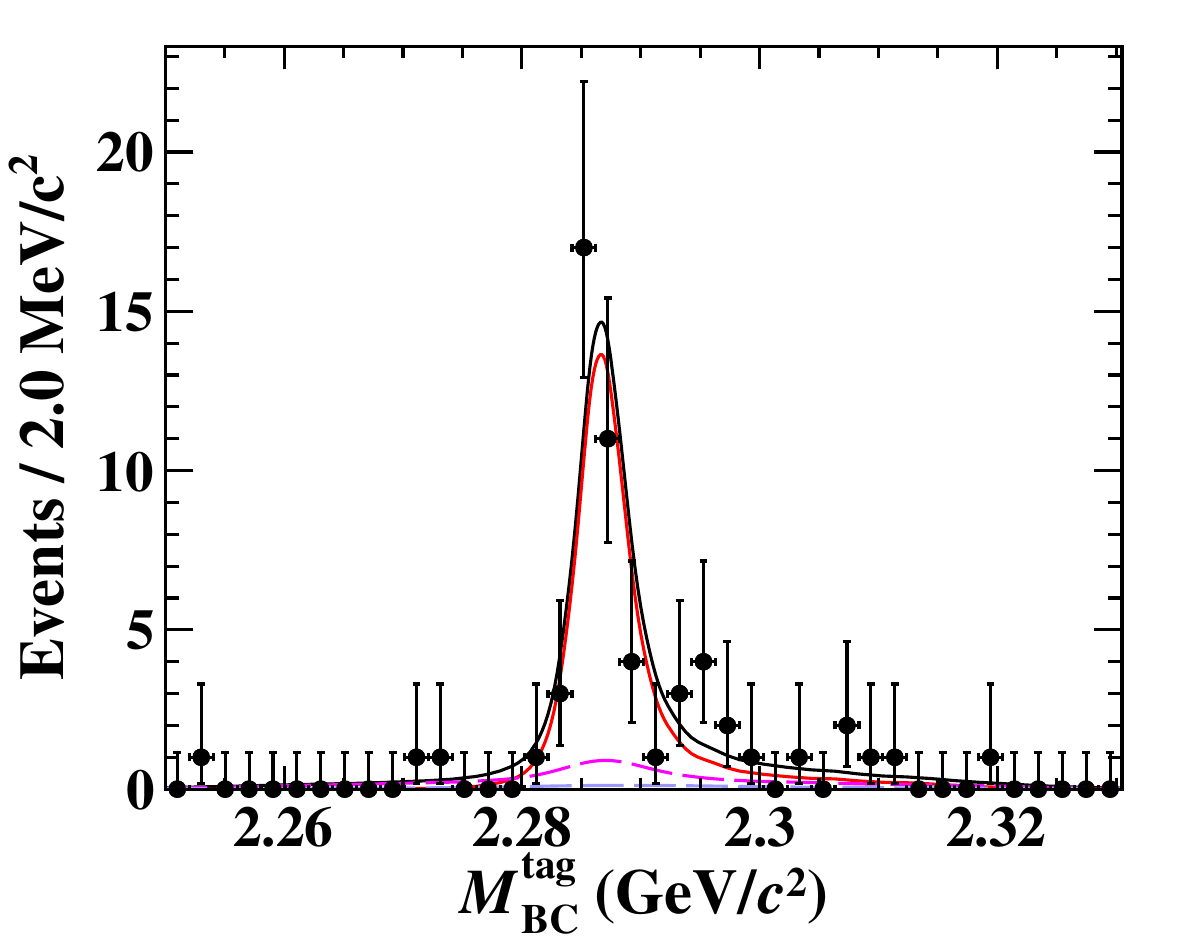}
  \includegraphics[width=0.24\textwidth]{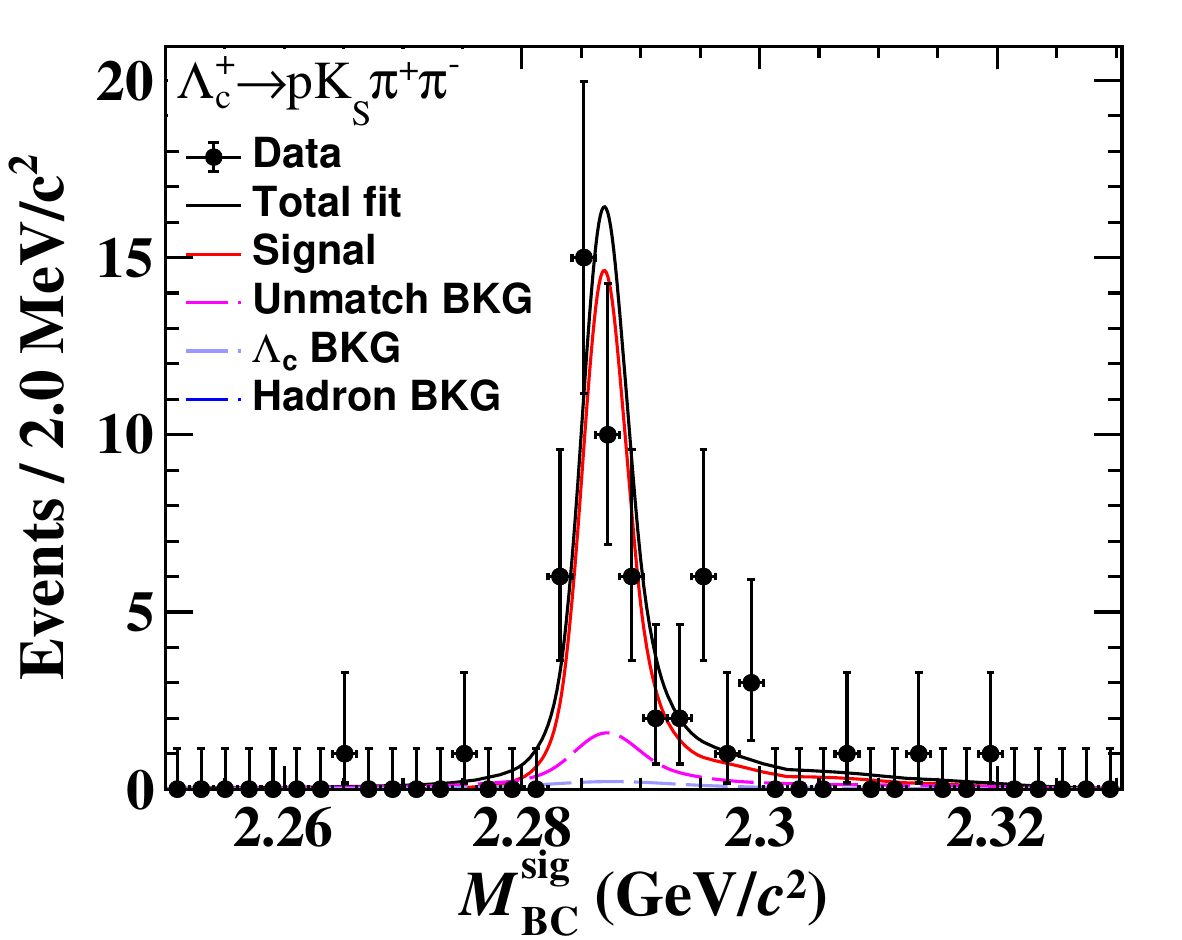}
  \includegraphics[width=0.24\textwidth]{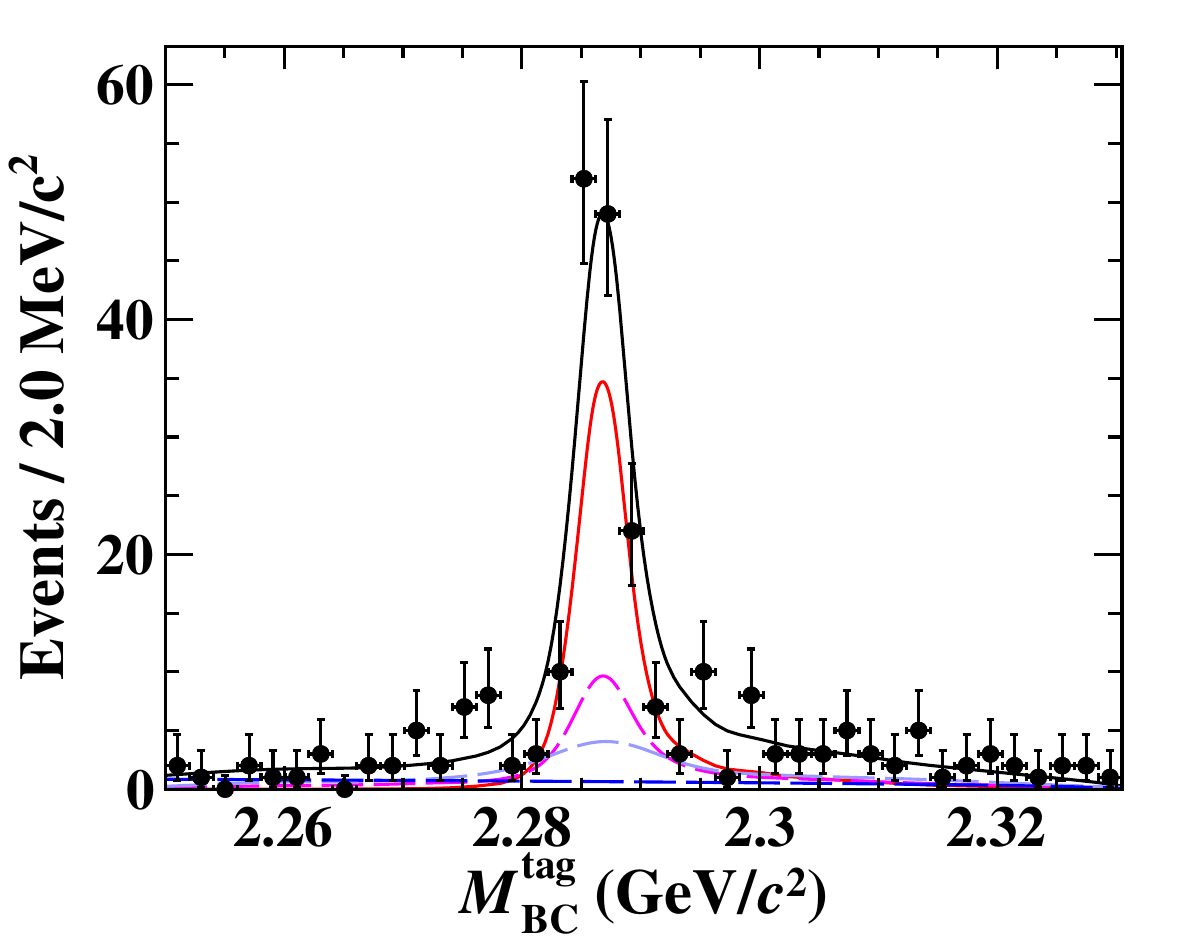}
  \includegraphics[width=0.24\textwidth]{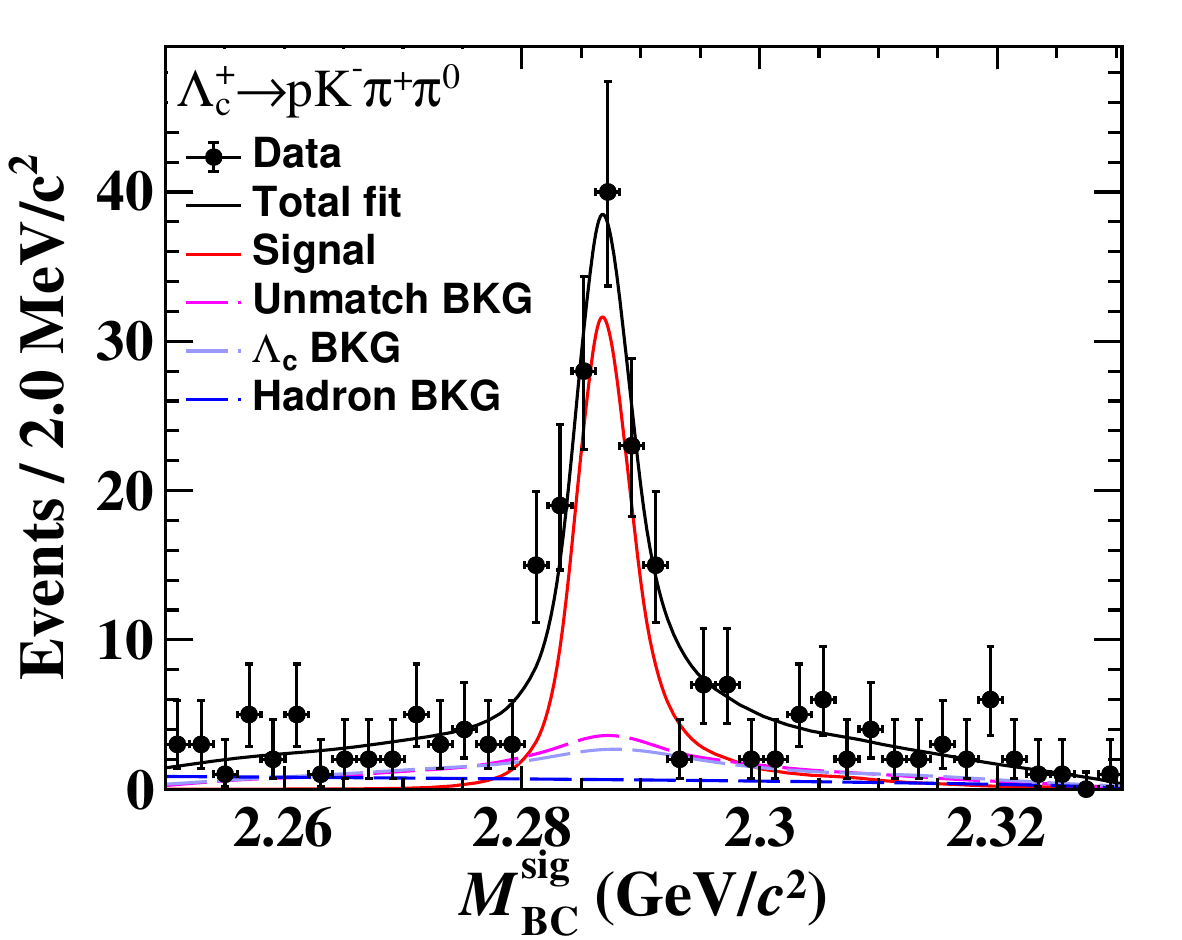}
  \includegraphics[width=0.24\textwidth]{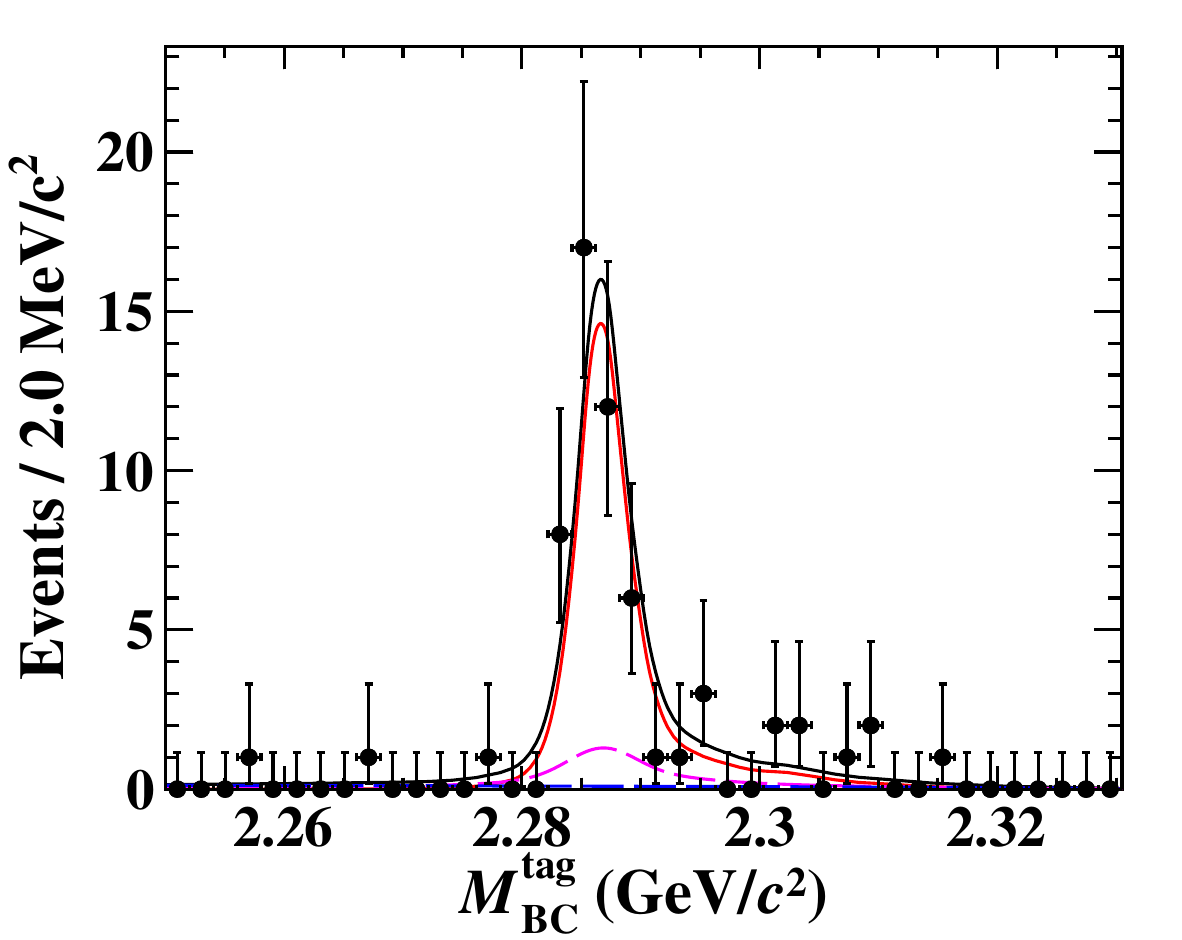}
  \includegraphics[width=0.24\textwidth]{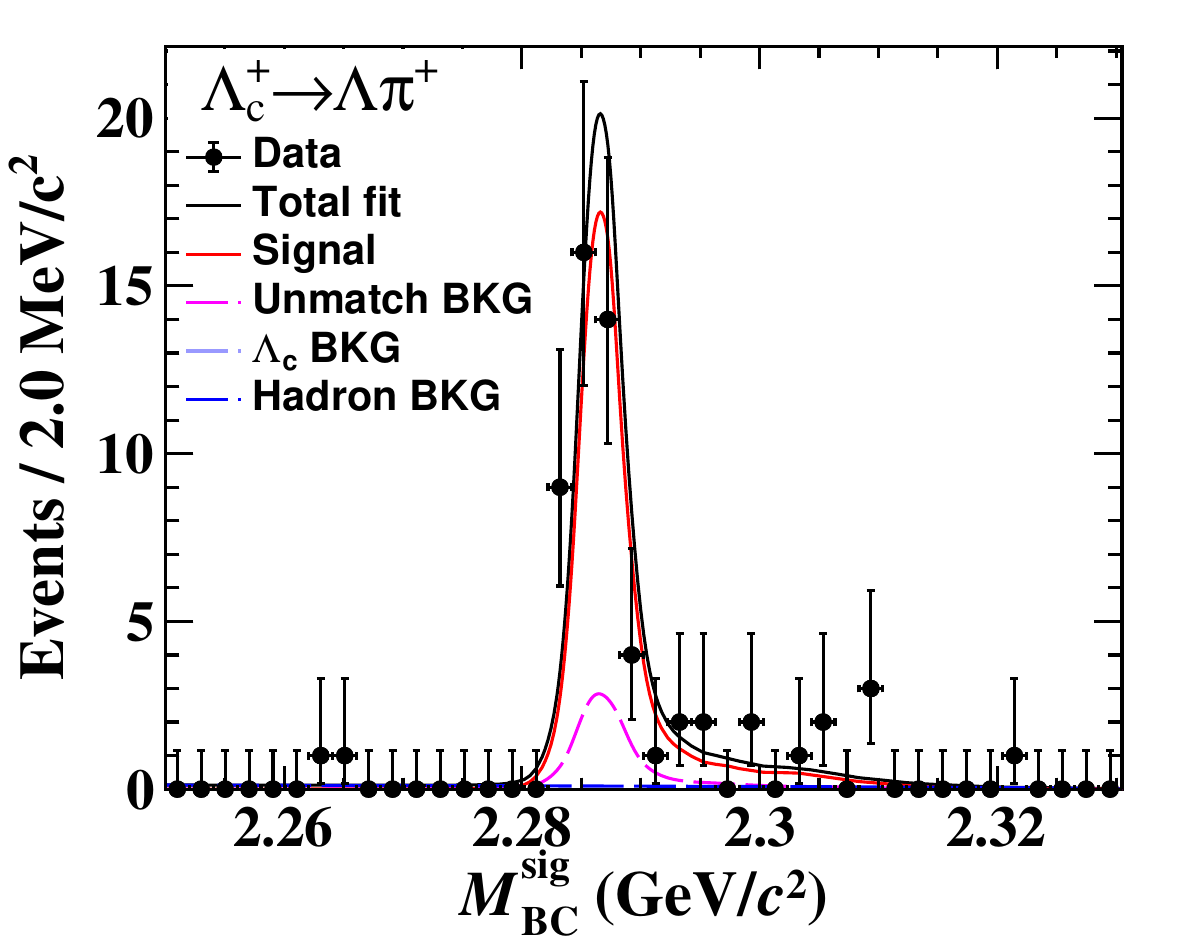}
  \includegraphics[width=0.24\textwidth]{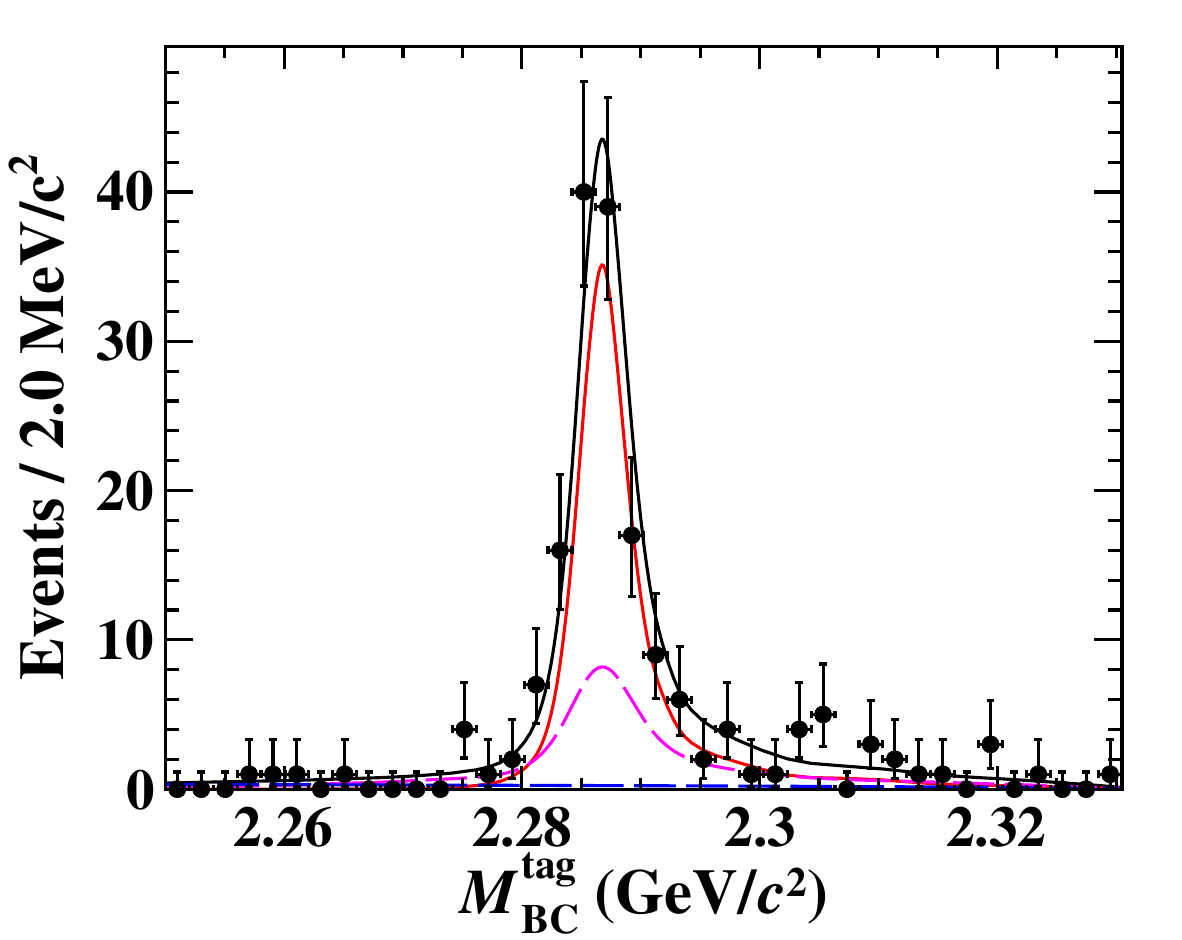}
  \includegraphics[width=0.24\textwidth]{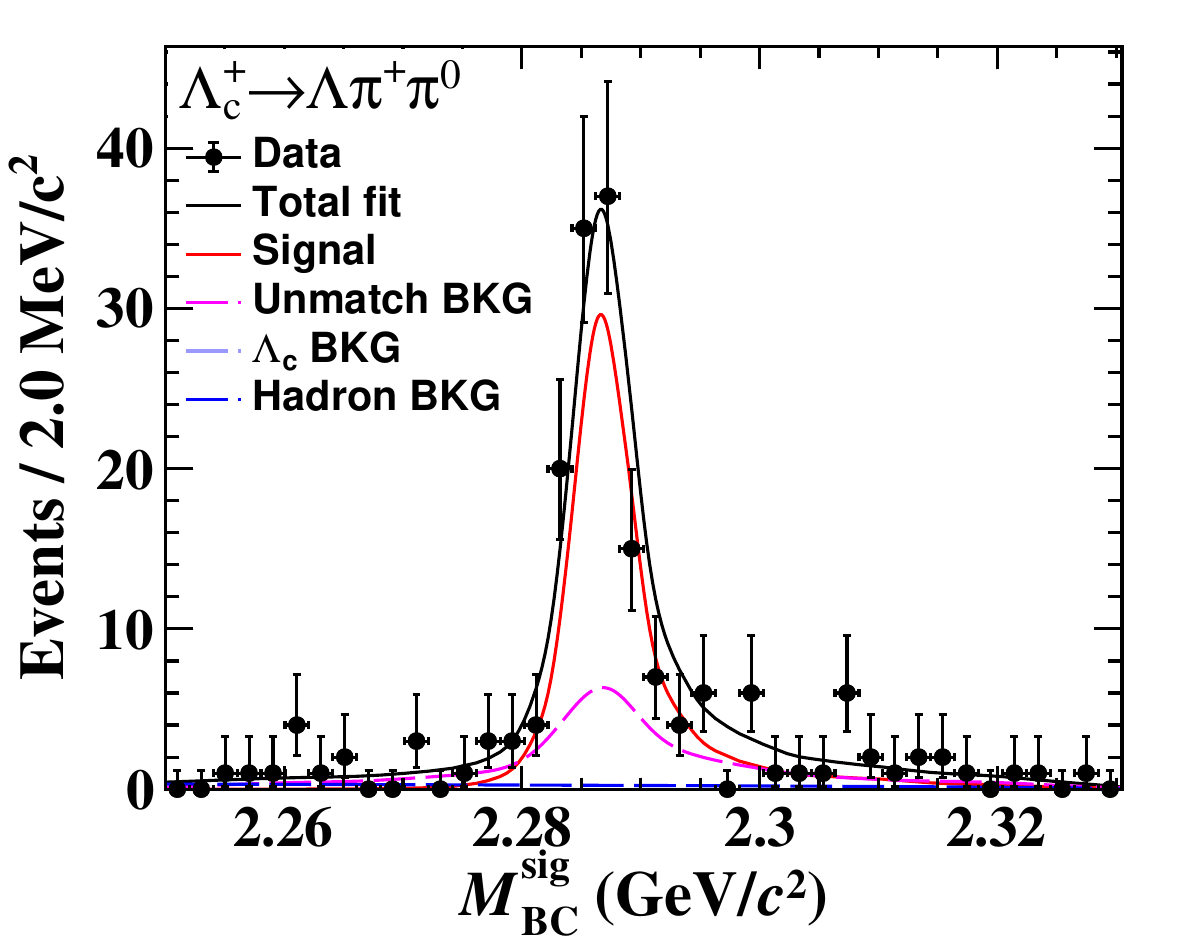}
  \includegraphics[width=0.24\textwidth]{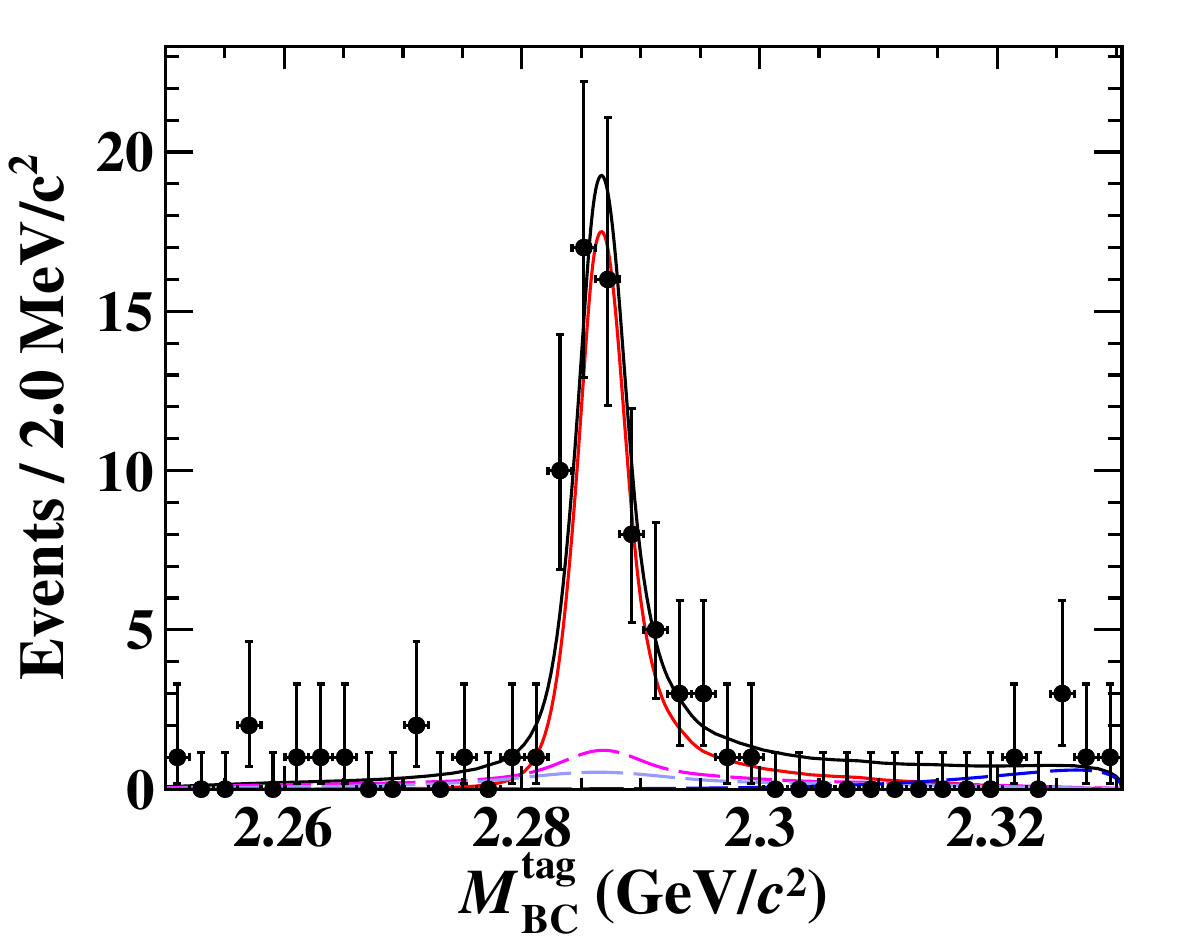}
  \includegraphics[width=0.24\textwidth]{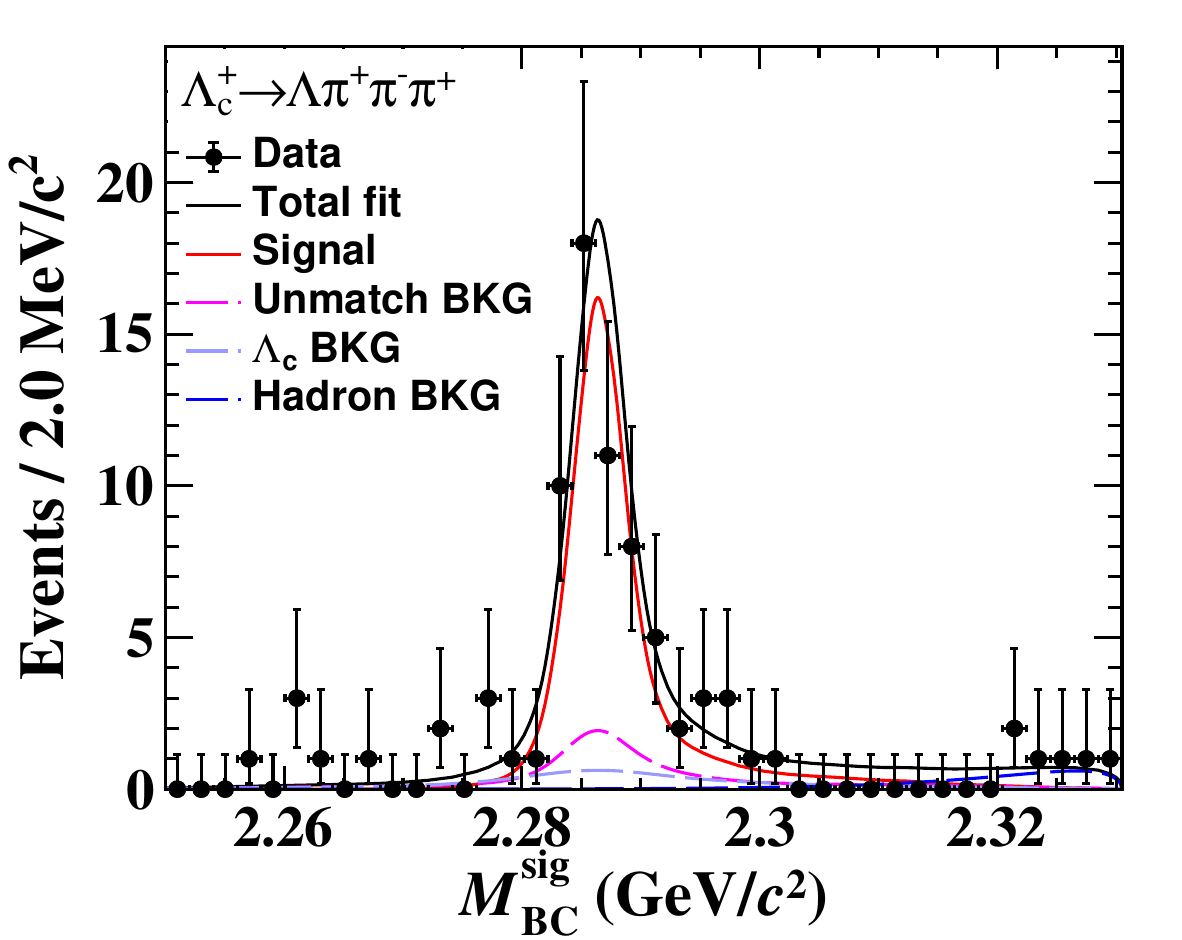}
  \includegraphics[width=0.24\textwidth]{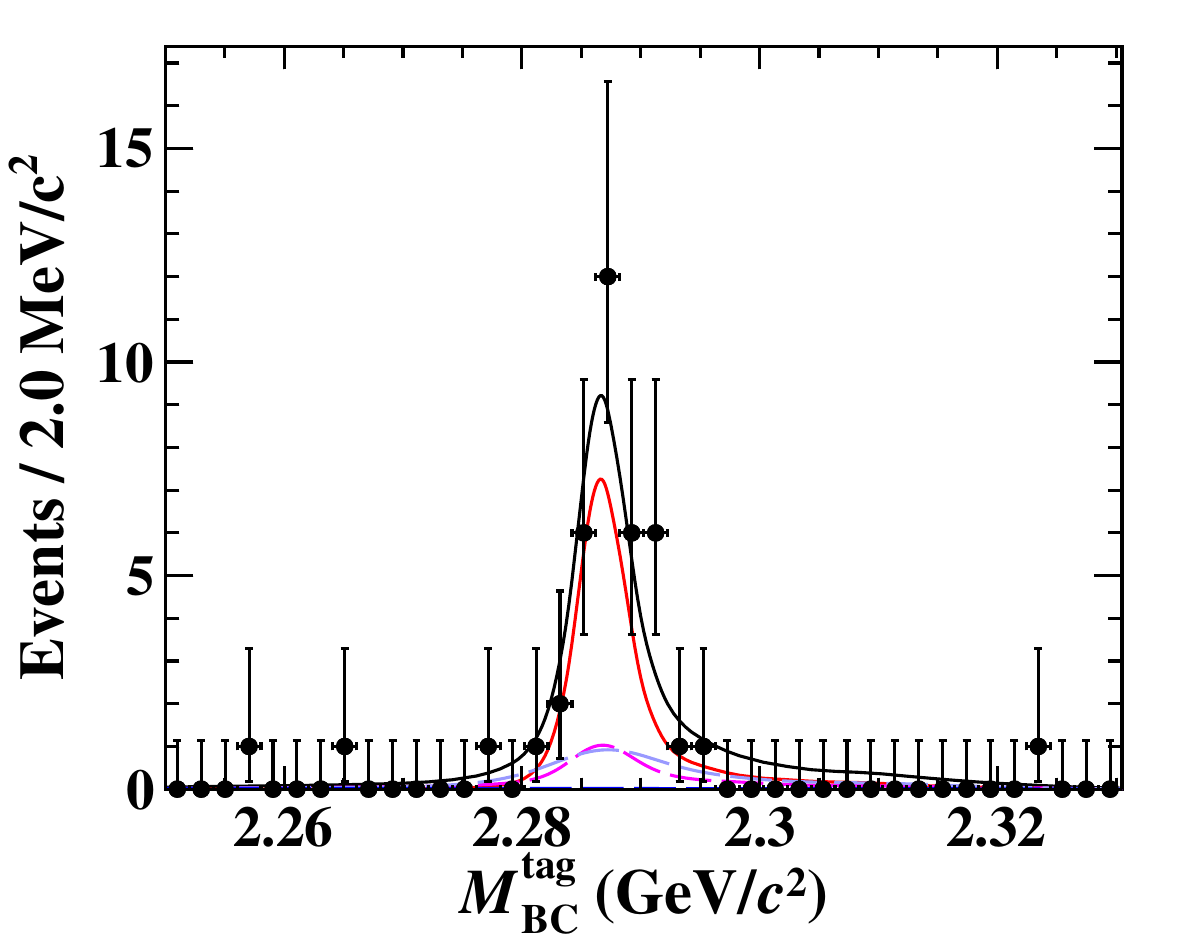}
  \includegraphics[width=0.24\textwidth]{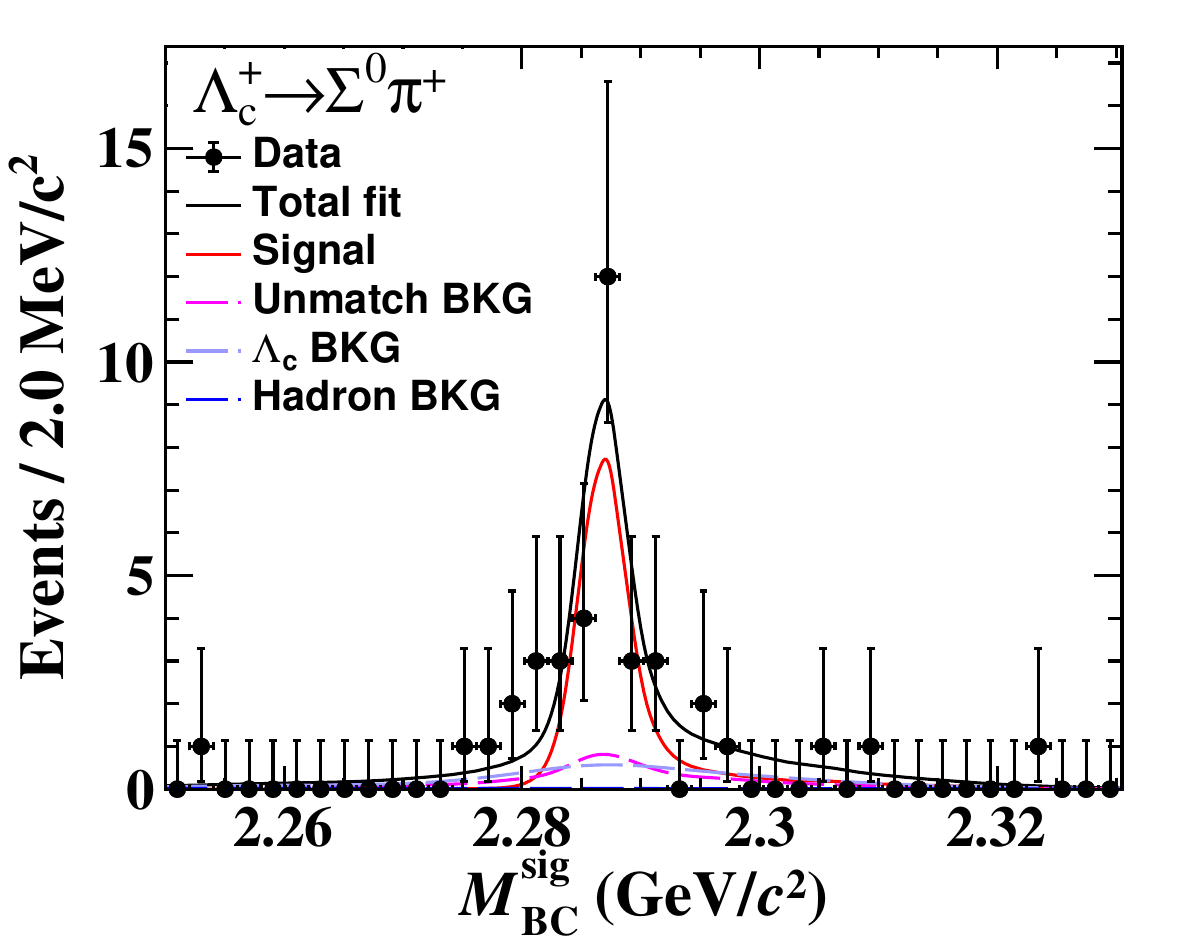}
  \includegraphics[width=0.24\textwidth]{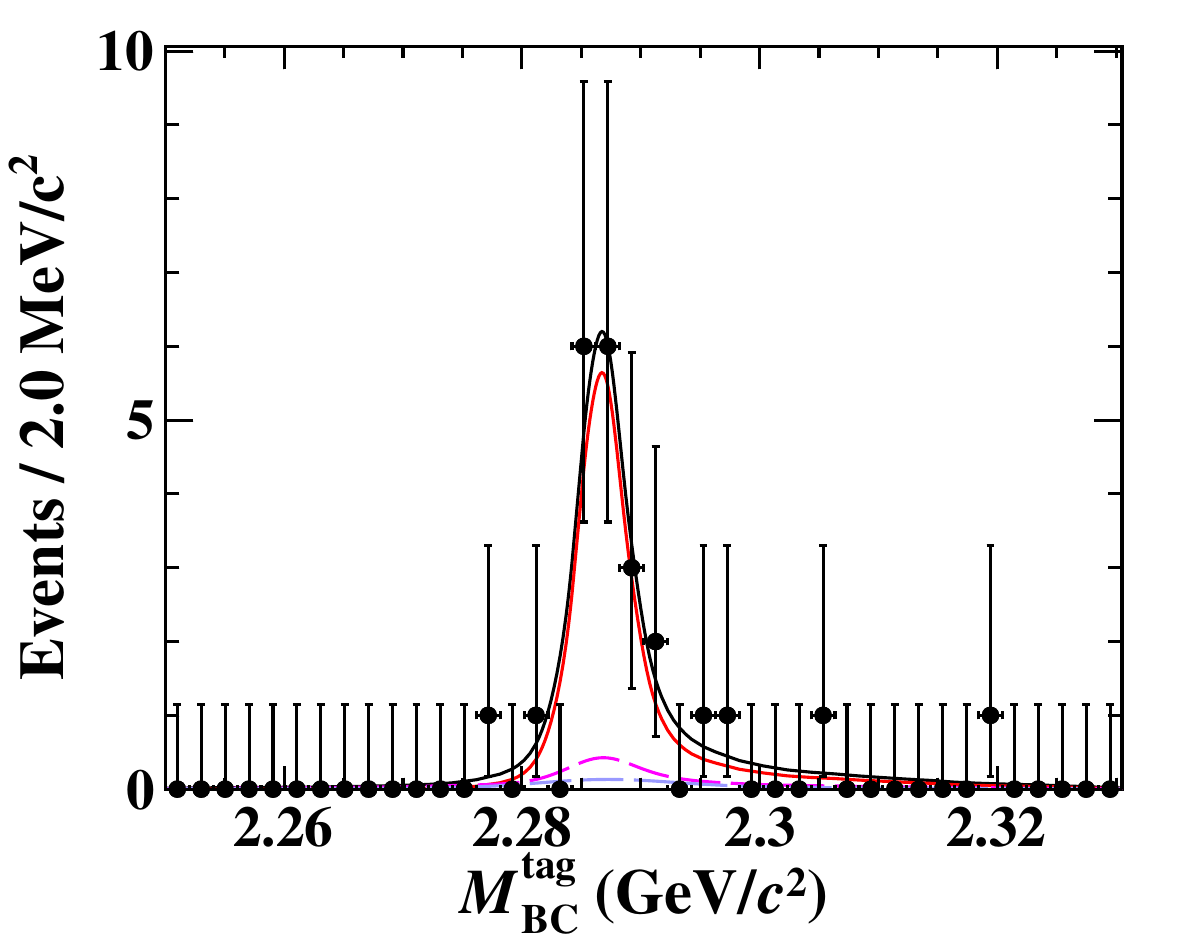}
  \includegraphics[width=0.24\textwidth]{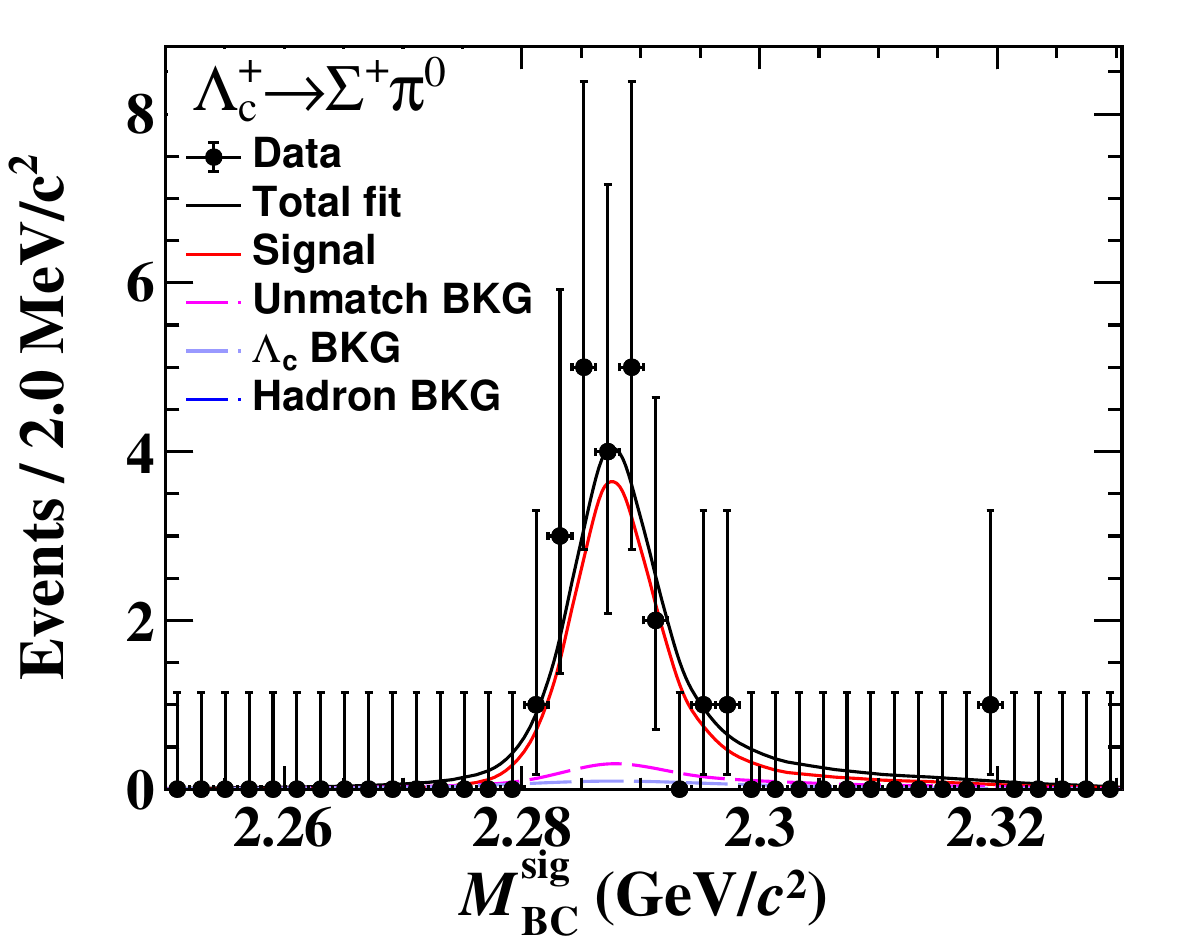}
  \includegraphics[width=0.24\textwidth]{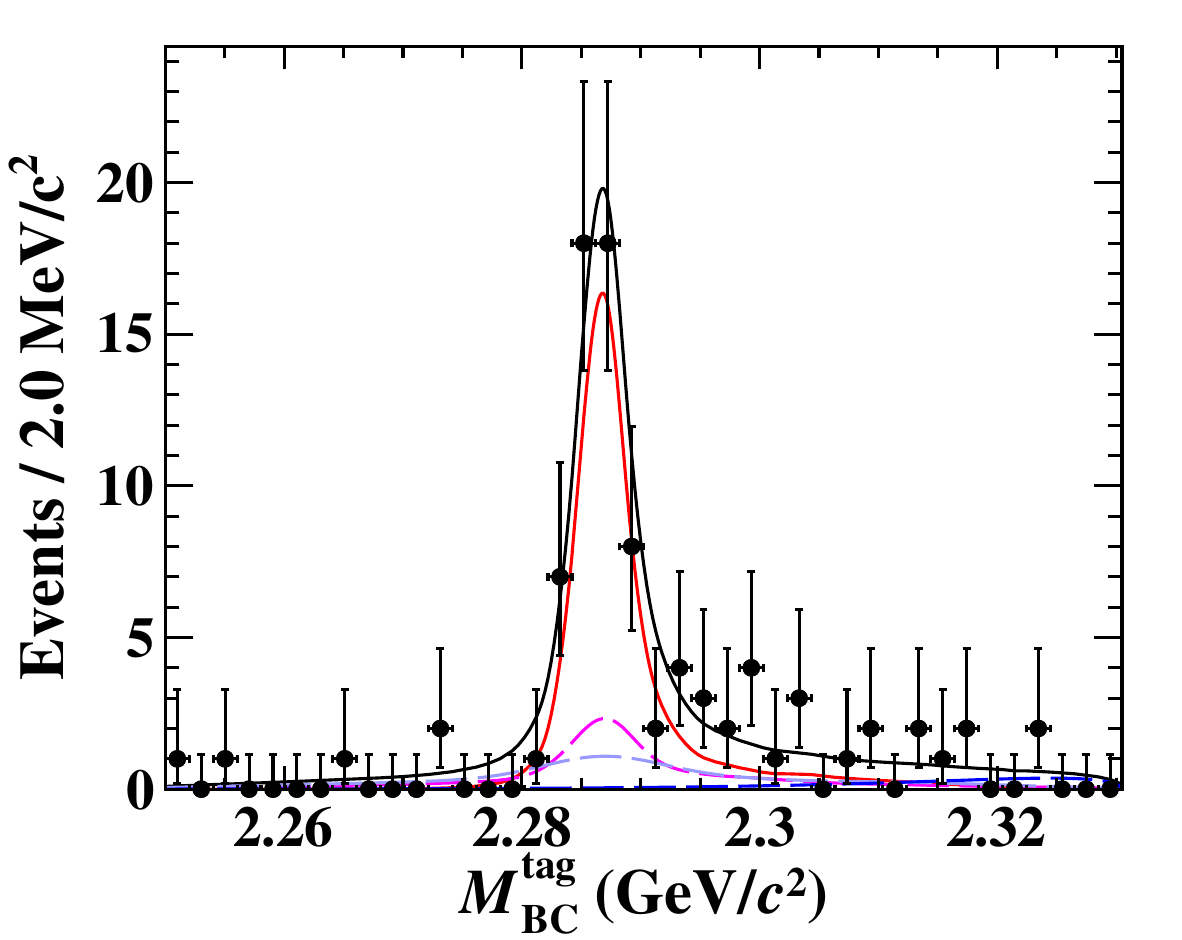}
  \includegraphics[width=0.24\textwidth]{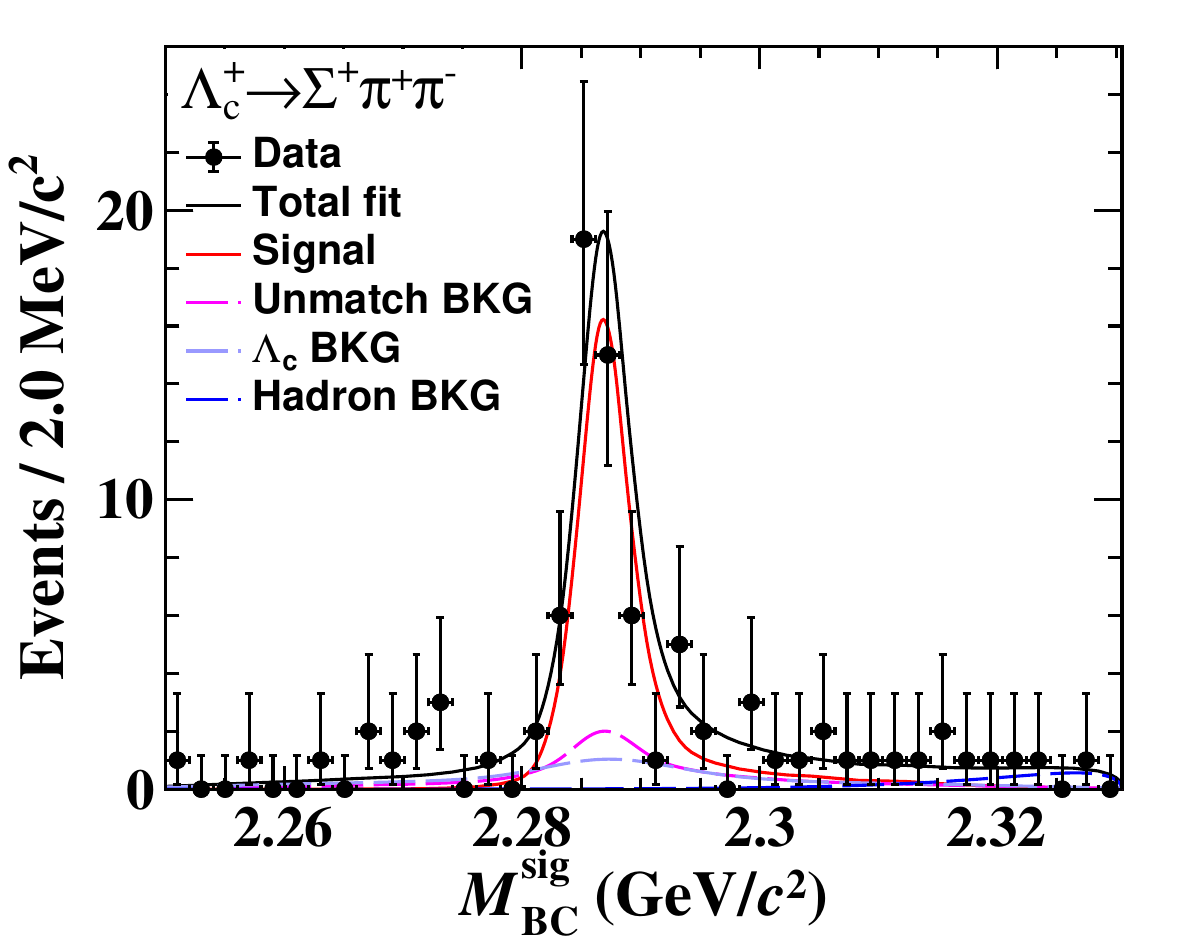}
  \includegraphics[width=0.24\textwidth]{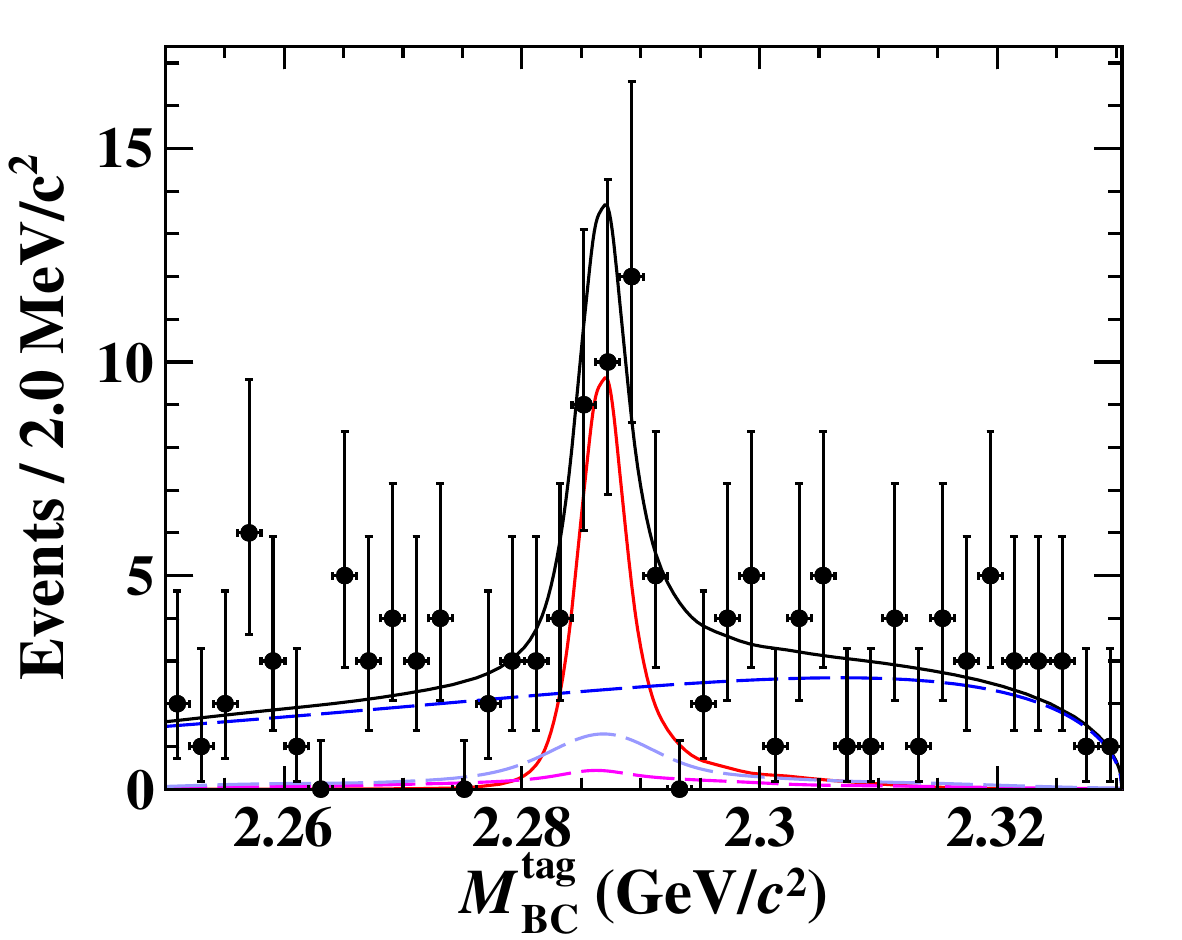}
  \includegraphics[width=0.24\textwidth]{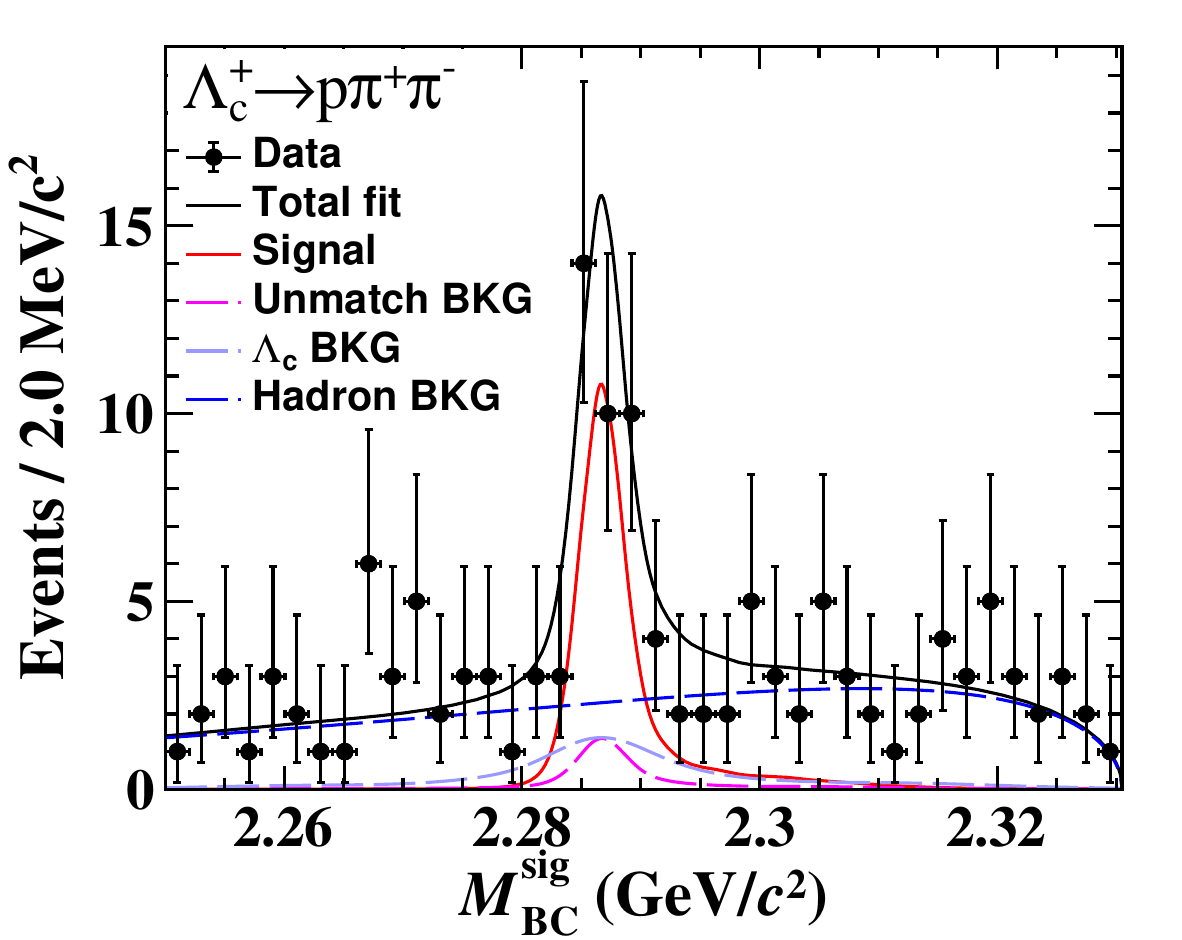}
     \vspace*{-0.5cm}
  \end{center}
\caption{The projections of the 2D fits on the $M_{\rm BC}^{\rm tag}$ and $M_{\rm BC}^{\rm sig}$ distributions of the accepted DT candidates at $\sqrt{s}=4640.91~\mev$. The plots in the first and third columns show the combined 12 tag modes for each signal mode.
The points with error bars are data, the black lines are the sum of fit functions, the red lines are the matched signal shapes, the pink dashed lines are the unmatched signal shapes, the lilac dashed lines are the non-signal $\lcp\lcm$ shapes, and the blue dashed lines are the ARGUS functions.}
\label{fig:DT_yield_4640}
\end{figure}

\begin{figure}[!htbp]
  \begin{center}
  
  \includegraphics[width=0.24\textwidth]{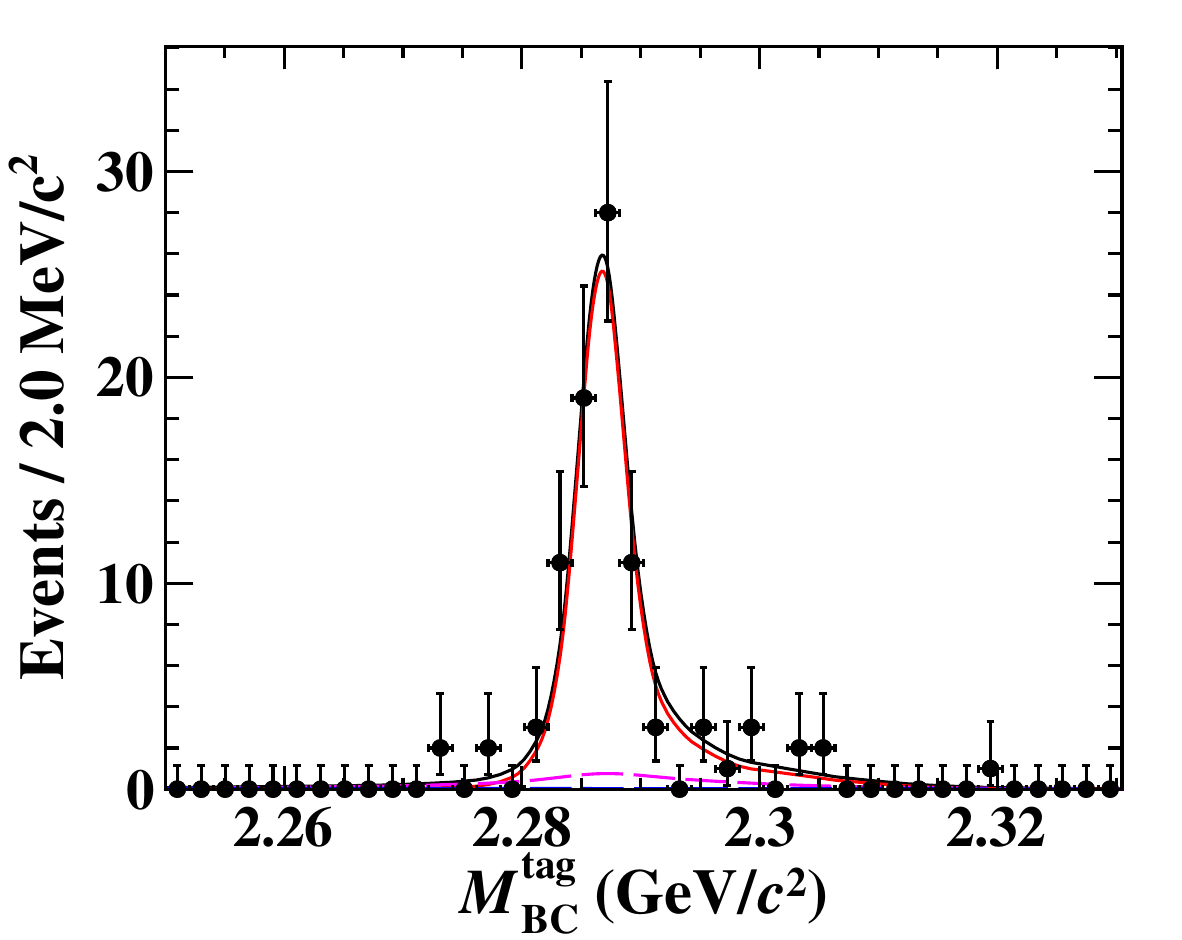}
  \includegraphics[width=0.24\textwidth]{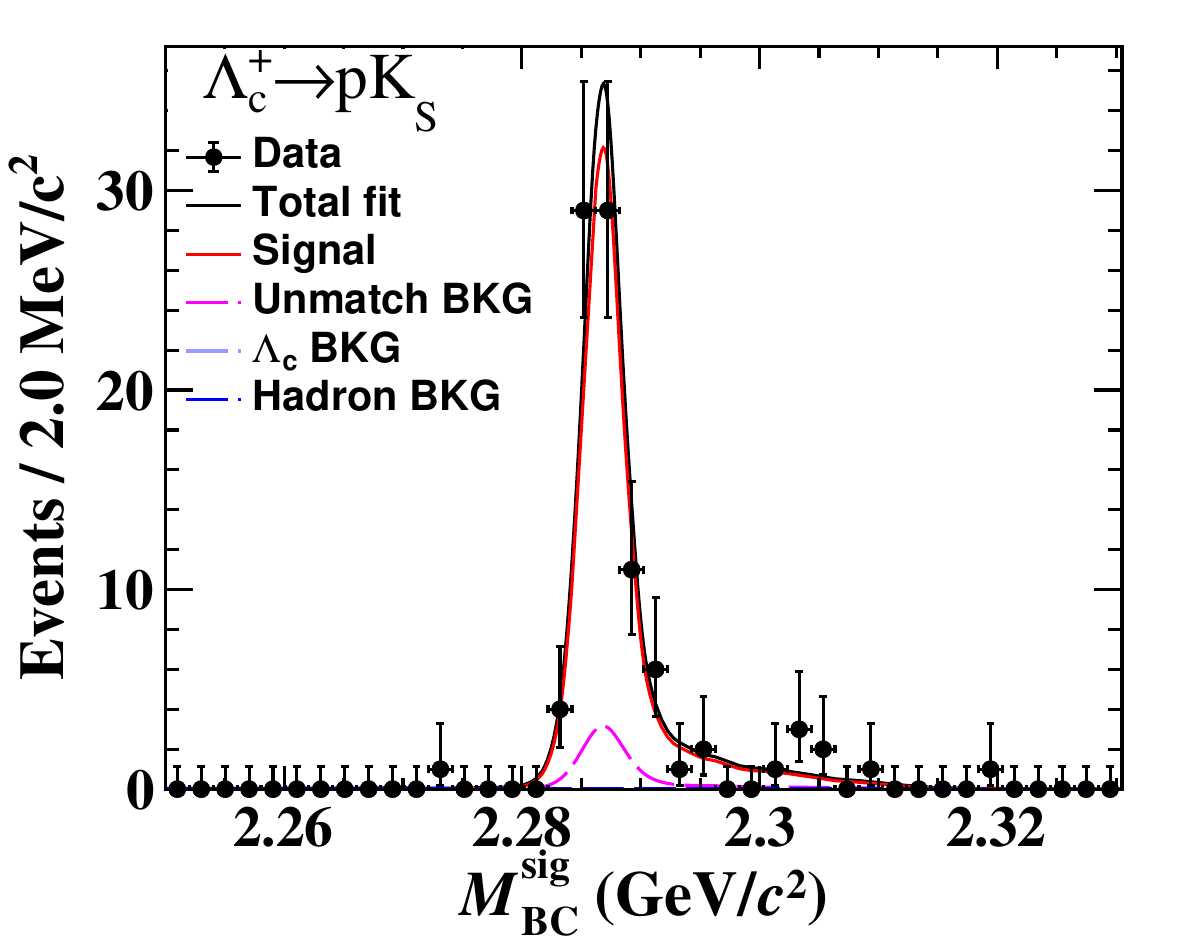}
  \includegraphics[width=0.24\textwidth]{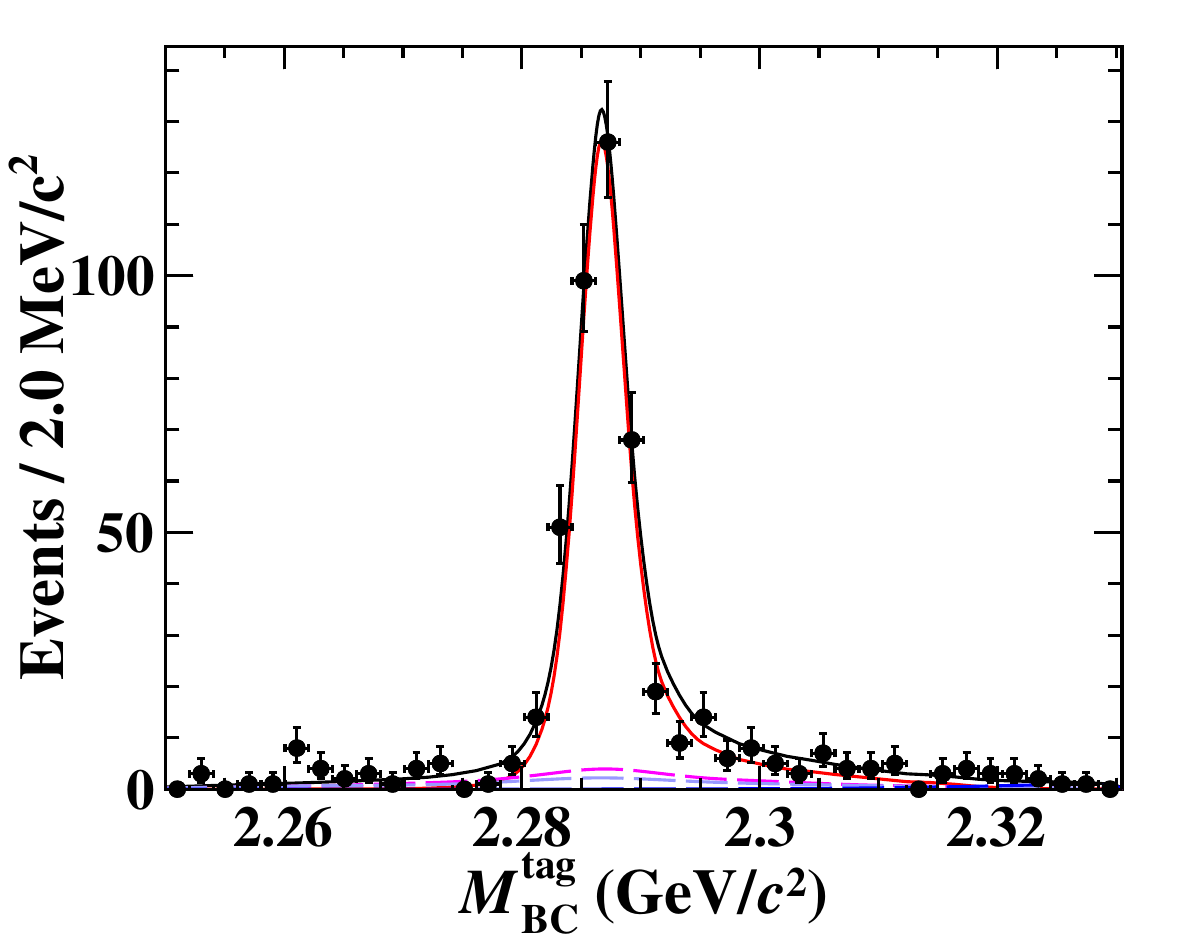}
  \includegraphics[width=0.24\textwidth]{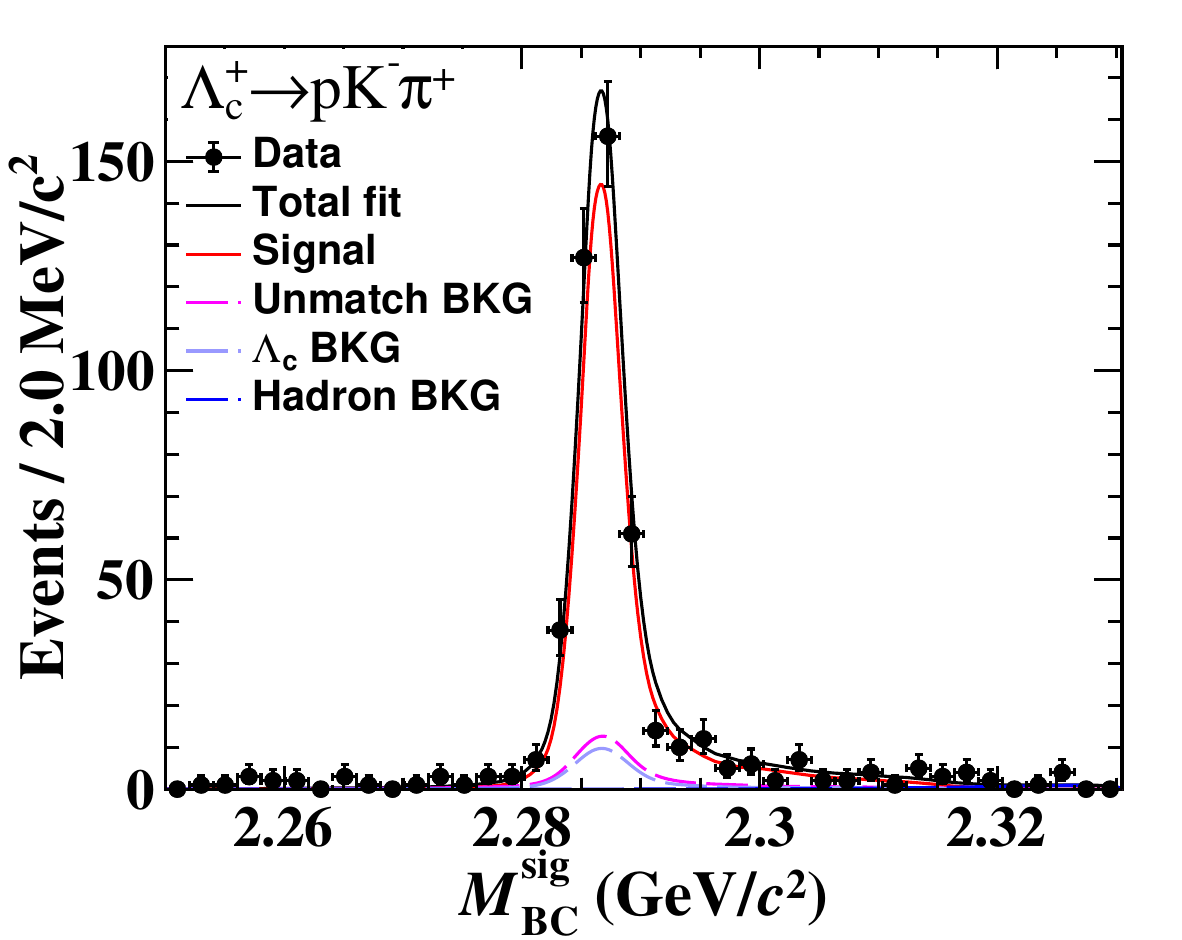}
  \includegraphics[width=0.24\textwidth]{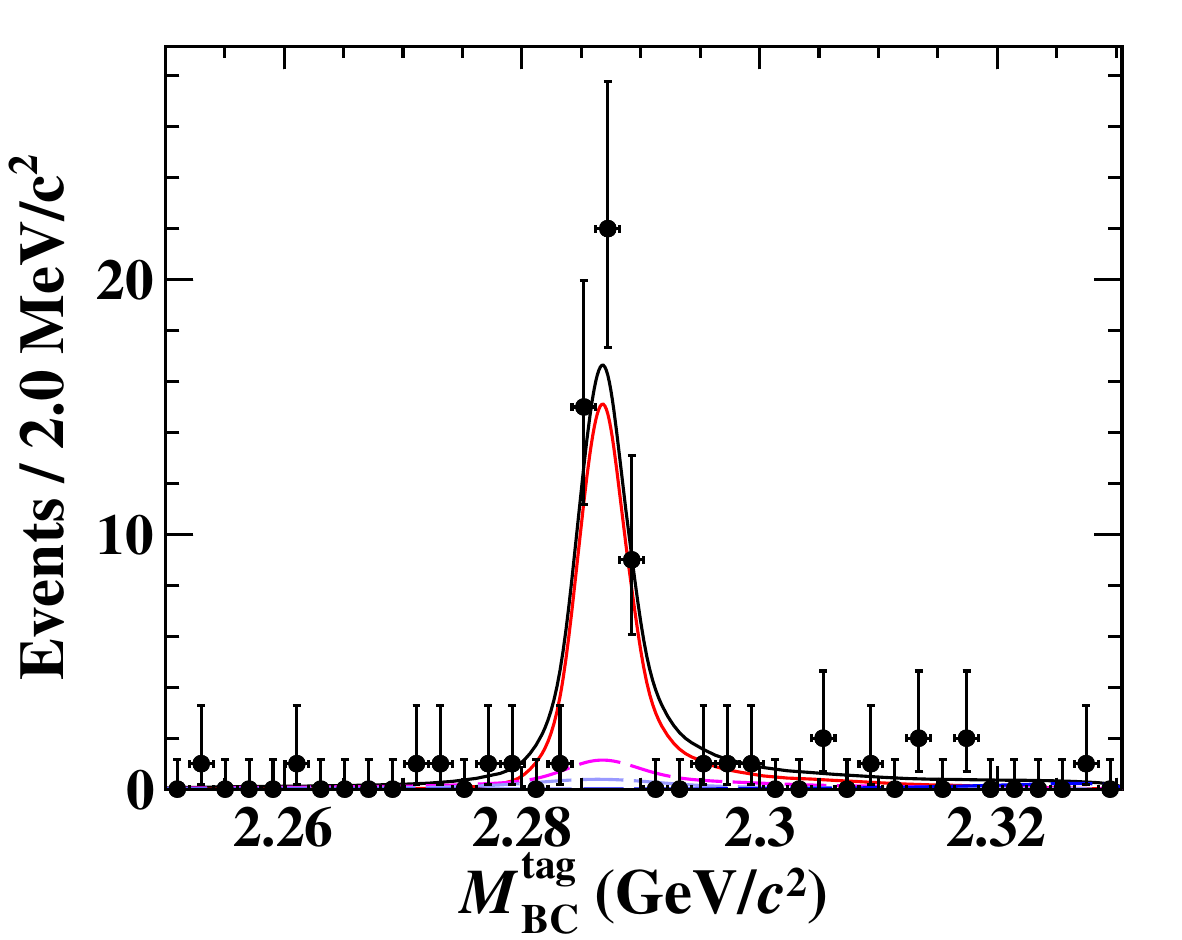}
  \includegraphics[width=0.24\textwidth]{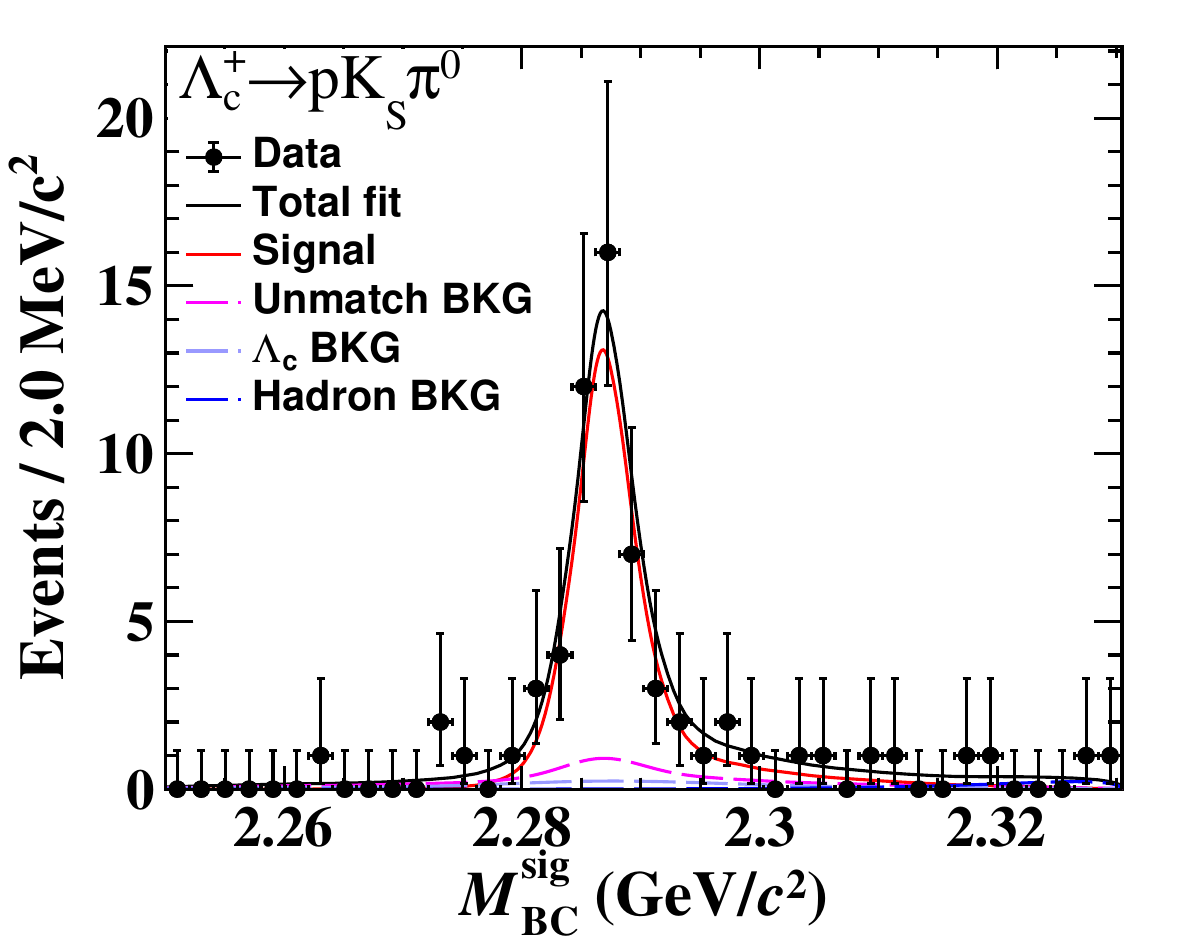}
  \includegraphics[width=0.24\textwidth]{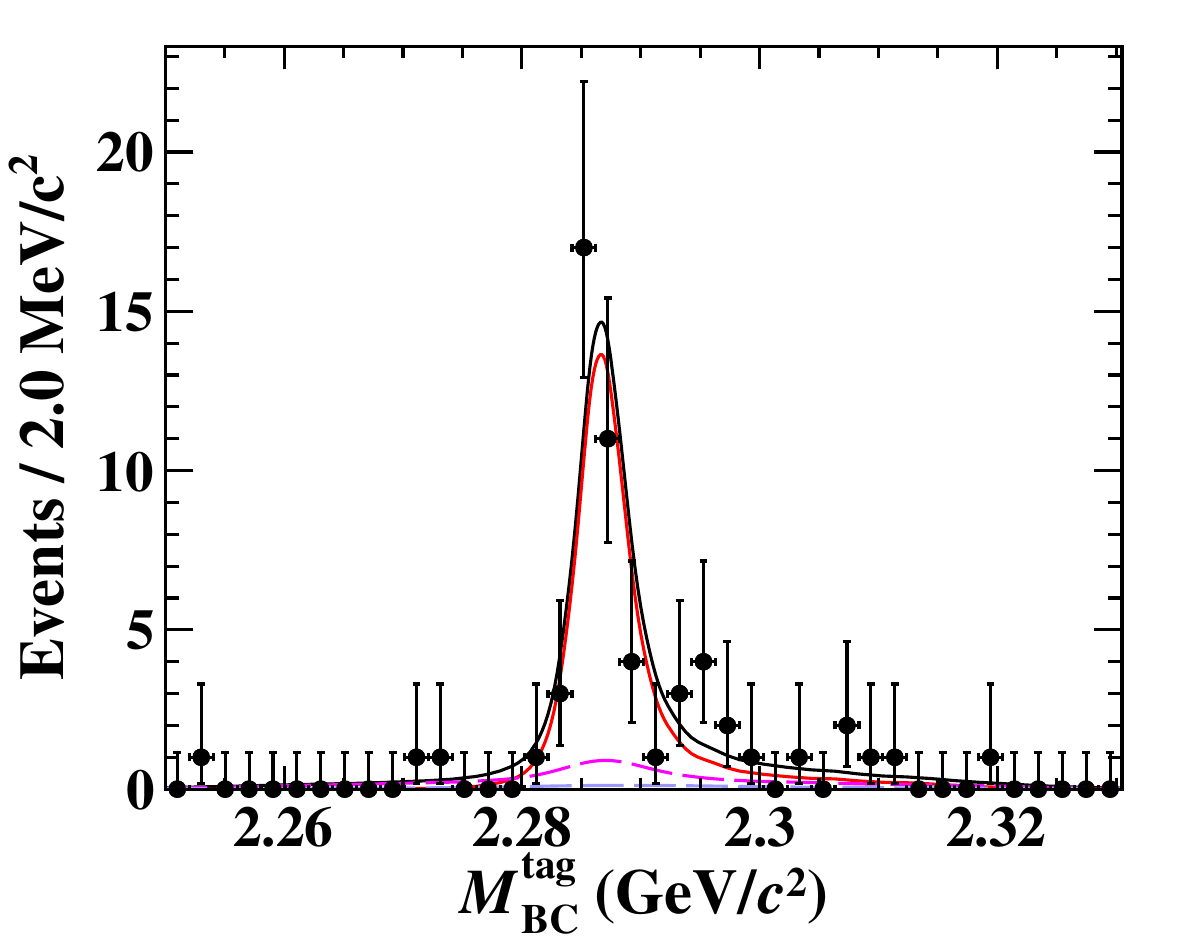}
  \includegraphics[width=0.24\textwidth]{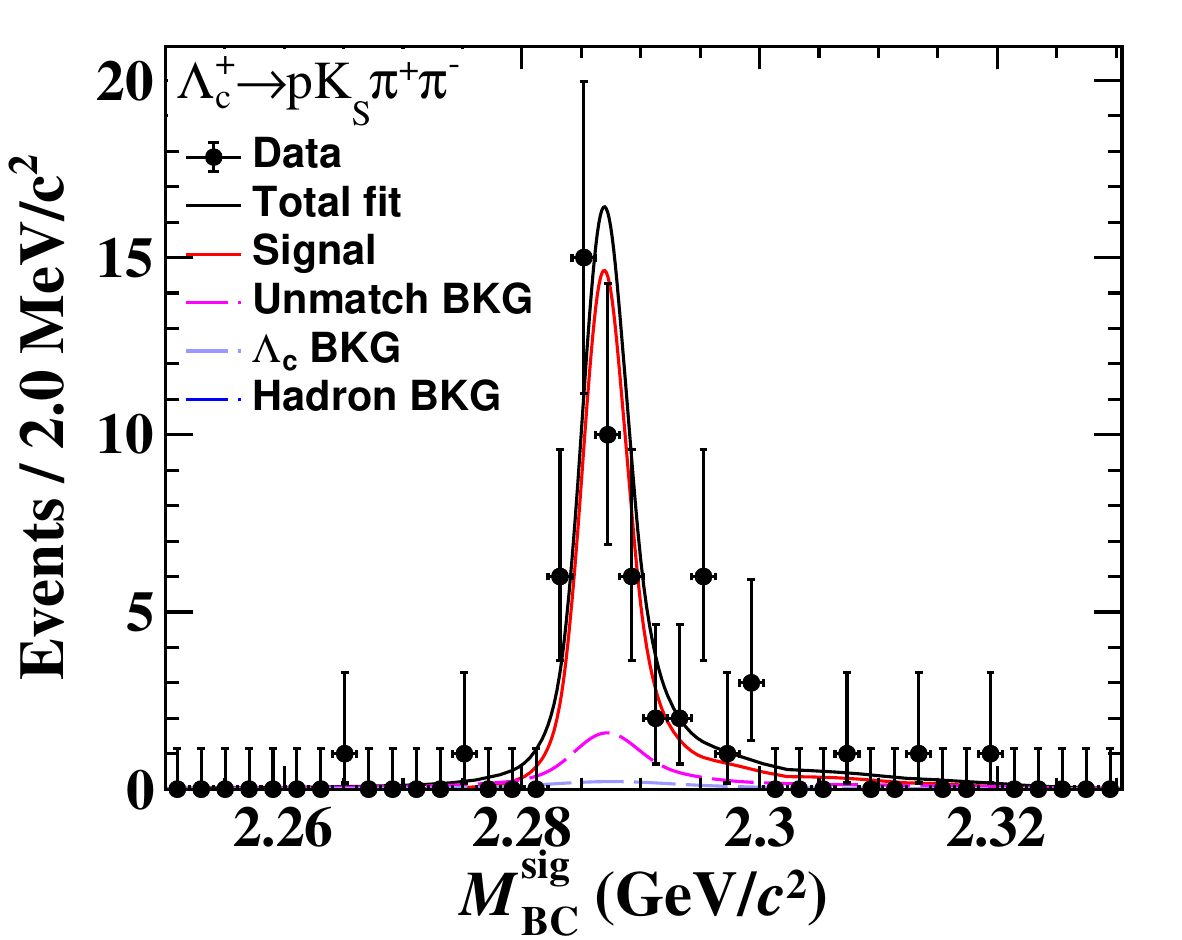}
  \includegraphics[width=0.24\textwidth]{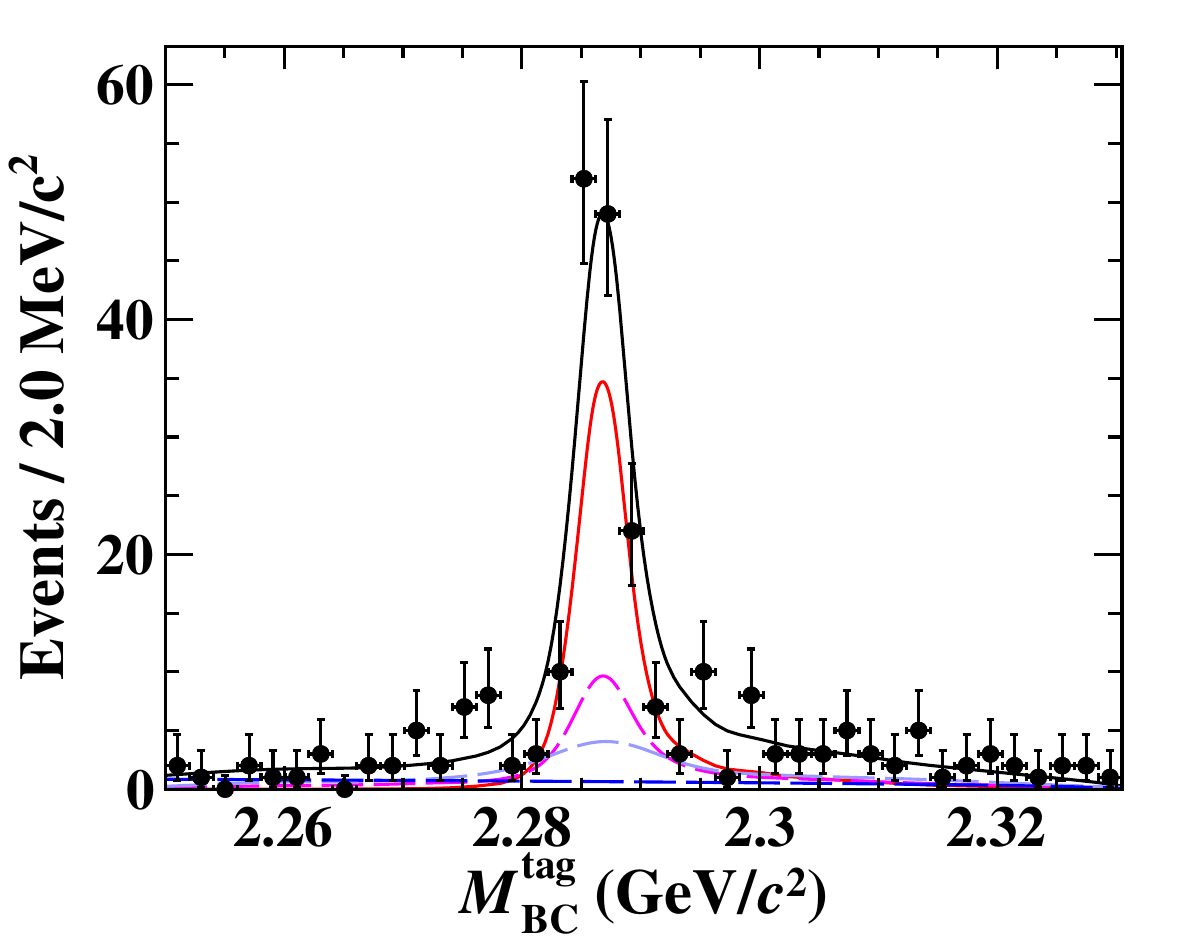}
  \includegraphics[width=0.24\textwidth]{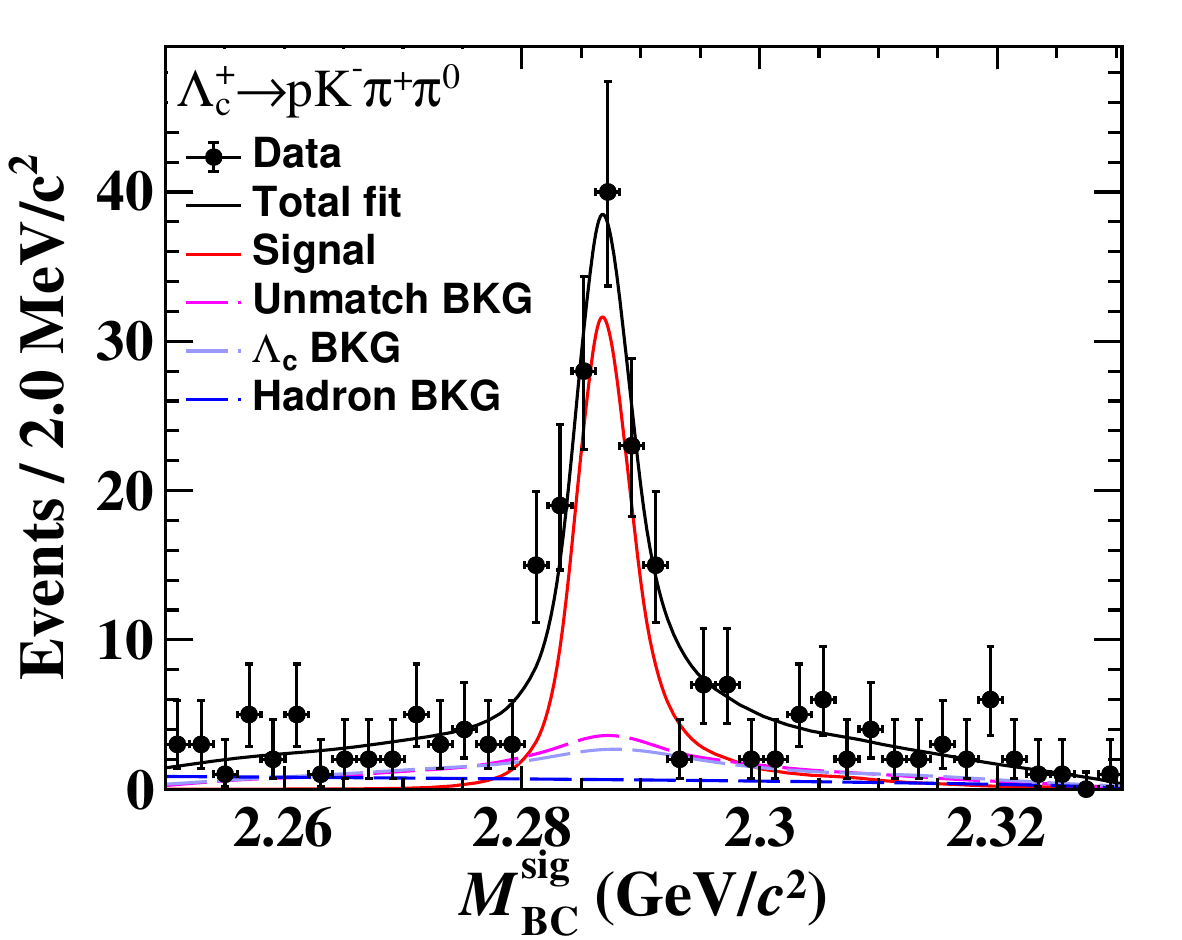}
  \includegraphics[width=0.24\textwidth]{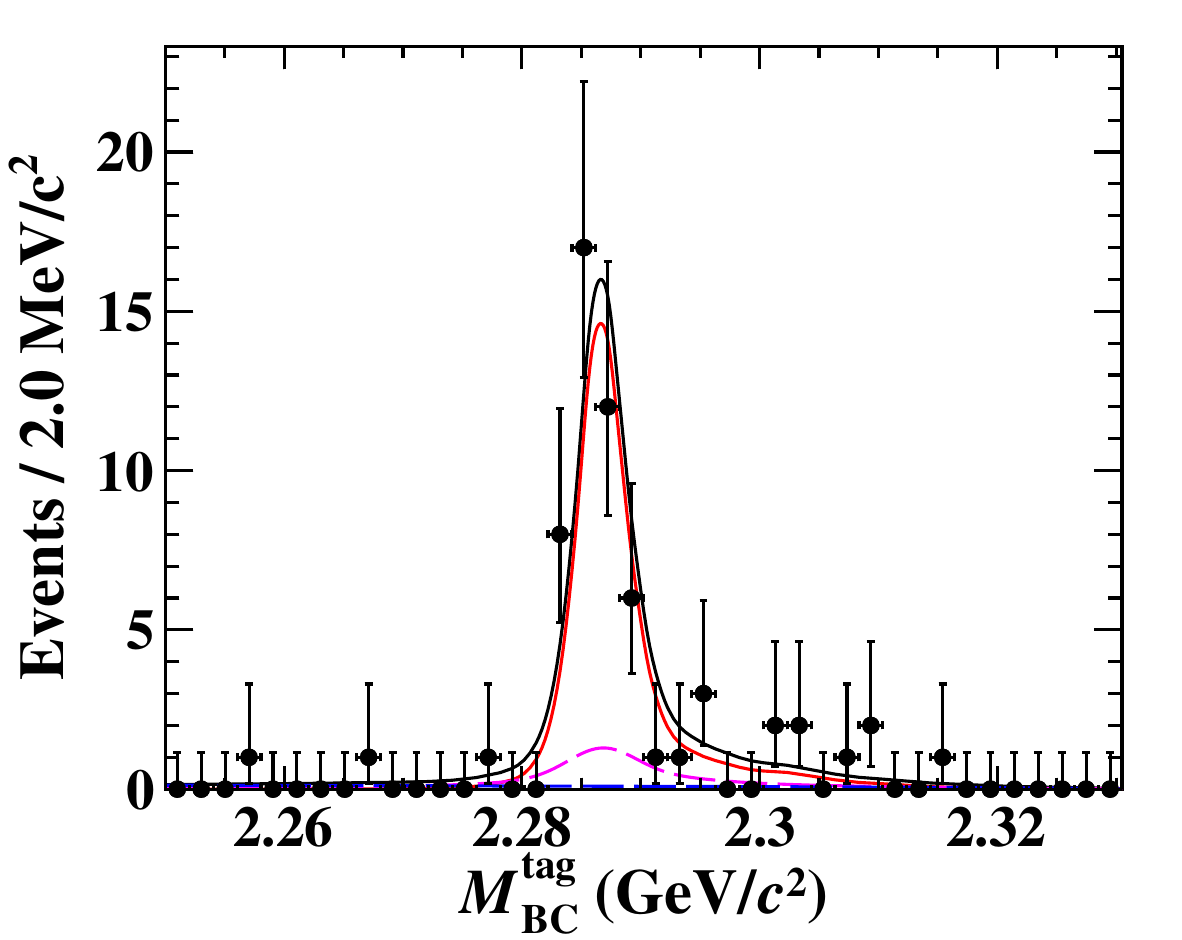}
  \includegraphics[width=0.24\textwidth]{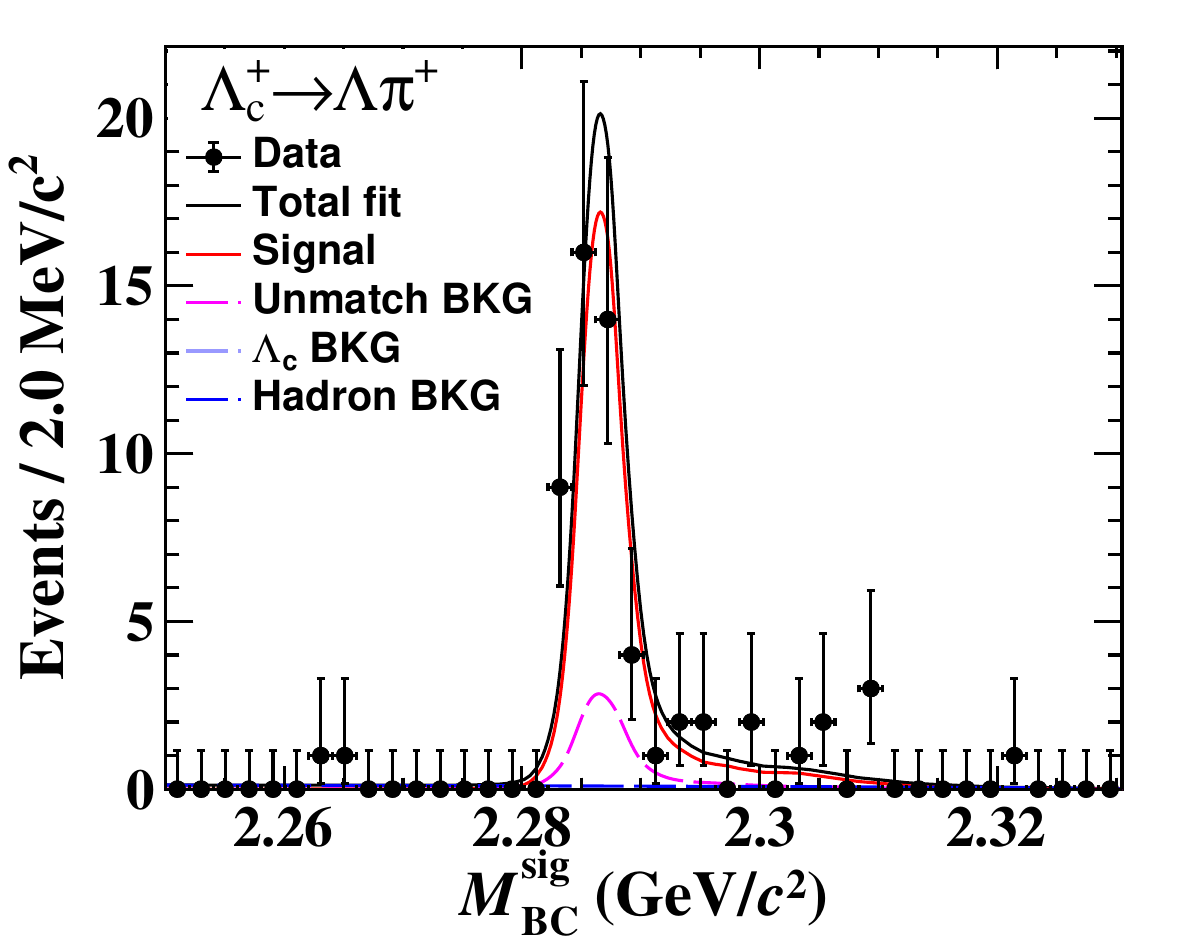}
  \includegraphics[width=0.24\textwidth]{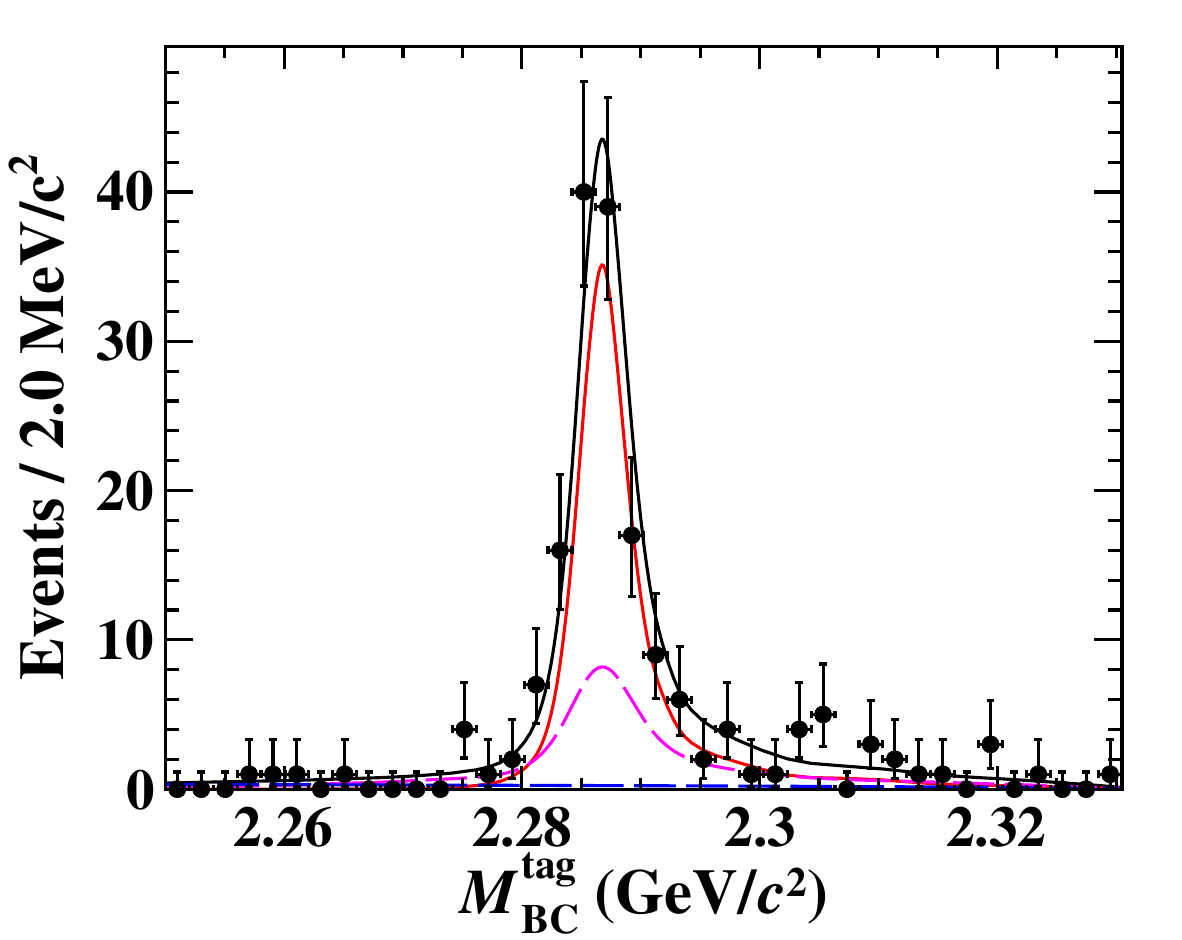}
  \includegraphics[width=0.24\textwidth]{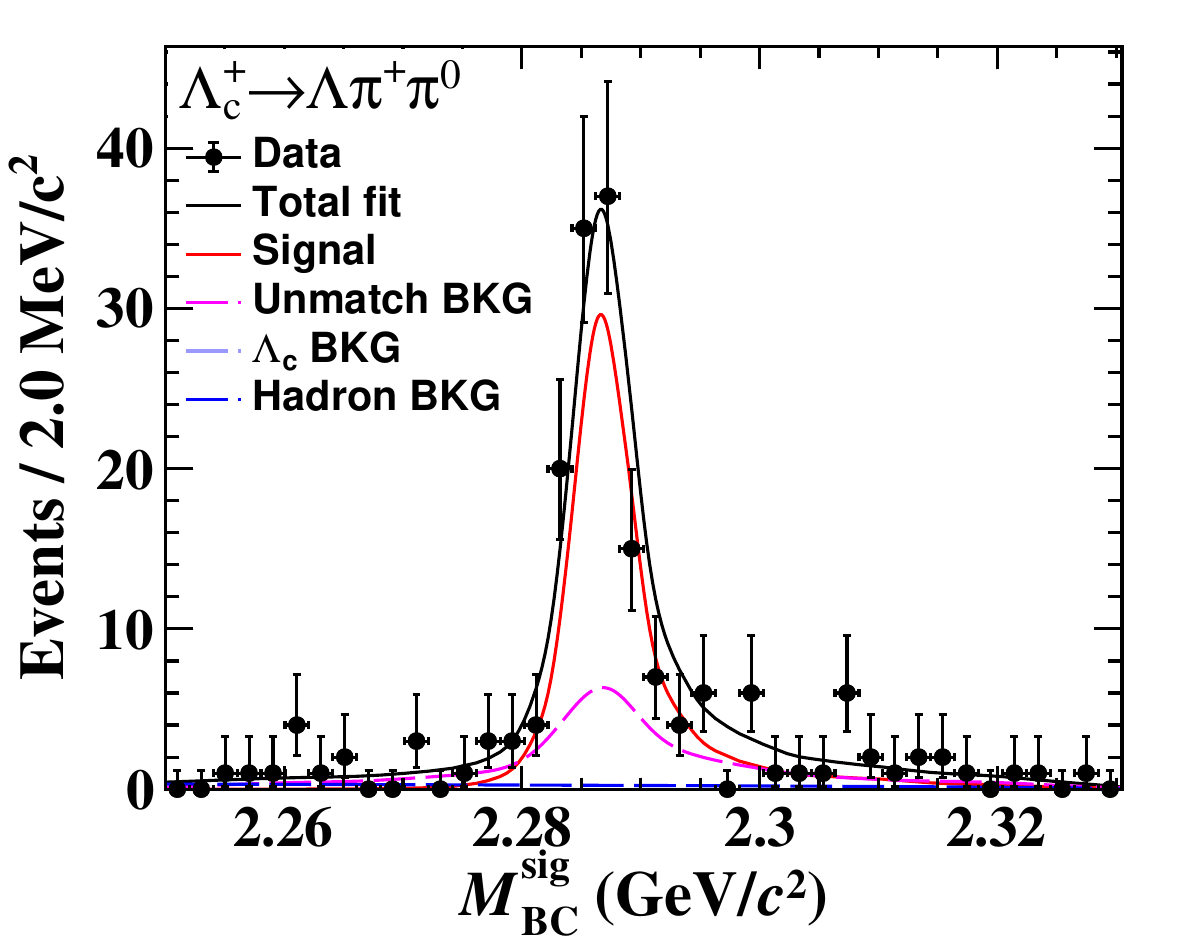}
  \includegraphics[width=0.24\textwidth]{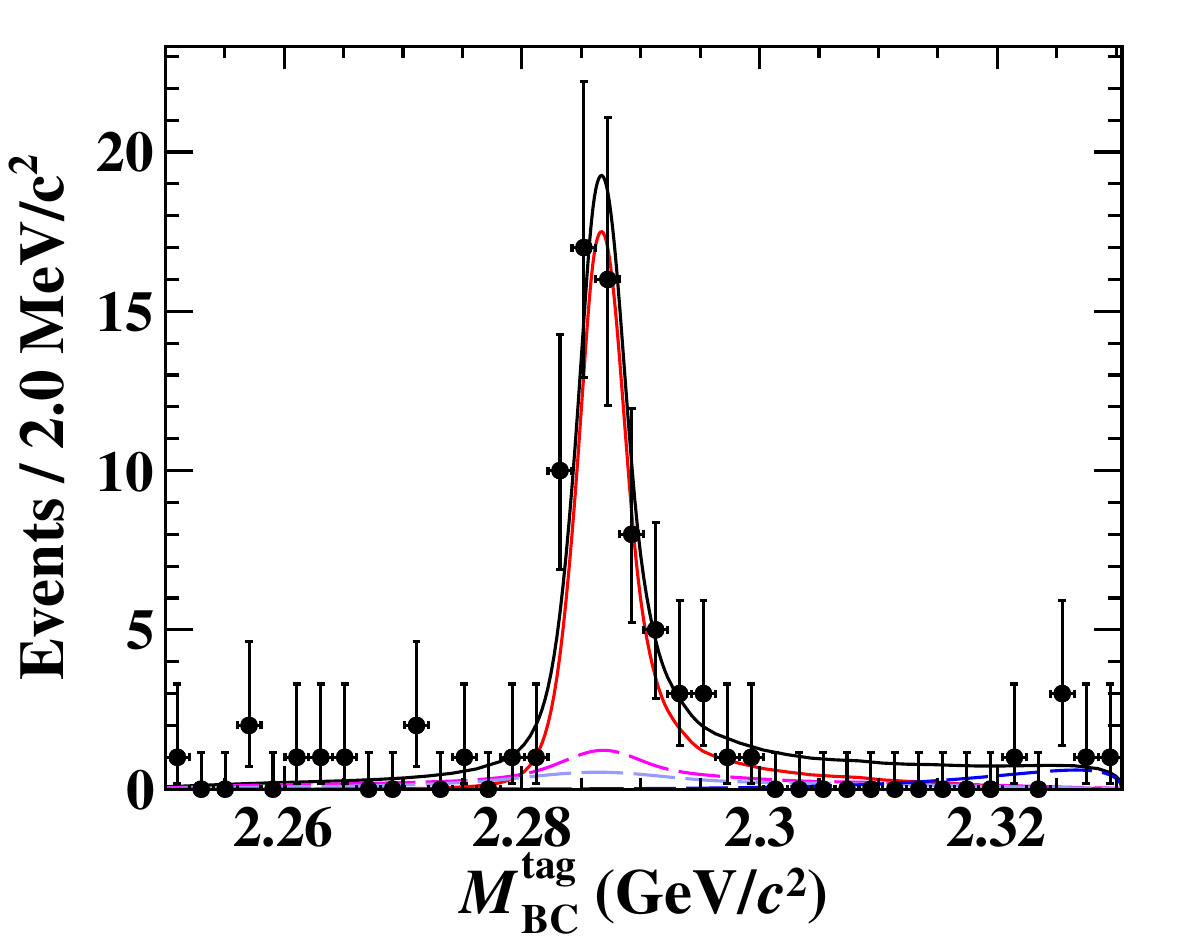}
  \includegraphics[width=0.24\textwidth]{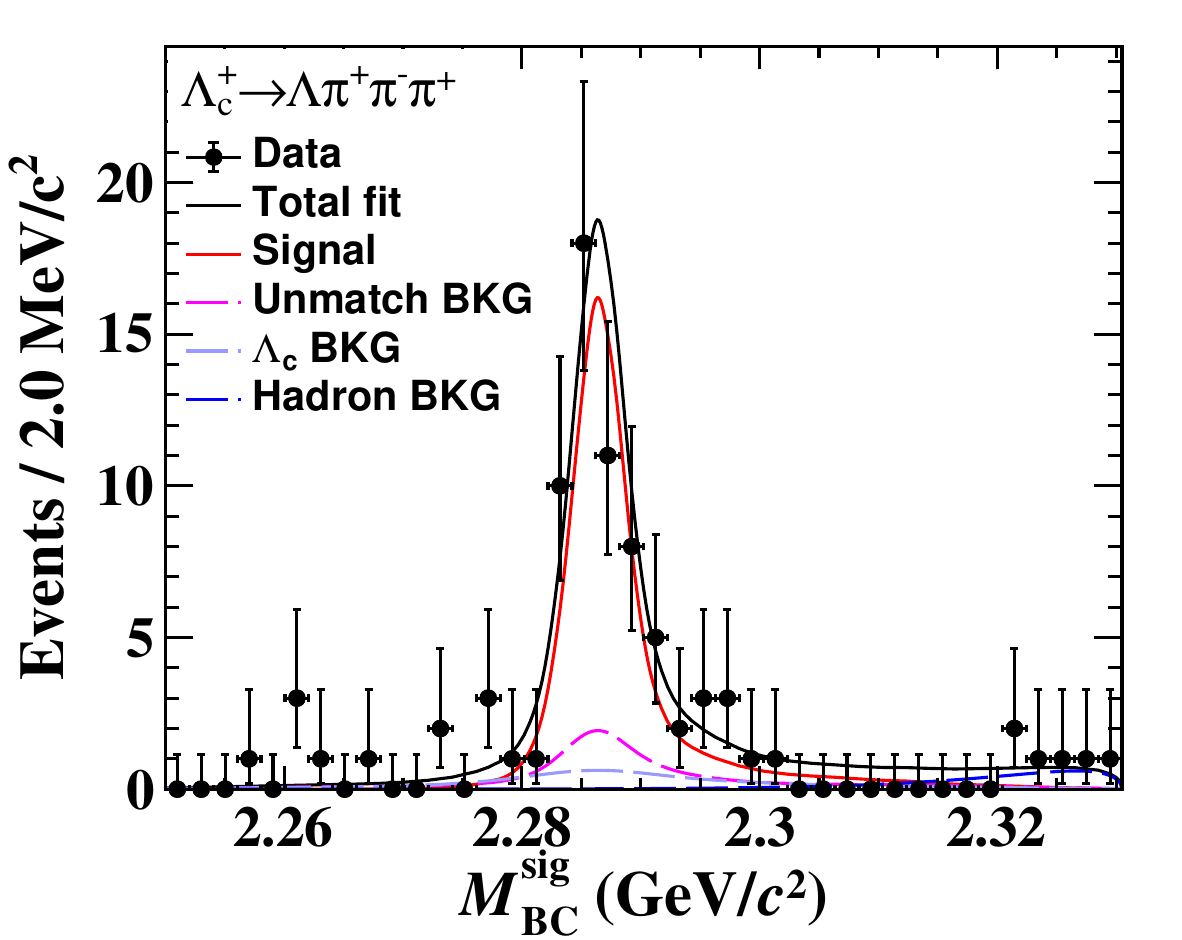}
  \includegraphics[width=0.24\textwidth]{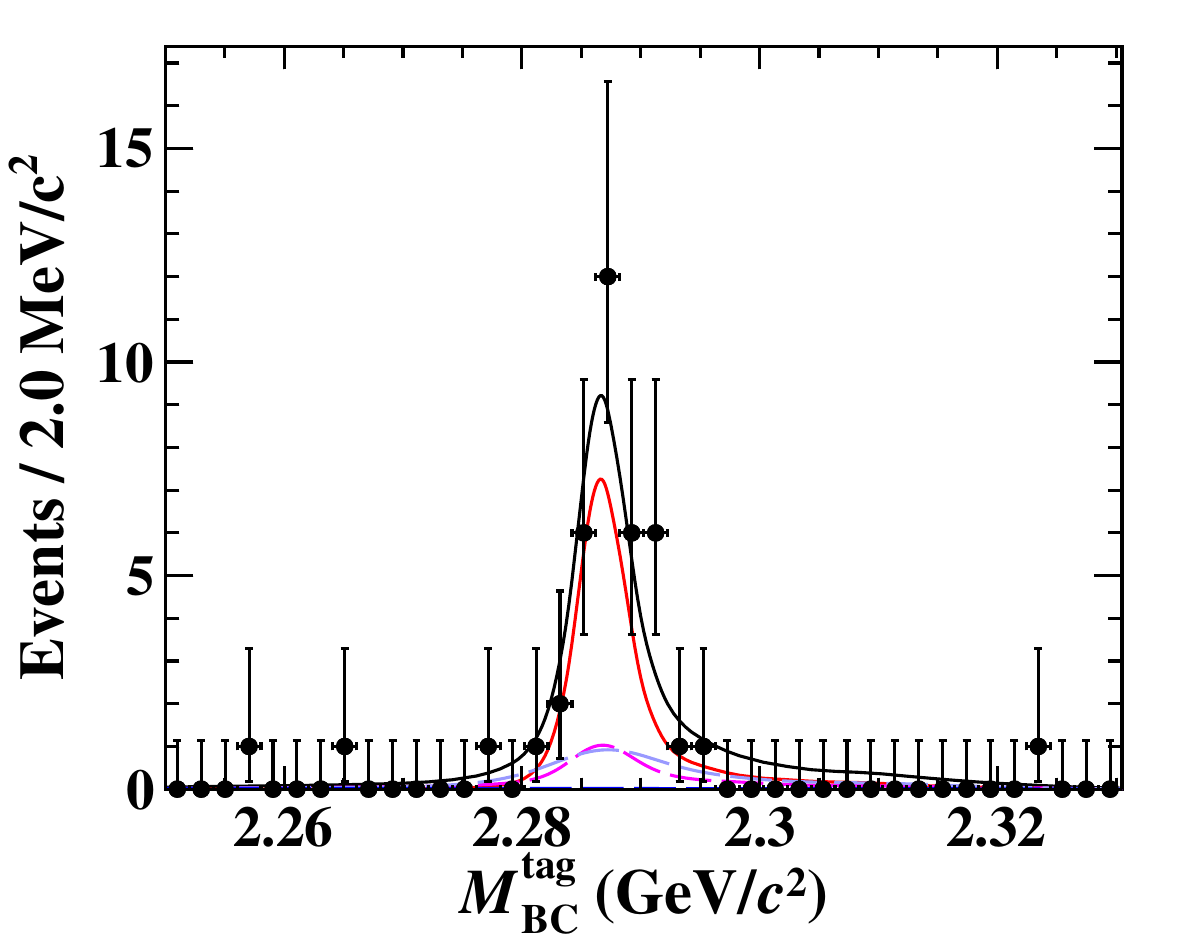}
  \includegraphics[width=0.24\textwidth]{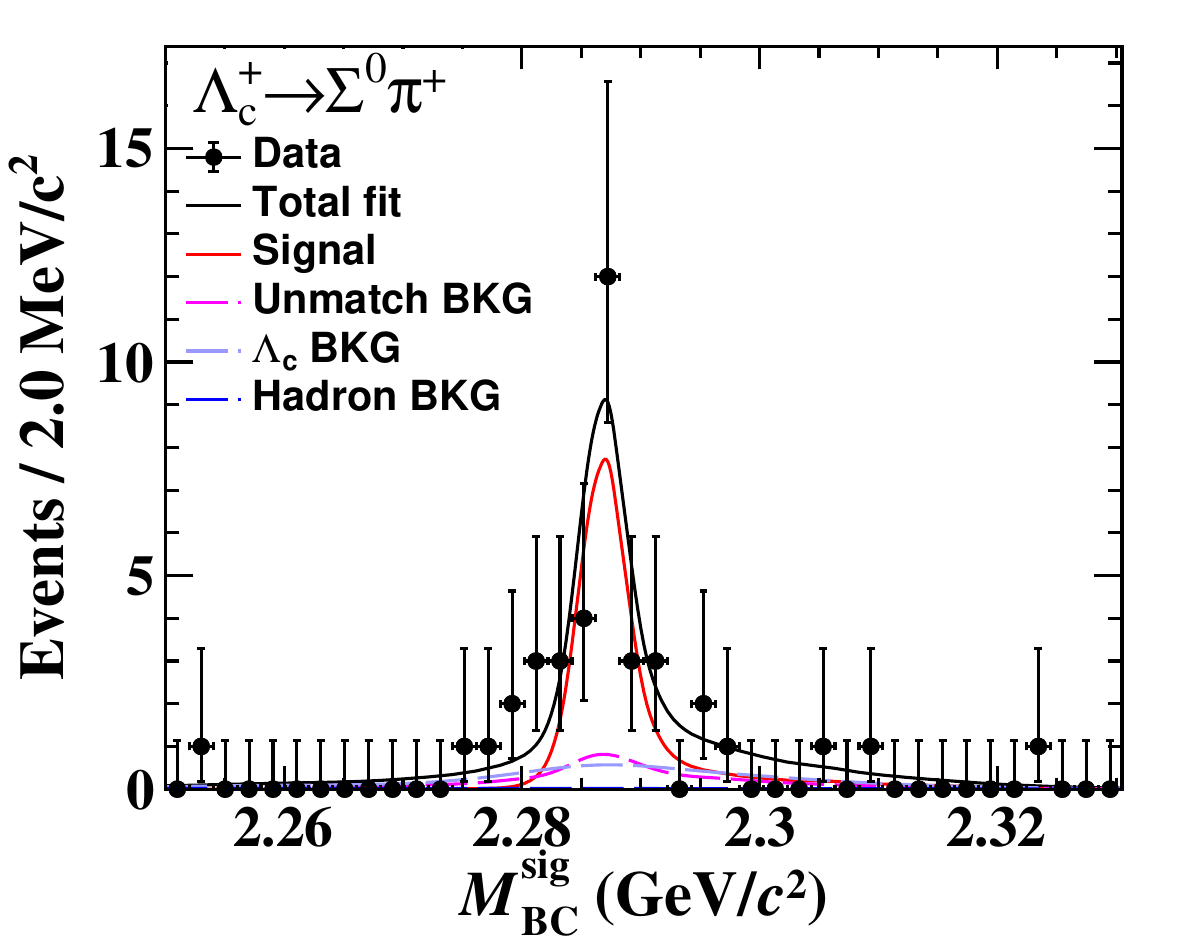}
  \includegraphics[width=0.24\textwidth]{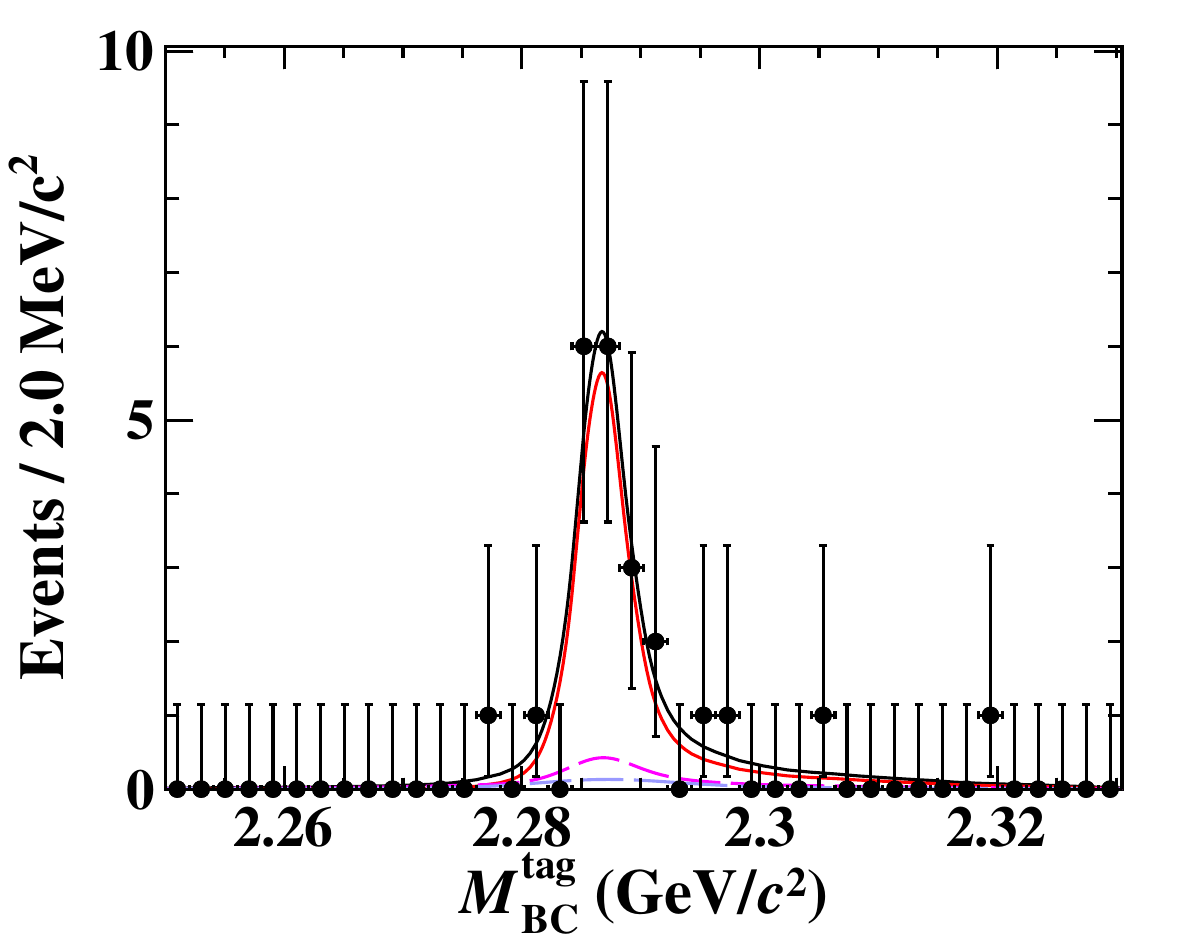}
  \includegraphics[width=0.24\textwidth]{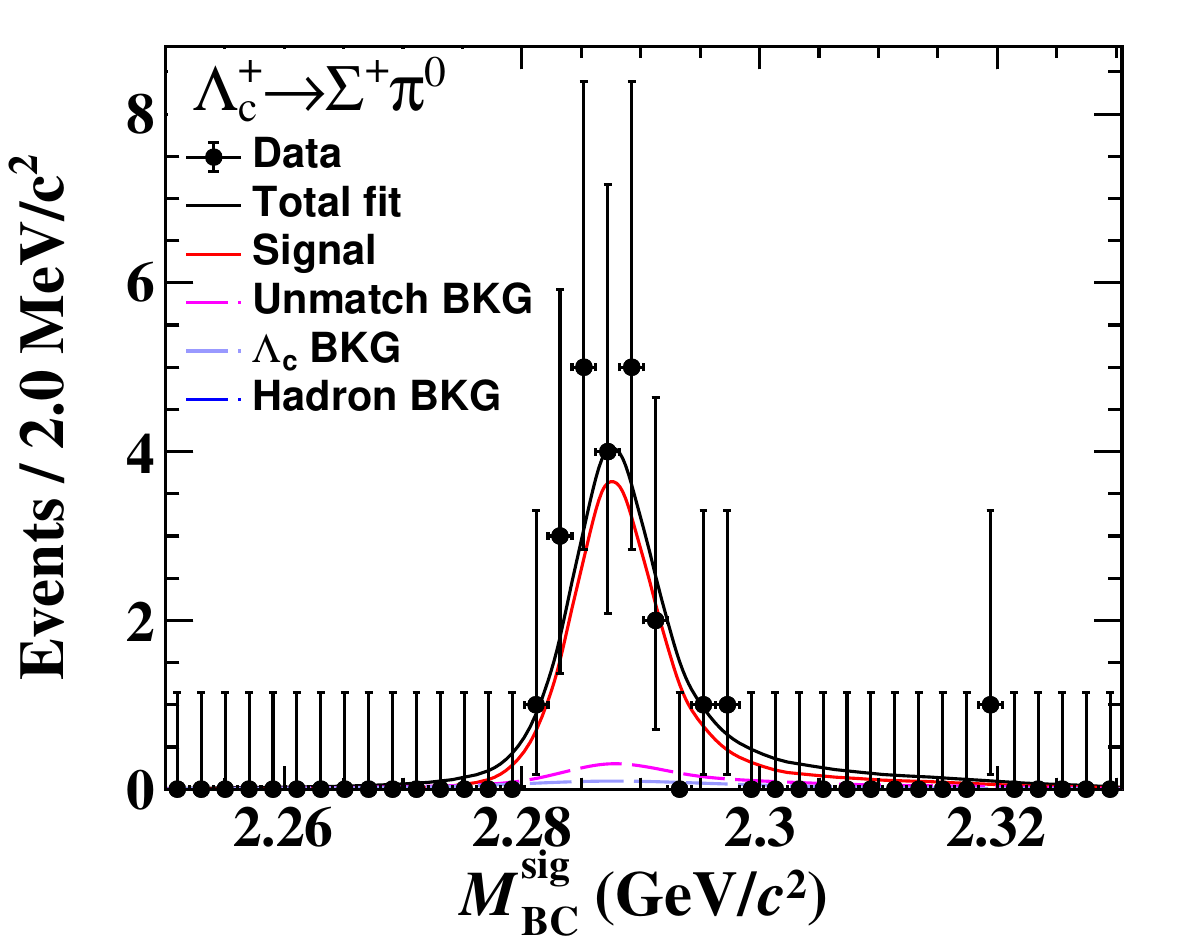}
  \includegraphics[width=0.24\textwidth]{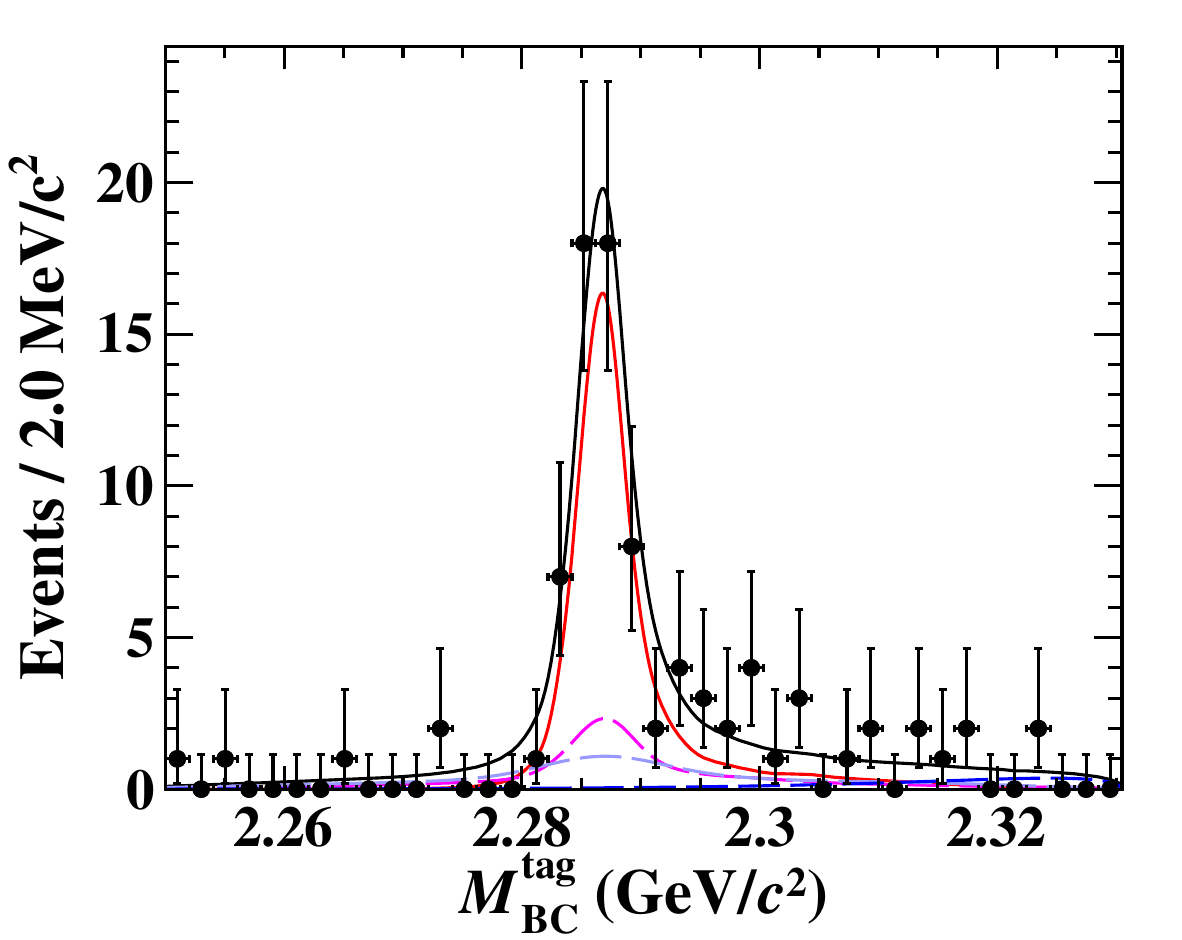}
  \includegraphics[width=0.24\textwidth]{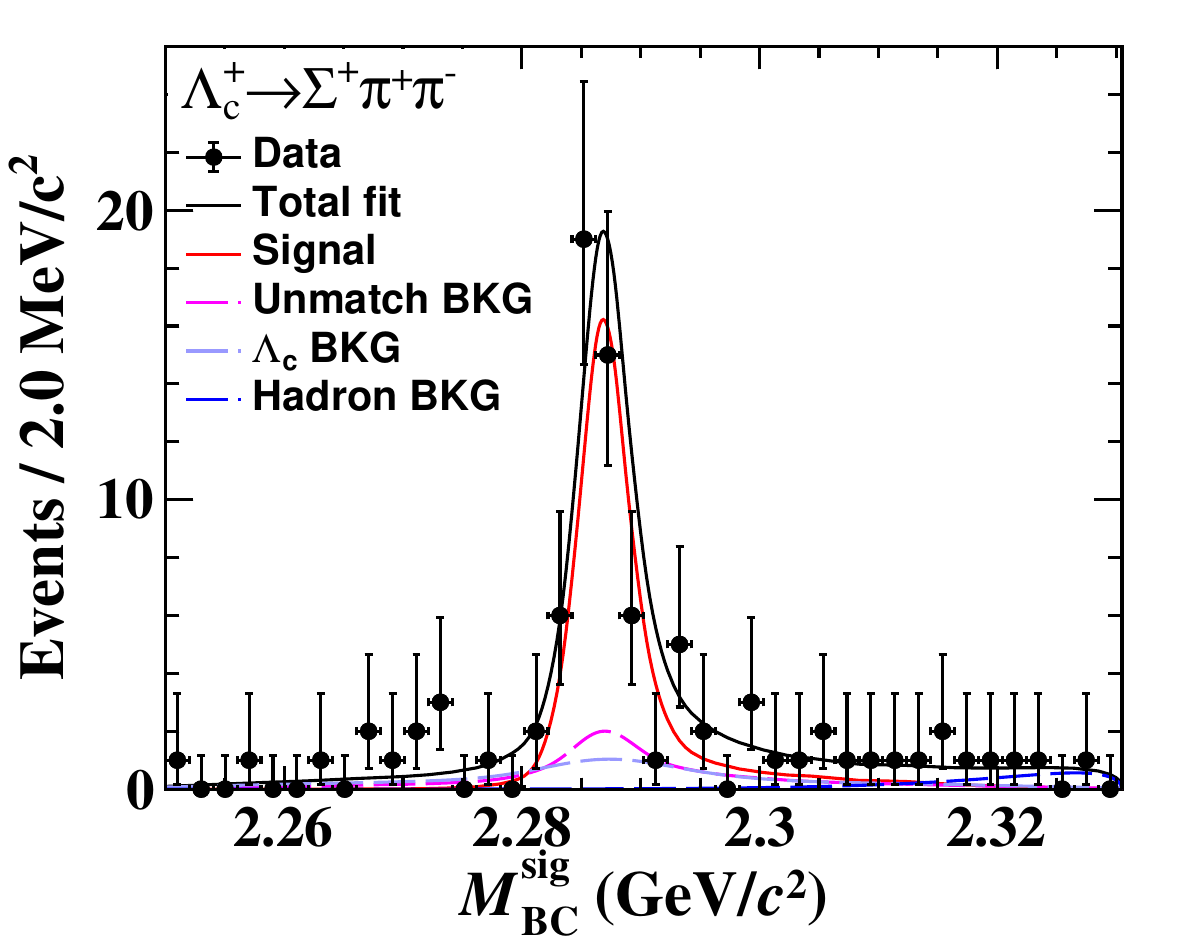}
  \includegraphics[width=0.24\textwidth]{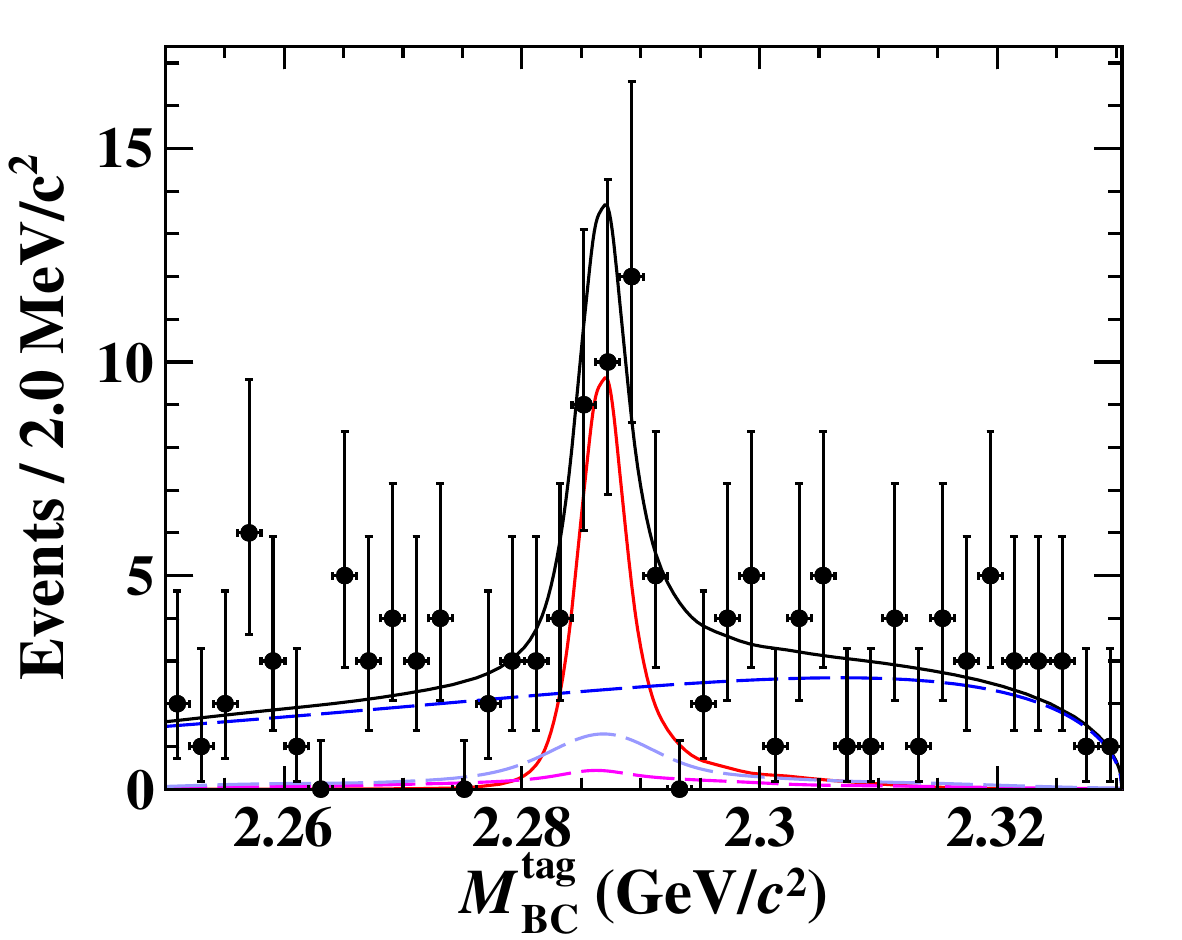}
  \includegraphics[width=0.24\textwidth]{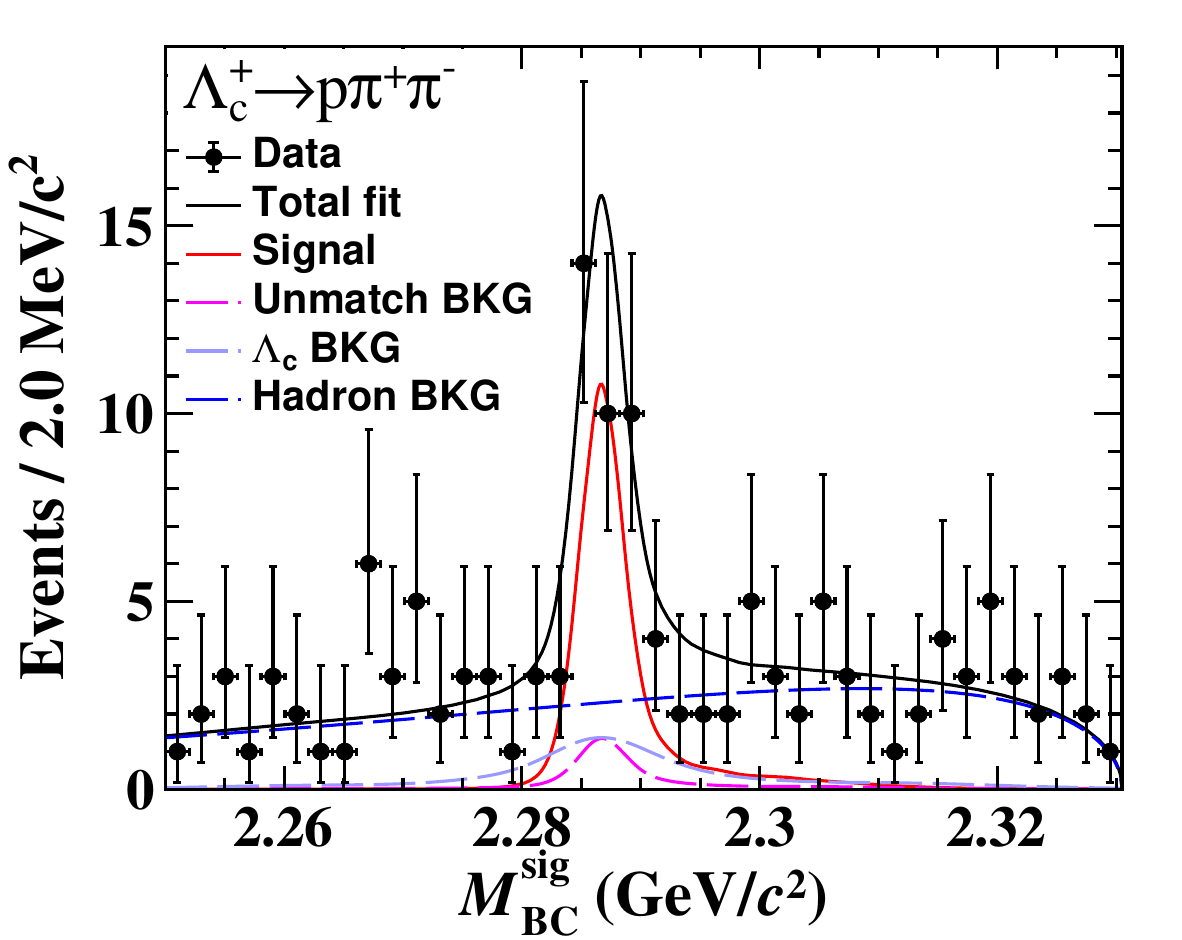}
     \vspace*{-0.5cm}
  \end{center}
\caption{The projections of the 2D fits on the $M_{\rm BC}^{\rm tag}$ and $M_{\rm BC}^{\rm sig}$ distributions of the accepted DT candidates at $\sqrt{s}=4661.24~\mev$. The plots in the first and third columns show the combined 12 tag modes for each signal mode. 
The points with error bars are data, the black lines are the sum of fit functions, the red lines are the matched signal shapes, the pink dashed lines are the unmatched signal shapes, the lilac dashed lines are the non-signal $\lcp\lcm$ shapes, and the blue dashed lines are the ARGUS functions.}
\label{fig:DT_yield_4660}
\end{figure}

\begin{figure}[!htbp]
  \begin{center}
  
  \includegraphics[width=0.24\textwidth]{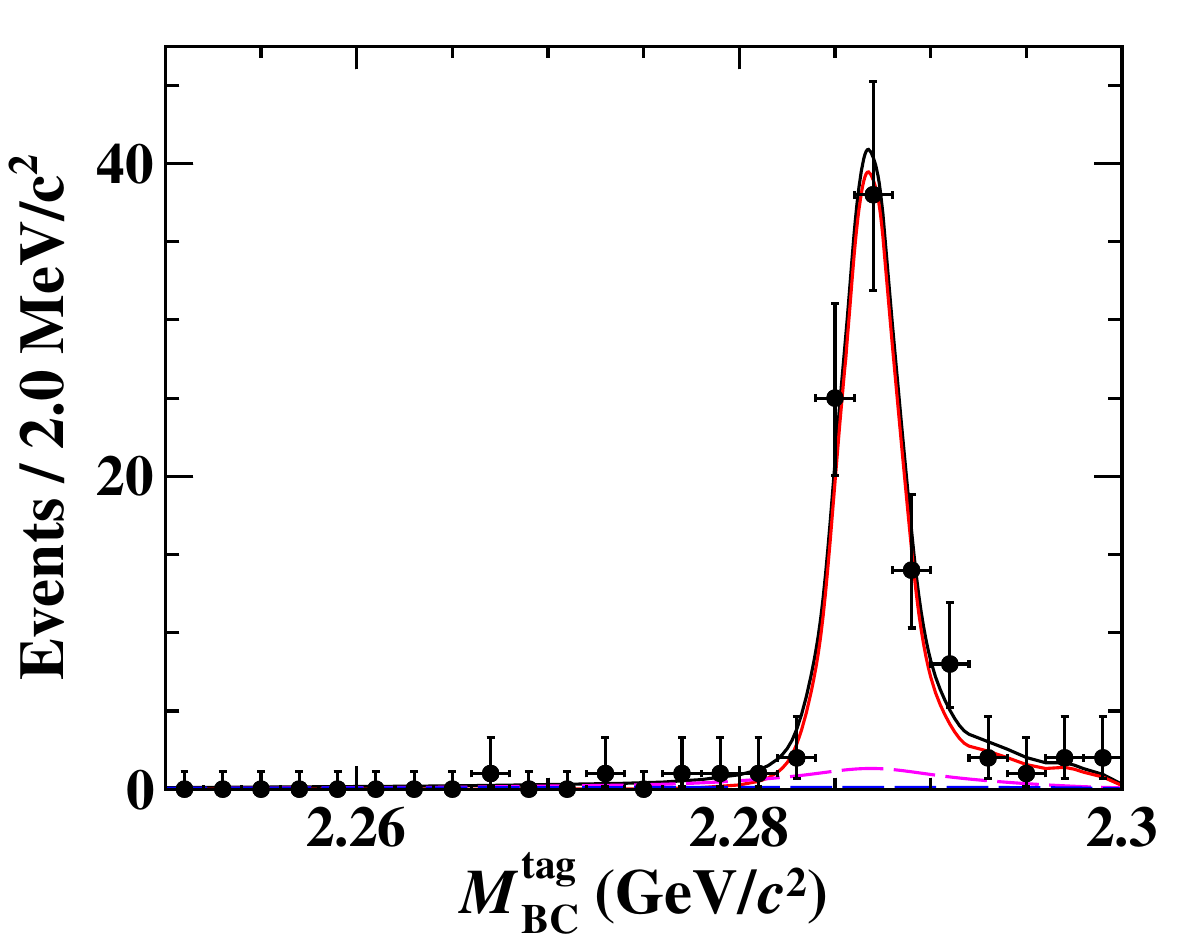}
  \includegraphics[width=0.24\textwidth]{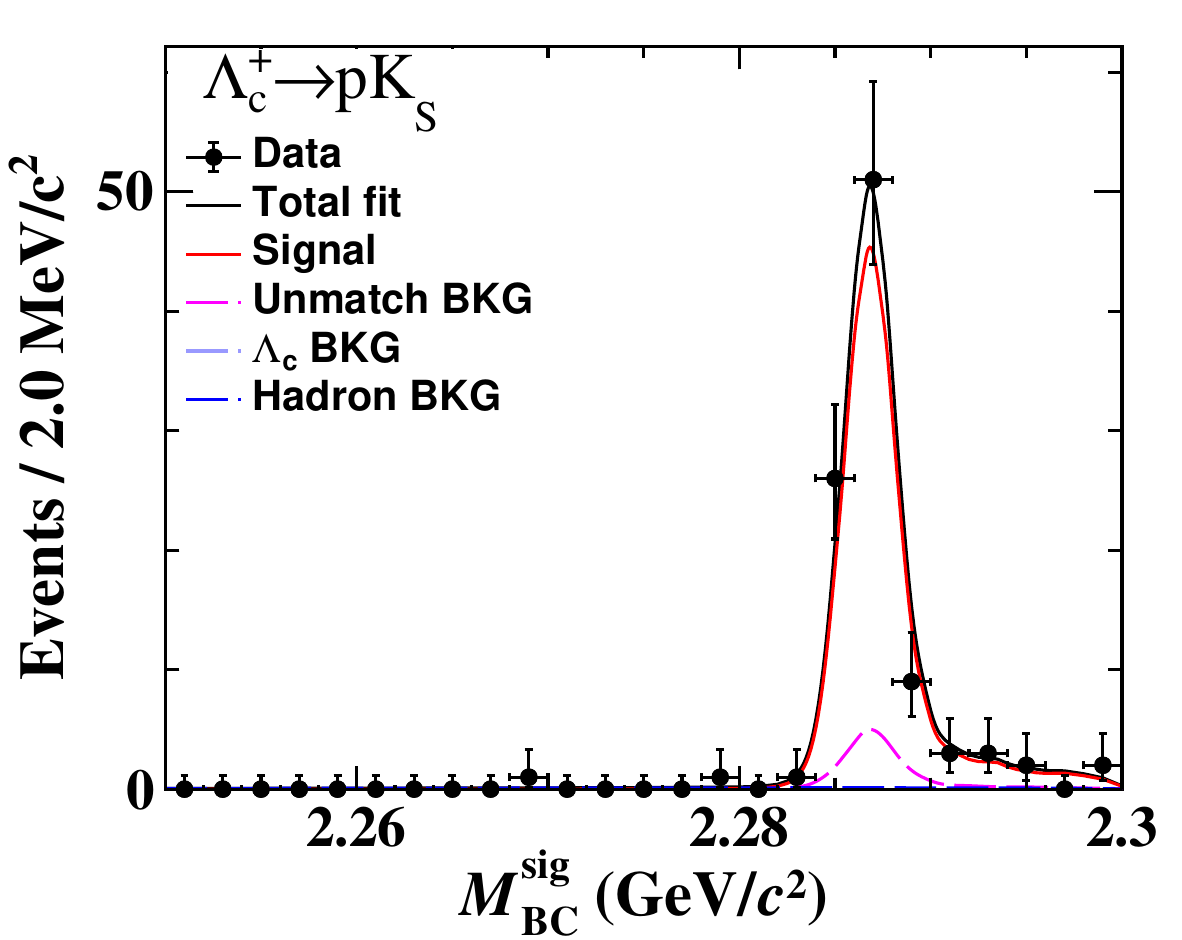}
  \includegraphics[width=0.24\textwidth]{Figure/4680/datamatch_tagmode1.pdf}
  \includegraphics[width=0.24\textwidth]{Figure/4680/datamatch_sigmode1.pdf}
  \includegraphics[width=0.24\textwidth]{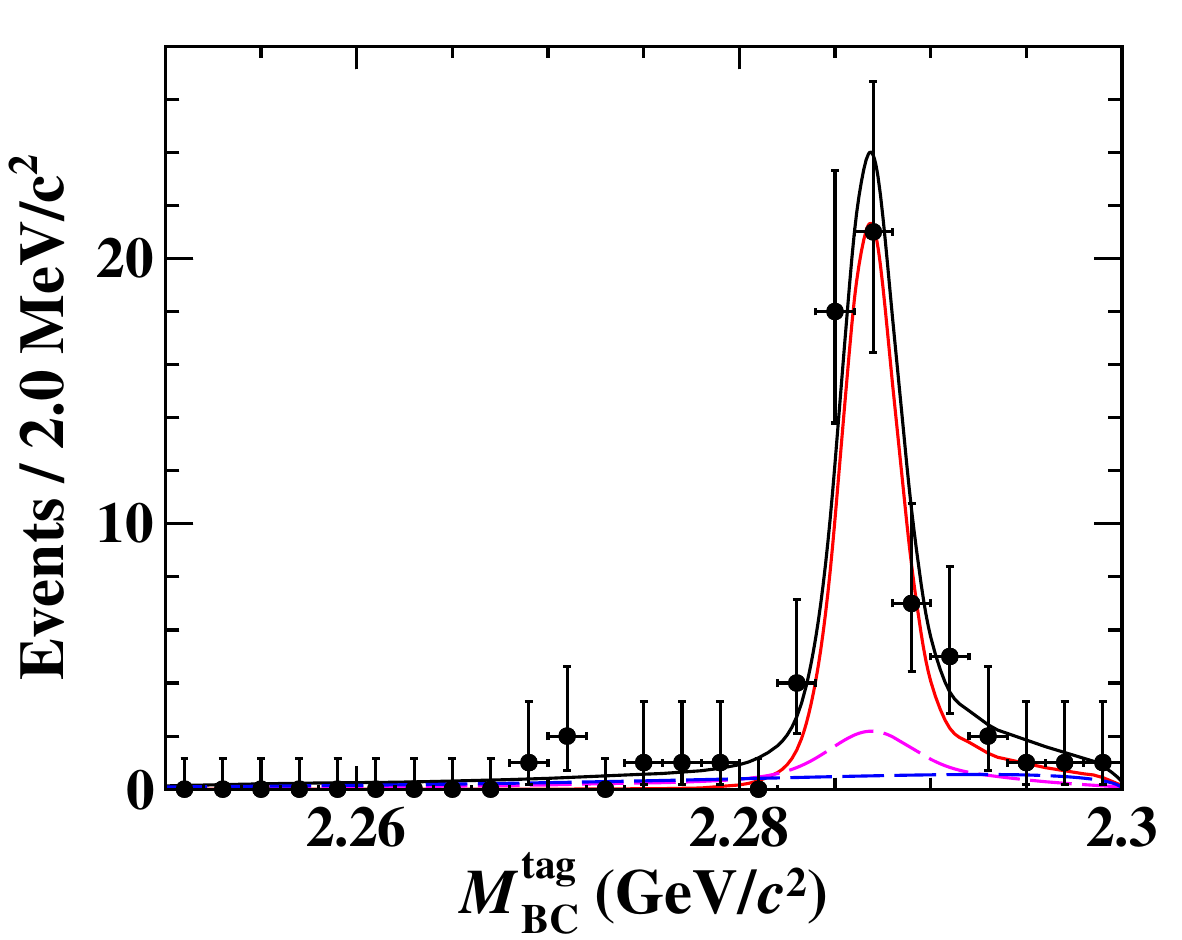}
  \includegraphics[width=0.24\textwidth]{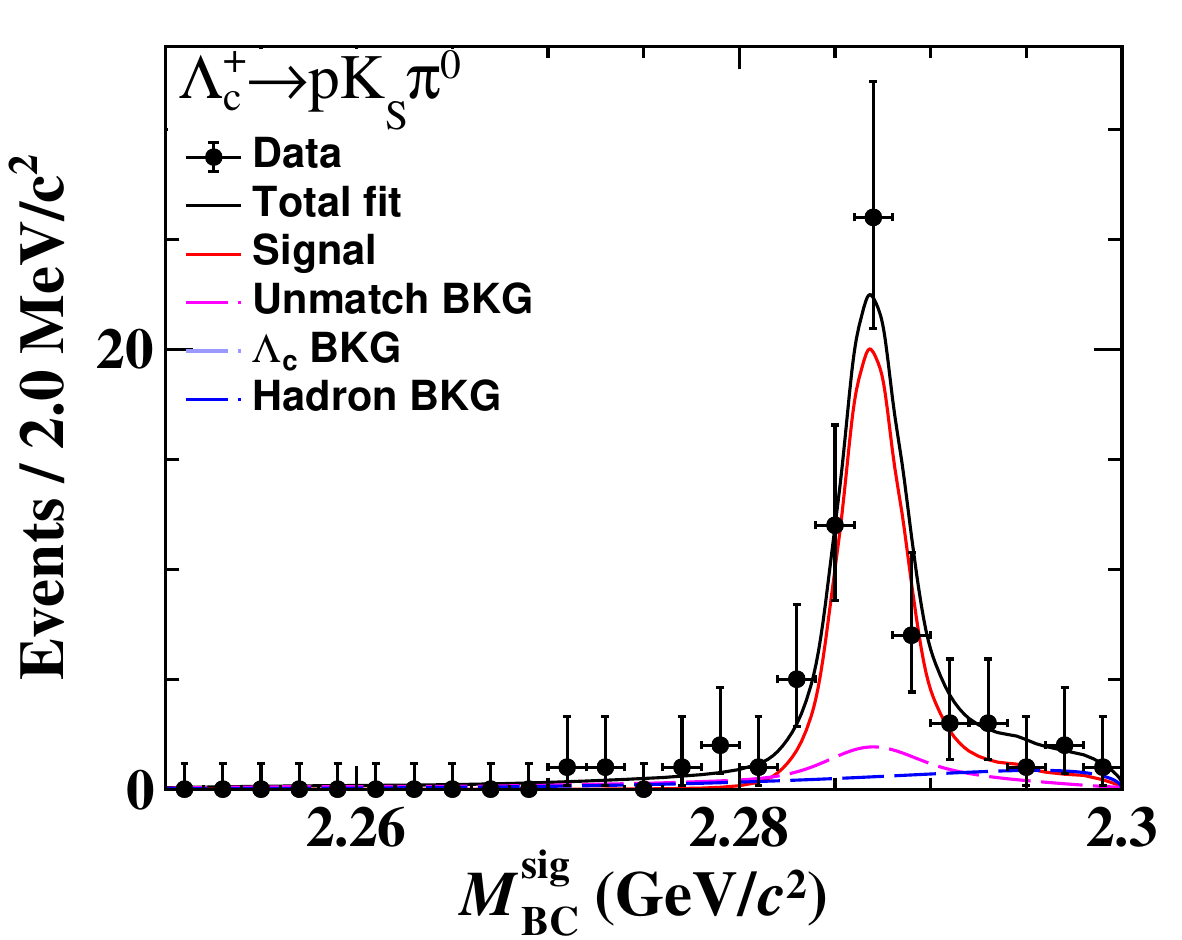}
  \includegraphics[width=0.24\textwidth]{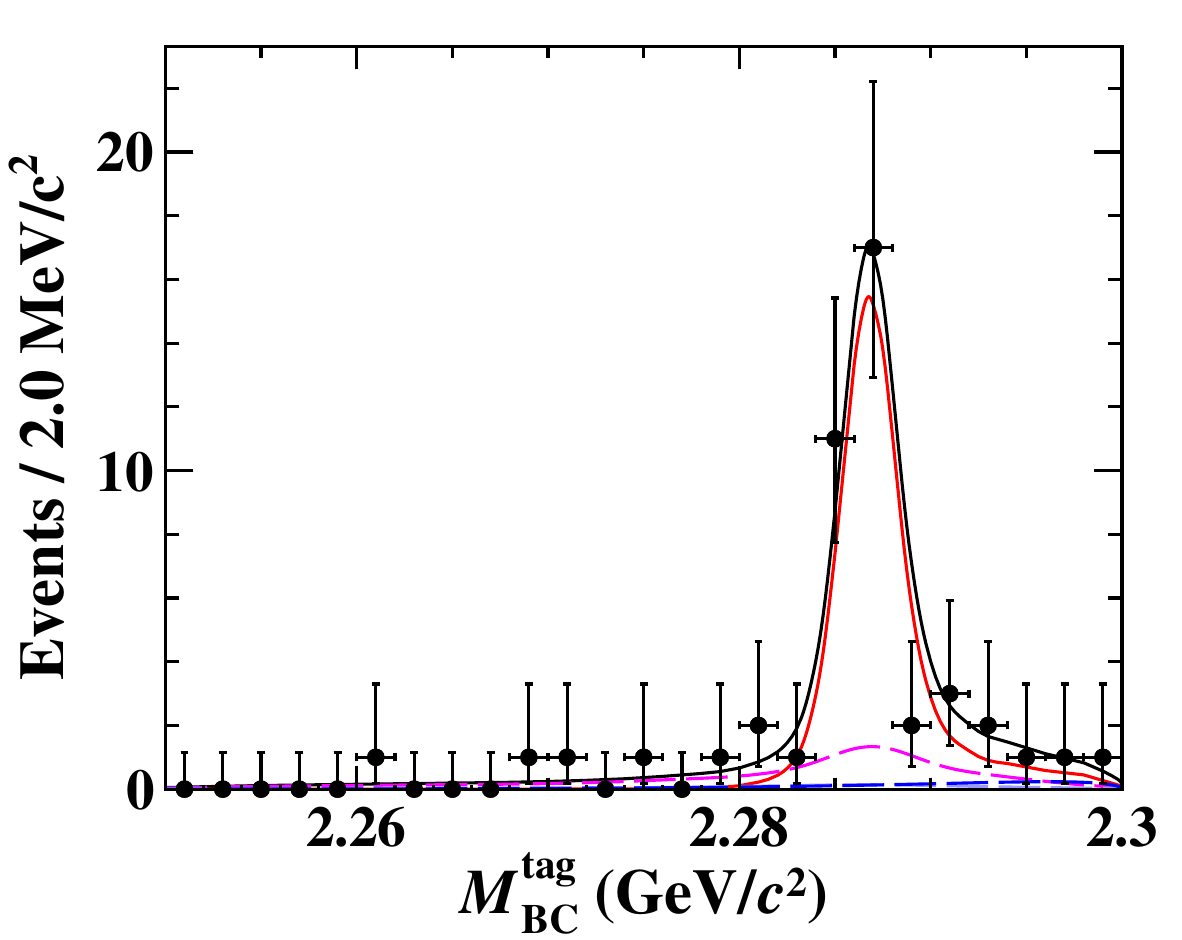}
  \includegraphics[width=0.24\textwidth]{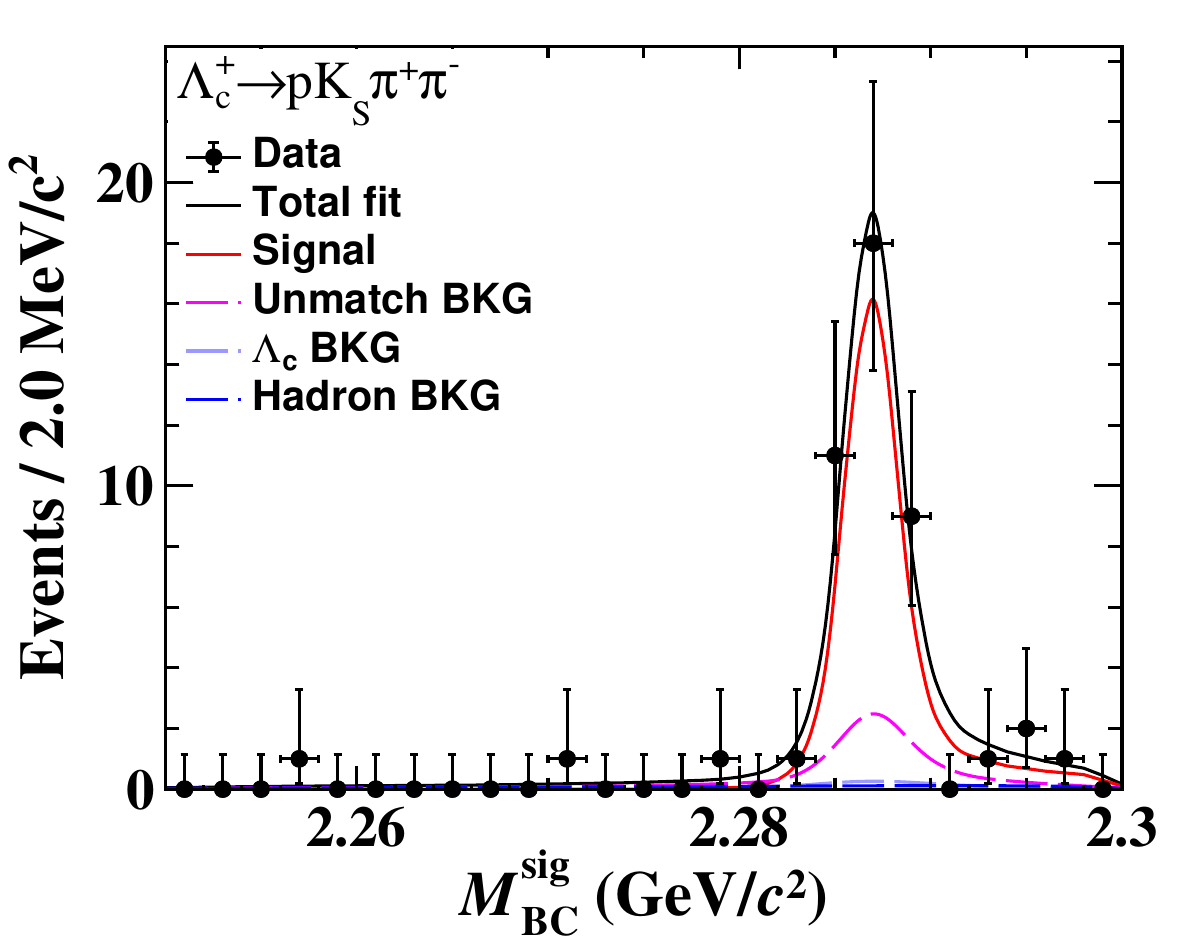}
  \includegraphics[width=0.24\textwidth]{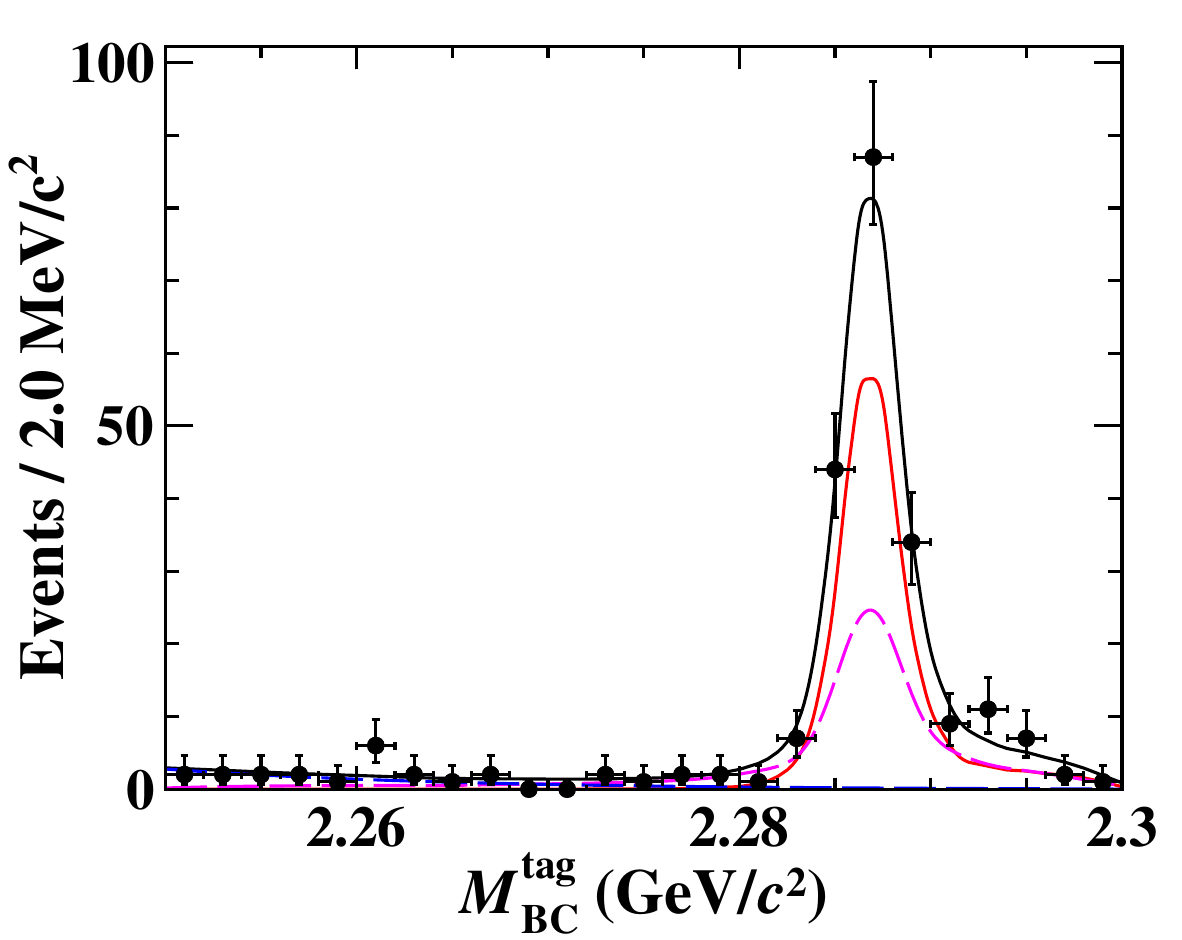}
  \includegraphics[width=0.24\textwidth]{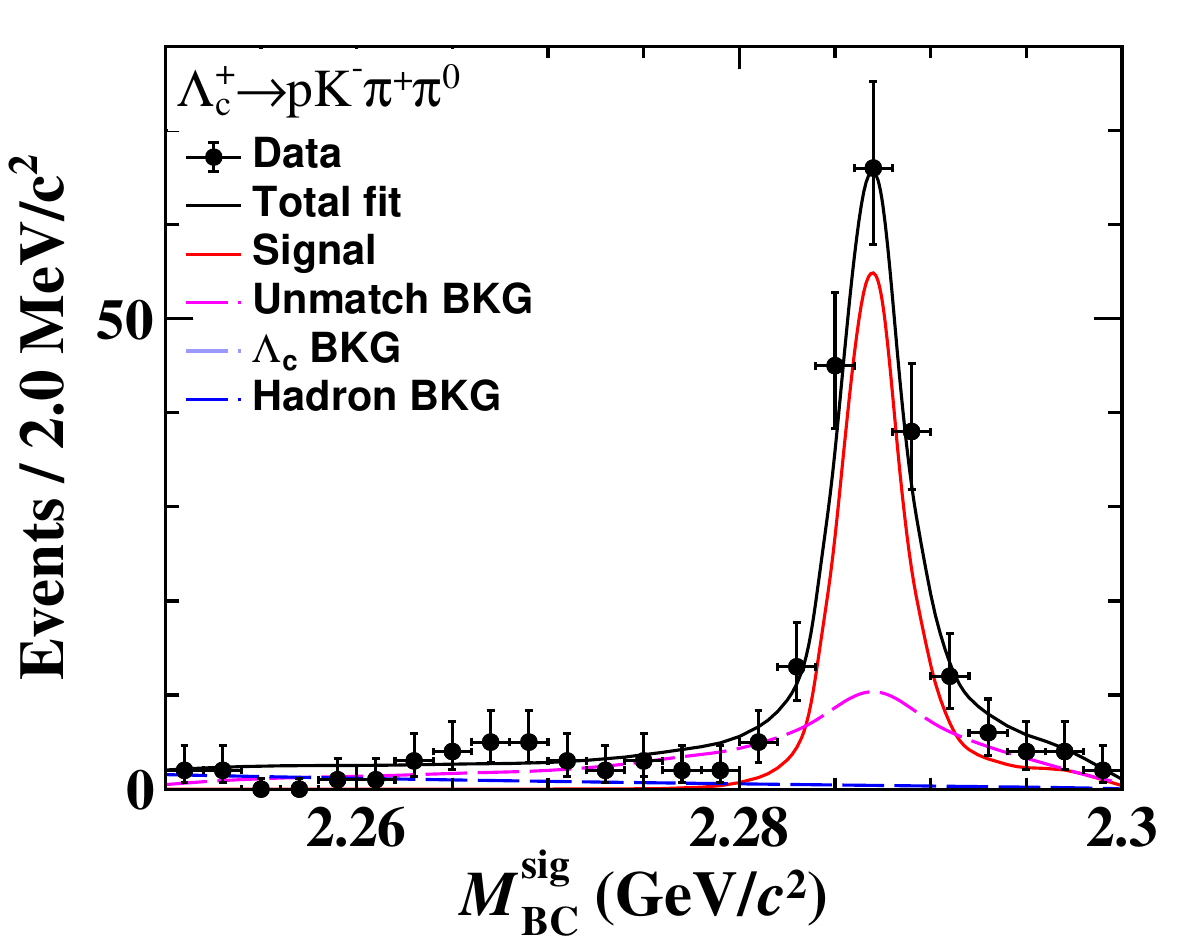}
  \includegraphics[width=0.24\textwidth]{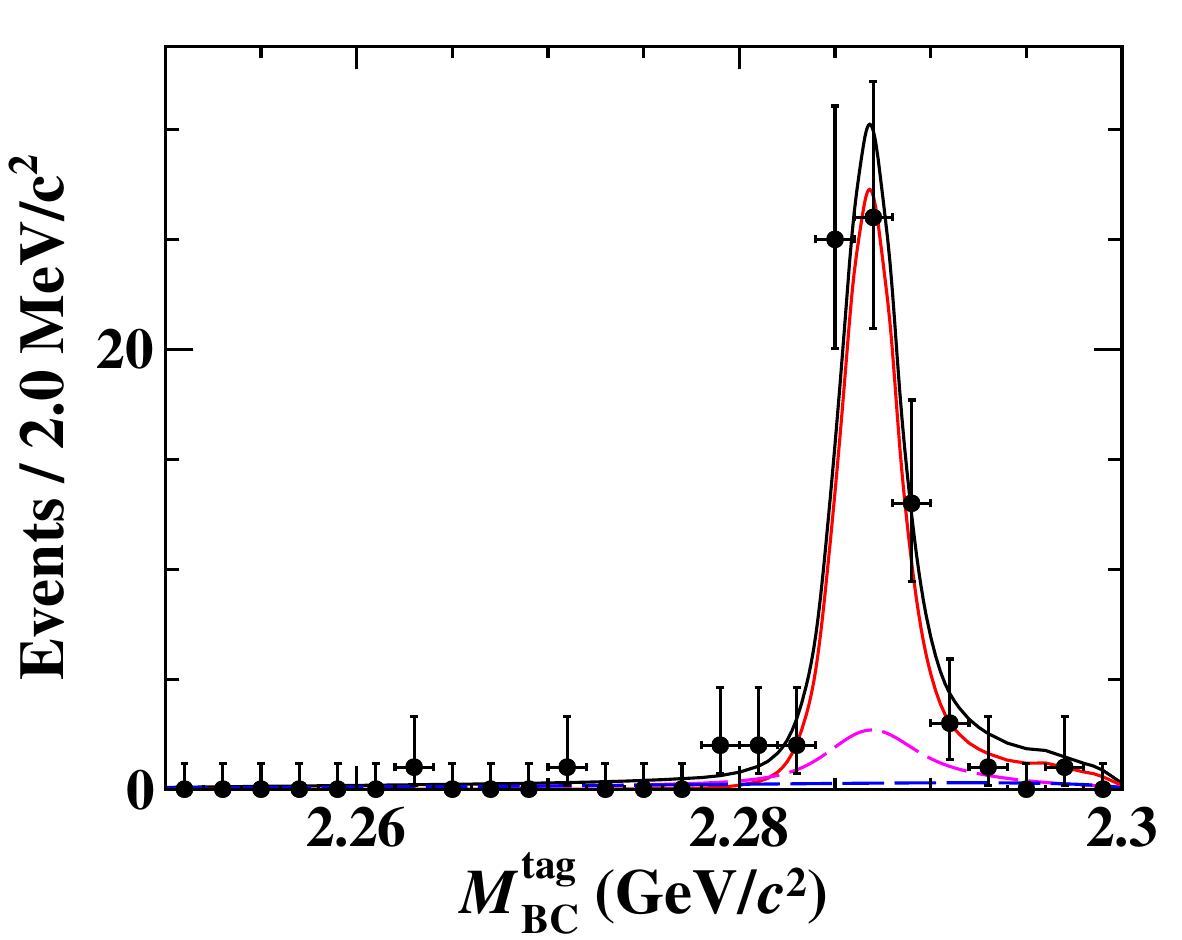}
  \includegraphics[width=0.24\textwidth]{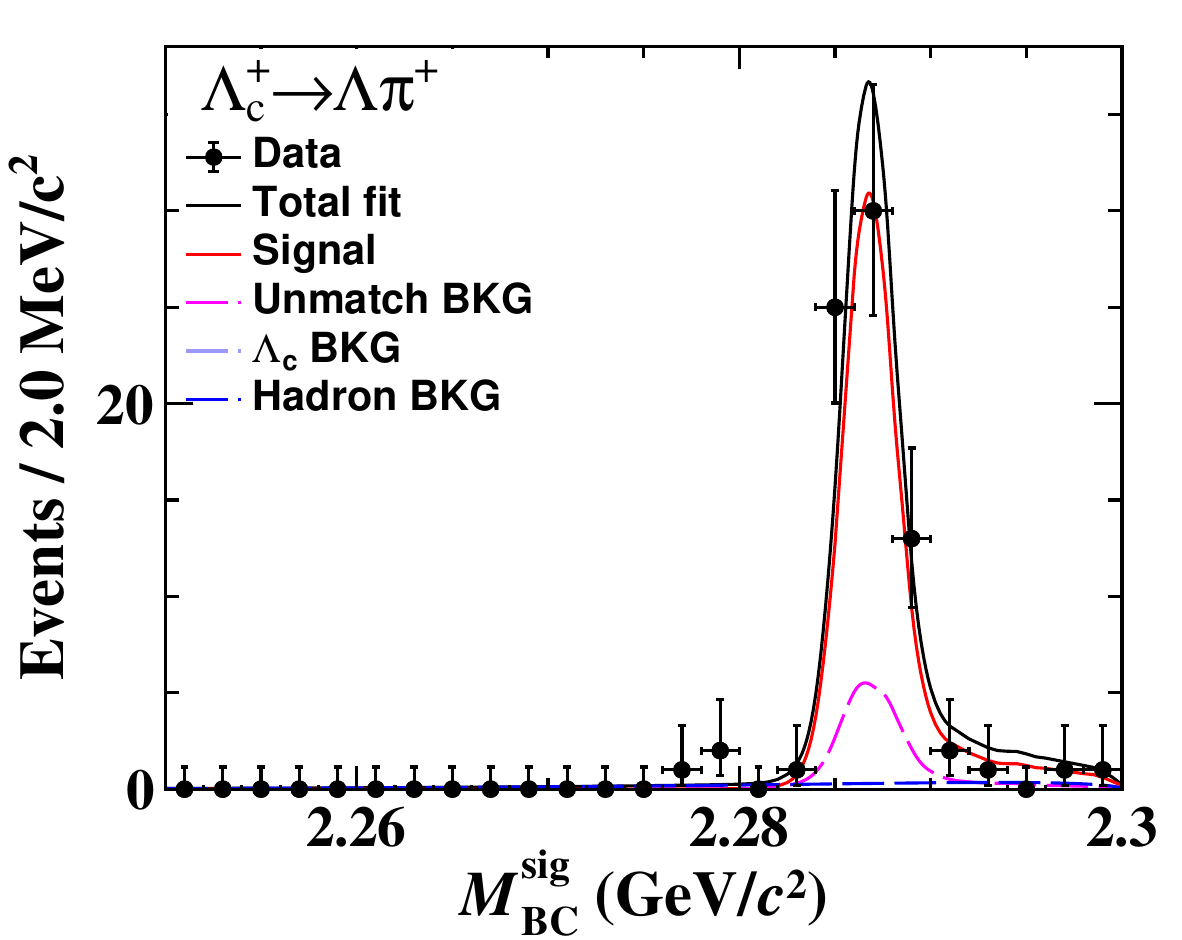}
  \includegraphics[width=0.24\textwidth]{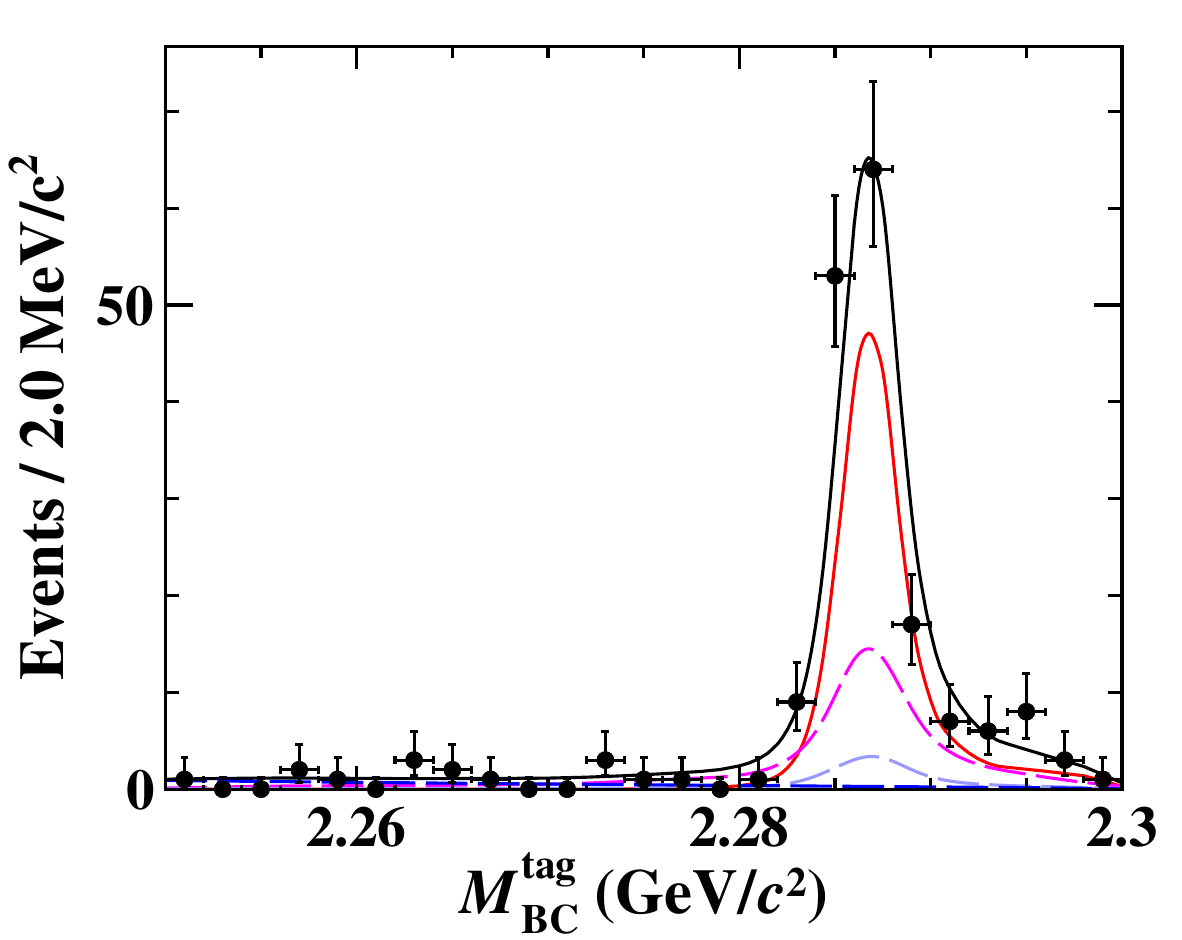}
  \includegraphics[width=0.24\textwidth]{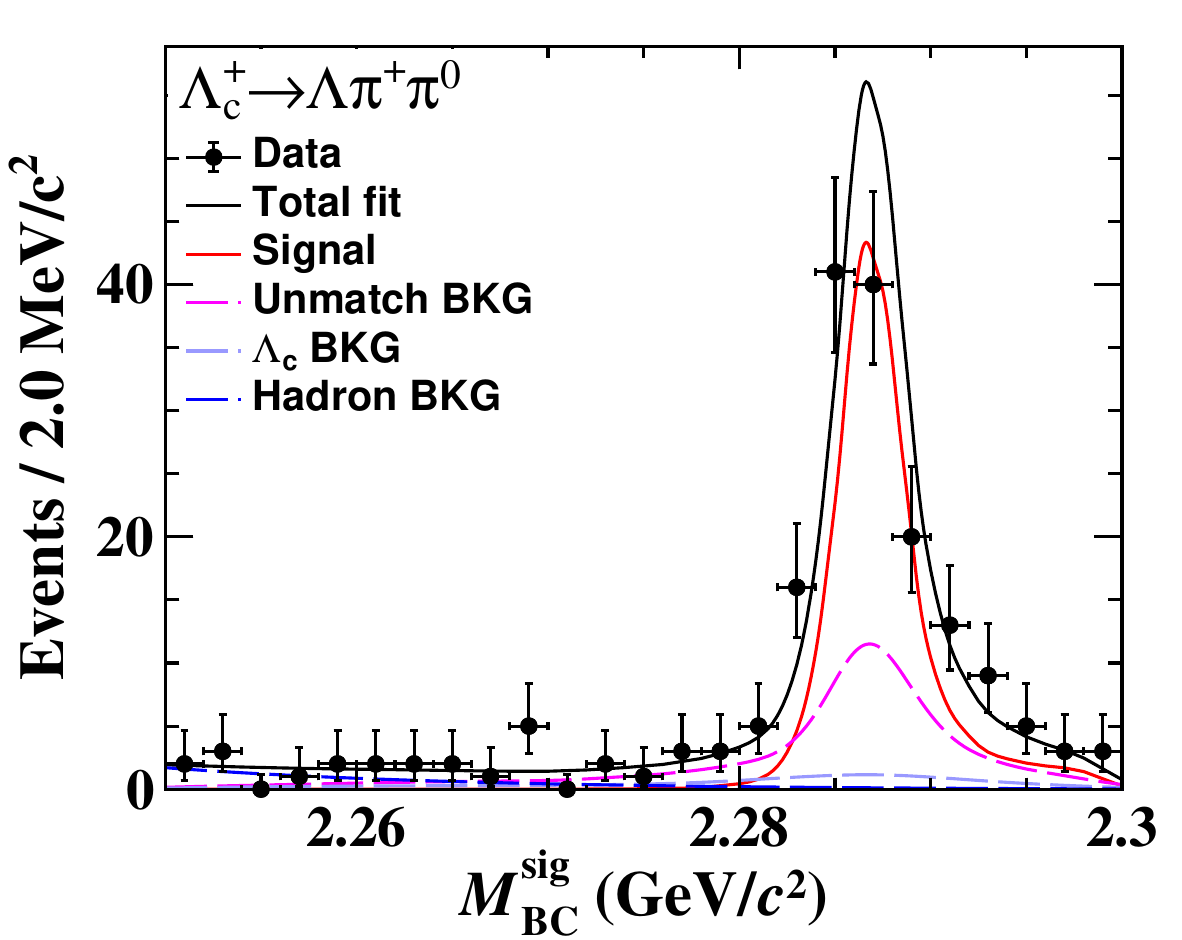}
  \includegraphics[width=0.24\textwidth]{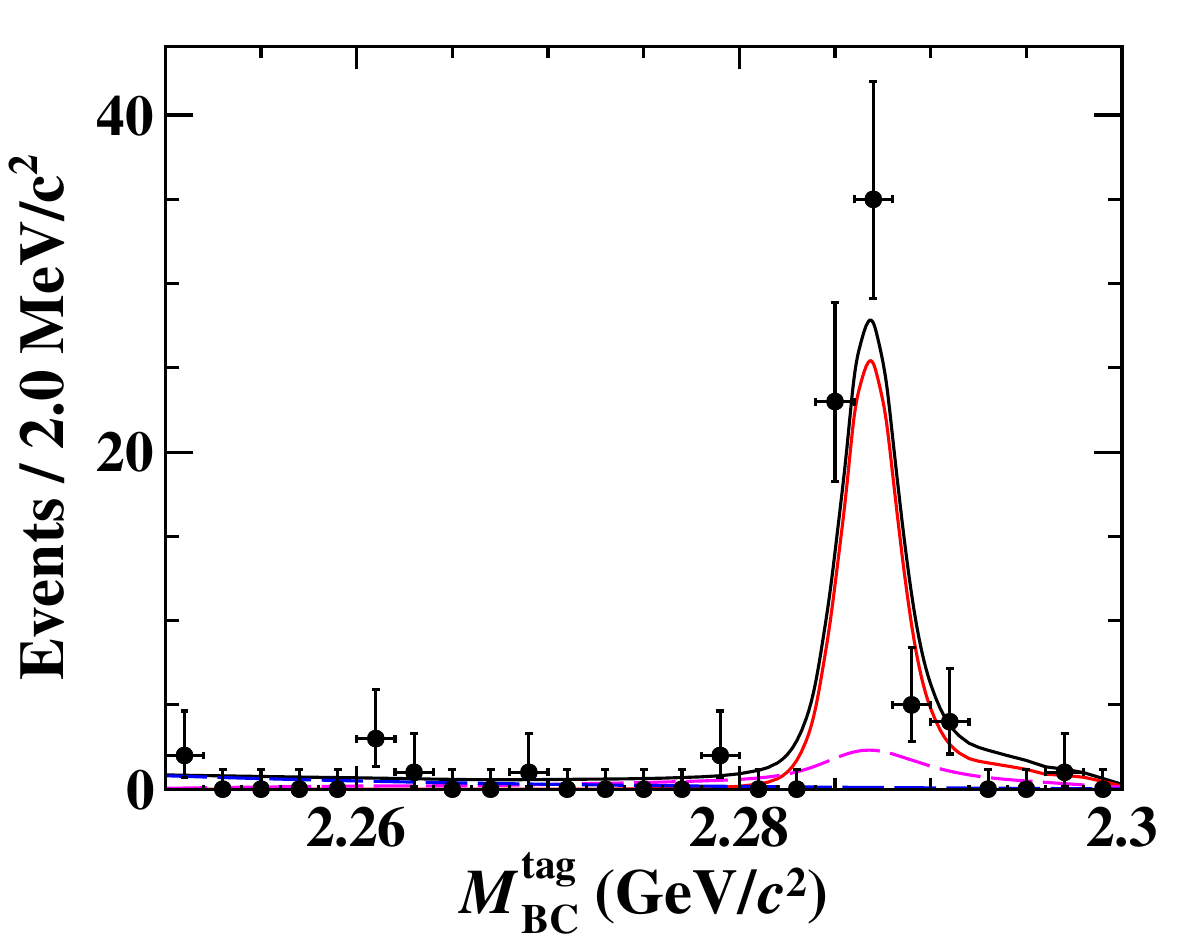}
  \includegraphics[width=0.24\textwidth]{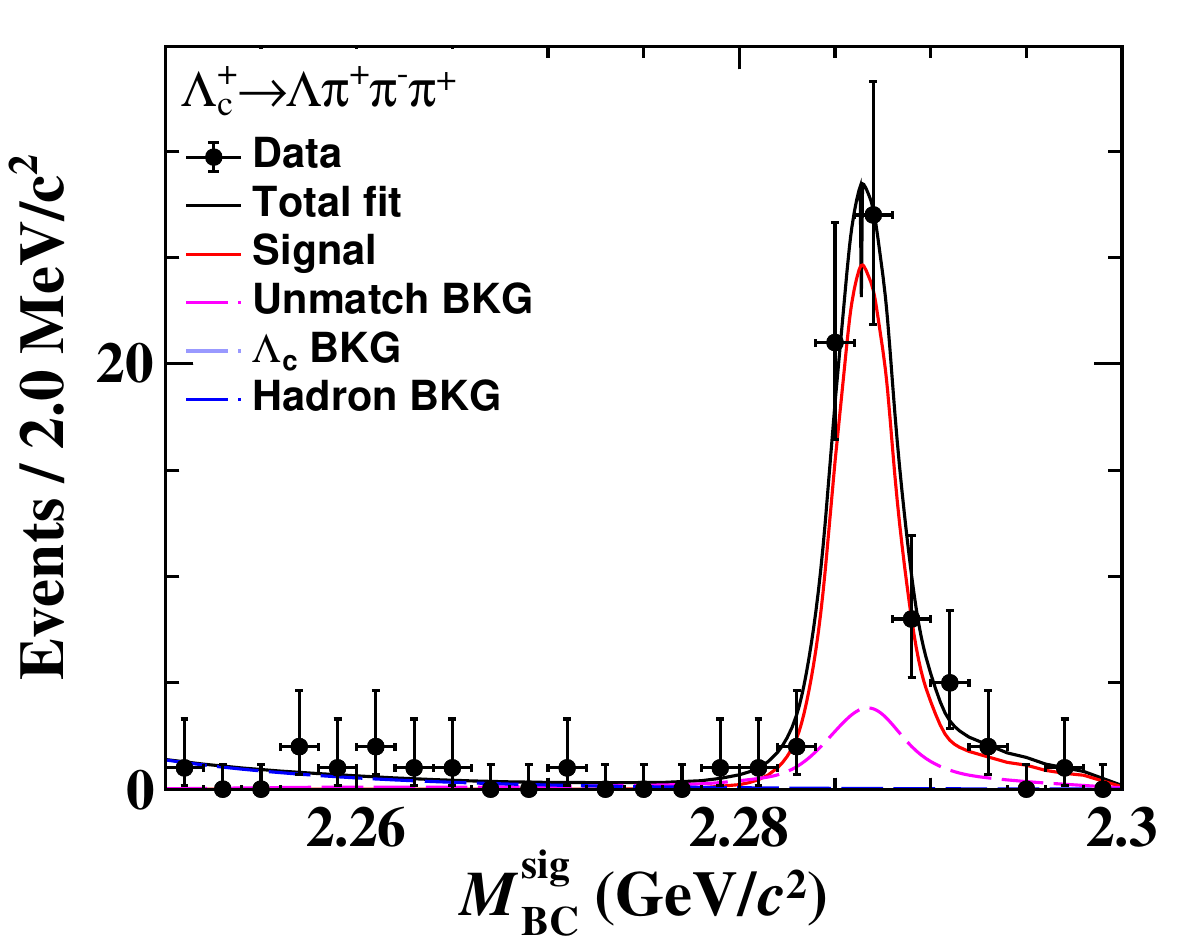}
  \includegraphics[width=0.24\textwidth]{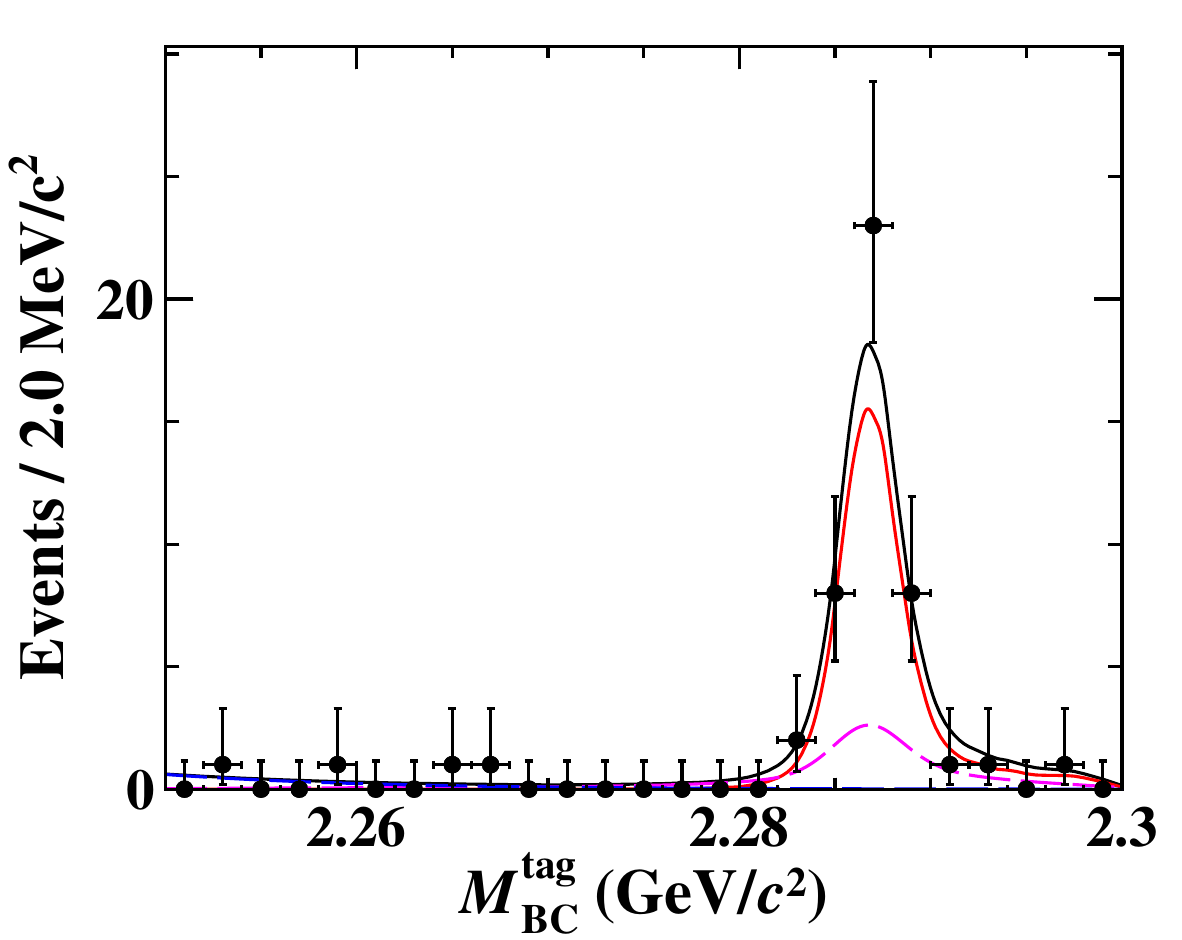}
  \includegraphics[width=0.24\textwidth]{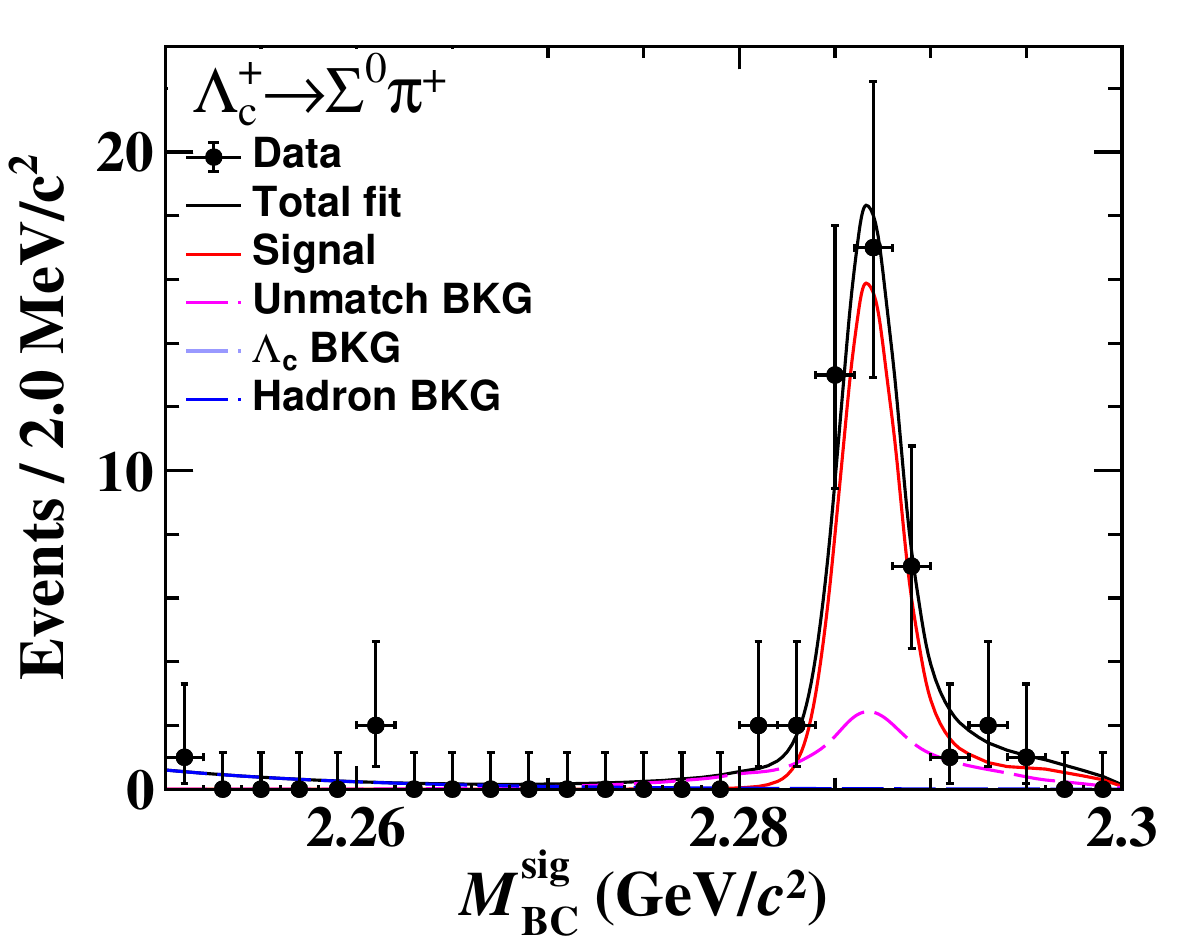}
  \includegraphics[width=0.24\textwidth]{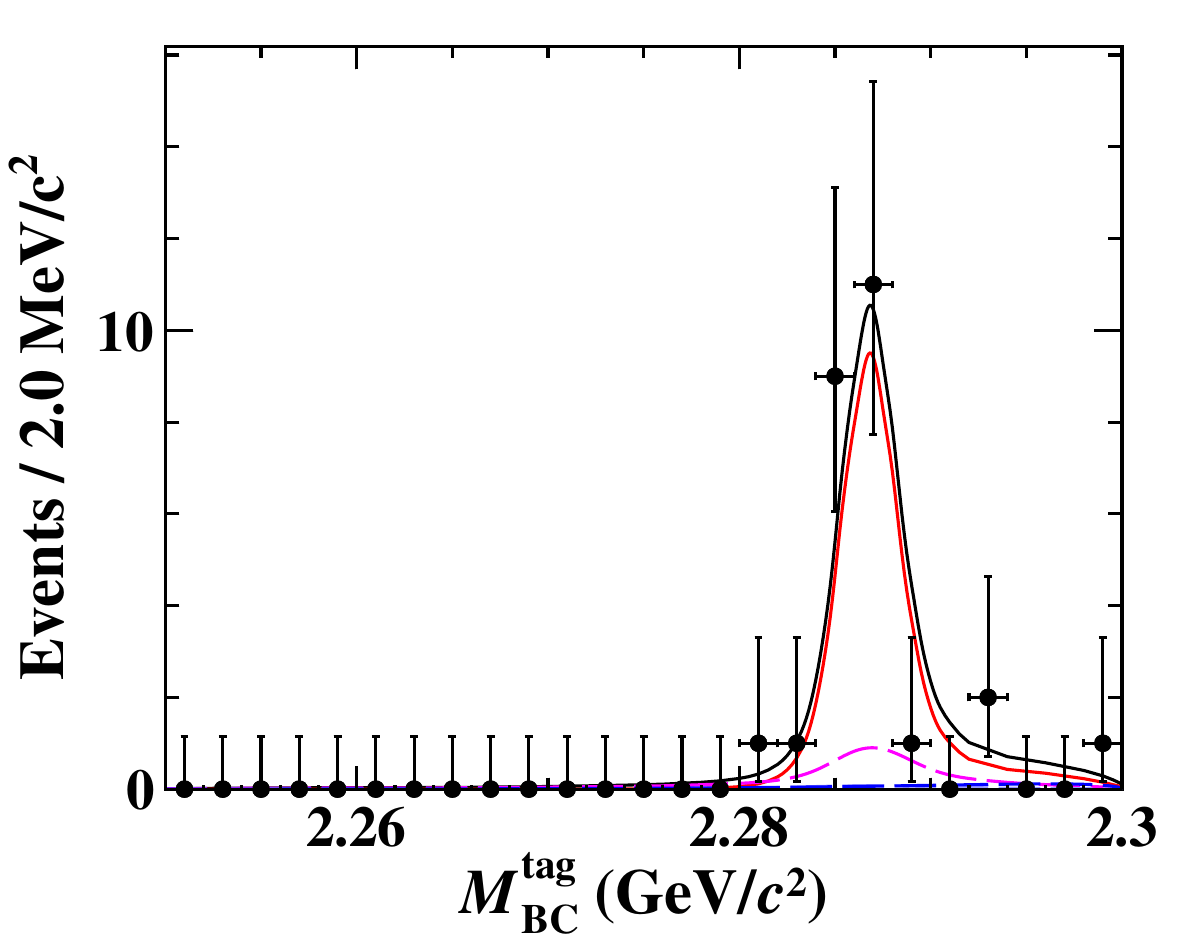}
  \includegraphics[width=0.24\textwidth]{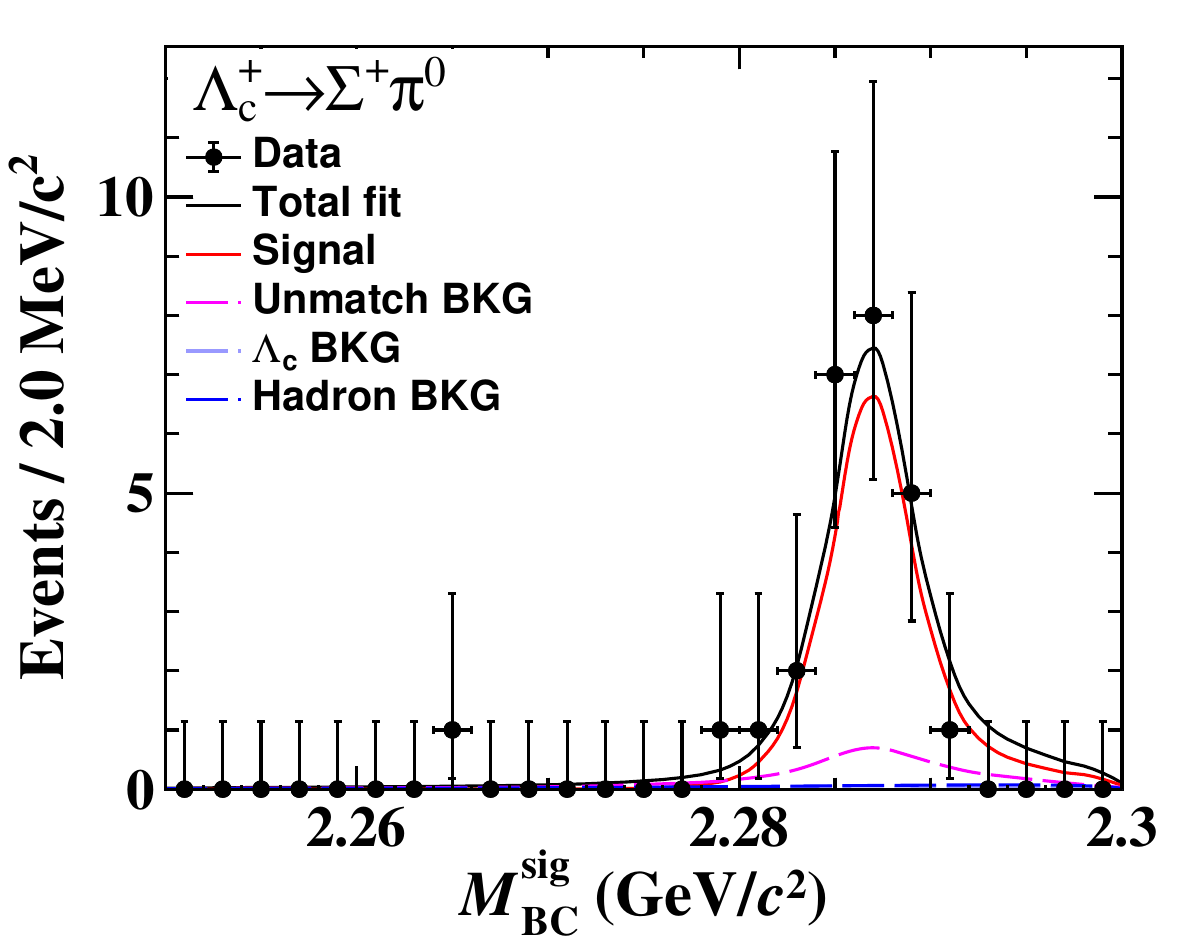}
  \includegraphics[width=0.24\textwidth]{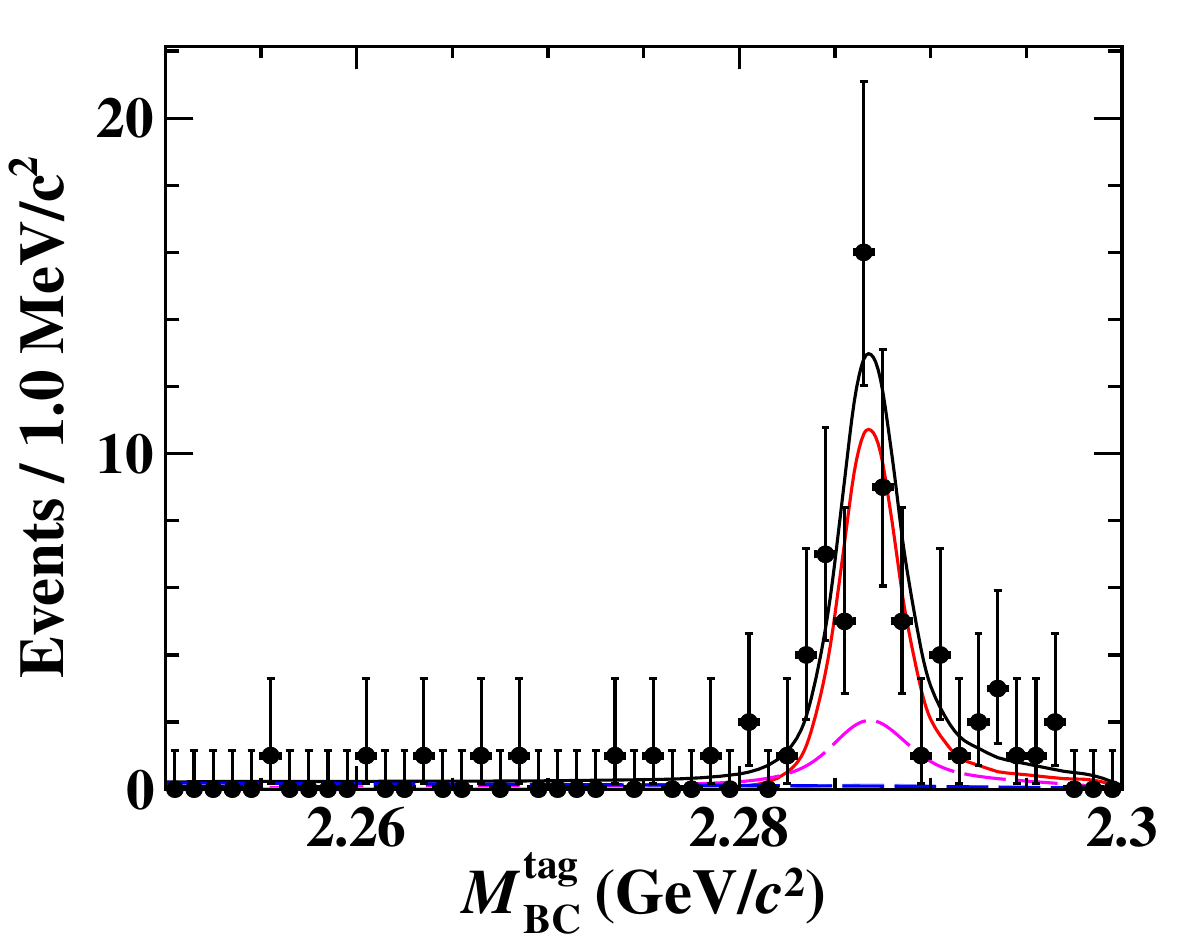}
  \includegraphics[width=0.24\textwidth]{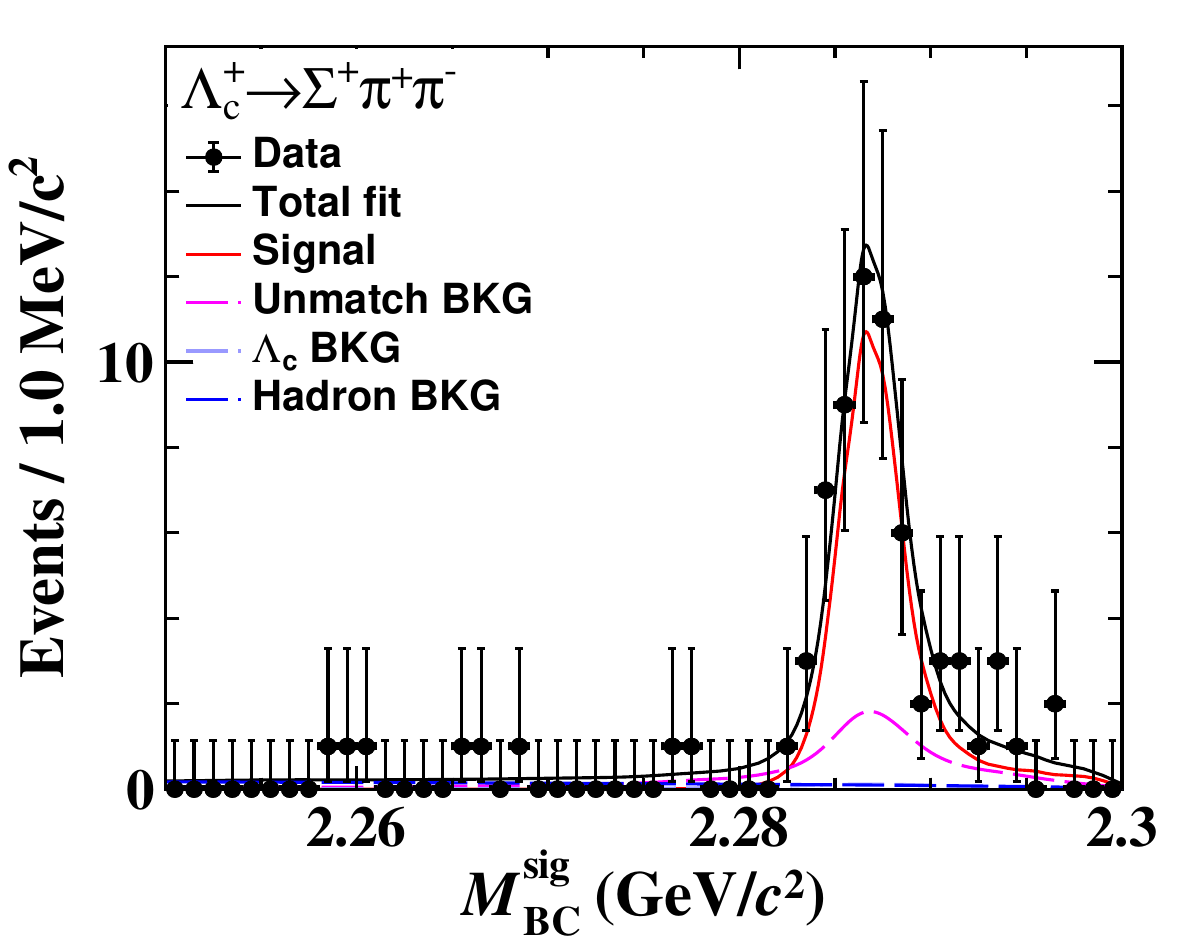}
  \includegraphics[width=0.24\textwidth]{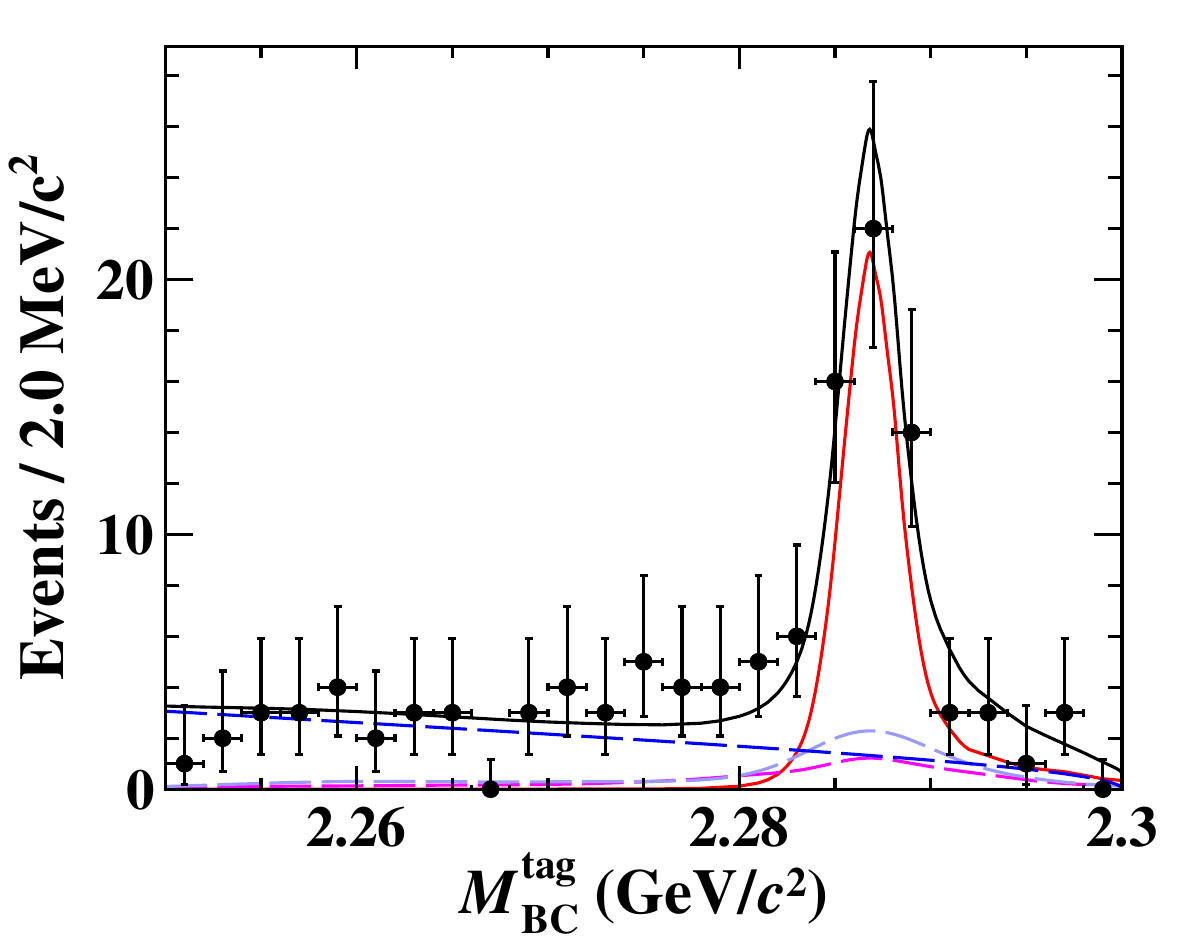}
  \includegraphics[width=0.24\textwidth]{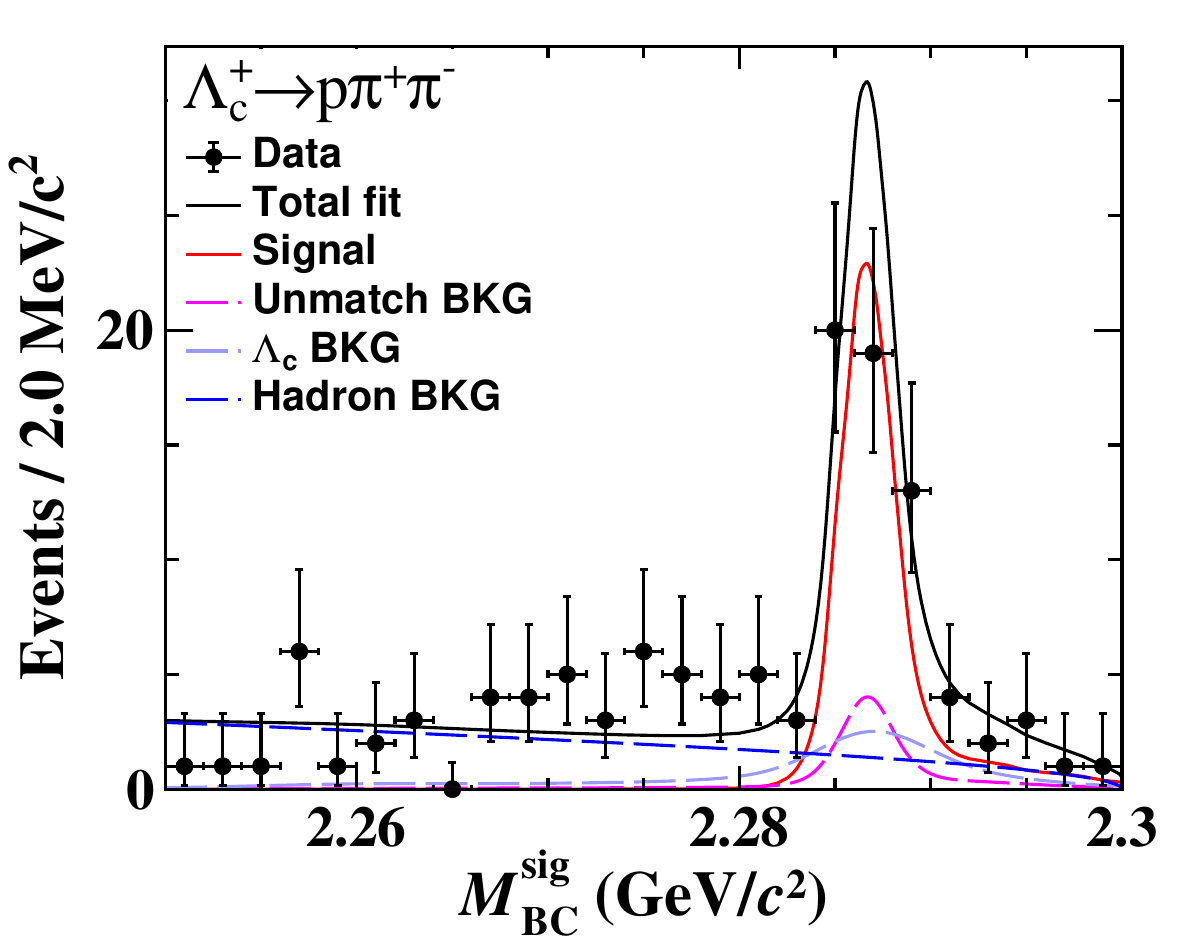}
     \vspace*{-0.5cm}
  \end{center}
\caption{The projections of the 2D fits on the $M_{\rm BC}^{\rm tag}$ and $M_{\rm BC}^{\rm sig}$ distributions of the accepted DT candidates at $\sqrt{s}=4681.92~\mev$. The plots in the first and third columns show the combined 12 tag modes for each signal mode.
The points with error bars are data, the black lines are the sum of fit functions, the red lines are the matched signal shapes, the pink dashed lines are the unmatched signal shapes, the lilac dashed lines are the non-signal $\lcp\lcm$ shapes, and the blue dashed lines are the ARGUS functions.}
\label{fig:DT_yield_4680}
\end{figure}
\begin{figure}[!htbp]
  \begin{center}
  
  \includegraphics[width=0.24\textwidth]{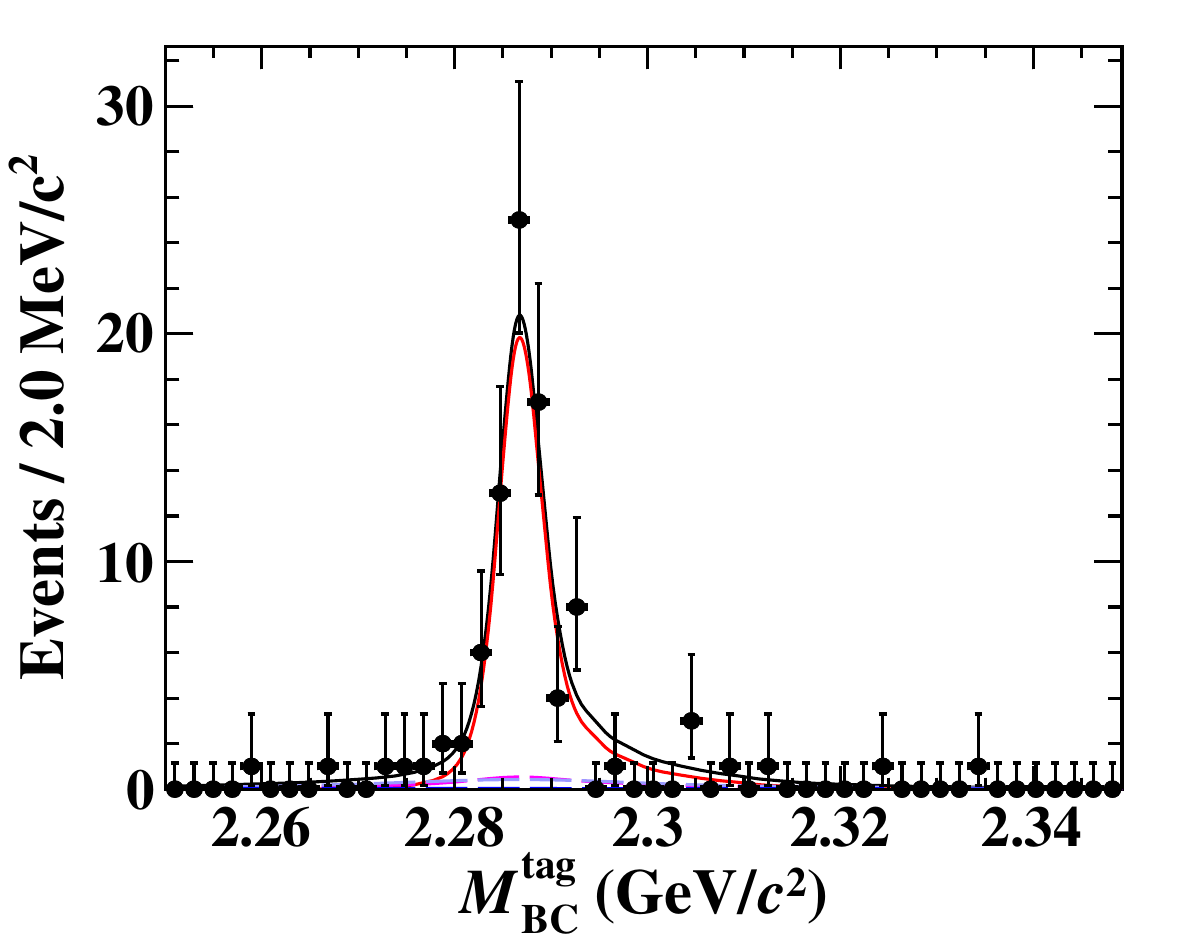}
  \includegraphics[width=0.24\textwidth]{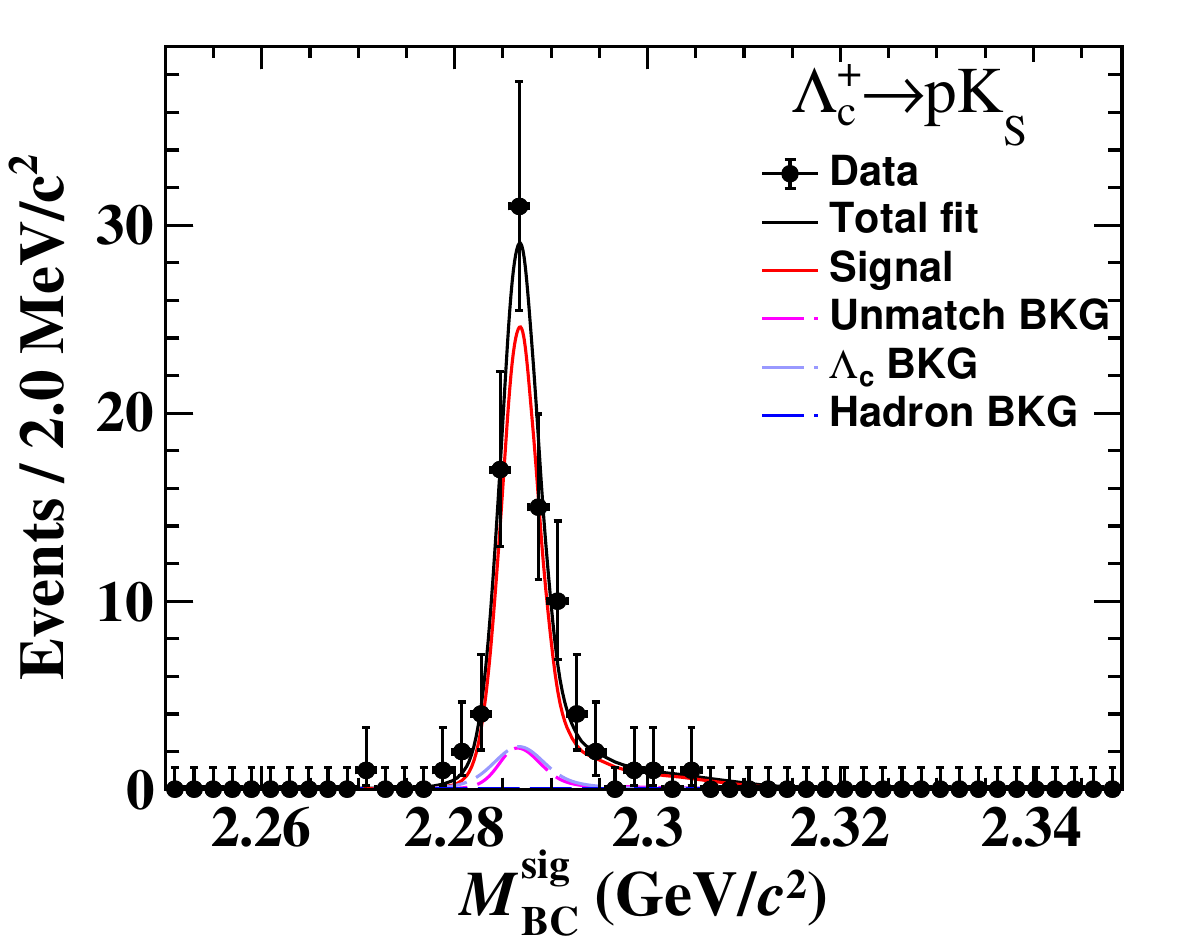}
  \includegraphics[width=0.24\textwidth]{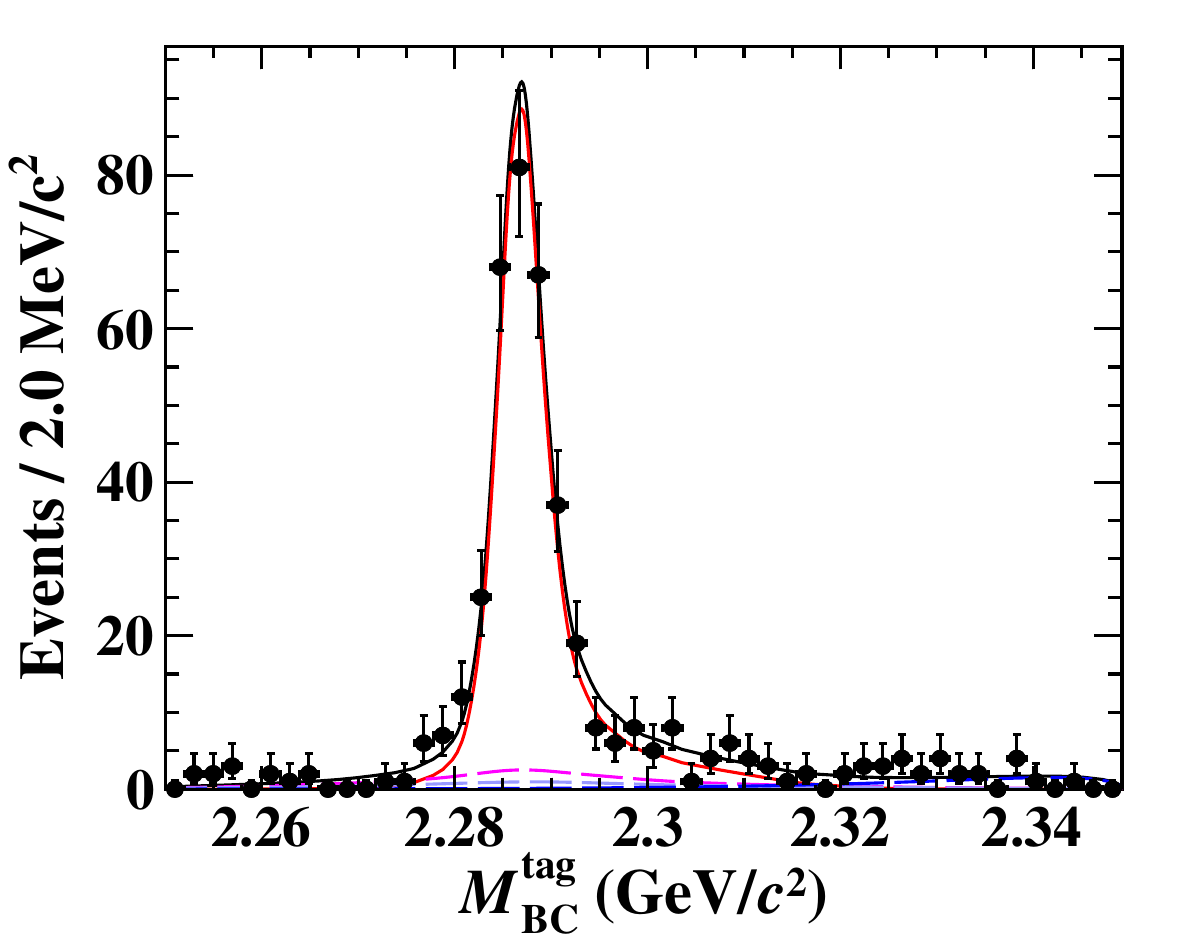}
  \includegraphics[width=0.24\textwidth]{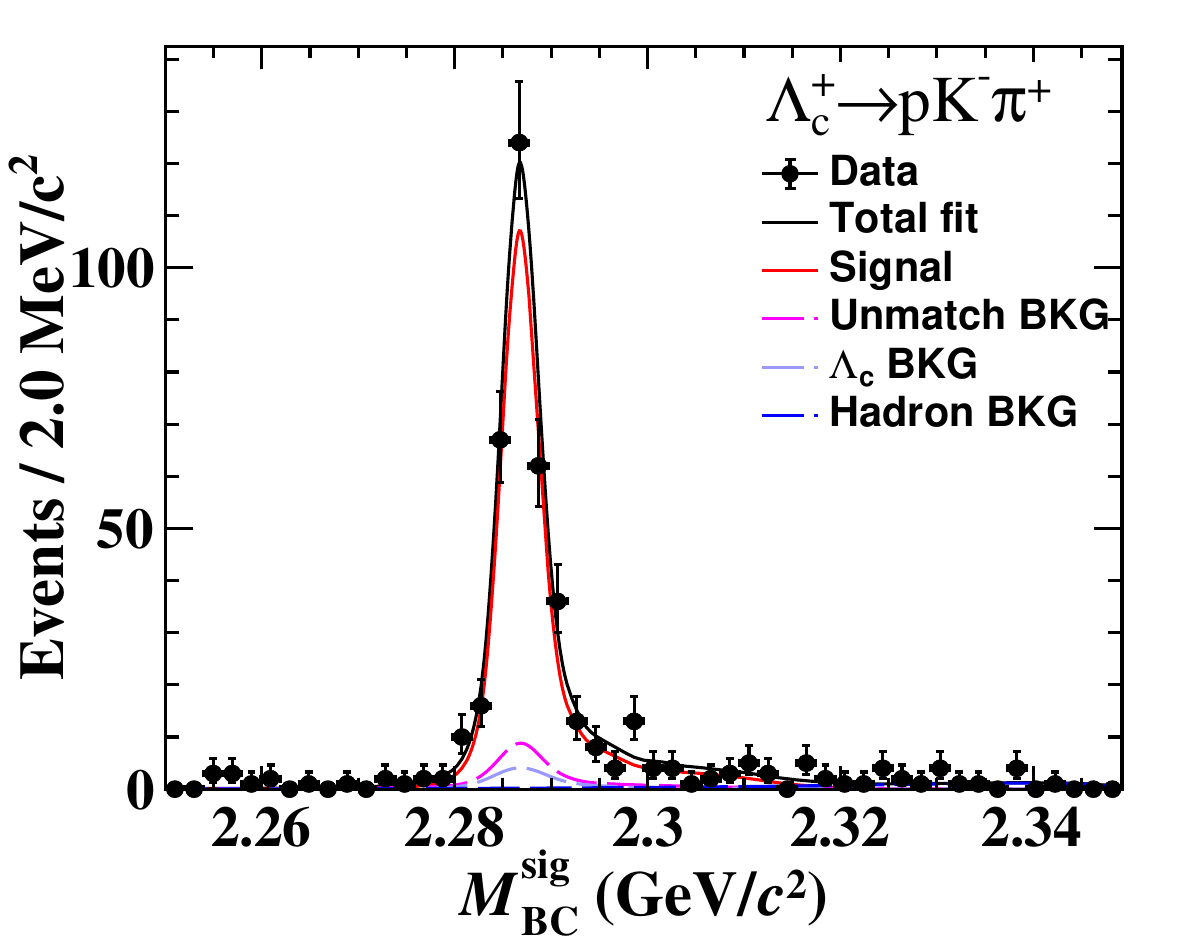}
  \includegraphics[width=0.24\textwidth]{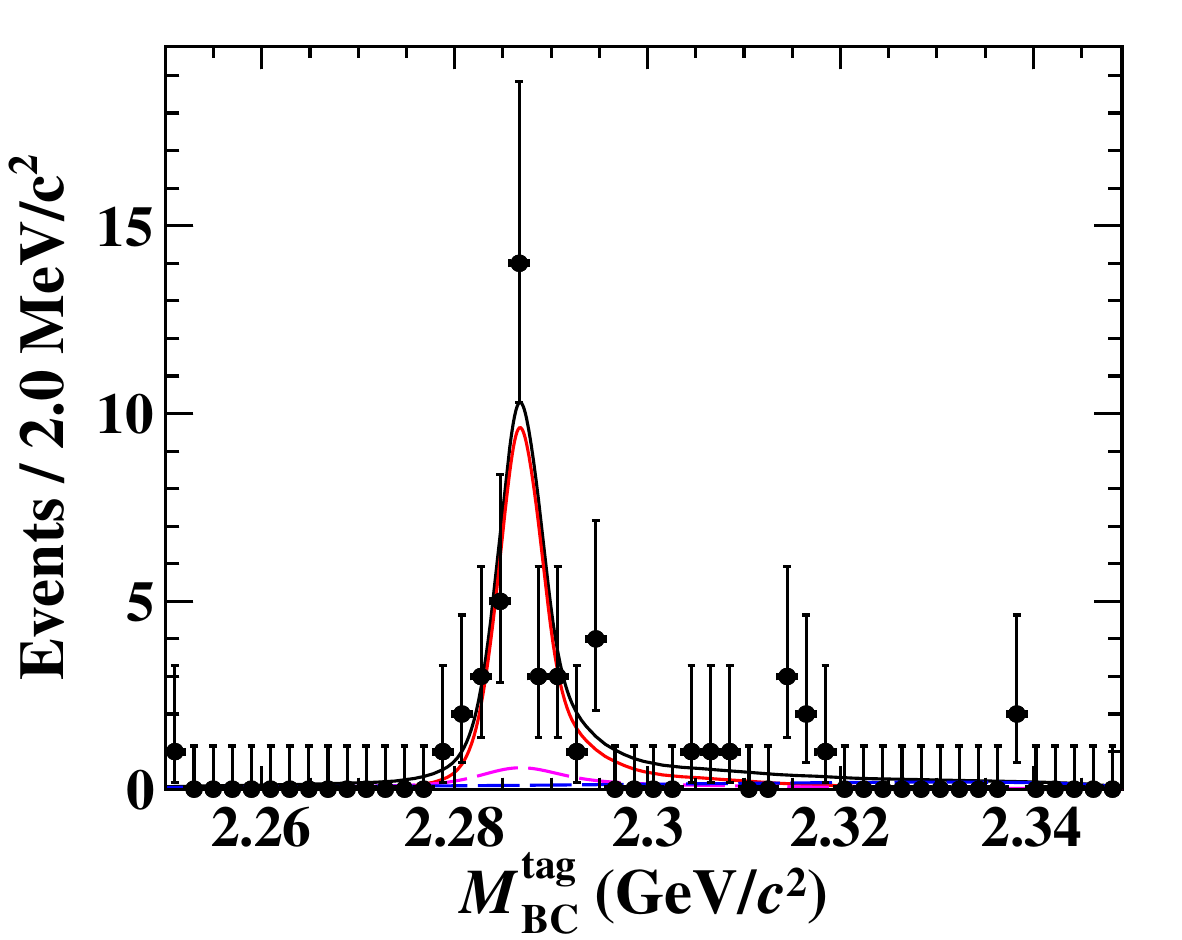}
  \includegraphics[width=0.24\textwidth]{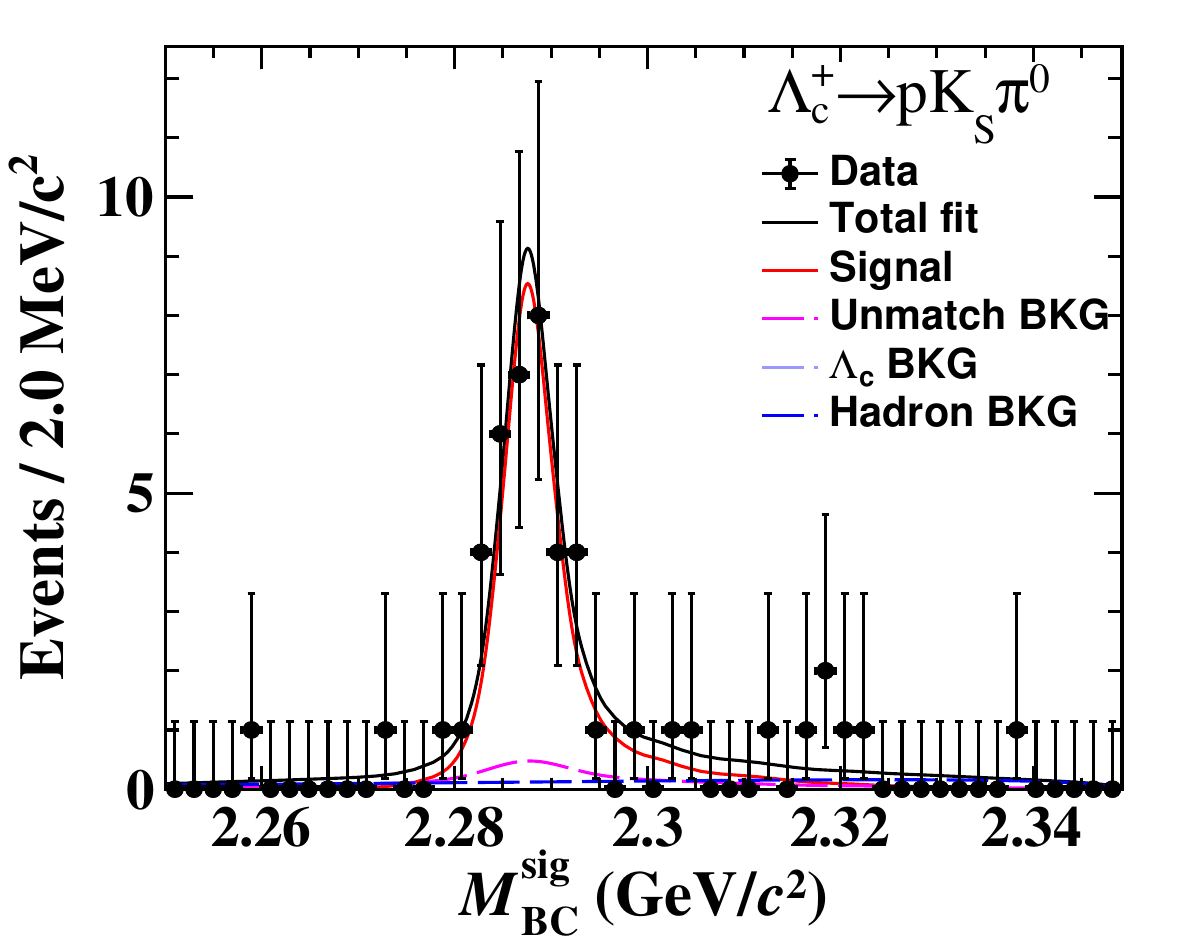}
  \includegraphics[width=0.24\textwidth]{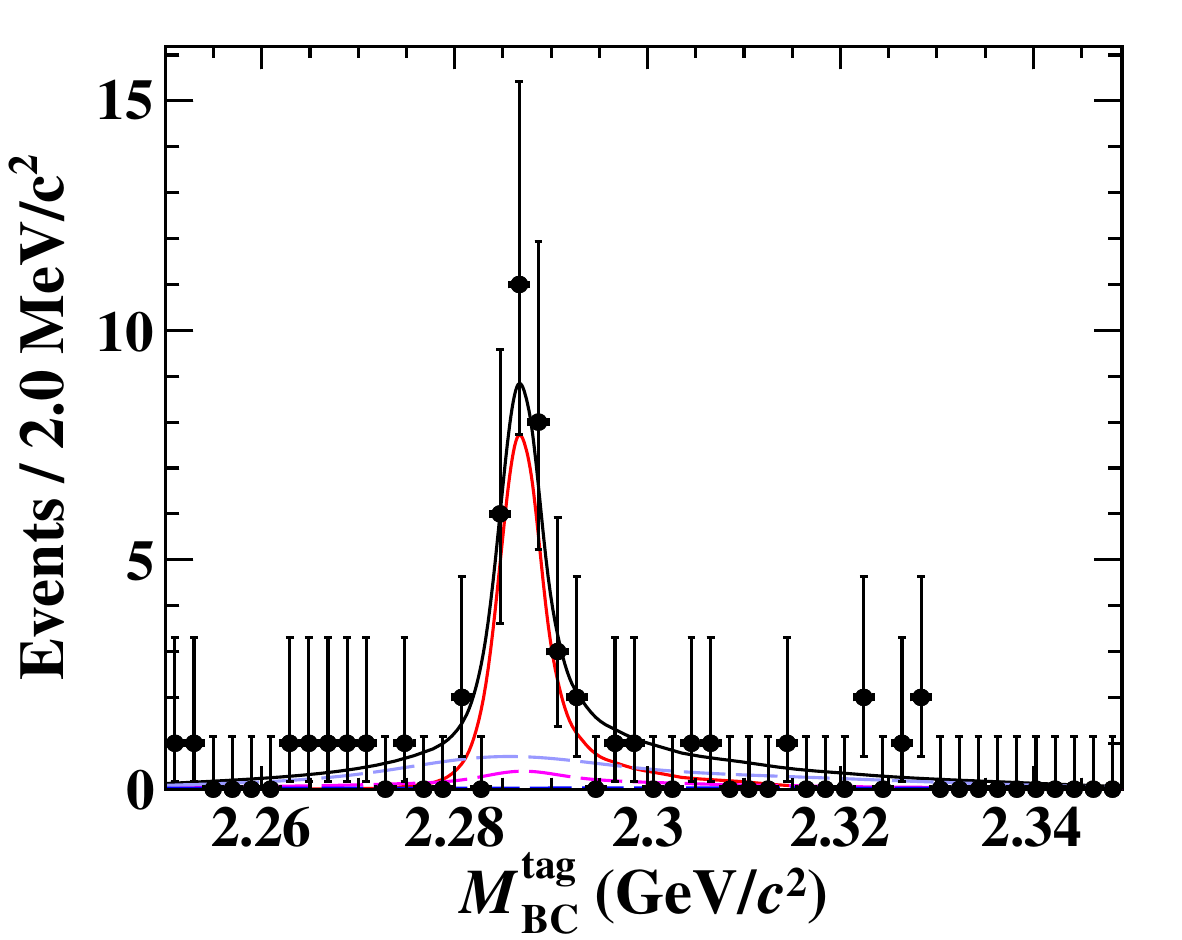}
  \includegraphics[width=0.24\textwidth]{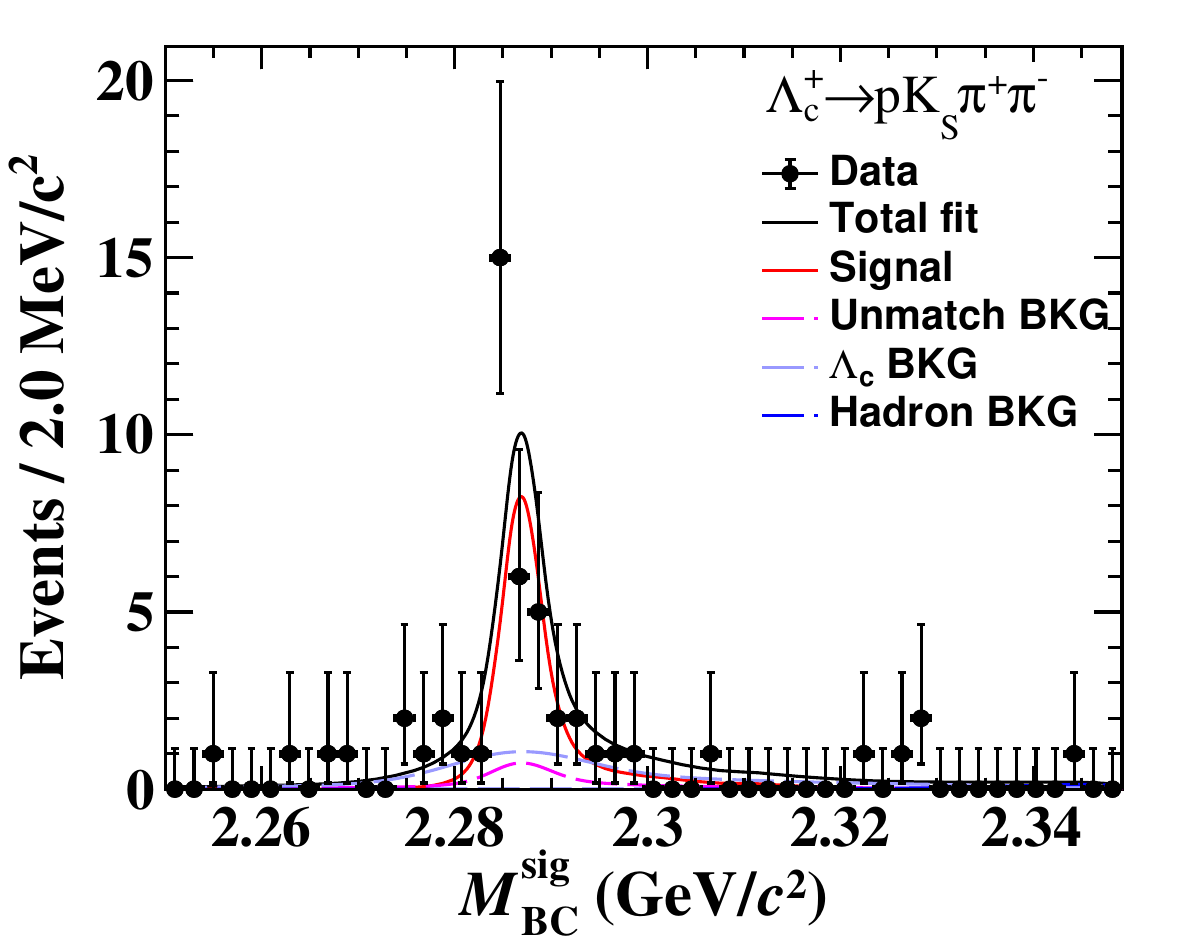}
  \includegraphics[width=0.24\textwidth]{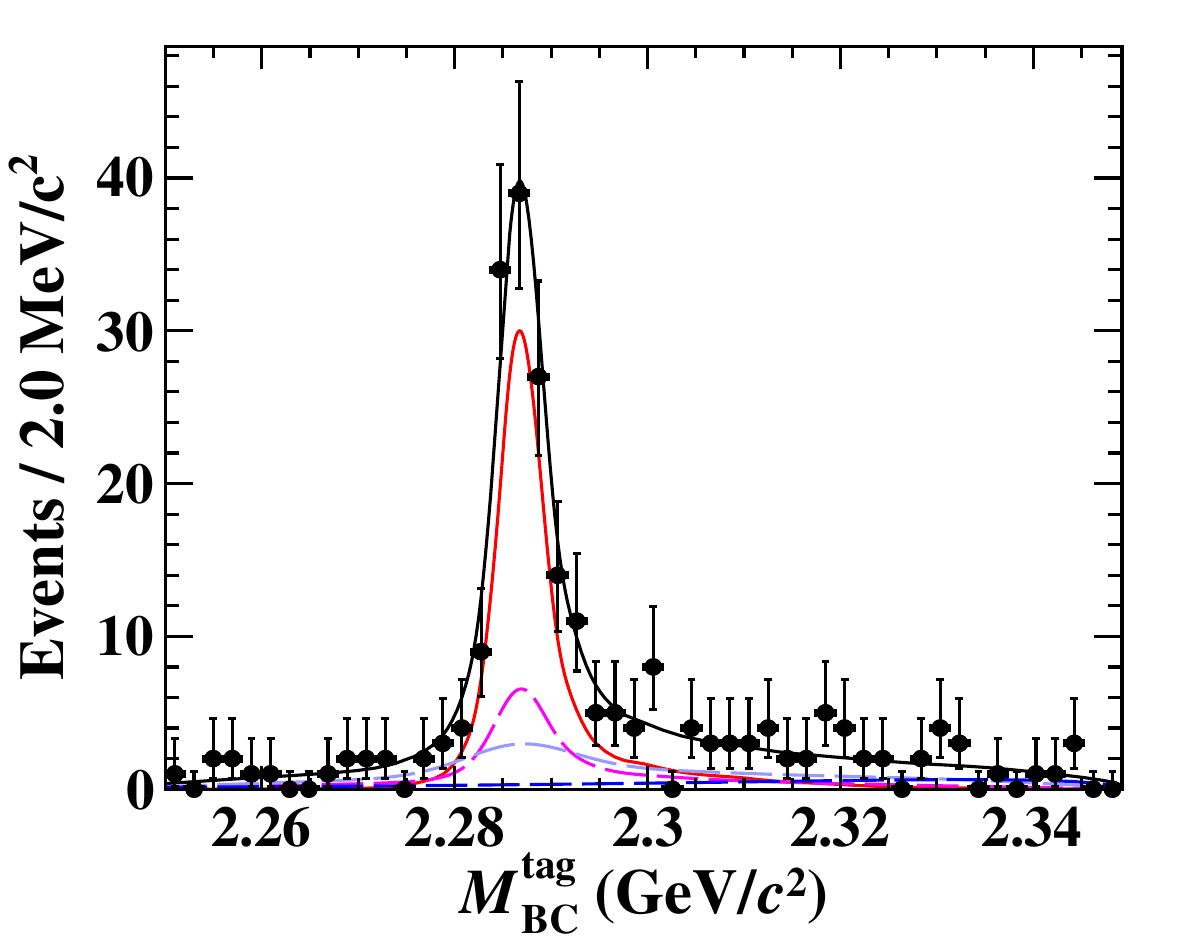}
  \includegraphics[width=0.24\textwidth]{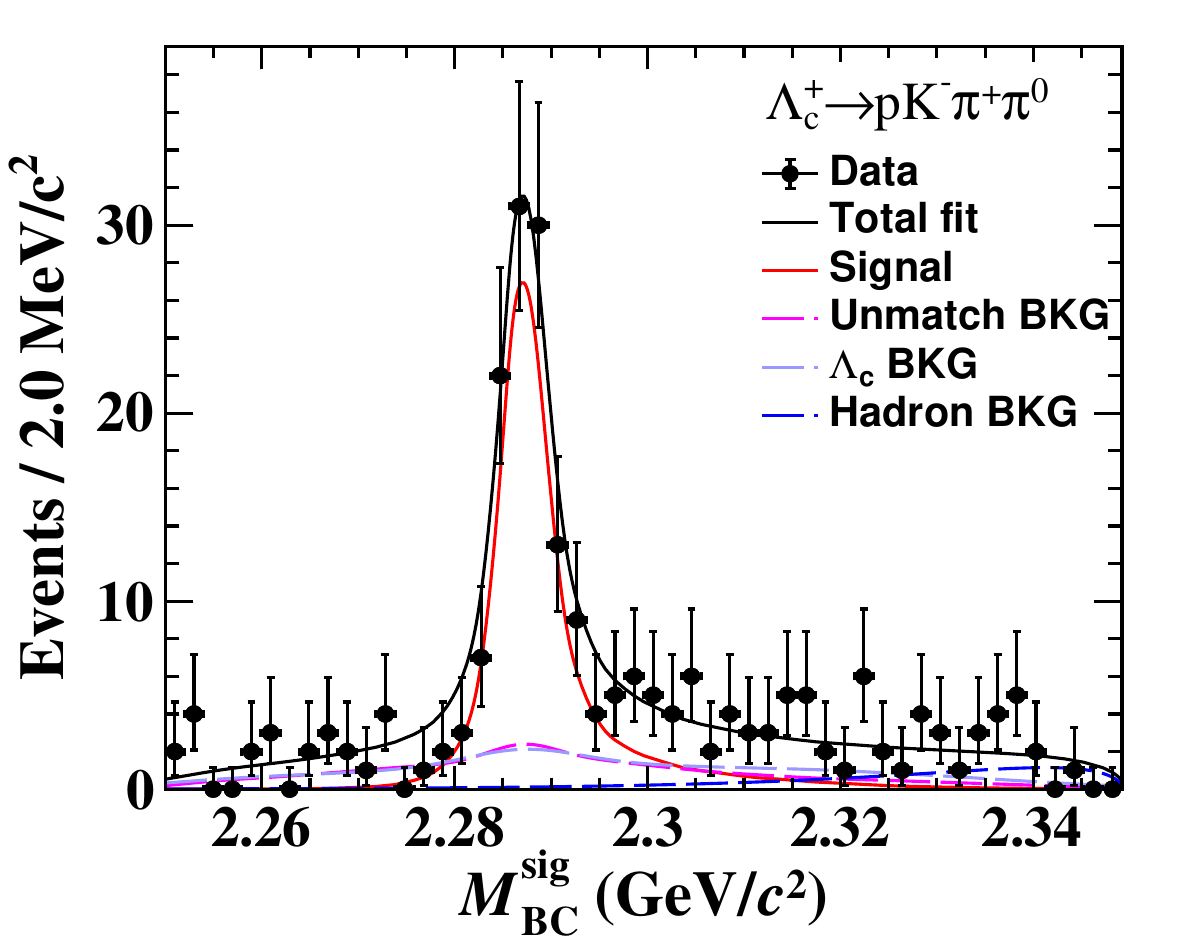}
  \includegraphics[width=0.24\textwidth]{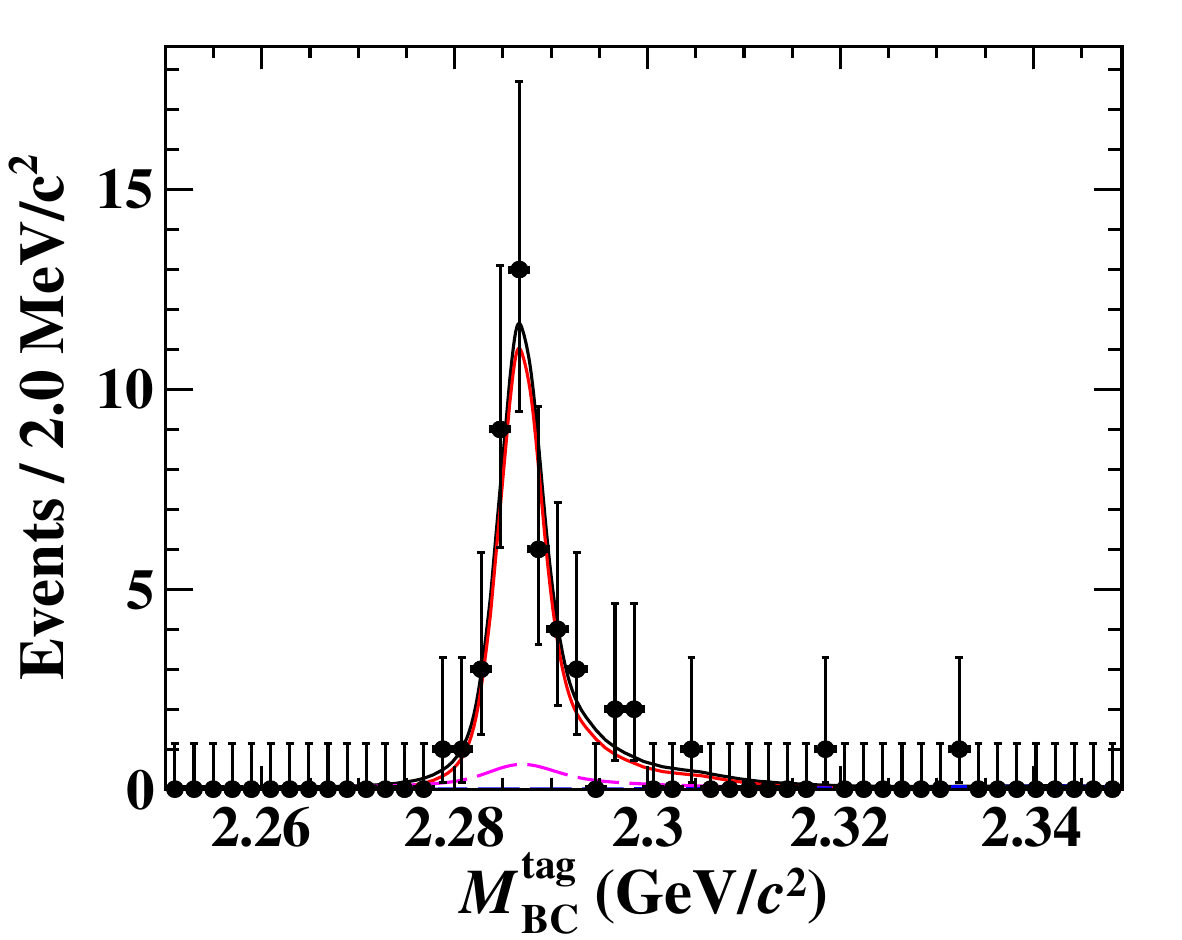}
  \includegraphics[width=0.24\textwidth]{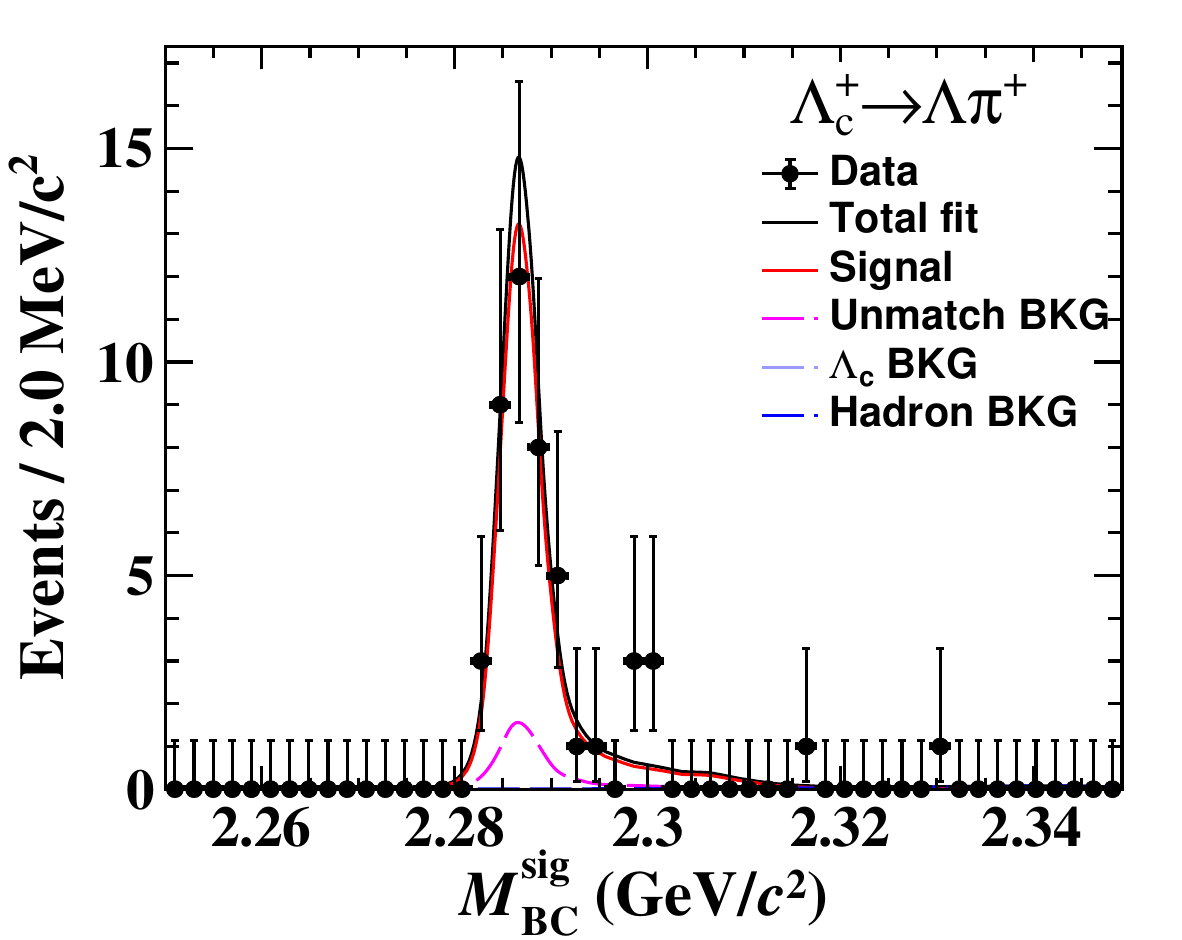}
  \includegraphics[width=0.24\textwidth]{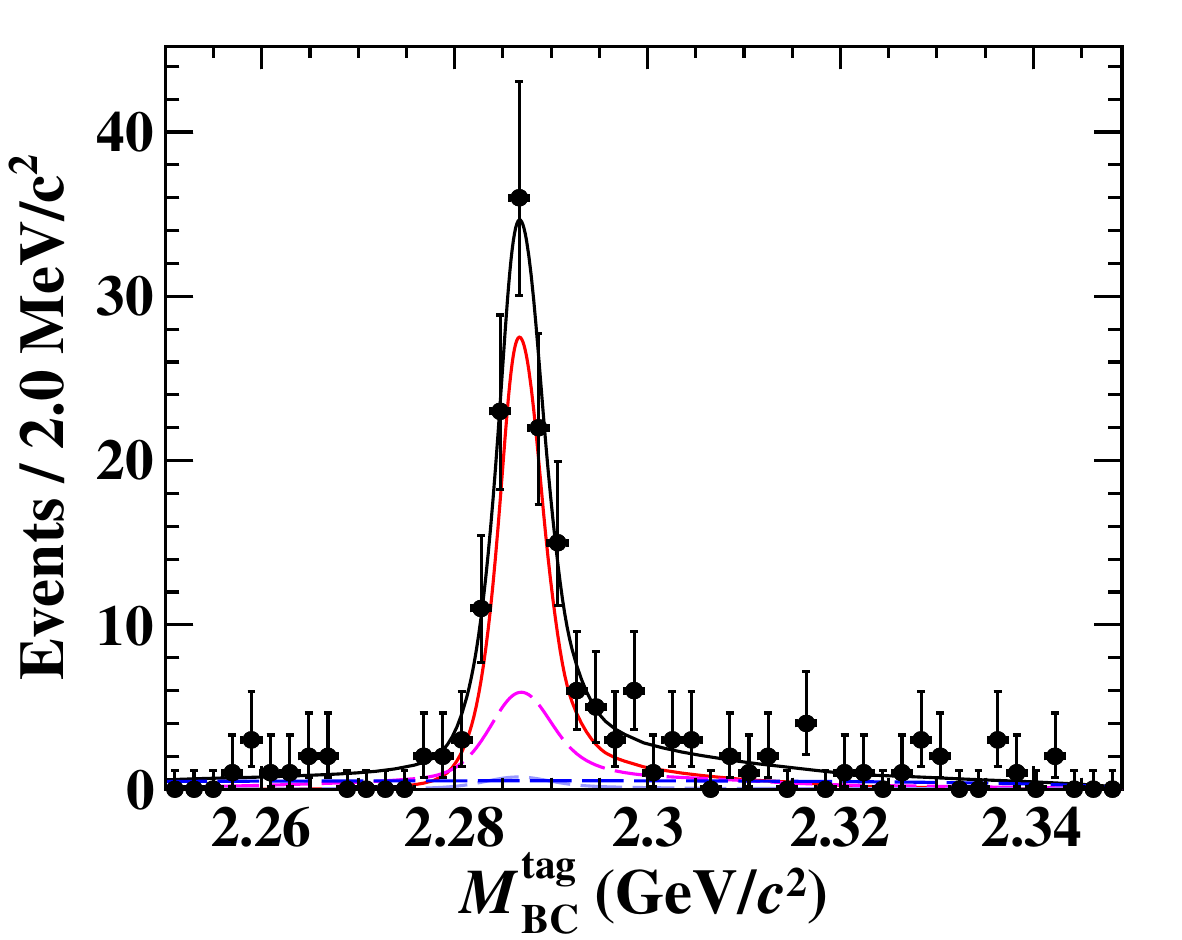}
  \includegraphics[width=0.24\textwidth]{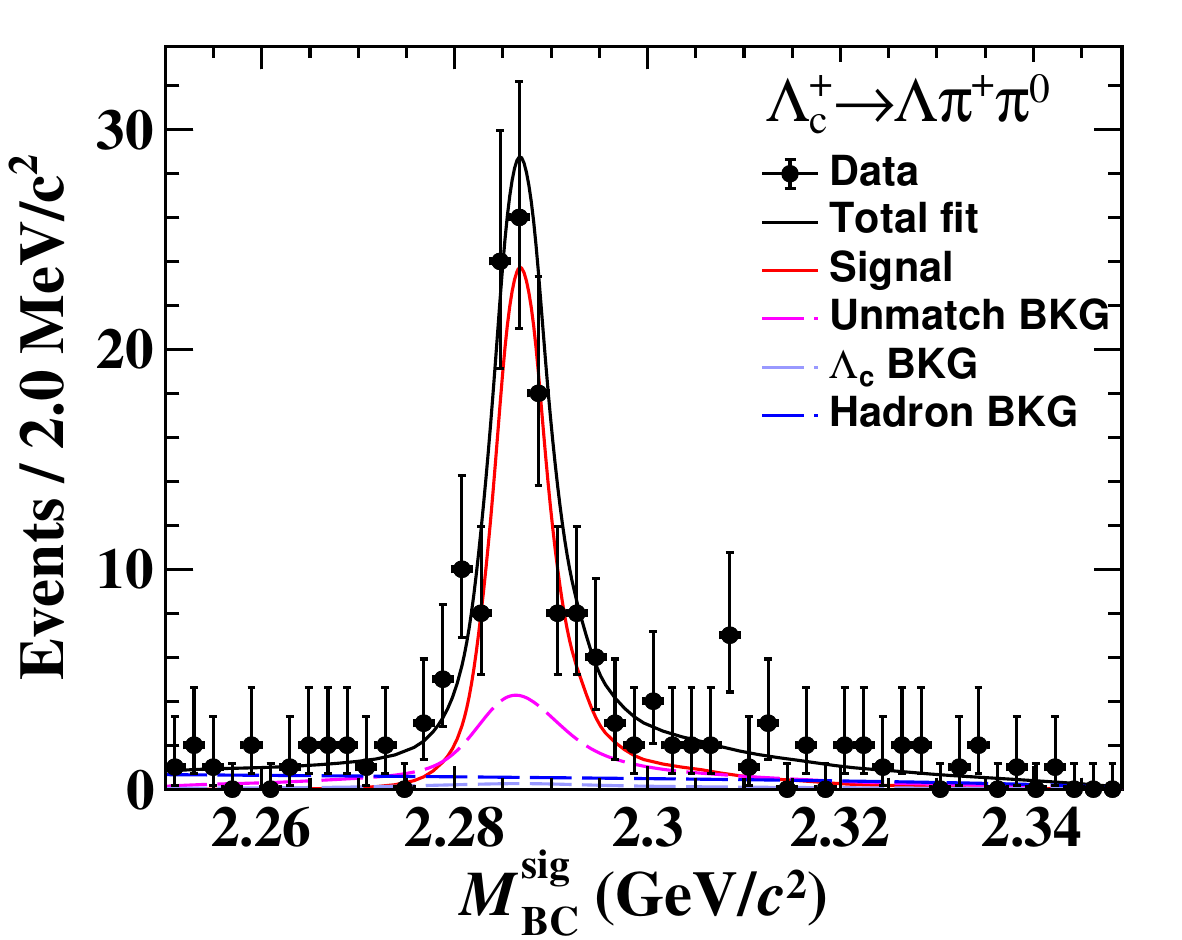}
  \includegraphics[width=0.24\textwidth]{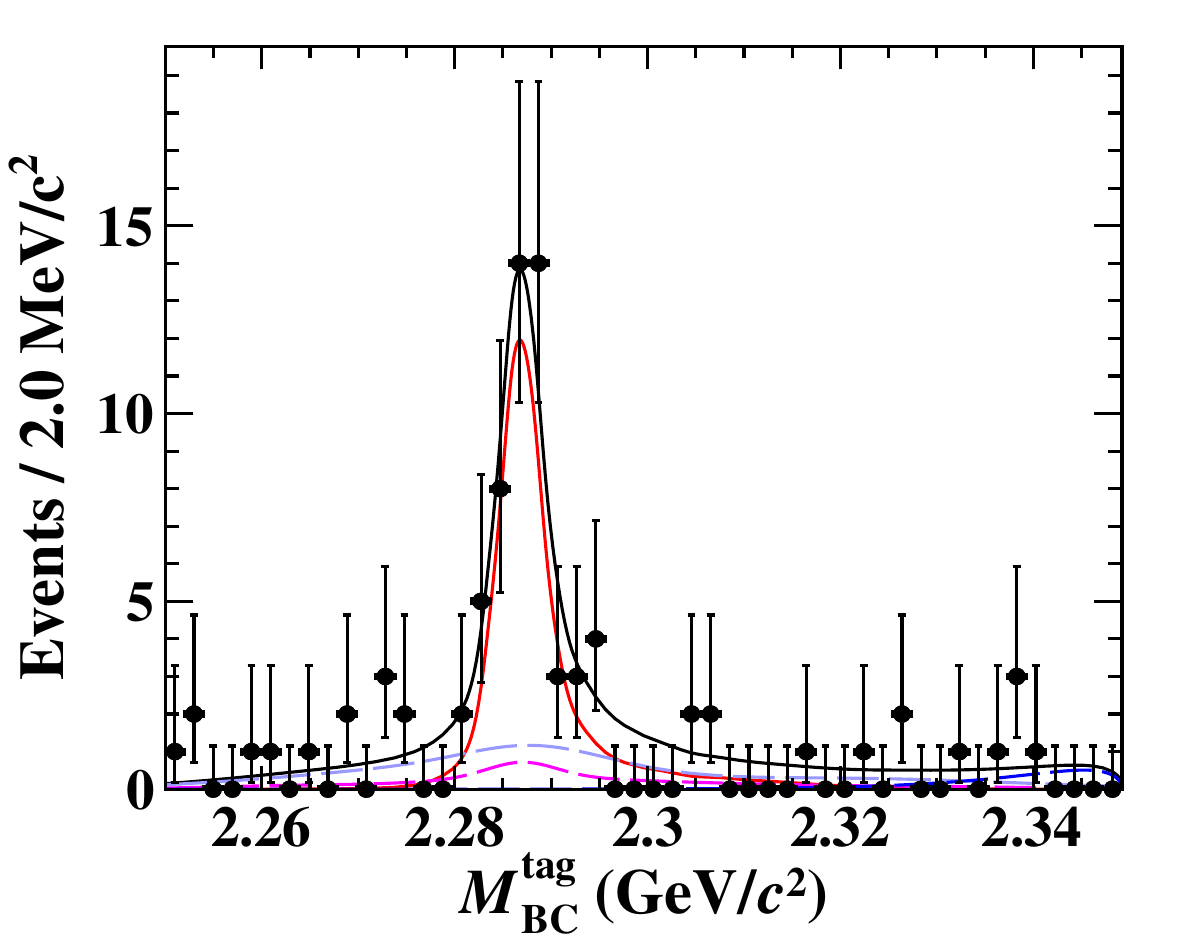}
  \includegraphics[width=0.24\textwidth]{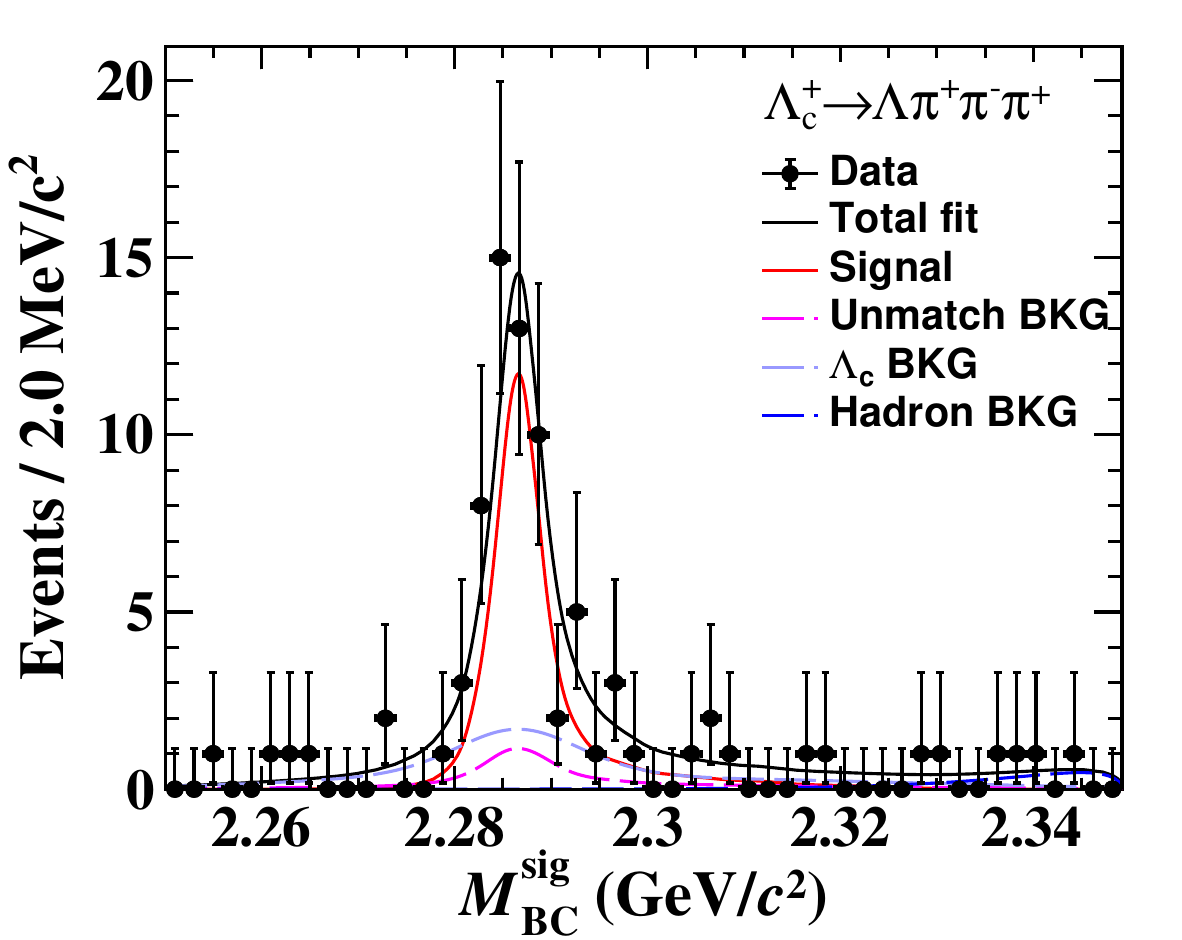}
  \includegraphics[width=0.24\textwidth]{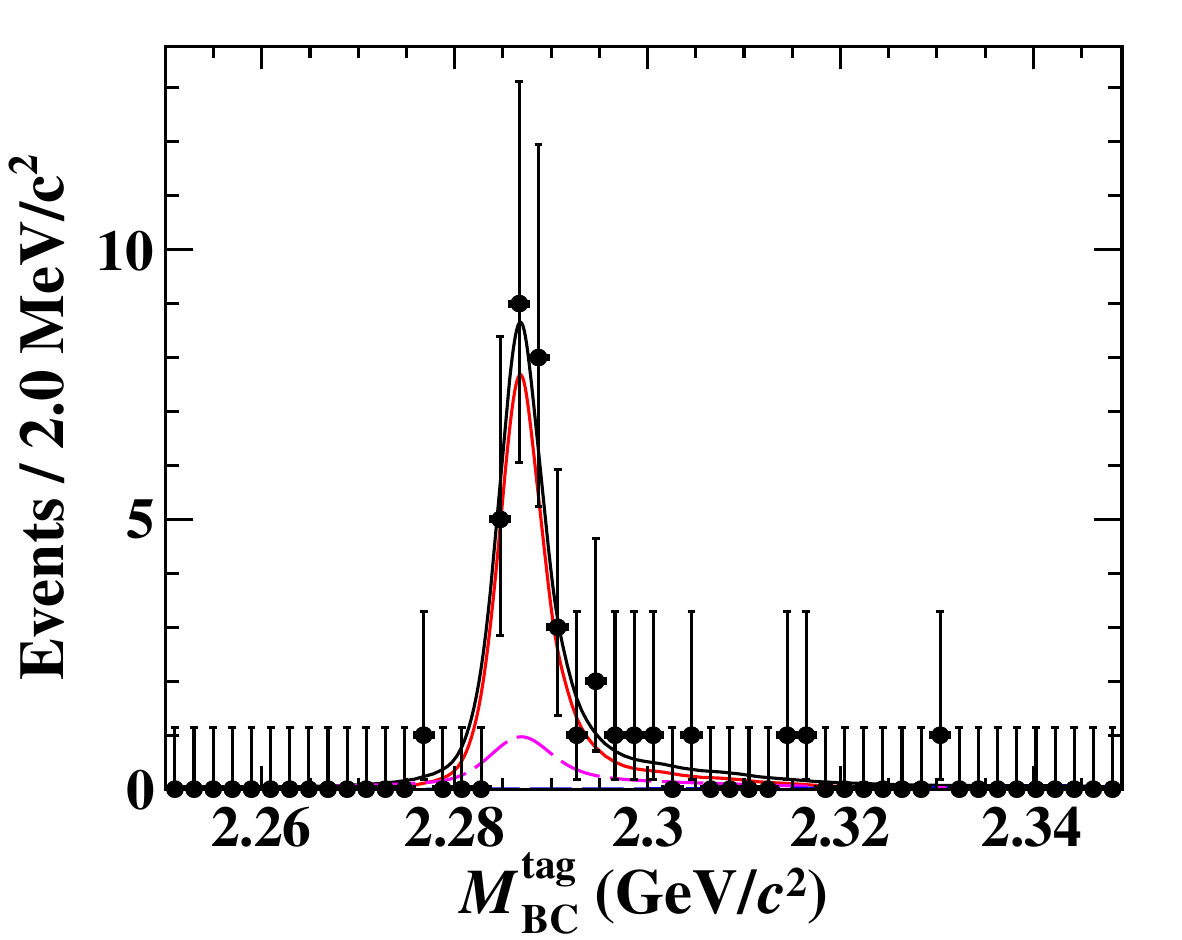}
  \includegraphics[width=0.24\textwidth]{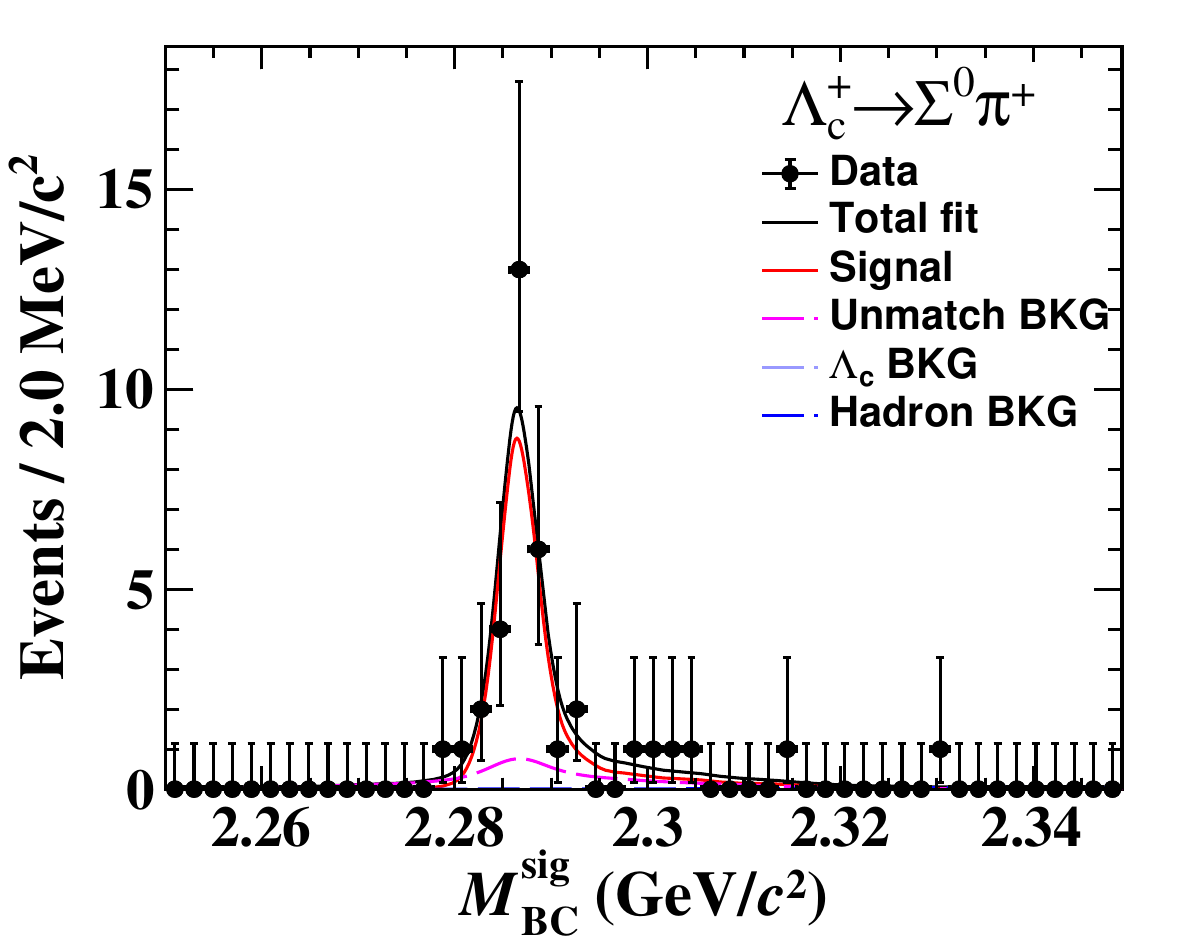}
  \includegraphics[width=0.24\textwidth]{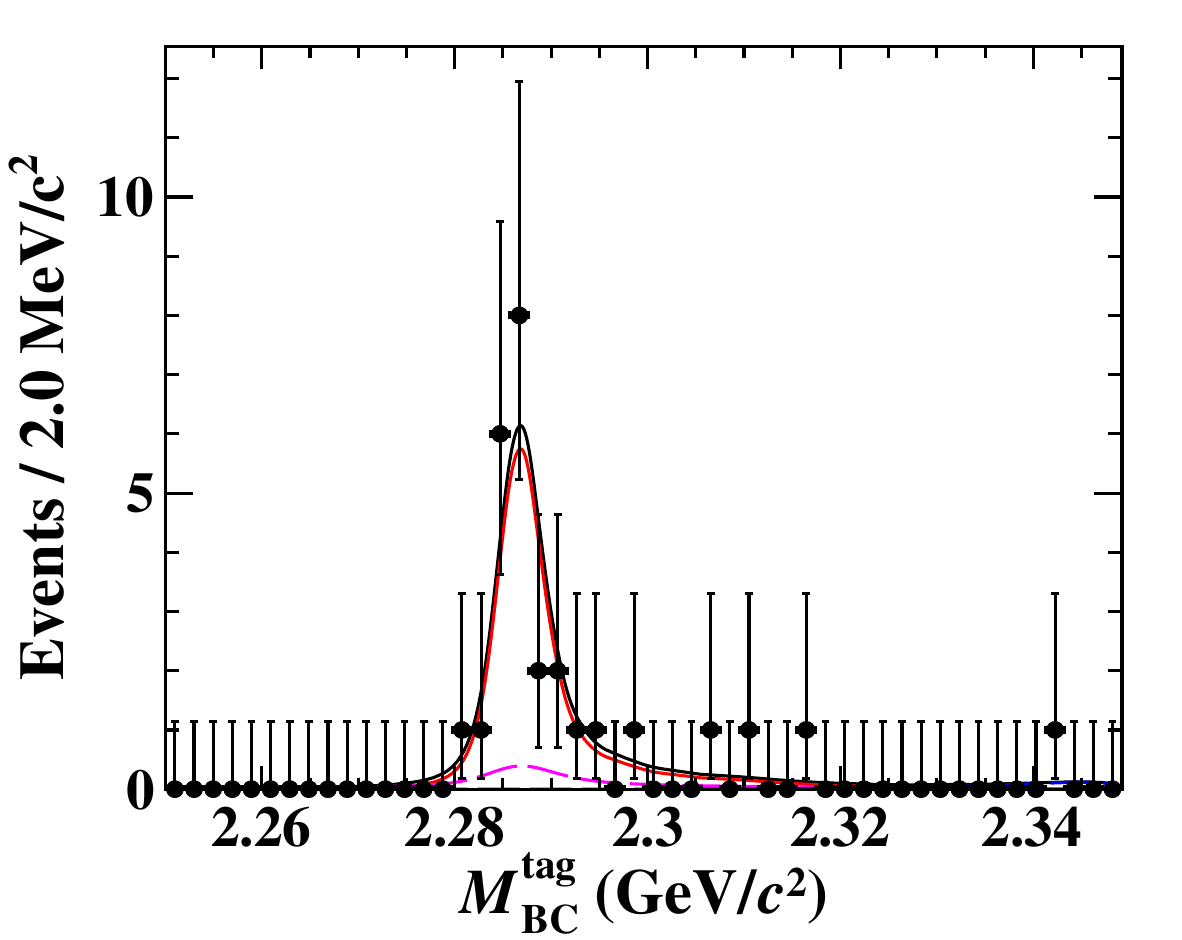}
  \includegraphics[width=0.24\textwidth]{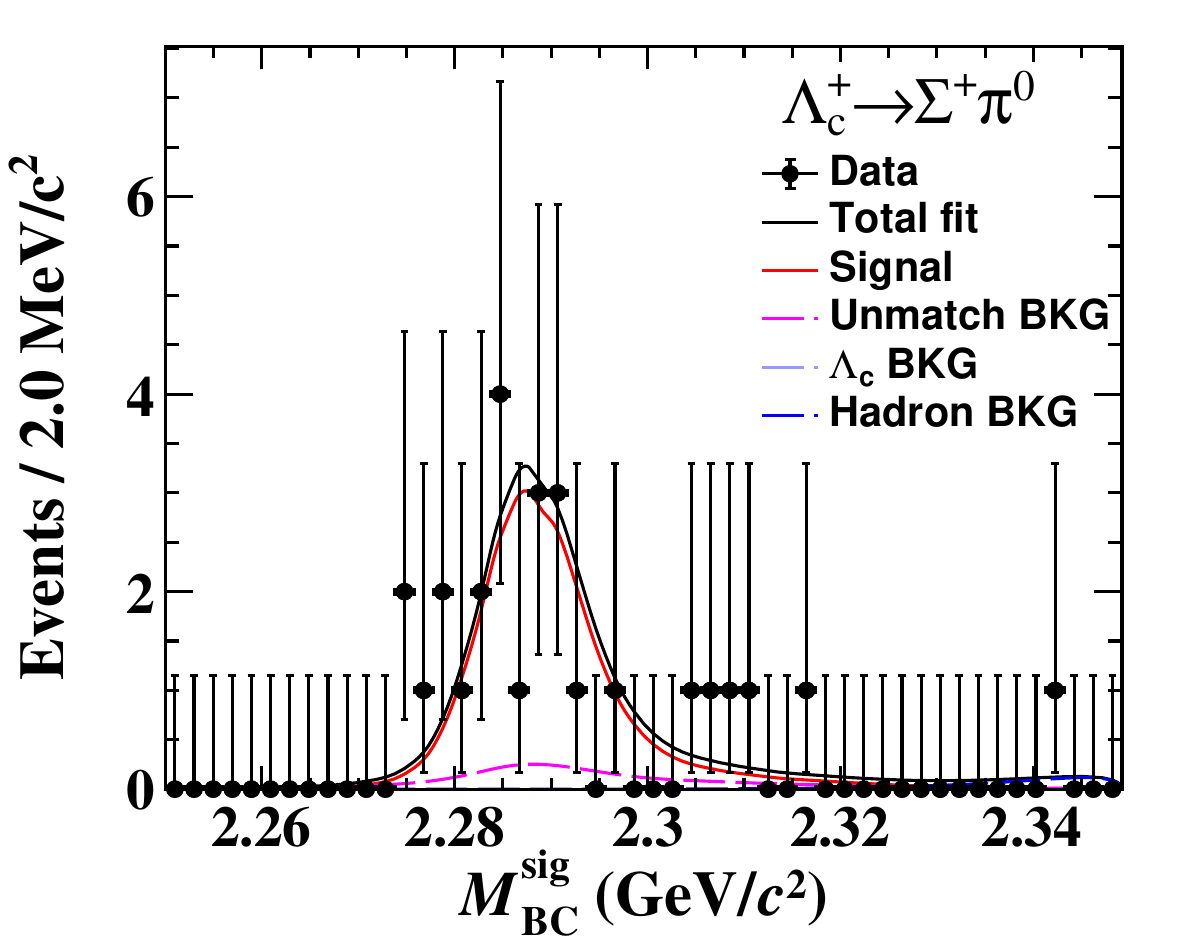}
  \includegraphics[width=0.24\textwidth]{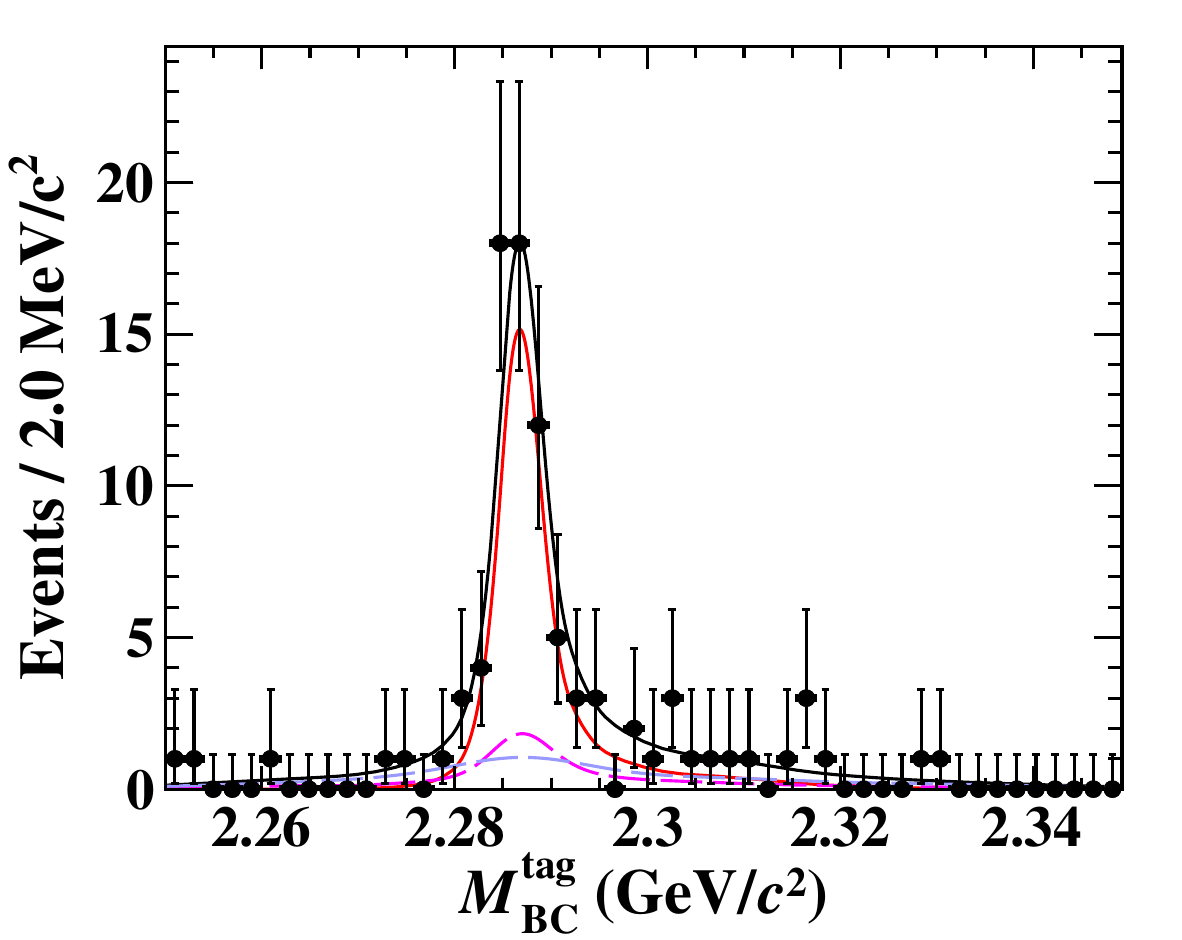}
  \includegraphics[width=0.24\textwidth]{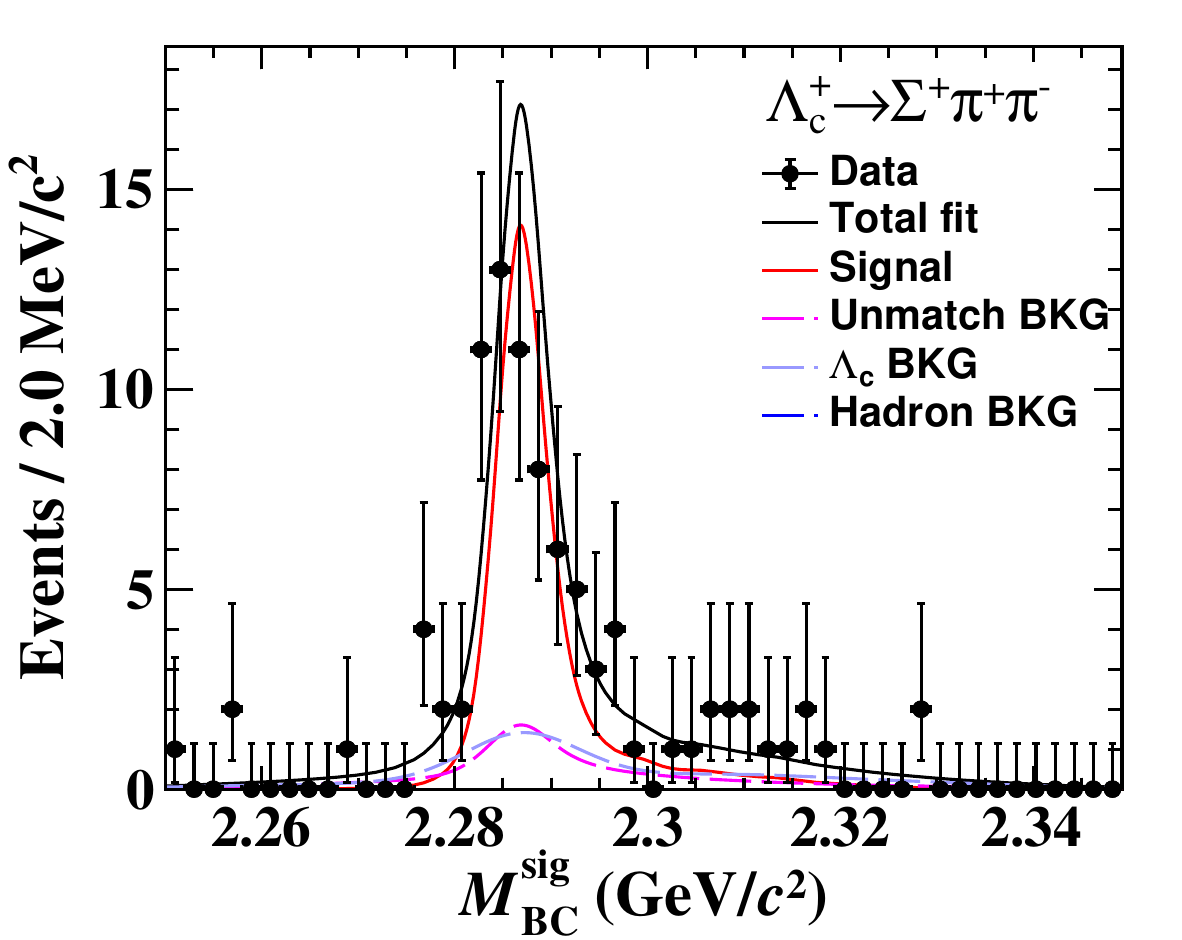}
  \includegraphics[width=0.24\textwidth]{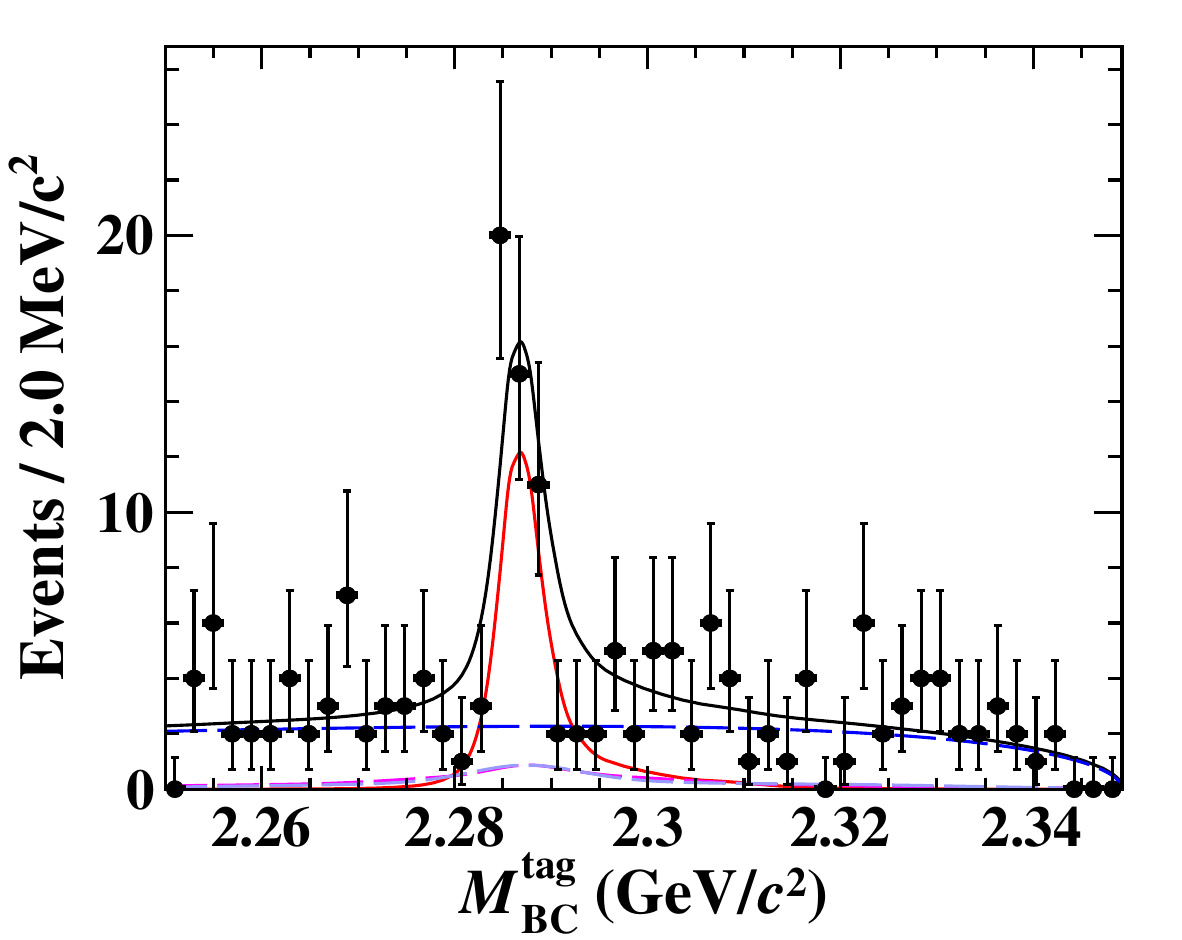}
  \includegraphics[width=0.24\textwidth]{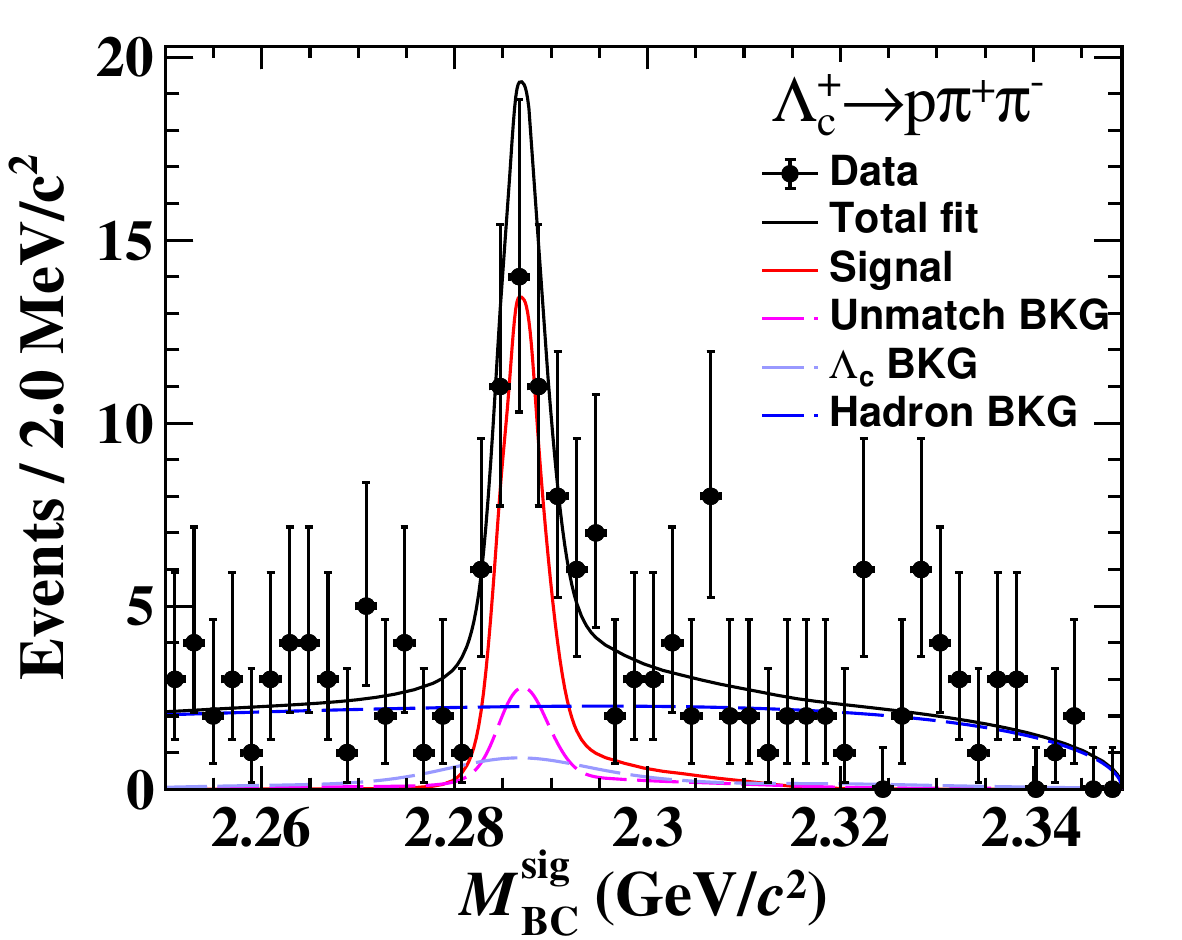}
     \vspace*{-0.5cm}
  \end{center}
\caption{The projections of the 2D fits on the $M_{\rm BC}^{\rm tag}$ and $M_{\rm BC}^{\rm sig}$ distributions of the accepted DT candidates at $\sqrt{s}=4698.82~\mev$. The plots in the first and third columns show the combined 12 tag modes for each signal mode. 
The points with error bars are data, the black lines are the sum of fit functions, the red lines are the matched signal shapes, the pink dashed lines are the unmatched signal shapes, the lilac dashed lines are the non-signal $\lcp\lcm$ shapes, and the blue dashed lines are the ARGUS functions.}
\label{fig:DT_yield_4700}
\end{figure}

\section{Systematic uncertainties}
\label{sys}
Tables~\ref{tab:sys_err_4600},~\ref{tab:sys_err_4612},~\ref{tab:sys_err_4626},~\ref{tab:sys_err_4640},~\ref{tab:sys_err_4660} and ~\ref{tab:sys_err_4700} show the systematic uncertainties at different energy points.
  \begin{table*}[!htbp]
  \begin{center}
  \footnotesize
    \caption{The systematic uncertainties in BF measurements(\%) at $\sqrt{s}=$ 4599.53 MeV.}
    \scriptsize
  \begin{tabular}{l|ccccccccc}
      \hline \hline
       Source                &  Tracking  &  PID  & $K_S^0$  & $\Lambda$ & $\pi^0$ &2D fit & Signal model & MC statistics  & Intermediate BF \\  \hline
      $\textbf{$\modea$}$   & 0.4        & 0.1    & 2.3   &           &       & 0.6 & 0.1 & 0.5 &  0.1 \\
      $\textbf{$\modeb$}$   & 1.7        & 0.7    &       &           &       & 0.8 & 0.3 & 0.4 &      \\
      $\textbf{$\modec$}$   & 0.6        & 0.4    & 1.4   &           &  1.0  & 0.7 & 0.2 & 0.6 &  0.1 \\
      $\textbf{$\moded$}$   & 2.1        & 1.4    & 2.7   &           &       & 2.5 & 0.7 & 0.6 &  0.1 \\
      $\textbf{$\modee$}$   & 1.7        & 1.2    &       &           &  0.6  & 0.1 & 0.3 & 0.7 &      \\
      $\textbf{$\modeaa$}$  & 0.6        & 0.6    &       &   1.7     &       & 0.1 & 0.1 & 0.6 &  0.8 \\
      $\textbf{$\modebb$}$  & 0.6        & 0.4    &       &   0.4     &  1.1  & 1.0 & 0.1 & 0.7 &  0.8 \\
      $\textbf{$\modedd$}$  & 1.7        & 0.6    &       &   1.8     &       & 0.9 & 0.1 & 0.6 &  0.8 \\
      $\textbf{$\modeaaa$}$ & 0.6        & 0.6    &       &   1.3     &       & 3.5 & 0.1 & 0.7 &  0.8 \\
      $\textbf{$\modeccc$}$ & 0.4        & 0.1    &       &           &  2.5  & 2.4 & 0.1 & 0.5 &  0.1 \\
      $\textbf{$\modeddd$}$ & 1.9        & 1.2    &       &           &  0.4  & 1.9 & 0.2 & 0.5 &  0.1 \\
      $\textbf{$\modef$}$   & 2.0        & 0.9    &       &           &       & 1.8 & 0.5 & 0.5 &       \\
      \hline\hline                           
    \end{tabular}
    \label{tab:sys_err_4600}
  \end{center}
  \end{table*}
   
  \begin{table*}[!htbp]
  \begin{center}
  \footnotesize
    \caption{The systematic uncertainties in BF measurements(\%) at $\sqrt{s}=$ 4611.86 MeV.}
  \begin{tabular}{l|ccccccccc}
      \hline \hline
      Source                &  Tracking  &  PID  & $K_S^0$  & $\Lambda$ & $\pi^0$ &2D fit & Signal model & MC statistics  & Intermediate BF \\ \hline
      $\textbf{$\modea$}$   & 0.3   &  0.1  & 2.4  &      &       & 0.1 & 0.1   & 0.6 &  0.1 \\
      $\textbf{$\modeb$}$   & 1.7   &  0.6  &      &      &       & 0.1 & 0.3   & 0.5 &      \\
      $\textbf{$\modec$}$   & 0.6   &  0.5  & 1.1  &      &  1.0  & 0.1 & 0.2   & 1.0 &  0.1 \\
      $\textbf{$\moded$}$   & 1.8   &  1.1  & 2.6  &      &       & 1.8 & 0.7   & 0.8 &  0.1 \\
      $\textbf{$\modee$}$   & 1.7   &  1.1  &      &      &  0.6  & 0.1 & 0.3   & 0.7 &      \\
      $\textbf{$\modeaa$}$  & 0.5   &  0.6  &      & 1.7  &       & 0.1 & 0.1   & 0.7 &  0.8 \\
      $\textbf{$\modebb$}$  & 0.5   &  0.4  &      & 0.4  &  1.2  & 5.2 & 0.1   & 0.8 &  0.8 \\
      $\textbf{$\modedd$}$  & 1.8   &  1.0  &      & 1.7  &       & 0.1 & 0.1   & 0.7 &  0.8 \\
      $\textbf{$\modeaaa$}$ & 0.6   &  0.6  &      & 1.3  &       & 0.1 & 0.2   & 0.8 &  0.8 \\
      $\textbf{$\modeccc$}$ & 0.4   &  0.1  &      &      &  2.5  & 0.0 & 0.1   & 0.6 &  0.1 \\
      $\textbf{$\modeddd$}$ & 1.6   &  0.7  &      &      &  0.5  & 6.0 & 0.3   & 0.6 &  0.1 \\
      $\textbf{$\modef$}$   & 1.4   &  1.2  &      &      &       & 0.5 & 0.5   & 0.6 &   \\
      \hline\hline
    \end{tabular}
    \label{tab:sys_err_4612}
  \end{center}
\end{table*}
  \begin{table*}[!htbp]
  \begin{center}
          \scriptsize
    \caption{The systematic uncertainties in BF measurements(\%) at $\sqrt{s}=$ 4628.00 MeV.}
  \begin{tabular}{l|ccccccccc}
      \hline \hline
       Source                &  Tracking  &  PID  & $K_S^0$  & $\Lambda$ & $\pi^0$ &2D fit & Signal model & MC statistics  & Intermediate BF \\  \hline
      $\textbf{$\modea$}$   & 0.4   &  0.1  &  2.4  &      &       & 1.3 &0.1 & 0.6 &  0.1 \\
      $\textbf{$\modeb$}$   & 1.8   &  0.7  &       &      &       & 0.8 &0.3 & 0.5 &      \\
      $\textbf{$\modec$}$   & 0.4   &  0.5  &  1.1  &      &  1.1  & 2.1 &0.2 & 0.7 &  0.1 \\
      $\textbf{$\moded$}$   & 2.1   &  1.1  &  2.7  &      &       & 1.1 &0.5 & 0.7 &  0.1 \\
      $\textbf{$\modee$}$   & 1.6   &  1.2  &       &      &  0.6  & 1.5 &0.3 & 0.7 &      \\
      $\textbf{$\modeaa$}$  & 0.6   &  0.6  &       & 1.7  &       & 1.1 &0.1 & 0.7 &  0.8 \\
      $\textbf{$\modebb$}$  & 0.3   &  0.4  &       & 0.4  &  1.1  & 1.4 &0.1 & 0.7 &  0.8 \\
      $\textbf{$\modedd$}$  & 1.5   &  0.8  &       & 1.4  &       & 2.7 &0.2 & 0.7 &  0.8 \\
      $\textbf{$\modeaaa$}$ & 0.5   &  0.6  &       & 1.3  &       & 0.2 &0.1 & 0.8 &  0.8 \\
      $\textbf{$\modeccc$}$ & 0.5   &  0.2  &       &      &  2.4  & 1.8 &0.1 & 0.6 &  0.1 \\
      $\textbf{$\modeddd$}$ & 1.5   &  1.2  &       &      &  0.5  & 0.6 &0.1 & 0.6 &  0.1 \\
      $\textbf{$\modef$}$   & 1.8   &  0.8  &       &      &       & 1.5 &0.7 & 0.5 &      \\
      \hline\hline
    \end{tabular}

    \label{tab:sys_err_4626}
  \end{center}
  \end{table*}
  \begin{table*}[!htbp]
  \begin{center}
    \scriptsize
    \caption{The systematic uncertainties in BF measurements(\%) at $\sqrt{s}=$ 4640.91 MeV.}
    
  \begin{tabular}{l|ccccccccc}
      \hline \hline
       Source                &  Tracking  &  PID  & $K_S^0$  & $\Lambda$ & $\pi^0$ &2D fit & Signal model & MC statistics  & Intermediate BF \\   \hline
      $\textbf{$\modea$}$   & 0.4 & 0.1   &  2.5 &     &     & 1.1& 0.1 & 0.6  &  0.1 \\
      $\textbf{$\modeb$}$   & 1.7 & 0.7   &      &     &     & 0.2& 0.2 & 0.5  &      \\
      $\textbf{$\modec$}$   & 0.6 & 0.3   &  1.7 &     & 1.1 & 1.9& 0.2 & 0.7  &  0.1 \\
      $\textbf{$\moded$}$   & 2.0 & 1.2   &  2.6 &     &     & 2.3& 0.1 & 0.7  &  0.1 \\
      $\textbf{$\modee$}$   & 1.7 & 1.3   &      &     & 0.7 & 0.3& 0.2 & 0.7  &      \\
      $\textbf{$\modeaa$}$  & 0.6 & 0.6   &      & 1.7 &     & 2.0& 0.1 & 0.7  &  0.8 \\
      $\textbf{$\modebb$}$  & 0.6 & 0.4   &      & 0.5 & 1.2 & 0.9& 0.1 & 0.7  &  0.8 \\
      $\textbf{$\modedd$}$  & 1.8 & 0.8   &      & 1.4 &     & 2.2& 0.1 & 0.7  &  0.8 \\
      $\textbf{$\modeaaa$}$ & 0.6 & 0.6   &      & 1.3 &     & 2.0& 0.1 & 0.9  &  0.8 \\
      $\textbf{$\modeccc$}$ & 0.7 & 0.2   &      &     & 2.4 & 2.3& 0.1 & 0.5  &  0.1 \\
      $\textbf{$\modeddd$}$ & 1.6 & 1.1   &      &     & 0.5 & 1.3& 0.3 & 0.6  &  0.1 \\
      $\textbf{$\modef$}$   & 1.6 & 0.9   &      &     &     & 4.6& 0.9 & 0.5  &               \\
      \hline\hline
    \end{tabular}

    \label{tab:sys_err_4640}
  \end{center}
  \end{table*}
  \begin{table*}[!htbp]
  \begin{center}
     \scriptsize
    \caption{The systematic uncertainties in BF measurements(\%) at $\sqrt{s}=$ 4661.24 MeV.}
  \begin{tabular}{l|ccccccccc}
      \hline \hline
       Source                &  Tracking  &  PID  & $K_S^0$  & $\Lambda$ & $\pi^0$ &2D fit & Signal model & MC statistics  & Intermediate BF \\ \hline
      $\textbf{$\modea$}$   & 0.4   & 0.1   & 2.6 &     &     & 1.2& 0.1 & 0.6 &  0.1 \\
      $\textbf{$\modeb$}$   & 1.6   & 0.7   &     &     &     & 0.2& 0.2 & 0.5 &      \\
      $\textbf{$\modec$}$   & 0.6   & 0.4   & 1.6 &     & 1.0 & 0.3& 0.2 & 0.6 &  0.1 \\
      $\textbf{$\moded$}$   & 2.0   & 1.1   & 2.6 &     &     & 0.8& 0.6 & 0.7 &  0.1 \\
      $\textbf{$\modee$}$   & 1.8   & 1.0   &     &     & 0.6 & 0.3& 0.3 & 0.7 &      \\
      $\textbf{$\modeaa$}$  & 0.6   & 0.6   &     & 1.7 &     & 1.2& 0.1 & 0.7 &  0.8 \\
      $\textbf{$\modebb$}$  & 0.6   & 0.4   &     & 0.4 & 1.2 & 1.2& 0.1 & 0.8 &  0.8 \\
      $\textbf{$\modedd$}$  & 2.0   & 0.9   &     & 1.1 &     & 1.0& 0.1 & 0.7 &  0.8 \\
      $\textbf{$\modeaaa$}$ & 0.6   & 0.6   &     & 1.5 & 2.4 & 1.7& 0.2 & 0.6 &  0.8 \\
      $\textbf{$\modeccc$}$ & 0.4   & 0.1   &     &     & 0.5 & 2.9& 0.1 & 0.5 &  0.1 \\
      $\textbf{$\modeddd$}$ & 1.8   & 1.2   &     &     &     & 2.3& 0.2 & 0.6 &  0.1 \\
      $\textbf{$\modef$}$   & 1.7   & 1.1   &     &     &     & 2.2& 1.0 & 0.5 &   \\
      \hline\hline
    \end{tabular}

    \label{tab:sys_err_4660}
  \end{center}
  \end{table*}

  \begin{table*}[!htbp]
  \begin{center}
    \scriptsize
    \caption{The systematic uncertainties in BF measurements(\%) at $\sqrt{s}=$ 4698.82 MeV.}
  \begin{tabular}{l|ccccccccc}
      \hline \hline
       Source                &  Tracking  &  PID  & $K_S^0$  & $\Lambda$ & $\pi^0$ &2D fit & Signal model & MC statistics  & Intermediate BF \\   \hline
      $\textbf{$\modea$}$   & 0.4   & 0.1   & 2.7  &     &     & 0.8  &  0.1  & 0.6 &  0.1 \\
      $\textbf{$\modeb$}$   & 1.7   & 0.7   &      &     &     & 0.1  &  0.2  & 0.5 &      \\
      $\textbf{$\modec$}$   & 0.7   & 0.3   & 1.6  &     & 1.1 & 2.3  &  0.2  & 0.7 &  0.1 \\
      $\textbf{$\moded$}$   & 2.1   & 1.2   & 2.5  &     &     & 1.1  &  0.5  & 0.7 &  0.1 \\
      $\textbf{$\modee$}$   & 1.7   & 1.1   &      &     & 0.6 & 1.0  &  0.4  & 0.7 &      \\
      $\textbf{$\modeaa$}$  & 0.5   & 0.6   &      & 1.7 &     & 1.3  &  0.1  & 0.6 &  0.8 \\
      $\textbf{$\modebb$}$  & 0.5   & 0.4   &      & 0.6 & 1.2 & 0.6  &  0.1  & 0.7 &  0.8 \\
      $\textbf{$\modedd$}$  & 0.8   & 0.8   &      & 0.8 &     & 1.5  &  0.1  & 0.7 &  0.8 \\
      $\textbf{$\modeaaa$}$ & 0.6   & 0.6   &      & 1.5 & 2.3 & 3.2  &  0.2  & 0.6 &  0.8 \\
      $\textbf{$\modeccc$}$ & 0.6   & 0.2   &      &     & 0.5 & 1.3  &  0.1  & 0.5 &  0.1 \\
      $\textbf{$\modeddd$}$ & 1.6   & 1.1   &      &     &     & 1.8  &  0.2  & 0.6 &  0.1 \\
      $\textbf{$\modef$}$   & 2.0   & 1.1   &      &     &     & 4.4  &  1.0  & 0.6 & 0.5    \\
      \hline\hline
    \end{tabular}

    \label{tab:sys_err_4700}
  \end{center}
  \end{table*}

\clearpage
\section{DT efficiencies}
Tables~\ref{tab:DTeff_4600},~\ref{tab:DTeff_4612},~\ref{tab:DTeff_4626},~\ref{tab:DTeff_4640},~\ref{tab:DTeff_4660}, ~\ref{tab:DTeff_4680} and ~\ref{tab:DTeff_4700} show the DT efficiencies at different energy points.
\begin{sidewaystable*}[hpt]
\caption{The DT efficiencies for 12 tag modes versus 12 signal modes at $\sqrt{s}=4599.53\mev$, where the uncertainties are statistical only.}
  \begin{center}
 \footnotesize
 \resizebox{1.0\textwidth}{!}{
  \begin{tabular}{l|c|c|c|c|c|c|c|c|c|c|c|c}
      \hline \hline
  Tag/signal mode  & $\textbf{$\modea$}$    & $\textbf{$\modeb$}$  & $\textbf{$\modec$}$  & $\textbf{$\moded$}$  & $\textbf{$\modee$}$  & $\textbf{$\modeaa$}$  & $\textbf{$\modebb$}$ & $\textbf{$\modedd$}$&$\textbf{$\modeaaa$}$&$\textbf{$\modeccc$}$ &$\textbf{$\modeddd$}$ &$\textbf{$\modef$}$ \\ \hline
 $\textbf{$\bmodea$}$   &$ 31.8\pm0.3$ &$29.0\pm0.3 $&$12.6\pm0.2 $&$11.9\pm0.1 $&$10.6\pm0.1 $&$25.4\pm0.3$ &$10.9\pm0.1 $ & $8.2\pm0.1 $&$15.5\pm0.2$ &$11.9\pm0.1$ &$10.4\pm0.1 $&$32.2\pm 0.3$\\
 $\textbf{$\bmodeb$}$   &$ 28.4\pm0.3$ &$26.5\pm0.3 $&$11.1\pm0.1 $&$10.5\pm0.1 $&$ 8.9\pm0.1 $&$22.4\pm0.3$ &$ 9.3\pm0.1 $ & $7.4\pm0.1 $&$13.7\pm0.2$ &$10.1\pm0.1$ &$ 9.3\pm0.1 $&$29.5\pm 0.3$\\
 $\textbf{$\bmodec$}$   &$ 12.3\pm0.1$ &$11.2\pm0.1 $&$ 4.4\pm0.1 $&$ 4.1\pm0.1 $&$ 3.5\pm0.1 $&$ 9.8\pm0.1$ &$ 3.9\pm0.1 $ & $3.0\pm0.0 $&$ 5.9\pm0.1$ &$ 4.1\pm0.1$ &$ 3.7\pm0.0 $&$13.1\pm 0.2$\\
 $\textbf{$\bmoded$}$   &$ 11.6\pm0.1$ &$10.6\pm0.1 $&$ 4.1\pm0.1 $&$ 3.3\pm0.1 $&$ 3.4\pm0.1 $&$ 9.1\pm0.1$ &$ 3.3\pm0.1 $ & $2.5\pm0.0 $&$ 5.1\pm0.1$ &$ 3.9\pm0.1$ &$ 3.6\pm0.0 $&$12.3\pm 0.2$\\
 $\textbf{$\bmodee$}$   &$ 10.2\pm0.1$ &$ 8.9\pm0.1 $&$ 3.3\pm0.1 $&$ 3.1\pm0.1 $&$ 2.7\pm0.1 $&$ 7.8\pm0.1$ &$ 2.7\pm0.1 $ & $2.2\pm0.0 $&$ 4.3\pm0.1$ &$ 3.1\pm0.0$ &$ 2.7\pm0.0 $&$10.3\pm 0.1$\\
 $\textbf{$\bmodeaa$}$  &$ 27.0\pm0.3$ &$24.7\pm0.3 $&$10.8\pm0.1 $&$10.1\pm0.1 $&$ 8.9\pm0.1 $&$21.3\pm0.3$ &$ 9.1\pm0.1 $ & $6.3\pm0.1 $&$12.0\pm0.1$ &$ 9.3\pm0.1$ &$ 9.1\pm0.1 $&$28.7\pm 0.3$\\
 $\textbf{$\bmodebb$}$  &$ 11.3\pm0.1$ &$10.1\pm0.1 $&$ 4.1\pm0.1 $&$ 3.5\pm0.1 $&$ 3.1\pm0.1 $&$ 8.6\pm0.1$ &$ 3.5\pm0.1 $ & $2.4\pm0.0 $&$ 5.2\pm0.1$ &$ 3.9\pm0.1$ &$ 3.2\pm0.0 $&$11.9\pm 0.1$\\
 $\textbf{$\bmodedd$}$  &$  8.0\pm0.1$ &$ 7.4\pm0.1 $&$ 3.0\pm0.0 $&$ 2.4\pm0.0 $&$ 2.4\pm0.0 $&$ 6.2\pm0.1$ &$ 2.4\pm0.0 $ & $1.6\pm0.0 $&$ 3.5\pm0.0$ &$ 2.8\pm0.0$ &$ 2.4\pm0.0 $&$ 8.5\pm 0.1$\\
 $\textbf{$\bmodeaaa$}$ &$ 15.2\pm0.2$ &$13.7\pm0.2 $&$ 5.9\pm0.1 $&$ 5.2\pm0.1 $&$ 4.5\pm0.1 $&$11.5\pm0.1$ &$ 4.9\pm0.1 $ & $3.6\pm0.0 $&$ 6.8\pm0.1$ &$ 5.3\pm0.1$ &$ 4.8\pm0.1 $&$15.8\pm 0.2$\\
 $\textbf{$\bmodeccc$}$ &$ 12.1\pm0.1$ &$10.1\pm0.1 $&$ 4.2\pm0.1 $&$ 3.9\pm0.1 $&$ 3.3\pm0.0 $&$ 9.3\pm0.1$ &$ 3.7\pm0.1 $ & $2.8\pm0.0 $&$ 5.6\pm0.1$ &$ 3.9\pm0.0$ &$ 3.5\pm0.0 $&$11.8\pm 0.1$\\
 $\textbf{$\bmodeddd$}$ &$ 10.2\pm0.1$ &$ 9.2\pm0.1 $&$ 3.9\pm0.1 $&$ 3.5\pm0.0 $&$ 3.0\pm0.0 $&$ 8.3\pm0.1$ &$ 3.3\pm0.0 $ & $2.4\pm0.0 $&$ 4.8\pm0.1$ &$ 3.5\pm0.0$ &$ 3.0\pm0.0 $&$10.7\pm 0.1$\\
 $\textbf{$\bmodef$}$   &$ 31.5\pm0.3$ &$29.5\pm0.3 $&$13.1\pm0.2 $&$12.2\pm0.1 $&$10.5\pm0.1 $&$26.2\pm0.3$ &$11.1\pm0.1 $ & $8.5\pm0.1 $&$15.8\pm0.2$ &$11.8\pm0.1$ &$10.7\pm0.1 $&$34.9\pm 0.3$\\
\hline \hline                                               
   \end{tabular}}
   \label{tab:DTeff_4600}
  \end{center}
\end{sidewaystable*}

\begin{sidewaystable}[hpt] 
\caption{The DT efficiencies for 12 tag modes versus 12 signal modes at $\sqrt{s}=4611.86\mev$, where the uncertainties are statistical only.}
\footnotesize
  \begin{center}
  \resizebox{1.0\textwidth}{!}{
   \begin{tabular}{l|c|c|c|c|c|c|c|c|c|c|c|c}
      \hline \hline
 Tag/signal mode  & $\textbf{$\modea$}$    & $\textbf{$\modeb$}$  & $\textbf{$\modec$}$  & $\textbf{$\moded$}$  & $\textbf{$\modee$}$  & $\textbf{$\modeaa$}$  & $\textbf{$\modebb$}$ & $\textbf{$\modedd$}$&$\textbf{$\modeaaa$}$&$\textbf{$\modeccc$}$ &$\textbf{$\modeddd$}$ &$\textbf{$\modef$}$ \\ \hline
 $\textbf{$\bmodea$}$   &$28.3\pm0.3$ &$27.3\pm0.3$ &$11.3\pm0.2$&$10.3\pm0.2$&$9.6\pm0.1$ &$23.0\pm0.3$&$9.8\pm0.2$ &$7.4\pm0.1$ &$14.0\pm0.2$ &$11.4\pm0.1$&$9.8\pm0.1$ & $30.7\pm0.3$    \\
 $\textbf{$\bmodeb$}$   &$26.8\pm0.3$ &$25.4\pm0.3$ &$10.4\pm0.2$&$9.4\pm0.1$ &$8.8\pm0.1$ &$20.7\pm0.3$&$8.6\pm0.1$ &$6.7\pm0.1$ &$12.4\pm0.2$ &$10.2\pm0.1$&$8.7\pm0.1$ & $28.4\pm0.3$     \\
 $\textbf{$\bmodec$}$   &$11.0\pm0.2$ &$10.5\pm0.2$ &$3.9\pm0.1$ &$3.7\pm0.1$ &$3.3\pm0.1$ &$9.2\pm0.1$ &$3.6\pm0.1$ &$2.7\pm0.0$ &$5.2\pm0.1$ &4$.1\pm0.1$  &$3.6\pm0.0$ & $12.1\pm0.2$         \\
 $\textbf{$\bmoded$}$   &$10.0\pm0.2$ &$9.4\pm0.1$  &$3.7\pm0.1$ &$2.8\pm0.1$ &$3.0\pm0.1$ &$8.1\pm0.1$ &$2.8\pm0.1$ &$2.0\pm0.0$ &$4.4\pm0.1$ &3$.6\pm0.1$  &$3.1\pm0.1$ & $11.0\pm0.2$          \\
 $\textbf{$\bmodee$}$   & $9.3\pm0.1$ &$8.8\pm0.1$  &$3.2\pm0.1$ &$2.7\pm0.1$ &$2.5\pm0.1$ &$7.4\pm0.1$ &$2.7\pm0.1$ &$2.0\pm0.0$ &$3.9\pm0.1$ &3$.0\pm0.0$  &$2.5\pm0.0$ & $9.7\pm0.1$           \\
 $\textbf{$\bmodeaa$}$  &$24.5\pm0.3$ &$22.5\pm0.3$ &$10.1\pm0.2$&$8.9\pm0.1$ &$8.3\pm0.1$ &$19.4\pm0.3$&$8.1\pm0.1$ &$5.6\pm0.1$ &$10.8\pm0.2$ &$8.9\pm0.1$ &$8.4\pm0.1$ & $26.6\pm0.3$      \\
 $\textbf{$\bmodebb$}$  &$10.3\pm0.2$ &$9.3\pm0.1$  &$3.8\pm0.1$ &$3.0\pm0.1$ &$3.1\pm0.1$ &$7.6\pm0.1$ &$3.0\pm0.1$ &$2.1\pm0.0$ &$4.6\pm0.1$ &3$.7\pm0.1$  &$2.8\pm0.0$ & $11.2\pm0.2$          \\
 $\textbf{$\bmodedd$}$  & $7.2\pm0.1$ &$6.8\pm0.1$  &$2.6\pm0.0$ &$2.0\pm0.0$ &$2.1\pm0.0$ &$5.5\pm0.1$ &$2.0\pm0.0$ &$1.3\pm0.0$ &$3.1\pm0.0$ &2$.6\pm0.0$  &$2.1\pm0.0$ & $7.7\pm0.1$           \\
 $\textbf{$\bmodeaaa$}$ &$13.8\pm0.2$ &$12.5\pm0.2$ &$5.2\pm0.1$ &$4.5\pm0.1$ &$4.0\pm0.1$ &$10.5\pm0.2$&$4.4\pm0.1$ &$3.2\pm0.0$ &$6.0\pm0.1$ &5$.0\pm0.1$  &$4.4\pm0.1$ & $14.4\pm0.2$        \\
 $\textbf{$\bmodeccc$}$ &$11.5\pm0.1$ &$10.2\pm0.1$ &$4.2\pm0.1$ &$3.6\pm0.1$ &$3.2\pm0.1$ &$8.9\pm0.1$ &$3.5\pm0.1$ &$2.6\pm0.0$ &$5.3\pm0.1$ &3$.9\pm0.0$  &$3.4\pm0.0$ & $11.6\pm0.1$         \\
 $\textbf{$\bmodeddd$}$ & $9.6\pm0.1$ &$8.7\pm0.1$  &$3.8\pm0.1$ &$3.1\pm0.0$ &$2.8\pm0.0$ &$7.7\pm0.1$ &$2.9\pm0.0$ &$2.1\pm0.0$ &$4.5\pm0.1$ &3$.5\pm0.0$  &$2.8\pm0.0$ & $10.1\pm0.1$           \\
 $\textbf{$\bmodef$}$   &$30.0\pm0.3$ &$28.4\pm0.3$ &$12.1\pm0.2$&$10.9\pm0.2$&$10.0\pm0.2$&$24.5\pm0.3$&$10.4\pm0.2$&$7.7\pm0.1$ &$14.3\pm0.2$ &$11.6\pm0.1$&$10.1\pm0.1$& $33.0\pm0.3$ \\
\hline \hline
   \end{tabular}}
   \label{tab:DTeff_4612}
  \end{center}
\end{sidewaystable}

\begin{sidewaystable}[hpt]
\caption{The DT efficiencies for 12 tag modes versus 12 signal modes at $\sqrt{s}=4628.00\mev$, where the uncertainties are statistical only.}
  \begin{center}
  \footnotesize
  \resizebox{1.0\textwidth}{!}{
  \begin{tabular}{l|c|c|c|c|c|c|c|c|c|c|c|c}
      \hline \hline
  Tag/signal mode  & $\textbf{$\modea$}$    & $\textbf{$\modeb$}$  & $\textbf{$\modec$}$  & $\textbf{$\moded$}$  & $\textbf{$\modee$}$  & $\textbf{$\modeaa$}$  & $\textbf{$\modebb$}$ & $\textbf{$\modedd$}$&$\textbf{$\modeaaa$}$&$\textbf{$\modeccc$}$ &$\textbf{$\modeddd$}$ &$\textbf{$\modef$}$ \\ \hline
 $\textbf{$\bmodea$}$   &$27.6\pm0.3$  & $25.4\pm0.3$ & $11.2\pm0.2$ & $10.0\pm0.1$ & $9.5\pm0.1$ & $22.2\pm0.3$ & $9.2\pm0.1$ & $7.2\pm0.1$ & $13.3\pm0.2$ & $10.8\pm0.1$ & $9.3\pm0.1$ & $29.5\pm0.3$    \\
 $\textbf{$\bmodeb$}$  & $24.9\pm0.3$ & $23.7\pm0.3$ & $10.2\pm0.1$ & $9.3\pm0.1$ & $8.4\pm0.1$ & $19.8\pm0.3$ & $8.3\pm0.1$ & $6.5\pm0.1$ & $11.8\pm0.2$ & $9.4\pm0.1$ & $8.3\pm0.1$ & $27.6\pm0.3$    \\
 $\textbf{$\bmodec$}$  & $11.0\pm0.1$ & $10.2\pm0.1$ & $3.8\pm0.1$ & $3.4\pm0.1$ & $3.2\pm0.1$ & $8.6\pm0.1$ & $3.4\pm0.1$ & $2.5\pm0.0$ & $5.1\pm0.1$ & $3.9\pm0.1$ & $3.4\pm0.0$ & $11.8\pm0.2$       \\
 $\textbf{$\bmoded$}$  & $9.7\pm0.1$ & $9.4\pm0.1$ & $3.4\pm0.1$ & $2.5\pm0.1$ & $2.9\pm0.1$ & $7.5\pm0.1$ & $2.7\pm0.1$ & $2.0\pm0.0$ & $4.2\pm0.1$ & $3.5\pm0.1$ & $3.0\pm0.0$ & $10.4\pm0.1$           \\
 $\textbf{$\bmodee$}$  &$ 9.2\pm0.1$ & $8.4\pm0.1$ & $3.0\pm0.1$ & $2.7\pm0.1$ & $2.5\pm0.1$ & $7.0\pm0.1$ & $2.6\pm0.1$ & $1.9\pm0.0$ & $3.9\pm0.1$ & $2.9\pm0.0$ & $2.4\pm0.0$ & $9.5\pm0.1$            \\
 $\textbf{$\bmodeaa$}$ &$ 23.7\pm0.3$ & $21.7\pm0.3$ & $9.3\pm0.1$ & $8.3\pm0.1$ & $7.9\pm0.1$ & $18.6\pm0.3$ & $7.8\pm0.1$ & $5.4\pm0.1$ & $10.2\pm0.1$ & $8.2\pm0.1$ & $7.8\pm0.1$ & $25.3\pm0.3$      \\
 $\textbf{$\bmodebb$}$ &$ 9.6\pm0.1$ & $8.9\pm0.1$ & $3.6\pm0.1$ & $2.9\pm0.1$ & $2.9\pm0.1$ & $7.3\pm0.1$ & $2.9\pm0.1$ & $2.0\pm0.0$ & $4.4\pm0.1$ & $3.5\pm0.1$ & $2.7\pm0.0$ & $10.4\pm0.1$      \\
 $\textbf{$\bmodedd$}$ &$ 7.0\pm0.1$ & $6.5\pm0.1$ & $2.4\pm0.0$ & $1.9\pm0.0$ & $2.0\pm0.0$ & $5.2\pm0.1$ & $2.0\pm0.0$ & $1.3\pm0.0$ & $3.0\pm0.0$ & $2.6\pm0.0$ & $2.1\pm0.0$ & $7.5\pm0.1$            \\
 $\textbf{$\bmodeaaa$}$&$ 13.1\pm0.2$ & $11.9\pm0.2$ & $5.1\pm0.1$ & $4.3\pm0.1$ & $4.1\pm0.1$ & $9.9\pm0.1$ & $4.1\pm0.1$ & $3.1\pm0.0$ & $5.8\pm0.1$ & $4.6\pm0.1$ & $4.2\pm0.1$ & $13.8\pm0.2$        \\
 $\textbf{$\bmodeccc$}$&$ 10.9\pm0.1$ & $9.4\pm0.1$ & $4.0\pm0.1$ & $3.5\pm0.1$ & $3.1\pm0.0$ & $8.2\pm0.1$ & $3.3\pm0.0$ & $2.6\pm0.0$ & $5.0\pm0.1$ & $3.8\pm0.0$ & $3.2\pm0.0$ & $11.1\pm0.1$          \\
 $\textbf{$\bmodeddd$}$&$ 4.7\pm0.1$ & $4.3\pm0.1$ & $1.8\pm0.0$ & $1.4\pm0.0$ & $1.3\pm0.0$ & $3.7\pm0.1$ & $1.4\pm0.0$ & $1.0\pm0.0$ & $2.2\pm0.0$ & $1.7\pm0.0$ & $1.4\pm0.0$ & $4.9\pm0.1$          \\
 $\textbf{$\bmodef$}$  &$ 28.9\pm0.3$ & $27.6\pm0.3$ & $11.7\pm0.2$ & $10.3\pm0.1$ & $9.8\pm0.1$ & $23.2\pm0.3$ & $9.7\pm0.1$ & $7.4\pm0.1$ & $13.8\pm0.2$ & $11.1\pm0.1$ & $9.8\pm0.1$ & $31.6\pm0.3$   \\
\hline \hline
   \end{tabular}}
   \label{tab:DTeff_4626}
  \end{center}
\end{sidewaystable}
\begin{sidewaystable}[hpt]
  \caption{The DT efficiencies for 12 tag modes versus 12 signal modes at $\sqrt{s}=4640.91\mev$, where the uncertainties are statistical only.}
  \begin{center}
  \footnotesize
  \resizebox{1.0\textwidth}{!}{
  \begin{tabular}{l|c|c|c|c|c|c|c|c|c|c|c|c}
      \hline \hline
  Tag/signal mode  & $\textbf{$\modea$}$    & $\textbf{$\modeb$}$  & $\textbf{$\modec$}$  & $\textbf{$\moded$}$  & $\textbf{$\modee$}$  & $\textbf{$\modeaa$}$  & $\textbf{$\modebb$}$ & $\textbf{$\modedd$}$&$\textbf{$\modeaaa$}$&$\textbf{$\modeccc$}$ &$\textbf{$\modeddd$}$ &$\textbf{$\modef$}$ \\ \hline
  $\textbf{$\bmodea$}$   &$26.6\pm0.3$  & $26.1\pm0.3$ & $10.9\pm0.1$ & $10.2\pm0.1$ & $9.6\pm0.1$ & $21.4\pm0.3$ & $9.3\pm0.1$ & $7.1\pm0.1$ & $13.0\pm0.2$ & $10.3\pm0.1$ & $9.3\pm0.1$ & $29.4\pm0.3$  \\
  $\textbf{$\bmodeb$}$   &$25.5\pm0.3$ & $23.7\pm0.3$ & $10.0\pm0.1$ & $9.0\pm0.1$ & $8.7\pm0.1$ & $19.8\pm0.3$ & $8.2\pm0.1$ & $6.5\pm0.1$ & $12.0\pm0.2$ & $9.3\pm0.1$ & $8.5\pm0.1$ & $27.0\pm0.3$     \\
  $\textbf{$\bmodec$}$   &$10.7\pm0.1$ & $10.0\pm0.1$ & $3.8\pm0.1$ & $3.5\pm0.1$ & $3.3\pm0.1$ & $8.2\pm0.1$ & $3.3\pm0.1$ & $2.6\pm0.0$ & $5.0\pm0.1$ & $3.7\pm0.1$ & $3.2\pm0.0$ & $11.5\pm0.2$        \\
  $\textbf{$\bmoded$}$   &$9.9\pm0.1$ & $9.0\pm0.1$ & $3.5\pm0.1$ & $2.5\pm0.1$ & $2.9\pm0.1$ & $7.5\pm0.1$ & $2.8\pm0.1$ & $2.0\pm0.0$ & $4.1\pm0.1$ & $3.5\pm0.0$ & $3.0\pm0.0$ & $10.8\pm0.1$          \\
  $\textbf{$\bmodee$}$   &$9.2\pm0.1$ & $8.7\pm0.1$ & $3.1\pm0.1$ & $2.6\pm0.1$ & $2.5\pm0.1$ & $6.9\pm0.1$ & $2.6\pm0.1$ & $2.0\pm0.0$ & $3.9\pm0.1$ & $2.9\pm0.0$ & $2.4\pm0.0$ & $9.4\pm0.1$           \\
  $\textbf{$\bmodeaa$}$  &$22.6\pm0.3$ & $21.5\pm0.3$ & $8.9\pm0.1$ & $8.3\pm0.1$ & $7.8\pm0.1$ & $18.2\pm0.3$ & $7.9\pm0.1$ & $5.2\pm0.1$ & $10.1\pm0.1$ & $8.2\pm0.1$ & $7.9\pm0.1$ & $24.1\pm0.3$      \\
  $\textbf{$\bmodebb$}$  &$9.6\pm0.1$ & $8.9\pm0.1$ & $3.4\pm0.1$ & $3.0\pm0.1$ & $3.0\pm0.1$ & $7.5\pm0.1$ & $2.8\pm0.1$ & $2.0\pm0.0$ & $4.5\pm0.1$ & $3.5\pm0.0$ & $2.7\pm0.0$ & $10.3\pm0.1$          \\
  $\textbf{$\bmodedd$}$  &$6.9\pm0.1$ & $6.5\pm0.1$ & $2.5\pm0.0$ & $1.9\pm0.0$ & $2.2\pm0.0$ & $5.1\pm0.1$ & $2.0\pm0.0$ & $1.3\pm0.0$ & $3.0\pm0.0$ & $2.5\pm0.0$ & $2.2\pm0.0$ & $7.5\pm0.1$           \\
  $\textbf{$\bmodeaaa$}$ &$12.7\pm0.2$ & $12.1\pm0.2$ & $5.0\pm0.1$ & $4.2\pm0.1$ & $4.1\pm0.1$ & $9.6\pm0.1$ & $4.2\pm0.1$ & $3.0\pm0.0$ & $5.5\pm0.1$ & $4.5\pm0.1$ & $4.1\pm0.1$ & $13.5\pm0.2$        \\
  $\textbf{$\bmodeccc$}$ &$10.4\pm0.1$ & $9.3\pm0.1$ & $3.8\pm0.1$ & $3.5\pm0.0$ & $3.1\pm0.0$ & $8.2\pm0.1$ & $3.3\pm0.0$ & $2.5\pm0.0$ & $4.8\pm0.1$ & $3.5\pm0.0$ & $3.2\pm0.0$ & $11.0\pm0.1$         \\
  $\textbf{$\bmodeddd$}$ &$9.1\pm0.1$ & $8.5\pm0.1$ & $3.4\pm0.0$ & $2.9\pm0.0$ & $2.7\pm0.0$ & $7.2\pm0.1$ & $2.8\pm0.0$ & $2.1\pm0.0$ & $4.1\pm0.1$ & $3.2\pm0.0$ & $2.7\pm0.0$ & $9.6\pm0.1$           \\
  $\textbf{$\bmodef$}$   &$28.6\pm0.3$ & $27.0\pm0.3$ & $11.5\pm0.2$ & $10.8\pm0.1$ & $9.6\pm0.1$ & $22.2\pm0.3$ & $9.7\pm0.1$ & $7.4\pm0.1$ & $13.5\pm0.2$ & $11.0\pm0.1$ & $9.6\pm0.1$ & $30.5\pm0.3$   \\
\hline \hline
   \end{tabular}}
   \label{tab:DTeff_4640}
  \end{center}
\end{sidewaystable}
  
\begin{sidewaystable}[hpt]
\caption{The DT efficiencies for 12 tag modes versus 12 signal modes at $\sqrt{s}=4661.24\mev$, where the uncertainties are statistical only.}
  \begin{center}
  \footnotesize
     \resizebox{1.0\textwidth}{!}{
  \begin{tabular}{l|c|c|c|c|c|c|c|c|c|c|c|c}
      \hline \hline
  Tag/signal mode  & $\textbf{$\modea$}$    & $\textbf{$\modeb$}$  & $\textbf{$\modec$}$  & $\textbf{$\moded$}$  & $\textbf{$\modee$}$  & $\textbf{$\modeaa$}$  & $\textbf{$\modebb$}$ & $\textbf{$\modedd$}$&$\textbf{$\modeaaa$}$&$\textbf{$\modeccc$}$ &$\textbf{$\modeddd$}$ &$\textbf{$\modef$}$ \\ \hline
  $\textbf{$\bmodea$}$    &$25.7\pm0.3$  & $25.4\pm0.3$ & $10.8\pm0.1$ & $9.7\pm0.1$ & $9.4\pm0.1$ & $20.7\pm0.3$ & $9.0\pm0.1$ & $7.1\pm0.1$ & $12.8\pm0.2$ & $10.2\pm0.1$ & $9.2\pm0.1$ & $28.4\pm0.3$ \\
  $\textbf{$\bmodeb$}$    &$24.9\pm0.3$ & $24.6\pm0.3$ & $10.0\pm0.1$ & $9.1\pm0.1$ & $8.4\pm0.1$ & $18.9\pm0.3$ & $8.2\pm0.1$ & $6.5\pm0.1$ & $11.6\pm0.1$ & $9.1\pm0.1$ & $8.3\pm0.1$ & $26.5\pm0.3$   \\
  $\textbf{$\bmodec$}$    &$10.6\pm0.1$ & $10.1\pm0.1$ & $3.7\pm0.1$ & $3.6\pm0.1$ & $3.2\pm0.1$ & $8.5\pm0.1$ & $3.2\pm0.1$ & $2.6\pm0.0$ & $4.9\pm0.1$ & $3.8\pm0.1$ & $3.3\pm0.0$ & $11.7\pm0.2$      \\
  $\textbf{$\bmoded$}$    &$9.5\pm0.1$ & $9.2\pm0.1$ & $3.5\pm0.1$ & $2.5\pm0.1$ & $3.0\pm0.1$ & $7.5\pm0.1$ & $2.8\pm0.1$ & $2.0\pm0.0$ & $4.3\pm0.1$ & $3.5\pm0.0$ & $3.1\pm0.0$ & $10.5\pm0.1$        \\
  $\textbf{$\bmodee$}$    &$9.0\pm0.1$ & $8.4\pm0.1$ & $3.0\pm0.1$ & $2.7\pm0.1$ & $2.4\pm0.1$ & $6.9\pm0.1$ & $2.4\pm0.1$ & $2.0\pm0.0$ & $3.8\pm0.1$ & $2.8\pm0.0$ & $2.5\pm0.0$ & $9.5\pm0.1$         \\
  $\textbf{$\bmodeaa$}$   &$22.1\pm0.3$ & $20.8\pm0.3$ & $9.2\pm0.1$ & $8.3\pm0.1$ & $7.8\pm0.1$ & $17.4\pm0.3$ & $7.5\pm0.1$ & $5.3\pm0.1$ & $9.6\pm0.1$ & $8.0\pm0.1$ & $7.5\pm0.1$ & $24.0\pm0.3$     \\
  $\textbf{$\bmodebb$}$   &$9.3\pm0.1$ & $8.9\pm0.1$ & $3.4\pm0.1$ & $3.0\pm0.1$ & $2.8\pm0.1$ & $7.1\pm0.1$ & $2.9\pm0.1$ & $2.1\pm0.0$ & $4.2\pm0.1$ & $3.4\pm0.0$ & $2.7\pm0.0$ & $10.2\pm0.1$        \\
  $\textbf{$\bmodedd$}$   &$6.9\pm0.1$ & $6.5\pm0.1$ & $2.5\pm0.0$ & $2.0\pm0.0$ & $2.2\pm0.0$ & $5.2\pm0.1$ & $2.0\pm0.0$ & $1.3\pm0.0$ & $3.0\pm0.0$ & $2.5\pm0.0$ & $2.2\pm0.0$ & $7.4\pm0.1$         \\
  $\textbf{$\bmodeaaa$}$  &$12.5\pm0.2$ & $11.7\pm0.2$ & $4.8\pm0.1$ & $4.3\pm0.1$ & $4.0\pm0.1$ & $9.3\pm0.1$ & $3.9\pm0.1$ & $3.1\pm0.0$ & $5.5\pm0.1$ & $4.5\pm0.1$ & $4.0\pm0.1$ & $13.3\pm0.2$      \\
  $\textbf{$\bmodeccc$}$  &$10.2\pm0.1$ & $9.1\pm0.1$ & $3.9\pm0.1$ & $3.5\pm0.0$ & $3.1\pm0.0$ & $8.0\pm0.1$ & $3.3\pm0.0$ & $2.5\pm0.0$ & $4.8\pm0.1$ & $3.6\pm0.0$ & $3.1\pm0.0$ & $10.6\pm0.1$       \\
  $\textbf{$\bmodeddd$}$  &$9.0\pm0.1$ & $8.3\pm0.1$ & $3.5\pm0.0$ & $3.0\pm0.0$ & $2.8\pm0.0$ & $6.9\pm0.1$ & $2.8\pm0.0$ & $2.2\pm0.0$ & $4.0\pm0.1$ & $3.1\pm0.0$ & $2.6\pm0.0$ & $9.3\pm0.1$         \\
  $\textbf{$\bmodef$}$    &$27.8\pm0.3$ & $26.5\pm0.3$ & $11.6\pm0.1$ & $10.4\pm0.1$ & $9.7\pm0.1$ & $22.1\pm0.3$ & $9.6\pm0.1$ & $7.4\pm0.1$ & $13.3\pm0.2$ & $10.6\pm0.1$ & $9.4\pm0.1$ & $30.1\pm0.3$ \\
\hline \hline
   \end{tabular}}
   \label{tab:DTeff_4660}
  \end{center}
\end{sidewaystable}

\begin{sidewaystable}[hpt]
 \caption{The DT efficiencies for 12 tag modes versus 12 signal modes at $\sqrt{s}=4681.92\mev$, where the uncertainties are statistical only.}
  \begin{center}
  \footnotesize
    \resizebox{1.0\textwidth}{!}{
  \begin{tabular}{l|c|c|c|c|c|c|c|c|c|c|c|c}
      \hline \hline
  Tag/signal mode  & $\textbf{$\modea$}$    & $\textbf{$\modeb$}$  & $\textbf{$\modec$}$  & $\textbf{$\moded$}$  & $\textbf{$\modee$}$  & $\textbf{$\modeaa$}$  & $\textbf{$\modebb$}$ & $\textbf{$\modedd$}$&$\textbf{$\modeaaa$}$&$\textbf{$\modeccc$}$ &$\textbf{$\modeddd$}$ &$\textbf{$\modef$}$ \\ \hline
  $\textbf{$\bmodea$}$   &$24.9\pm0.3$  & $24.8\pm0.3$ & $10.6\pm0.1$ & $9.5\pm0.1$ & $9.3\pm0.1$ & $19.7\pm0.3$ & $8.8\pm0.1$ & $7.2\pm0.1$ & $12.4\pm0.2$ & $9.9\pm0.1$ & $8.8\pm0.1$ & $27.4\pm0.3$  \\
  $\textbf{$\bmodeb$}$   &$24.3\pm0.3$ & $23.8\pm0.3$ & $9.7\pm0.1$ & $9.3\pm0.1$ & $8.5\pm0.1$ & $18.7\pm0.3$ & $8.2\pm0.1$ & $6.6\pm0.1$ & $11.1\pm0.1$ & $9.2\pm0.1$ & $8.1\pm0.1$ & $25.8\pm0.3$    \\
  $\textbf{$\bmodec$}$   &$10.3\pm0.1$ & $9.8\pm0.1$ & $3.7\pm0.1$ & $3.5\pm0.1$ & $3.1\pm0.1$ & $8.2\pm0.1$ & $3.1\pm0.1$ & $2.5\pm0.0$ & $4.8\pm0.1$ & $3.6\pm0.0$ & $3.2\pm0.0$ & $11.2\pm0.1$       \\
  $\textbf{$\bmoded$}$   &$9.2\pm0.1$ & $9.3\pm0.1$ & $3.4\pm0.1$ & $2.6\pm0.1$ & $3.0\pm0.1$ & $7.1\pm0.1$ & $2.8\pm0.1$ & $2.0\pm0.0$ & $4.4\pm0.1$ & $3.4\pm0.0$ & $3.0\pm0.0$ & $10.1\pm0.1$        \\
  $\textbf{$\bmodee$}$   &$8.9\pm0.1$ & $8.5\pm0.1$ & $2.9\pm0.1$ & $2.7\pm0.1$ & $2.5\pm0.1$ & $6.5\pm0.1$ & $2.5\pm0.1$ & $2.0\pm0.0$ & $3.8\pm0.1$ & $2.8\pm0.0$ & $2.4\pm0.0$ & $9.4\pm0.1$         \\
  $\textbf{$\bmodeaa$}$  &$21.0\pm0.3$ & $20.5\pm0.3$ & $8.9\pm0.1$ & $7.9\pm0.1$ & $7.4\pm0.1$ & $17.2\pm0.2$ & $7.3\pm0.1$ & $5.3\pm0.1$ & $9.5\pm0.1$ & $7.6\pm0.1$ & $7.4\pm0.1$ & $22.9\pm0.3$     \\
  $\textbf{$\bmodebb$}$  &$9.1\pm0.1$ & $8.9\pm0.1$ & $3.3\pm0.1$ & $3.0\pm0.1$ & $2.9\pm0.1$ & $6.8\pm0.1$ & $2.8\pm0.1$ & $2.1\pm0.0$ & $4.1\pm0.1$ & $3.3\pm0.0$ & $2.5\pm0.0$ & $9.5\pm0.1$         \\
  $\textbf{$\bmodedd$}$  &$7.0\pm0.1$ & $6.6\pm0.1$ & $2.5\pm0.0$ & $2.0\pm0.0$ & $2.2\pm0.0$ & $5.2\pm0.1$ & $2.1\pm0.0$ & $1.3\pm0.0$ & $3.0\pm0.0$ & $2.5\pm0.0$ & $2.2\pm0.0$ & $7.3\pm0.1$         \\
  $\textbf{$\bmodeaaa$}$ &$12.2\pm0.1$ & $11.1\pm0.1$ & $4.8\pm0.1$ & $4.5\pm0.1$ & $4.0\pm0.1$ & $9.1\pm0.1$ & $3.9\pm0.1$ & $3.1\pm0.0$ & $5.5\pm0.1$ & $4.3\pm0.1$ & $3.9\pm0.1$ & $13.1\pm0.2$      \\
  $\textbf{$\bmodeccc$}$ &$10.1\pm0.1$ & $9.2\pm0.1$ & $3.7\pm0.0$ & $3.4\pm0.0$ & $3.0\pm0.0$ & $7.6\pm0.1$ & $3.1\pm0.0$ & $2.5\pm0.0$ & $4.6\pm0.1$ & $3.4\pm0.0$ & $3.1\pm0.0$ & $10.1\pm0.1$       \\
  $\textbf{$\bmodeddd$}$ &$8.5\pm0.1$ & $8.1\pm0.1$ & $3.4\pm0.0$ & $2.9\pm0.0$ & $2.7\pm0.0$ & $6.8\pm0.1$ & $2.6\pm0.0$ & $2.2\pm0.0$ & $3.9\pm0.1$ & $3.1\pm0.0$ & $2.6\pm0.0$ & $9.3\pm0.1$         \\
  $\textbf{$\bmodef$}$   &$26.8\pm0.3$ & $25.8\pm0.3$ & $11.2\pm0.1$ & $10.0\pm0.1$ & $9.6\pm0.1$ & $21.0\pm0.3$ & $8.9\pm0.1$ & $7.3\pm0.1$ & $13.0\pm0.2$ & $10.1\pm0.1$ & $9.3\pm0.1$ & $27.9\pm0.3$ \\
\hline \hline
   \end{tabular}}
   \label{tab:DTeff_4680}
  \end{center}
\end{sidewaystable}

\begin{sidewaystable}[hpt]
\caption{The DT efficiencies for 12 tag modes versus 12 signal modes at $\sqrt{s}=4698.82\mev$, where the uncertainties are statistical only.}
  \begin{center}
  \footnotesize
    \resizebox{1.0\textwidth}{!}{
  \begin{tabular}{l|c|c|c|c|c|c|c|c|c|c|c|c}
      \hline \hline
  Tag/signal mode & $\textbf{$\modea$}$    & $\textbf{$\modeb$}$  & $\textbf{$\modec$}$  & $\textbf{$\moded$}$  & $\textbf{$\modee$}$  & $\textbf{$\modeaa$}$  & $\textbf{$\modebb$}$ & $\textbf{$\modedd$}$&$\textbf{$\modeaaa$}$&$\textbf{$\modeccc$}$ &$\textbf{$\modeddd$}$ &$\textbf{$\modef$}$ \\ \hline
 $\textbf{$\bmodea$}$   &$24.9\pm0.3$  & $24.5\pm0.3$ & $10.3\pm0.1$ & $9.5\pm0.1$ & $8.8\pm0.1$ & $19.5\pm0.3$ & $8.1\pm0.1$ & $6.9\pm0.1$ & $11.9\pm0.1$ & $9.3\pm0.1$ & $8.5\pm0.1$ & $26.8\pm0.3$ \\
 $\textbf{$\bmodeb$}$   &$23.9\pm0.3$ & $23.2\pm0.3$ & $9.5\pm0.1$ & $9.1\pm0.1$ & $8.4\pm0.1$ & $17.9\pm0.2$ & $8.0\pm0.1$ & $6.6\pm0.1$ & $11.0\pm0.1$ & $8.9\pm0.1$ & $8.0\pm0.1$ & $25.5\pm0.3$   \\
 $\textbf{$\bmodec$}$   &$10.1\pm0.1$ & $9.6\pm0.1$ & $3.7\pm0.1$ & $3.5\pm0.1$ & $3.1\pm0.1$ & $8.1\pm0.1$ & $3.2\pm0.1$ & $2.6\pm0.0$ & $4.7\pm0.1$ & $3.6\pm0.0$ & $3.2\pm0.0$ & $10.8\pm0.1$      \\
 $\textbf{$\bmoded$}$   &$9.3\pm0.1$ & $9.2\pm0.1$ & $3.4\pm0.1$ & $2.5\pm0.1$ & $3.0\pm0.1$ & $7.0\pm0.1$ & $2.7\pm0.1$ & $2.1\pm0.0$ & $4.2\pm0.1$ & $3.4\pm0.0$ & $2.9\pm0.0$ & $9.9\pm0.1$        \\
 $\textbf{$\bmodee$}$   &$8.5\pm0.1$ & $8.3\pm0.1$ & $2.9\pm0.1$ & $2.6\pm0.1$ & $2.5\pm0.1$ & $6.4\pm0.1$ & $2.4\pm0.0$ & $2.0\pm0.0$ & $3.8\pm0.1$ & $2.7\pm0.0$ & $2.3\pm0.0$ & $8.8\pm0.1$        \\
 $\textbf{$\bmodeaa$}$  &$20.8\pm0.3$ & $19.6\pm0.3$ & $8.8\pm0.1$ & $7.7\pm0.1$ & $7.3\pm0.1$ & $15.5\pm0.2$ & $7.1\pm0.1$ & $5.2\pm0.1$ & $9.2\pm0.1$ & $7.4\pm0.1$ & $7.2\pm0.1$ & $22.3\pm0.3$    \\
 $\textbf{$\bmodebb$}$  &$8.6\pm0.1$ & $8.7\pm0.1$ & $3.3\pm0.1$ & $2.9\pm0.1$ & $2.8\pm0.1$ & $6.7\pm0.1$ & $2.8\pm0.1$ & $2.0\pm0.0$ & $4.0\pm0.1$ & $3.3\pm0.0$ & $2.5\pm0.0$ & $9.5\pm0.1$        \\
 $\textbf{$\bmodedd$}$  &$6.7\pm0.1$ & $6.7\pm0.1$ & $2.6\pm0.0$ & $2.0\pm0.0$ & $2.2\pm0.0$ & $5.1\pm0.1$ & $2.0\pm0.0$ & $1.4\pm0.0$ & $3.1\pm0.0$ & $2.5\pm0.0$ & $2.1\pm0.0$ & $7.3\pm0.1$        \\
 $\textbf{$\bmodeaaa$}$ &$11.6\pm0.1$ & $11.1\pm0.1$ & $4.7\pm0.1$ & $4.3\pm0.1$ & $4.0\pm0.1$ & $8.8\pm0.1$ & $3.8\pm0.1$ & $3.2\pm0.0$ & $5.2\pm0.1$ & $4.2\pm0.1$ & $3.9\pm0.1$ & $12.6\pm0.2$     \\
 $\textbf{$\bmodeccc$}$ &$9.4\pm0.1$ & $8.9\pm0.1$ & $3.7\pm0.0$ & $3.4\pm0.0$ & $2.9\pm0.0$ & $7.4\pm0.1$ & $3.1\pm0.0$ & $2.5\pm0.0$ & $4.5\pm0.1$ & $3.4\pm0.0$ & $2.9\pm0.0$ & $10.0\pm0.1$       \\
 $\textbf{$\bmodeddd$}$ &$8.3\pm0.1$ & $8.0\pm0.1$ & $3.4\pm0.0$ & $2.9\pm0.0$ & $2.6\pm0.0$ & $6.5\pm0.1$ & $2.7\pm0.0$ & $2.1\pm0.0$ & $3.9\pm0.1$ & $3.0\pm0.0$ & $2.5\pm0.0$ & $8.7\pm0.1$        \\
 $\textbf{$\bmodef$}$   &$26.2\pm0.3$ & $25.4\pm0.3$ & $10.7\pm0.1$ & $9.9\pm0.1$ & $9.1\pm0.1$ & $20.2\pm0.3$ & $9.0\pm0.1$ & $7.3\pm0.1$ & $12.5\pm0.2$ & $10.0\pm0.1$ & $8.7\pm0.1$ & $28.4\pm0.3$ \\
\hline \hline
   \end{tabular} }
   \label{tab:DTeff_4700}
  \end{center}
\end{sidewaystable}

\begin{sidewaystable}[hpt]
  \begin{center}
   \caption{Correlation coefficients among $19$ fit parameters including statistical and systematic uncertainties.}
 \resizebox{\linewidth}{!}{
   \begin{tabular}{lcccccccccccccccccccc}
      \hline \hline
  & $N_{\lcp\lcm}(4600)$ &  $N_{\lcp\lcm}(4612)$ & $N_{\lcp\lcm}(4628)$ & $N_{\lcp\lcm}(4640)$ & $N_{\lcp\lcm}(4660)$ & $N_{\lcp\lcm}(4680)$ & $N_{\lcp\lcm}(4700)$ &$\mathcal{B}(\textbf{$\modea$})$ & $\mathcal{B}(\textbf{$\modeb$})$   & $\mathcal{B}(\textbf{$\modec$})$ & $\mathcal{B}(\textbf{$\moded$})$ & $\mathcal{B}(\textbf{$\modee$})$ & $\mathcal{B}(\textbf{$\modeaa$})$ & $\mathcal{B}(\textbf{$\modebb$})$ & $\mathcal{B}(\textbf{$\modedd$})$ & $\mathcal{B}(\textbf{$\modeaaa$})$ & $\mathcal{B}(\textbf{$\modeccc$})$ & $\mathcal{B}(\textbf{$\modeddd$})$ & $\mathcal{B}(\textbf{$\modef$})$   \\ \hline
 $N_{\lcp\lcm}(4600)$    &  1&  0.51&  0.78&  0.73&  0.73&   0.8&  0.71&  $-0.7$& $-0.64$& $-0.42$& $-0.32$& $-0.47$& $-0.52$& $-0.47$&  $-0.5$& $-0.51$&  $-0.3$& $-0.49$& $-0.41$ \\
 $N_{\lcp\lcm}(4612)$    &     &     1&  0.53&   0.5&   0.5&  0.55&  0.48& $-0.48$& $-0.44$&  $-0.3$& $-0.25$& $-0.32$& $-0.36$& $-0.34$& $-0.34$& $-0.34$& $-0.21$& $-0.34$& $-0.28$ \\
 $N_{\lcp\lcm}(4628)$    &     &      &     1&  0.77&  0.77&  0.85&  0.75& $-0.72$&  $-0.6$& $-0.42$&$ -0.28$& $-0.42$& $-0.52$& $-0.46$& $-0.45$& $-0.52$& $-0.31$& $-0.45$& $-0.36$ \\
 $N_{\lcp\lcm}(4640)$    &     &      &      &     1&  0.71&  0.79&  0.69&  $-0.7$& $-0.65$& $-0.43$& $-0.35$& $-0.49$& $-0.53$& $-0.47$& $-0.51$& $-0.52$&  $-0.3$& $-0.51$& $-0.42$ \\
 $N_{\lcp\lcm}(4660)$    &     &      &      & &     1&  0.79&   0.7& $-0.68$& $-0.61$& $-0.41$& $-0.31$& $-0.45$& $-0.51$& $-0.46$& $-0.48$&  $-0.5$&  $-0.3$& $-0.46$& $-0.38$ \\
 $N_{\lcp\lcm}(4680)$    &     &      &      & & &     1&  0.77& $-0.77$& $-0.68$& $-0.47$& $-0.36$&  $-0.5$& $-0.57$&  $-0.5$& $-0.55$& $-0.56$& $-0.33$& $-0.52$& $-0.43$ \\
 $N_{\lcp\lcm}(4700)$    &     &      &      & & & &     1& $-0.67$&  $-0.6$&  $-0.4$&  $-0.3$& $-0.43$& $-0.49$& $-0.44$& $-0.48$& $-0.48$& $-0.28$& $-0.44$& $-0.37$ \\
 $\mathcal{B}(\textbf{$\modea$})$  &    &      &      & & & & &     1&  0.65&  0.52&  0.53&   0.5&   0.5&  0.45&  0.49&  0.49&  0.29&  0.52&  0.42 \\
 $\mathcal{B}(\textbf{$\modeb$})$  &    &      &      & & & & &  &     1&   0.5&   0.6&  0.79&  0.58&  0.56&   0.7&  0.54&  0.31&  0.83&  0.62 \\
 $\mathcal{B}(\textbf{$\modec$})$  &    &      &      & & & & &  &  &     1&  0.48&  0.45&  0.34&  0.43&  0.36&  0.33&   0.4&  0.44&  0.31 \\
 $\mathcal{B}(\textbf{$\moded$})$  &    &      &      & & & & &  &  &  &     1&  0.51&  0.34&  0.33&  0.42&  0.32&  0.17&  0.54&  0.39 \\
 $\mathcal{B}(\textbf{$\modee$})$  &    &      &      & & & & &  &  &  &  &     1&  0.47&  0.52&  0.55&  0.44&  0.36&  0.72&  0.51 \\
 $\mathcal{B}(\textbf{$\modeaa$})$ &    &      &      & & & & &  &  &  &  &  &     1&  0.51&  0.62&  0.66&  0.22&  0.48&  0.37 \\
 $\mathcal{B}(\textbf{$\modebb$})$ &   &      &      & & & & &  &  &  &  &  &  &     1&  0.48&  0.46&  0.45&   0.5&  0.36 \\
 $\mathcal{B}(\textbf{$\modedd$})$ &   &      &      & & & & &  &  &  &  &  &  &  &     1&  0.55&  0.23&  0.58&  0.44 \\
 $\mathcal{B}(\textbf{$\modeaaa$})$&   &      &      & & & & &  &  &  &  &  &  &  &  &     1&  0.21&  0.45&  0.35 \\
 $\mathcal{B}(\textbf{$\modeccc$})$&   &      &      & & & & &  &  &  &  &  &  &  &  &  &     1&   0.3&  0.19 \\
 $\mathcal{B}(\textbf{$\modeddd$})$&   &      &      & & & & &  &  &  &  &  &  &  &  &  &  &     1&  0.53 \\
 $\mathcal{B}(\textbf{$\modef$})$  &   &      &      & & & & &  &  &  &  &  &  &  &  &  &  &  &     1 \\
 \hline
   \end{tabular}}
   \label{tab:correlation}
 \end{center}
\end{sidewaystable}
\clearpage
\acknowledgments
\input{acknowledgement_2025-08-29}
\clearpage
\bibliographystyle{JHEP}
\bibliography{biblio.bib}
\clearpage
\section*{The BESIII collaboration}
\addcontentsline{toc}{section}{The BESIII collaboration}
\input{authorlist_2025-08-29}
\end{document}

%% file: acknowledgement_2025-08-29.tex

The BESIII Collaboration thanks the staff of BEPCII (https://cstr.cn/31109.02.BEPC) and the IHEP computing center for their strong support. This work is supported in part by National Key R\&D Program of China under Contracts Nos. 2023YFA1606000, 2023YFA1606704; National Natural Science Foundation of China (NSFC) under Contracts Nos. 11635010, 11935015, 11935016, 11935018, 12025502, 12035009, 12035013, 12061131003, 12192260, 12192261, 12192262, 12192263, 12192264, 12192265, 12221005, 12225509, 12235017, 12361141819, 12422504, 12365015, 124B2097; the Chinese Academy of Sciences (CAS) Large-Scale Scientific Facility Program; the Strategic Priority Research Program of Chinese Academy of Sciences under Contract No. XDA0480600; CAS under Contract No. YSBR-101; 100 Talents Program of CAS; Fundamental Research Funds for the Central Universities, Lanzhou University under Contracts Nos. lzujbky-2025-ytB01, lzujbky-2023-stlt01, University of Chinese Academy of Sciences; The Natural Science Foundation of Inner Mongolia Autonomous Region No. 2023QN01011; The Institute of Nuclear and Particle Physics (INPAC) and Shanghai Key Laboratory for Particle Physics and Cosmology; ERC under Contract No. 758462; German Research Foundation DFG under Contract No. FOR5327; Istituto Nazionale di Fisica Nucleare, Italy; Knut and Alice Wallenberg Foundation under Contracts Nos. 2021.0174, 2021.0299, 2023.0315; Ministry of Development of Turkey under Contract No. DPT2006K-120470; National Research Foundation of Korea under Contract No. NRF-2022R1A2C1092335; National Science and Technology fund of Mongolia; Polish National Science Centre under Contract No. 2024/53/B/ST2/00975; STFC (United Kingdom); Swedish Research Council under Contract No. 2019.04595; U. S. Department of Energy under Contract No. DE-FG02-05ER41374



%% file: authorlist_2025-08-29.tex
M.~Ablikim$^{1}$\BESIIIorcid{0000-0002-3935-619X},
M.~N.~Achasov$^{4,c}$\BESIIIorcid{0000-0002-9400-8622},
P.~Adlarson$^{81}$\BESIIIorcid{0000-0001-6280-3851},
X.~C.~Ai$^{86}$\BESIIIorcid{0000-0003-3856-2415},
C.~S.~Akondi$^{31A,31B}$\BESIIIorcid{0000-0001-6303-5217},
R.~Aliberti$^{39}$\BESIIIorcid{0000-0003-3500-4012},
A.~Amoroso$^{80A,80C}$\BESIIIorcid{0000-0002-3095-8610},
Q.~An$^{77,64,\dagger}$,
Y.~Bai$^{62}$\BESIIIorcid{0000-0001-6593-5665},
O.~Bakina$^{40}$\BESIIIorcid{0009-0005-0719-7461}
% ...
\\
\vspace{0.2cm}
(BESIII Collaboration)\\
\vspace{0.2cm} {\it
$^{1}$ Institute of High Energy Physics, Beijing 100049, People's Republic of China\\
% ...
$^{4}$ Budker Institute of Nuclear Physics SB RAS (BINP), Novosibirsk 630090, Russia\\
% ...
$^{31}$ INFN Sezione di Ferrara, (A)INFN Sezione di Ferrara, I-44122, Ferrara, Italy; (B)University of Ferrara, I-44122, Ferrara, Italy\\
% ...
$^{39}$ Johannes Gutenberg University of Mainz, Johann-Joachim-Becher-Weg 45, D-55099 Mainz, Germany\\
$^{40}$ Joint Institute for Nuclear Research, 141980 Dubna, Moscow region, Russia\\
% ...
$^{62}$ Southeast University, Nanjing 211100, People's Republic of China\\
% ...
$^{64}$ State Key Laboratory of Particle Detection and Electronics, Beijing 100049, Hefei 230026, People's Republic of China\\
% ...
$^{77}$ University of Science and Technology of China, Hefei 230026, People's Republic of China\\
% ...
$^{80}$ University of Turin and INFN, (A)University of Turin, I-10125, Turin, Italy; (B)University of Eastern Piedmont, I-15121, Alessandria, Italy; (C)INFN, I-10125, Turin, Italy\\
$^{81}$ Uppsala University, Box 516, SE-75120 Uppsala, Sweden\\
% ...
$^{86}$ Zhengzhou University, Zhengzhou 450001, People's Republic of China\\

\vspace{0.2cm}
$^{\dagger}$ Deceased\\
% ...
$^{c}$ Also at the Novosibirsk State University, Novosibirsk, 630090, Russia\\
% ...

}

%% file: biblio.bib
@article{BESIII:2020nme,
    author = "Ablikim, M. and others",
    collaboration = "BESIII",
    title = "{Future Physics Programme of BESIII}",
    eprint = "1912.05983",
    archivePrefix = "arXiv",
    primaryClass = "hep-ex",
    reportNumber = "HEP-Physics-Report-BESIII-2019-12-13",
    doi = "10.1088/1674-1137/44/4/040001",
    journal = "Chin. Phys. C",
    volume = "44",
    number = "4",
    pages = "040001",
    year = "2020"
}

@article{Korner:1992wi,
    author = "Korner, J. G. and Kramer, M.",
    title = "{Exclusive nonleptonic charm baryon decays}",
    reportNumber = "DESY-92-049, MZ-TH-91-07",
    doi = "10.1007/BF01561305",
    journal = "Z. Phys. C",
    volume = "55",
    pages = "659--670",
    year = "1992"
}

@article{Ivanov:1997ra,
    author = "Ivanov, Mikhail A. and Korner, J. G. and Lyubovitskij, Valery E. and Rusetsky, A. G.",
    title = "{Exclusive nonleptonic decays of bottom and charm baryons in a relativistic three quark model: Evaluation of nonfactorizing diagrams}",
    eprint = "hep-ph/9709372",
    archivePrefix = "arXiv",
    reportNumber = "MZ-TH-97-15",
    doi = "10.1103/PhysRevD.57.5632",
    journal = "Phys. Rev. D",
    volume = "57",
    pages = "5632--5652",
    year = "1998"
}

@article{Cheng:1991sn,
    author = "Cheng, Hai-Yang and Tseng, B.",
    title = "{Nonleptonic weak decays of charmed baryons}",
    reportNumber = "IP-ASTP-17-91",
    doi = "10.1103/PhysRevD.46.1042",
    journal = "Phys. Rev. D",
    volume = "46",
    pages = "1042",
    year = "1992",
    note = "[Erratum: Phys.Rev.D 55, 1697 (1997)]"
}

@article{Xu:1992vc,
    author = "Xu, Q. P. and Kamal, A. N.",
    title = "{Cabibbo favored nonleptonic decays of charmed baryons}",
    reportNumber = "ALBERTA-THY-8-92",
    doi = "10.1103/PhysRevD.46.270",
    journal = "Phys. Rev. D",
    volume = "46",
    pages = "270--278",
    year = "1992"
}

@article{Cheng:1993gf,
    author = "Cheng, Hai-Yang and Tseng, B.",
    title = "{Cabibbo allowed nonleptonic weak decays of charmed baryons}",
    eprint = "hep-ph/9304286",
    archivePrefix = "arXiv",
    reportNumber = "IP-ASTP-10-93, ITP-SB-93-20",
    doi = "10.1103/PhysRevD.48.4188",
    journal = "Phys. Rev. D",
    volume = "48",
    pages = "4188--4202",
    year = "1993"
}

@article{Xu:1992sw,
    author = "Xu, Q. P. and Kamal, A. N.",
    title = "{The Nonleptonic charmed baryon decays: $B_c\to B (\frac{3}{2}^+,~\rm decuplet) + P(0^-)~ \rm or ~V(1^-)$}",
    reportNumber = "ALBERTA-THY-20-92",
    doi = "10.1103/PhysRevD.46.3836",
    journal = "Phys. Rev. D",
    volume = "46",
    pages = "3836--3844",
    year = "1992"
}

@article{Zenczykowski:1993jm,
    author = "Zenczykowski, Piotr",
    title = "{Nonleptonic charmed baryon decays: Symmetry properties of parity violating amplitudes}",
    reportNumber = "INP-1655-PH",
    doi = "10.1103/PhysRevD.50.5787",
    journal = "Phys. Rev. D",
    volume = "50",
    pages = "5787--5792",
    year = "1994"
}

@article{Sharma:1998rd,
    author = "Sharma, K. K. and Verma, R. C.",
    title = "{A Study of weak mesonic decays of $\Lambda(c)$ and $\Xi(c)$ baryons on the basis of HQET results}",
    eprint = "hep-ph/9803302",
    archivePrefix = "arXiv",
    doi = "10.1007/s100529801008",
    journal = "Eur. Phys. J. C",
    volume = "7",
    pages = "217--224",
    year = "1999"
}

@article{Korner:1978ec,
    author = {K\"orner, J. G. and Kramer, G. and Willrodt, J.},
    title = "{Weak Decays of the Charmed Baryon C$_0^+$ and the Inclusive Yield of $\Lambda$ and $p$}",
    reportNumber = "DESY-78-13",
    doi = "10.1016/0370-2693(78)90495-1",
    journal = "Phys. Lett. B",
    volume = "78",
    pages = "492",
    year = "1978",
    note = "[Erratum: Phys.Lett.B 81, 419--419 (1979)]"
}

@article{Uppal:1994pt,
    author = "Uppal, T. and Verma, R. C. and Khanna, M. P.",
    title = "{Constituent quark model analysis of weak mesonic decays of charm baryons}",
    doi = "10.1103/PhysRevD.49.3417",
    journal = "Phys. Rev. D",
    volume = "49",
    pages = "3417--3425",
    year = "1994"
}

@article{Geng:2019xbo,
    author = "Geng, C. Q. and Liu, Chia-Wei and Tsai, Tien-Hsueh",
    title = "{Asymmetries of anti-triplet charmed baryon decays}",
    eprint = "1902.06189",
    archivePrefix = "arXiv",
    primaryClass = "hep-ph",
    doi = "10.1016/j.physletb.2019.05.024",
    journal = "Phys. Lett. B",
    volume = "794",
    pages = "19--28",
    year = "2019"
}

@article{Sharma:1996sc,
    author = "Sharma, K. K. and Verma, R. C.",
    title = "{SU(3) flavor analysis of two-body weak decays of charmed baryons}",
    eprint = "hep-ph/9704391",
    archivePrefix = "arXiv",
    reportNumber = "PRINT-97-113 (PANJAB)",
    doi = "10.1103/PhysRevD.55.7067",
    journal = "Phys. Rev. D",
    volume = "55",
    pages = "7067--7074",
    year = "1997"
}

@article{Geng:2017esc,
    author = "Geng, C. Q. and Hsiao, Y. K. and Lin, Yu-Heng and Liu, Liang-Liang",
    title = "{Non-leptonic two-body weak decays of $\Lambda_c(2286)$}",
    eprint = "1708.02460",
    archivePrefix = "arXiv",
    primaryClass = "hep-ph",
    doi = "10.1016/j.physletb.2017.11.062",
    journal = "Phys. Lett. B",
    volume = "776",
    pages = "265--269",
    year = "2018"
}

@article{Geng:2017mxn,
    author = "Geng, C. Q. and Hsiao, Y. K. and Liu, Chia-Wei and Tsai, Tien-Hsueh",
    title = "{Charmed Baryon Weak Decays with SU(3) Flavor Symmetry}",
    eprint = "1709.00808",
    archivePrefix = "arXiv",
    primaryClass = "hep-ph",
    doi = "10.1007/JHEP11(2017)147",
    journal = "JHEP",
    volume = "11",
    pages = "147",
    year = "2017"
}

@article{Geng:2018plk,
    author = "Geng, C. Q. and Hsiao, Y. K. and Liu, Chia-Wei and Tsai, Tien-Hsueh",
    title = "{Antitriplet charmed baryon decays with SU(3) flavor symmetry}",
    eprint = "1801.03276",
    archivePrefix = "arXiv",
    primaryClass = "hep-ph",
    doi = "10.1103/PhysRevD.97.073006",
    journal = "Phys. Rev. D",
    volume = "97",
    number = "7",
    pages = "073006",
    year = "2018"
}

@article{Lu:2016ogy,
    author = {L\"u, Cai-Dian and Wang, Wei and Yu, Fu-Sheng},
    title = "{Test flavor SU(3) symmetry in exclusive $\Lambda_c$ decays}",
    eprint = "1601.04241",
    archivePrefix = "arXiv",
    primaryClass = "hep-ph",
    doi = "10.1103/PhysRevD.93.056008",
    journal = "Phys. Rev. D",
    volume = "93",
    number = "5",
    pages = "056008",
    year = "2016"
}

@article{Hsiao:2019yur,
    author = "Hsiao, Y. K. and Yao, Yu and Zhao, H. J.",
    title = "{Two-body charmed baryon decays involving vector meson with $SU(3)$ flavor symmetry}",
    eprint = "1902.08783",
    archivePrefix = "arXiv",
    primaryClass = "hep-ph",
    doi = "10.1016/j.physletb.2019.03.031",
    journal = "Phys. Lett. B",
    volume = "792",
    pages = "35--39",
    year = "2019"
}

@article{He:2018joe,
    author = "He, Xiao-Gang and Shi, Yu-Ji and Wang, Wei",
    title = "{Unification of Flavor SU(3) Analyses of Heavy Hadron Weak Decays}",
    eprint = "1811.03480",
    archivePrefix = "arXiv",
    primaryClass = "hep-ph",
    doi = "10.1140/epjc/s10052-020-7862-5",
    journal = "Eur. Phys. J. C",
    volume = "80",
    number = "5",
    pages = "359",
    year = "2020"
}

@article{Cheng:2021qpd,
    author = "Cheng, Hai-Yang",
    title = "{Charmed baryon physics circa 2021}",
    eprint = "2109.01216",
    archivePrefix = "arXiv",
    primaryClass = "hep-ph",
    doi = "10.1016/j.cjph.2022.06.021",
    journal = "Chin. J. Phys.",
    volume = "78",
    pages = "324--362",
    year = "2022"
}

@article{Li:2021iwf,
    author = "Li, Hai-Bo and Lyu, Xiao-Rui",
    title = "{Study of the standard model with weak decays of charmed hadrons at BESIII}",
    eprint = "2103.00908",
    archivePrefix = "arXiv",
    primaryClass = "hep-ex",
    doi = "10.1093/nsr/nwab181",
    journal = "Natl. Sci. Rev.",
    volume = "8",
    number = "11",
    pages = "nwab181",
    year = "2021"
}

@article{Li:2025nzx,
    author = "Li, Pei-Rong and Lyu, Xiao-Rui and Zheng, Yangheng",
    title = "{Experimental overview on the charmed baryon decays}",
    eprint = "2509.19141",
    archivePrefix = "arXiv",
    primaryClass = "hep-ex",
    doi = "10.1088/1674-1137/ae1187",
    journal = "Chin. Phys. C",
    volume = "50",
    pages = "022002",
    year = "2026"
}

@article{BESIII:2015ysy,
    author = "Ablikim, M. and others",
    collaboration = "BESIII",
    title = "{Measurement of the absolute branching fraction for $\Lambda^+_{c}\to \Lambda e^+\nu_e$}",
    eprint = "1510.02610",
    archivePrefix = "arXiv",
    primaryClass = "hep-ex",
    doi = "10.1103/PhysRevLett.115.221805",
    journal = "Phys. Rev. Lett.",
    volume = "115",
    number = "22",
    pages = "221805",
    year = "2015"
}

@article{MARK-III:1985hbd,
    author = "Baltrusaitis, R. M. and others",
    collaboration = "MARK-III",
    title = "{Direct Measurements of Charmed d Meson Hadronic Branching Fractions}",
    reportNumber = "SLAC-PUB-3861",
    doi = "10.1103/PhysRevLett.56.2140",
    journal = "Phys. Rev. Lett.",
    volume = "56",
    pages = "2140",
    year = "1986"
}

@article{LHCb:2015eia,
    author = "Aaij, Roel and others",
    collaboration = "LHCb",
    title = "{Determination of the quark coupling strength $|V_{ub}|$ using baryonic decays}",
    eprint = "1504.01568",
    archivePrefix = "arXiv",
    primaryClass = "hep-ex",
    reportNumber = "CERN-PH-EP-2015-084, LHCB-PAPER-2015-013",
    doi = "10.1038/nphys3415",
    journal = "Nat. Phys.",
    volume = "11",
    pages = "743--747",
    year = "2015"
}

@article{LHCb:2022vns,
    author = "Aaij, Roel and others",
    collaboration = "LHCb",
    title = "{Study of the $B^{-} \to \Lambda_{c}^{+} \bar{\Lambda}_{c}^{-} K^{-}$ decay}",
    eprint = "2211.00812",
    archivePrefix = "arXiv",
    primaryClass = "hep-ex",
    reportNumber = "CERN-EP-2022-196, LHCb-PAPER-2022-028",
    doi = "10.1103/PhysRevD.108.012020",
    journal = "Phys. Rev. D",
    volume = "108",
    pages = "012020",
    year = "2023"
}

@article{Belle:2017jrt,
    author = "Li, Y. B. and others",
    collaboration = "Belle",
    title = "{Observation of $\Xi_{c}(2930)^0$ and updated measurement of $B^{-} \to K^{-} \Lambda_{c}^{+} \bar{\Lambda}_{c}^{-}$ at Belle}",
    eprint = "1712.03612",
    archivePrefix = "arXiv",
    primaryClass = "hep-ex",
    reportNumber = "Belle Preprint \# 2017-24; KEK Preprint \# 2017-35, BELLE-PREPRINT-\#-2017-24, KEK-PREPRINT-\#-2017-35",
    doi = "10.1140/epjc/s10052-018-5720-5",
    journal = "Eur. Phys. J. C",
    volume = "78",
    number = "3",
    pages = "252",
    year = "2018"
}

@article{LHCb:2020iby,
    author = "Aaij, Roel and others",
    collaboration = "LHCb",
    title = "{Observation of New $\Xi_c^0$ Baryons Decaying to $\Lambda_c^+ K^-$}",
    eprint = "2003.13649",
    archivePrefix = "arXiv",
    primaryClass = "hep-ex",
    reportNumber = "LHCb-PAPER-2020-004, CERN-EP-2020-038",
    doi = "10.1103/PhysRevLett.124.222001",
    journal = "Phys. Rev. Lett.",
    volume = "124",
    number = "22",
    pages = "222001",
    year = "2020"
}

@article{BaBar:2009eml,
    author = "Aubert, Bernard and others",
    collaboration = "BaBar",
    title = "{Observation of the baryonic B-decay $\bar{B}^0 \to \Lambda_c^+ \bar{p} K^- \pi^+$}",
    eprint = "0907.4566",
    archivePrefix = "arXiv",
    primaryClass = "hep-ex",
    reportNumber = "SLAC-PUB-13675, BABAR-PUB-09-014",
    doi = "10.1103/PhysRevD.80.051105",
    journal = "Phys. Rev. D",
    volume = "80",
    pages = "051105",
    year = "2009"
}

@article{Belle:2018yob,
    author = "Li, Y. B. and others",
    collaboration = "Belle",
    title = "{Evidence of a structure in $\bar{K}^{0} \Lambda _{c}^{+}$ consistent with a charged $\Xi _c(2930)^{+}$ , and updated measurement of $\bar{B}^{0} \rightarrow \bar{K}^{0} \Lambda _{c}^{+} \bar{\Lambda }_{c}^{-}$ at Belle}",
    eprint = "1806.09182",
    archivePrefix = "arXiv",
    primaryClass = "hep-ex",
    reportNumber = "Belle Preprint \# 2018-10; KEK Preprint \# 2018-16, BELLE-PREPRINT-2018-10, KEK-PREPRINT-2018-16",
    doi = "10.1140/epjc/s10052-018-6425-5",
    journal = "Eur. Phys. J. C",
    volume = "78",
    number = "11",
    pages = "928",
    year = "2018"
}

@article{LHCb:2022sck,
    author = "Aaij, Roel and others",
    collaboration = "LHCb",
    title = "{Amplitude analysis of the $\Lambda_c^+ \to p K^-\pi^+$ decay and $\Lambda_c^+$ baryon polarization measurement in semileptonic beauty hadron decays}",
    eprint = "2208.03262",
    archivePrefix = "arXiv",
    primaryClass = "hep-ex",
    reportNumber = "LHCb-PAPER-2022-002, CERN-EP-2022-124",
    doi = "10.1103/PhysRevD.108.012023",
    journal = "Phys. Rev. D",
    volume = "108",
    number = "1",
    pages = "012023",
    year = "2023"
}

@article{BESIII:2022udq,
    author = "Ablikim, Medina and others",
    collaboration = "BESIII",
    title = "{Partial wave analysis of the charmed baryon hadronic decay $ {\Lambda}_c^{+} $\textrightarrow{} \ensuremath{\Lambda}\ensuremath{\pi}$^{+}$\ensuremath{\pi}$^{0}$}",
    eprint = "2209.08464",
    archivePrefix = "arXiv",
    primaryClass = "hep-ex",
    doi = "10.1007/JHEP12(2022)033",
    journal = "JHEP",
    volume = "12",
    pages = "033",
    year = "2022"
}

@article{BESIII:2019odb,
    author = "Ablikim, Medina and others",
    collaboration = "BESIII",
    title = "{Measurements of Weak Decay Asymmetries of $\Lambda_c^+\to pK_S^0$, $\Lambda\pi^+$, $\Sigma^+\pi^0$, and $\Sigma^0\pi^+$}",
    eprint = "1905.04707",
    archivePrefix = "arXiv",
    primaryClass = "hep-ex",
    doi = "10.1103/PhysRevD.100.072004",
    journal = "Phys. Rev. D",
    volume = "100",
    number = "7",
    pages = "072004",
    year = "2019"
}

@article{BESIII:2025zbz,
    author = "Ablikim, Medina and others",
    collaboration = "BESIII",
    title = "{The Production and Decay Dynamics of the Charmed Baryon $\Lambda_c^+$ in $e^+e^-$ Annihilations near Threshold}",
    eprint = "2508.11400",
    archivePrefix = "arXiv",
    primaryClass = "hep-ex",
    month = "8",
    year = "2025"
}

@article{BESIII:2015bjk,
    author = "Ablikim, M. and others",
    collaboration = "BESIII",
    title = "{Measurements of absolute hadronic branching fractions of $\Lambda_{c}^{+}$ baryon}",
    eprint = "1511.08380",
    archivePrefix = "arXiv",
    primaryClass = "hep-ex",
    doi = "10.1103/PhysRevLett.116.052001",
    journal = "Phys. Rev. Lett.",
    volume = "116",
    number = "5",
    pages = "052001",
    year = "2016"
}

@article{BESIII:2016ozn,
    author = "Ablikim, Medina and others",
    collaboration = "BESIII",
    title = "{Measurement of Singly Cabibbo Suppressed Decays $\Lambda_c^{+}\to p\pi^{+}\pi^{-}$ and $\Lambda_c^{+}\to pK^{+}K^{-}$}",
    eprint = "1608.00407",
    archivePrefix = "arXiv",
    primaryClass = "hep-ex",
    doi = "10.1103/PhysRevLett.117.232002",
    journal = "Phys. Rev. Lett.",
    volume = "117",
    number = "23",
    pages = "232002",
    year = "2016",
    note = "[Addendum: Phys.Rev.Lett. 120, 029903 (2018)]"
}

@article{BESIII:2023rwv,
    author = "Ablikim, Medina and others",
    collaboration = "BESIII",
    title = "{Measurement of Energy-Dependent Pair-Production Cross Section and Electromagnetic Form Factors of a Charmed Baryon}",
    eprint = "2307.07316",
    archivePrefix = "arXiv",
    primaryClass = "hep-ex",
    doi = "10.1103/PhysRevLett.131.191901",
    journal = "Phys. Rev. Lett.",
    volume = "131",
    number = "19",
    pages = "191901",
    year = "2023"
}

@article{Belle:2008xmh,
    author = "Pakhlova, G. and others",
    collaboration = "Belle",
    title = "{Observation of a near-threshold enhancement in the $e^+e^-\to \Lambda_c^+\bar{\Lambda}_c^-$ cross section using initial-state radiation}",
    eprint = "0807.4458",
    archivePrefix = "arXiv",
    primaryClass = "hep-ex",
    doi = "10.1103/PhysRevLett.101.172001",
    journal = "Phys. Rev. Lett.",
    volume = "101",
    pages = "172001",
    year = "2008"
}

@article{BESIII:2022ulv,
    author = "Ablikim, M. and others",
    collaboration = "BESIII",
    title = "{Luminosities and energies of  $e^{+}e^{−}$ collision data taken between $\sqrt{s}=4.61$ GeV and $4.95$ GeV at BESIII}",
    eprint = "2205.04809",
    archivePrefix = "arXiv",
    primaryClass = "hep-ex",
    doi = "10.1088/1674-1137/ac84cc",
    journal = "Chin. Phys. C",
    volume = "46",
    number = "11",
    pages = "113003",
    year = "2022"
}

@article{BESIII:2009fln,
    author = "Ablikim, M. and others",
    collaboration = "BESIII",
    title = "{Design and Construction of the BESIII Detector}",
    eprint = "0911.4960",
    archivePrefix = "arXiv",
    primaryClass = "physics.ins-det",
    doi = "10.1016/j.nima.2009.12.050",
    journal = "Nucl. Instrum. Meth. A",
    volume = "614",
    pages = "345--399",
    year = "2010"
}

@inproceedings{Yu:2016cof,
    author = "Yu, Chenghui and others",
    title = "{BEPCII Performance and Beam Dynamics Studies on Luminosity}",
    booktitle = "{7th International Particle Accelerator Conference}",
    doi = "10.18429/JACoW-IPAC2016-TUYA01",
    pages = "TUYA01",
    year = "2016"
}

@article{Guo:2017sjt,
    author = "Guo, Ying-Xiao and others",
    title = "{The study of time calibration for upgraded end cap TOF of BESIII}",
    doi = "10.1007/s41605-017-0012-4",
    journal = "Radiat. Detect. Technol. Methods",
    volume = "1",
    pages = "15",
    year = "2017"
}

@article{Li:2017jpg,
    author = "Li, Xin and others",
    title = "{Study of MRPC technology for BESIII endcap-TOF upgrade}",
    doi = "10.1007/s41605-017-0014-2",
    journal = "Radiat. Detect. Technol. Methods",
    volume = "1",
    pages = "13",
    year = "2017"
}

@article{Cao:2020ibk,
    author = "Cao, P. and others",
    title = "{Design and construction of the new BESIII endcap Time-of-Flight system with MRPC Technology}",
    doi = "10.1016/j.nima.2019.163053",
    journal = "Nucl. Instrum. Meth. A",
    volume = "953",
    pages = "163053",
    year = "2020"
}

@article{GEANT4:2002zbu,
    author = "Agostinelli, S. and others",
    collaboration = "GEANT4",
    title = "{GEANT4--a simulation toolkit}",
    reportNumber = "SLAC-PUB-9350, FERMILAB-PUB-03-339, CERN-IT-2002-003",
    doi = "10.1016/S0168-9002(03)01368-8",
    journal = "Nucl. Instrum. Meth. A",
    volume = "506",
    pages = "250--303",
    year = "2003"
}

@article{Huang:2022wuo,
    author = "Huang, Kai-Xuan and Li, Zhi-Jun and Qian, Zhen and Zhu, Jiang and Li, Hao-Yuan and Zhang, Yu-Mei and Sun, Sheng-Sen and You, Zheng-Yun",
    title = "{Method for detector description transformation to Unity and application in BESIII}",
    eprint = "2206.10117",
    archivePrefix = "arXiv",
    primaryClass = "physics.ins-det",
    doi = "10.1007/s41365-022-01133-8",
    journal = "Nucl. Sci. Tech.",
    volume = "33",
    number = "11",
    pages = "142",
    year = "2022"
}

@article{Jadach:2000ir,
    author = "Jadach, S. and Ward, B. F. L. and Was, Z.",
    title = "{Coherent exclusive exponentiation for precision Monte Carlo calculations}",
    eprint = "hep-ph/0006359",
    archivePrefix = "arXiv",
    reportNumber = "CERN-TH-2000-087, UTHEP-99-09-01",
    doi = "10.1103/PhysRevD.63.113009",
    journal = "Phys. Rev. D",
    volume = "63",
    pages = "113009",
    year = "2001"
}

@article{Jadach:1999vf,
    author = "Jadach, S. and Ward, B. F. L. and Was, Z.",
    title = "{The Precision Monte Carlo event generator K K for two fermion final states in $e^+ e^-$ collisions}",
    eprint = "hep-ph/9912214",
    archivePrefix = "arXiv",
    reportNumber = "DESY-99-106, CERN-TH-99-235, UTHEP-99-08-01",
    doi = "10.1016/S0010-4655(00)00048-5",
    journal = "Comput. Phys. Commun.",
    volume = "130",
    pages = "260--325",
    year = "2000"
}

@article{Lange:2001uf,
    author = "Lange, D. J.",
    editor = "Erhan, S. and Schlein, P. and Rozen, Y.",
    title = "{The EvtGen particle decay simulation package}",
    doi = "10.1016/S0168-9002(01)00089-4",
    journal = "Nucl. Instrum. Meth. A",
    volume = "462",
    pages = "152--155",
    year = "2001"
}

@article{Ping:2008zz,
    author = "Ping, Rong-Gang",
    title = "{Event generators at BESIII}",
    doi = "10.1088/1674-1137/32/8/001",
    journal = "Chin. Phys. C",
    volume = "32",
    pages = "599",
    year = "2008"
}

@article{ParticleDataGroup:2024cfk,
    author = "Navas, S. and others",
    collaboration = "Particle Data Group",
    title = "{Review of particle physics}",
    doi = "10.1103/PhysRevD.110.030001",
    journal = "Phys. Rev. D",
    volume = "110",
    number = "3",
    pages = "030001",
    year = "2024 and 2025 update"
}

@article{Chen:2000tv,
    author = "Chen, J. C. and Huang, G. S. and Qi, X. R. and Zhang, D. H. and Zhu, Y. S.",
    title = "{Event generator for $J/\psi$ and $\psi(2S)$ decay}",
    doi = "10.1103/PhysRevD.62.034003",
    journal = "Phys. Rev. D",
    volume = "62",
    pages = "034003",
    year = "2000"
}

@article{Yang:2014vra,
    author = "Yang, Rui-Ling and Ping, Rong-Gang and Chen, Hong",
    title = "{Tuning and Validation of the Lundcharm Model with $J/\psi$ Decays}",
    doi = "10.1088/0256-307X/31/6/061301",
    journal = "Chin. Phys. Lett.",
    volume = "31",
    pages = "061301",
    year = "2014"
}

@article{Richter-Was:1992hxq,
    author = "Richter-Was, E.",
    title = "{QED bremsstrahlung in semileptonic B and leptonic tau decays}",
    reportNumber = "CERN-TH-6746-92",
    doi = "10.1016/0370-2693(93)90062-M",
    journal = "Phys. Lett. B",
    volume = "303",
    pages = "163--169",
    year = "1993"
}

@article{Guan:2013jua,
    author = {Guan, Yinghui and L\"u, Xiao-Rui and Zheng, Yangheng and Wang, Yi-Fang},
    title = "{Study of the efficiency of event start time determination at BESIII}",
    eprint = "1304.6177",
    archivePrefix = "arXiv",
    primaryClass = "physics.ins-det",
    doi = "10.1088/1674-1137/38/1/016201",
    journal = "Chin. Phys. C",
    volume = "38",
    number = "1",
    pages = "016201",
    year = "2014"
}

@article{BESIII:2022xne,
    author = "Ablikim, Medina and others",
    collaboration = "BESIII",
    title = "{Observations of the Cabibbo-Suppressed decays $\Lambda_{c}^{+}\to n\pi^{+}\pi^{0}$, $n\pi^{+}\pi^{-}\pi^{+}$ and the Cabibbo-Favored decay $\Lambda_{c}^{+}\to nK^{-}\pi^{+}\pi^{+}$}",
    eprint = "2210.03375",
    archivePrefix = "arXiv",
    primaryClass = "hep-ex",
    doi = "10.1088/1674-1137/ac9d29",
    journal = "Chin. Phys. C",
    volume = "47",
    number = "2",
    pages = "023001",
    year = "2023"
}

@article{Guan:2013hua,
    author = "Guan, Yinghui and Lu, Xiao-Rui and Zheng, Yangheng and Zhu, Yong-Sheng",
    title = "{Simultaneous least squares fitter based on the Lagrange multiplier method}",
    eprint = "1304.6170",
    archivePrefix = "arXiv",
    primaryClass = "hep-ex",
    doi = "10.1088/1674-1137/37/10/106201",
    journal = "Chin. Phys. C",
    volume = "37",
    pages = "106201",
    year = "2013"
}

@article{BESIII:2010ank,
    author = "Ablikim, M. and others",
    collaboration = "BESIII",
    title = "{Branching fraction measurements of $\chi_{c0}$ and $\chi_{c2}$ to $\pi^0\pi^0$ and $\eta\eta$}",
    eprint = "1001.5360",
    archivePrefix = "arXiv",
    primaryClass = "hep-ex",
    doi = "10.1103/PhysRevD.81.052005",
    journal = "Phys. Rev. D",
    volume = "81",
    pages = "052005",
    year = "2010"
}

@article{BESIII:2015jmz,
    author = "Ablikim, M. and others",
    collaboration = "BESIII",
    title = "{Study of decay dynamics and $CP$ asymmetry in $D^+ \to K^0_L e^+ \nu_e$ decay}",
    eprint = "1510.00308",
    archivePrefix = "arXiv",
    primaryClass = "hep-ex",
    doi = "10.1103/PhysRevD.92.112008",
    journal = "Phys. Rev. D",
    volume = "92",
    number = "11",
    pages = "112008",
    year = "2015"
}
